\documentclass[11pt]{article}
\usepackage{graphicx,psfrag}
\usepackage[utf8]{inputenc}
\usepackage{tabularx}
\usepackage[DIV13]{typearea}
\usepackage{ragged2e}
\usepackage{amsmath,amsfonts,bbm,mathrsfs,amssymb}
\usepackage{slashed,cancel}
\usepackage[usenames,dvipsnames]{xcolor}
\usepackage{cite}
\usepackage{amsmath}
\usepackage{bm} 
\usepackage{soul}
\usepackage{hyperref}
\hypersetup{colorlinks=true,urlcolor=blue,anchorcolor=blue,citecolor=blue,filecolor=blue,
            linkcolor=Magenta,menucolor=blue, linktocpage=true,pdfproducer=medialab}

\usepackage[english]{babel}
\usepackage{catchfile}
\usepackage{indentfirst}
\usepackage{graphicx}
\usepackage{float}
\usepackage{enumerate}
\usepackage{subcaption}
\usepackage{multirow}
\usepackage{physics}
\usepackage{makecell}
\usepackage{subcaption}

\usepackage{booktabs}
\usepackage{tabularx}
\usepackage{array}

\usepackage[normalem]{ulem}
\usepackage{tcolorbox}\usepackage{adjustbox}
\usepackage[compat=1.1.0]{tikz-feynman}

\newcommand{\email}[1]{\href{mailto:#1}{\tt #1}}
\renewcommand{\theequation}{\thesection.\arabic{equation}}
\numberwithin{equation}{section}

\newcommand{\purple}[1]{\color{purple} #1 \color{black}}

\definecolor{vierde}{rgb}{0.0, 0.5, 0.0}
\definecolor{OliveGreen}{rgb}{0,0.6,0}

\newcommand{\be}{\begin{equation}}
\newcommand{\ee}{\end{equation}}
\newcommand{\bea}{\begin{eqnarray}}
\newcommand{\eea}{\end{eqnarray}}
\newcommand{\B}{\beta}
\newcommand{\e}{\eta}
\def\ba{\begin{eqnarray}}
\def\ea{\end{eqnarray}}

\def\bt{\bar{t}}
\def\bp{\bar{p}}
\def\bq{\bar{q}}

\def\bs{\bar{s}}

\begin{document}
\renewcommand*{\thefootnote}{\fnsymbol{footnote}}
\begin{titlepage}

\vspace*{-1cm}

\begin{center}
\bf\LARGE{
Quantum detection of CP violation in the $t\bar{t}$ system: production\\[4mm]}
\centering
\vskip .3cm
\end{center}
\vskip 0.5  cm
\begin{center}
{\bf Priyanka Lamba}${}^{a}$~\footnote{\email{priyanka.lamba@uclouvain.be}},
{\bf Fabio Maltoni}${}^{a,b}$~\footnote{\email{fabio.maltoni@unibo.it}},
{\bf Olimpia Miniati}${}^{b}$~\footnote{\email{olimpia.miniati2@unibo.it}}
and
{\bf Eleni Vryonidou}${}^{c,d}$~\footnote{\email{vryonidou.eleni@ucy.ac.cy}}
\vskip .5cm
{\small
\vskip .2cm
${}^{a)}$ Centre for Cosmology, Particle Physics and Phenomenology (CP3), \\Universit\'{e} Catholique de Louvain, B-1348 Louvain-la-Neuve, Belgium\\
${}^{b)}$ Dipartimento di Fisica e Astronomia, Universit\`{a} di Bologna and INFN, Sezione di Bologna,\\ Via Irnerio 46, 40126 Bologna, Italy\\
${}^{c)}$ {Department of Physics and Astronomy, University of Manchester,\\ Oxford Road, Manchester M13~9PL, United Kingdom}\\
${}^{d)}$ {Department of Physics, University of Cyprus, P.O. Box 20537, 1678 Nicosia, Cyprus}

}
\end{center}
\vskip 2cm
\begin{abstract}
We investigate how possible new CP-violating top-quark interactions are
encoded in the quantum state of a produced $t\bar t$ pair. We derive analytic
expressions for the production density matrix in several benchmark channels
relevant to hadron, lepton and photon colliders. In a common spin basis, we
identify two characteristic CP-odd structures in the Fano--Bloch
decomposition: a difference between the top and antitop polarisation vectors
and an antisymmetric component of the spin-correlation matrix. We construct
observables that directly probe these structures and study how quantum
information measures, including discord, concurrence, magic and trace
distance, respond to CP-even and CP-odd SMEFT contributions. Finally, using
current measurements and future collider projections, we assess the
sensitivity of these observables to possible new sources of CP violation in
top-quark production. This establishes the production-level framework whose
experimental reconstruction is developed in a companion paper.
\end{abstract}
\vskip 3cm

\end{titlepage}
\setcounter{footnote}{0}

\renewcommand*{\thefootnote}{\arabic{footnote}}

\tableofcontents
\newpage 


\section{Introduction}
\label{sec:intro}

Among the quarks of the Standard Model (SM), the top quark provides the most direct access to spin information. Its lifetime is so short that it decays before hadronisation, preventing strong-interaction effects from depolarising its spin. The angular and momentum distributions of its decay products therefore retain information about the spin state in which it was produced. Accordingly, top-quark pair production provides a naturally occurring bipartite spin-$1/2$ system at collider energies, whose spin state can be represented by a two-qubit density matrix.  The study of top polarisation and spin correlations has a long history, both theoretically~\cite{Barger:1988jj,Mahlon:1995zn,Stelzer:1995gc,Parke:1996pr,Mahlon:1996pn,Mahlon:1997uc,Brandenburg:2002xr} and experimentally at the Tevatron~\cite{D0:2015kta} and at the LHC~\cite{CMS:2013roq,ATLAS:2014aus,CMS:2016piu,ATLAS:2016bac,CMS:2019nrx,ATLAS:2019zrq}. More recently, these measurements have been reinterpreted as providing direct information on the quantum state of the pair. In particular, recent ATLAS and CMS analyses have established entanglement between the top and antitop spins in selected regions of phase space~\cite{ATLAS:2023fsd,CMS:2024zkc,CMS:2024pts}.

This observation has motivated a systematic use of quantum information language in collider physics. The relevant object is the $t\bar t$ spin density matrix, or equivalently its Fano--Bloch decomposition~\cite{Fano:1957zz,Fano:1983zz}. In this decomposition the two single-particle polarisation vectors and the spin-correlation matrix provide a complete parametrisation of the spin state. These coefficients are the same quantities that appear in the conventional analysis of top spin correlations, but the density-matrix formulation makes it possible to ask different questions. One may ask whether the state is entangled, whether it violates a Bell inequality, whether it carries discord, whether it is close to a stabilizer state, or how far it is from another reference density matrix. Such questions have been explored in a rapidly growing literature on quantum observables at colliders~\cite{2003.02280,Fabbrichesi:2021npl,Severi:2021cnj,Severi:2022qjy,Aoude:2022imd,Afik:2022kwm,Aguilar-Saavedra:2022uye,Afik:2022dgh,Cheng:2023qmz,Han:2023fci,Dong:2023xiw,Cheng:2024btk,Aguilar-Saavedra:2024hwd,Aguilar-Saavedra:2023lwb,DelGratta:2025qyp,Lamba:2026qnk,Afik:2025ejh,A:2026pug,LoChiatto:2024dmx,Aguilar-Saavedra:2024fig,Cheng:2024rxi,Aguilar-Saavedra:2024vpd,Han:2024ugl,Maltoni:2024tul,Maltoni:2024csn,Aoude:2025jzc,Fabbrichesi:2025psr,Altakach:2026fpl,Choi:2026omc,Arai:2026jtc,Fang:2026ddi,Antozzi:2026vdi,Afik:2026pxv,Guo:2026yhz,Aoude:2026eeg}. 

The question addressed in this paper is whether this quantum information description can be used to diagnose beyond-the-standard-model sources of CP violation in $t\bar t$ production. CP violation is present in the SM through the complex phase of the CKM matrix, whose predictions successfully describe the CP-violating effects observed so far in quark-flavour processes. Nevertheless, the amount and structure of CP violation available in the SM are insufficient to account for the observed baryon asymmetry of the Universe within standard baryogenesis scenarios. Independently, the unexplained smallness of strong CP violation constitutes another indication that our understanding of CP symmetry may be incomplete. These considerations motivate broad searches for new CP-violating interactions.
One obvious direction is to search for CP violation in the lepton sector, which the new generation of neutrino experiments aims to probe. Another is to look for new interactions involving the third generation. The top quark is especially interesting in this respect. The large top mass makes it a privileged probe of electroweak symmetry breaking, and many extensions of the SM generate dipole, Yukawa or four-fermion interactions involving top quarks. More generally, CP violation can be studied systematically within the Standard Model Effective Field Theory (SMEFT), as discussed, e.g., in Ref.~\cite{Bonnefoy:2021tbt}. In the specific context of top-quark production, several dimension-six operators can generate CP-odd contributions~\cite{Kane:1991bg,Atwood:1991ka,Bernreuther:1992be,Atwood:2000tu,Zhang:2010dr,Aguilar-Saavedra:2008nuh,Gupta:2009wu,Bernreuther:2013aga,Bernreuther:2015yna,Cirigliano:2016nyn,Cirigliano:2016njn,Aguilar-Saavedra:2018ksv,deBeurs:2018pvs,Degrande:2021zpv,Bernreuther:2024ltu}. Searches for dimension-six interactions at high momentum transfer are an integral part of the LHC programme and will remain central at future lepton and hadron colliders, see e.g.,~\cite{deBlas:2025xhe,Armadillo:2026mvp}. 

The density-matrix viewpoint is particularly useful because CP symmetry acts directly on the spin state. For the unpolarised production processes considered here, after comparing CP-related kinematic configurations and expressing the top and antitop spins in a common basis, CP invariance implies a simple set of relations among the Fano--Bloch coefficients: the top and antitop polarisation vectors coincide, while the spin-correlation matrix is symmetric. The corresponding production-level CP markers are therefore the polarisation difference and the antisymmetric part of the correlation matrix. These relations are properties of the density matrix itself and do not depend on the particular observables used to reconstruct it. Conventional CP-odd spin asymmetries and triple-product correlations~\cite{Godbole:2006tq,Boudjema:2009fz,Rahaman:2021fcz,Baumgart:2012ay,Fischer:2018lme,Bernreuther:2017cyi} can be understood as specific projections of these CP-odd density-matrix components.

The quantum information formulation raises a further question. The presence of CP-odd components in the density matrix is not equivalent to a change in entanglement, discord, magic, or any other global quantum property. A CP-violating interaction may modify the state while producing only a small change in an entanglement measure; conversely, a state may remain maximally entangled for both CP-conserving and CP-violating choices of the underlying couplings. One must therefore determine which quantum observables are sensitive to CP violation, in which regions of phase space, and for which production mechanisms. This motivates analysing the production density matrix directly, before specifying a particular decay channel or tomographic reconstruction.

In this work, we compute the spin density matrix for several benchmark production mechanisms: the decay of a spin-zero state into \(t\bar t\), \(e^+e^- \to t\bar t\), photon fusion \(\gamma\gamma \to t\bar t\), and the partonic channels \(q\bar q \to t\bar t\) and \(gg \to t\bar t\) relevant to hadron colliders. For each channel, we identify the CP-even and CP-odd components of the Fano--Bloch coefficients and relate the anomalous couplings that generate them to the corresponding SMEFT operators. This provides a unified comparison among spin-zero decays, electroweak annihilation, photon fusion, and QCD production.

We therefore consider two complementary classes of observables. The first
class consists of direct CP markers constructed from the Fano--Bloch
coefficients themselves, such as polarisation asymmetries and antisymmetric
spin-correlation observables, together with the trace distance between the
state and its CP transform. The second class consists of quantum information
observables, such as discord, concurrence, and magic, which probe different
global properties of the same density matrix.

The present paper is the first part of a two-paper study. Here the object of study is the produced $t\bar t$ state itself: how it is generated from the hard interaction, how CP violation is encoded in it, and which quantum quantities respond to it. The reconstruction of the same state from the decay products, including the possibility that the $Wtb$ decay vertex also contains anomalous CP-violating interactions, is treated in the companion paper~\cite{Lamba:2026sni}. This separation is useful conceptually. The production problem determines the target density matrix and its CP-odd structures; the tomography problem determines how those structures are inferred experimentally and how possible decay-side effects can be disentangled.

The paper is organised as follows. In Section~\ref{sec:SMEFT} we introduce the CP-violating top-quark interactions and the SMEFT conventions used throughout the paper. In Section~\ref{sec:general_decomposition} we define the production density matrix, the Fano--Bloch coefficients, and the spin basis in which the CP relations take their simplest form. Section~\ref{sec:observables} discusses the CP transformation properties of the density matrix and introduces both  direct CP markers and the quantum information observables used later. In Section~\ref{sec:production_channels} we derive and analyse the density matrices for the benchmark production channels: spin-zero decay, $e^+e^-$ annihilation, photon fusion, and the $q\bar q$ and $gg$ partonic subprocesses. Section~\ref{sec:pheno} presents the phenomenological analysis and discusses the expected sensitivities at the LHC and at a future $e^+e^-$ collider. Finally, in Section~\ref{sec:conclusions} we draw our conclusions.


\section{CP-violating top-quark interactions and the SMEFT}
\label{sec:SMEFT}

We parametrise new physics interactions in terms of anomalous top-quark couplings. This choice is convenient for the explicit computation of production amplitudes and for the discussion of CP-violating observables. The relation between this anomalous-coupling description and the SMEFT framework is presented explicitly in the relevant sections below, where the corresponding mappings are derived. In the present section, we introduce the SMEFT operators relevant for our study and fix the conventions used throughout the paper.

The effects of heavy new physics are parametrised in the SMEFT through higher-dimensional operators added to the SM Lagrangian and suppressed by powers of the new physics scale $\Lambda$. Restricting ourselves to dimension-six terms, the effective Lagrangian can be written as
\begin{equation}
\mathcal{L}_{\text{SMEFT}}
=
\mathcal{L}_{\text{SM}}
+
\sum_i \frac{C_i}{\Lambda^2} O_i
+
\sum_j \left(
\frac{C_j}{\Lambda^2}\, {}^\ddagger O_j
+ \mathrm{h.c.}
\right)\, ,
\label{Lagrangian}
\end{equation}
where $C_i$ and $C_j$ denote the corresponding Wilson coefficients. Non-hermitian operators are marked with a double dagger ($^\ddagger$). This Lagrangian is used to study the impact of the operators listed in Tables~\ref{table: CP-odd operators}-~\ref{table: CP-even operators}  on the production processes considered in this work, with particular emphasis on CP-violating observables. In all cases, only tree-level SM contributions are included.

The operators introducing CP-violating top interactions are listed in Table~\ref{table: CP-odd operators}, following Refs.~\cite{Aguilar-Saavedra:2018ksv,Degrande:2021zpv}. For completeness, we also include in Table~\ref{table: CP-even operators} the CP-even operators that modify the Standard Model interaction vertices relevant for $t\bar t$ production and decay. Throughout this work, we restrict ourselves to flavour-conserving operators. In the operator definitions, $Q_L$ denotes the third-generation left-handed quark doublet, while $t_R$ and $b_R$ denote the right-handed top- and bottom-quark singlets. When explicit flavour indices are shown, they are written as $q_L^a$, $u_R^a$, and $d_R^a$, with $a$ indexing the generation (a=1, 2, 3). The convention adopted for the covariant derivative is
\begin{equation}
D_\mu = \partial_\mu
- i g_s T^A G^A_\mu
- i g \frac{\sigma^I}{2}  W^I_\mu
- i g' \frac{Y}{2} B_\mu \, ,
\label{partialD}
\end{equation} with $Q=T_3+Y/2$.

In Table~\ref{table: CP-odd operators}, the non-hermitian operators can in general have complex Wilson coefficients. For non-hermitian CP-odd operators, CP-violating effects in observables are proportional to the imaginary part of the relevant Wilson coefficient, or of an appropriate combination of Wilson coefficients. By contrast, the Wilson coefficients of hermitian operators are real as a consequence of the hermiticity of the SMEFT Lagrangian. These statements hold in the absence of absorptive phases. Since in this paper we only consider tree-level production, absorptive phases will be neglected.

For $2\to2$ scattering, we neglect operators inducing electric or magnetic dipole moments for the electron or for the incoming partons, since they are not relevant to the top-quark interactions that are the focus of this work. For the same reason, we do not include the CP-odd triple-gluon operator in our analysis. Although it is in principle relevant for CP-violating effects in QCD processes, it does not parameterise a direct anomalous interaction of the top quark and it is expected to be better constrained by QCD jet production similarly to its CP-even counterpart \cite{Krauss:2016ely,Hirschi:2018etq}. 

Four-fermion operators in Table~\ref{table: CP-odd operators} are  not included in the phenomenological analysis. In particular, the semileptonic operators $O_{\ell equ}^{(1,3)}$ do not interfere with the SM amplitude for $e^+e^- \to t\bar t$ and therefore modify the SM production matrices only at quadratic order in the EFT expansion. CP-violating contributions at quadratic order can be induced by the interference of the two operators, but we do not consider these in our analysis. Likewise, the four-quark operators $O_{quqd}^{(1,8)}$ are not retained in the analysis. In addition to being outside our flavour-conserving setup, they are forbidden if one imposes the symmetry $U(2)_q \times U(2)_u \times U(2)_d$, as motivated by the minimal-flavour-violation ansatz~\cite{Aguilar-Saavedra:2018ksv}. More generally, in the results presented below we only keep operators that can interfere with the SM amplitudes and contribute to the CP-violating observables under consideration.

Within the production channels and operator set considered here,
\(O_{bW}\) and \(O_{\phi tb}\) affect the top-quark decay vertex but not the
production amplitudes. Their effects are studied in the companion
paper~\cite{Lamba:2026sni}.
The operator $O_{t\phi}$ modifies the top-quark Yukawa interaction. 
It contributes to $t\bar t$ production in $e^+e^-$ and $q\bar q$ channels through Higgs exchange. 
However, in the massless limit for the initial fermions, $m_e=m_q=0$ (with $q=u,d,s,c$), the corresponding $s$-channel Higgs contributions vanish and are therefore neglected in our analysis of $t\bar t$ production in $e^+e^-$ and $q\bar q$ channels. 
\begin{table}[t]
\centering
\begin{tabular}{|c|c|c|c|}
\hline  
 \multicolumn{2}{|c|}{Four fermions} & \multicolumn{2}{|c|}{Two quarks}\\
\hline
   $^\ddagger O_{\ell equ}^{1(1133)}$ & $(\overline{\ell}_L e_R)\epsilon(\overline{Q}_L t_R)$ & $^\ddagger O_{tW}$ & $(\overline{Q}_L \sigma^{\mu\nu} t_R)\sigma^I \widetilde{\phi}W^I_{\mu\nu}$ \\
  $^\ddagger O_{\ell equ}^{3(1133)}$ & $(\overline{\ell}_L \sigma^{\mu\nu} e_R)\epsilon(\overline{Q}_L \sigma_{\mu\nu} t_R)$  & $^\ddagger O_{tB}$ & $(\overline{Q}_L \sigma^{\mu\nu} t_R)\widetilde{\phi}B_{\mu\nu}$ \\
 $^\ddagger O_{quqd}^{1(abcd)}$ & $(\overline{q}_L^a u_R^b)\epsilon(\overline{q}_L^c d_R^d)$ & $^\ddagger O_{bW}$ & $(\overline{Q}_L \sigma^{\mu\nu} b_R)\sigma^I \phi W^I_{\mu\nu}$ \\
 $^\ddagger O_{quqd}^{8(abcd)}$ & $(\overline{q}_L^a T^A u_R^b)\epsilon(\overline{q}_L^c T^A d_R^d)$ & $^\ddagger O_{\phi tb}$ & $i(\widetilde{\phi}^\dagger D_\mu \phi)(\overline{t}_{R} \gamma^\mu b_R)$\\ 
& & $^\ddagger O_{tG}$ & $(\overline{Q}_L \sigma^{\mu\nu} T^A t_R)\widetilde{\phi}G^A_{\mu\nu}$ \\
& & $^\ddagger O_{t\phi}$ &$(\overline{Q}_L t_R\widetilde{\phi})(\phi^\dagger\phi)$  \\
\hline
\end{tabular}
\caption{SMEFT operators whose complex Wilson coefficients can induce CP violation relevant to $t\bar t$ production and decay, following Refs.~\cite{Aguilar-Saavedra:2018ksv,Degrande:2021zpv}. The combination $O_{tZ}=-\sin\theta_W O_{tB}+\cos\theta_W O_{tW}$ can also be defined after electroweak symmetry breaking.}
\label{table: CP-odd operators}
\end{table}

\begin{table}[t]
\centering
\begin{tabular}{|c|c|}
\hline  
  \multicolumn{2}{|c|}{Two quarks}\\
\hline
  $O^{3}_{\phi Q}$&$  \left( \phi^\dagger \sigma^I \overleftrightarrow{iD}_\mu \phi \right)
\left( \overline{Q}_{L} \gamma^\mu \sigma^I Q_{L} \right) $\\
$ O^{1}_{\phi Q}$&$  \left( \phi^\dagger \overleftrightarrow{iD}_\mu \phi \right)
\left( \overline{Q}_{L} \gamma^\mu Q_{L} \right)$ \\
$ O_{\phi t}$ &$ (\phi^\dagger \overleftrightarrow{iD}_\mu \phi)
(\bar{t}_R \gamma^\mu t_R)$ \\
\hline
\end{tabular}
\caption{Dimension-six operators that modify the SM interactions of third-generation quarks with massive vector bosons that are relevant for our analysis. Specifically, $O^{3}_{\phi Q}$ contributes to the $V-A$ structure of $Wtb$ vertex, while all the three operators modify the SM-like interaction of a $t\bt$ pair with a $Z$ boson. The combinations $O_{\phi Q}^{(\pm)}=O_{\phi Q}^{1}\pm O_{\phi Q}^{3}$ are also commonly used in experimental constraints. }
\label{table: CP-even operators}
\end{table}

Before turning to the individual production channels, we summarize in
Table~\ref{table:SMEFT limits} representative current constraints on the
Wilson coefficients relevant to our analysis. These results are obtained
from different experimental measurements and global-fit frameworks and,
in some cases, employ different normalisation conventions. They should
therefore not be interpreted as a statistically combined set of limits.
Instead, we use them as reference ranges to motivate the benchmark values
adopted in the numerical analysis. Unless stated otherwise, the quoted
constraints correspond to a new physics scale of
$\Lambda=1~\mathrm{TeV}$.

\begin{table}[t]
\centering
\begin{tabular}{|c|c|c|}
\hline  
  Quantity & Limits & Reference\\
\hline
\hline
 $\operatorname{Re}(C_{tW})$& [-0.208,0.161]&\cite{terHoeve:2025gey} 95\% C.I. (w/ RGE)\\
  $\operatorname{Re}(C_{tW})$& [0.02, 0.34]  &\cite{ATLAS:2025adk} 95\% C.I. 
  \\
   $\operatorname{Im}(C_{tW})$& [-0.08, 0.12]  &\cite{ATLAS:2025adk} 95\% C.I. 
   \\
 \hline
 \hline
$\operatorname{Re}(C_{tB})$ &[-1.7, 1.6] & \cite{ATLAS:2023eld} 95$\%$ C.I. (obs.)\\ 
 $\operatorname{Im}(C_{tB})$& [-1.9, 1.9]& \cite{ATLAS:2023eld} 95$\%$ C.I. (obs.)\\ 
 \hline
 \hline
 $\operatorname{Re}(C_{tG})/g_s$&[0.002,0.184]&\cite{terHoeve:2025gey} 95\% C.I. (w/ RGE) \\
  $\operatorname{Re}(C_{tG})/y_tg_s$ & [-0.24, 0.07]&\cite{CMS:2019nrx} 95\% C.I. 
  \\
   $\operatorname{Im}(C_{tG})/y_tg_s$ &[-0.33, 0.20]& \cite{CMS:2019nrx} 95\% C.I. 
   \\
 \hline\hline
  $C^{3}_{\phi Q}$& [-0.723,0.406]&\cite{terHoeve:2025gey} 95\% C.I. (w/ RGE)\\
    $C^{3}_{\phi Q}$& [-0.75,0.41]&\cite{ATLAS:2025adk} 95\% C.I. \\ 
\hline
   $C^{(-)}_{\phi Q}$& [-0.775,1.545]&\cite{terHoeve:2025gey} 95\% C.I. (w/ RGE)\\
    $C^{(-)}_{\phi Q}$& [-2.68, 2.94] &\cite{CMS:2023xyc}$2\sigma$ C.I.\\
\hline
  $C_{\phi t}$&  [-2.241,0.104]&\cite{terHoeve:2025gey} 95\% C.I. (w/ RGE)\\
$C_{\phi t}$&[-2.2, 1.6] & \cite{ATLAS:2023eld} 95\% C.I. (obs)\\
\hline
\end{tabular}
\caption{Representative current constraints on the Wilson coefficients
relevant to this study. The intervals are taken from the references listed
in the last column and are used as reference ranges for selecting the
benchmark values adopted in the numerical analysis. Unless stated otherwise,
the limits correspond to $\Lambda=1~\mathrm{TeV}$.}
\label{table:SMEFT limits}
\end{table}


\section{Production density matrix: quantum state of the $t\bar t$ pair}
\label{sec:general_decomposition}

Since the top quark and the antitop are spin-$1/2$ particles, their spin degrees of freedom define a bipartite two-level system. The spin (or helicity) state of the produced $t\bar t$ pair can therefore be described by a two-qubit density matrix. A general two-qubit density matrix can be expanded in the tensor-product basis built from the identity and Pauli matrices as
\begin{equation}
\rho = \frac{1}{4}\left[
\mathbb{I}\otimes\mathbb{I}
+ B_i\,\sigma^i\otimes\mathbb{I}
+ \bar{B}_j\,\mathbb{I}\otimes\sigma^j
+ C_{ij}\,\sigma^i\otimes\sigma^j
\right],
\label{eq:rho2_qbit}
\end{equation}
where summation over the repeated indices \(i,j=1,2,3\) is understood. 
The coefficients \(B_i\), \(\bar B_j\), and \(C_{ij}\) are real and are given by
\begin{equation}
    B_i
    = \operatorname{Tr}\!\left[
        (\sigma_i \otimes \mathbb{I})\,\rho
      \right],
    \qquad
    \bar B_j
    = \operatorname{Tr}\!\left[
        (\mathbb{I} \otimes \sigma_j)\,\rho
      \right],
    \qquad
    C_{ij}
    = \operatorname{Tr}\!\left[
        (\sigma_i \otimes \sigma_j)\,\rho
      \right].
\end{equation}
The polarisation vectors
\begin{equation}
    \mathbf{B}=(B_1,B_2,B_3),
    \qquad
    \bar{\mathbf{B}}=(\bar B_1,\bar B_2,\bar B_3),
\end{equation}
describe the single-particle polarisations of the top quark and antiquark,
respectively, while the spin-correlation matrix $\mathbf C=(C_{ij})$ encodes their spin correlations.

Let us consider the production of a $t\bar t$ pair from an initial state $I(a_i)$, where $a_i$ denotes the independent parameters needed to characterise the production process. It is convenient to write the spin production matrix in terms of the helicity amplitudes as
\begin{equation}
\label{eq:Rmatrix}
R^I_{\lambda\lambda'\mu\mu'}
=
\frac{1}{N_I}
\sum_{\text{$I$-d.o.f.}}
\mathcal{M}^P(\lambda,\mu)\,
\mathcal{M}^{P\dagger}(\lambda',\mu') ,
\end{equation}
where $\mathcal{M}^P(\lambda,\mu)$ is the production amplitude for a top quark with helicity $\lambda$ and an antitop with helicity $\mu$ and $N_I$ accounts for the multiplicity of the initial state, i.e. the product of the spin, colour and other averaging
factors associated with the initial state. The sum runs over all
unobserved quantum numbers, such as initial-state colours and helicities
when they are not fixed, and final-state colour degrees of freedom. Independently of the underlying production mechanism, the matrix $R^I$ admits the general decomposition
\begin{equation}
R^I_{\lambda\lambda'\mu\mu'}
=
\widetilde{A}^I\,
\delta_{\lambda\lambda'}\delta_{\mu\mu'}
+
\tilde{B}_i^I\,
\sigma^i_{\lambda\lambda'}\delta_{\mu\mu'}
+
\tilde{\bar{B}}_j^{I}\,
\delta_{\lambda\lambda'}\sigma^j_{\mu\mu'}
+
\tilde{C}_{ij}^I\,
\sigma^i_{\lambda\lambda'}\sigma^j_{\mu\mu'} ,
\label{eq: R matrix decomposition}
\end{equation}
with $\tilde{A}^I=\mathrm{Tr}[R^I]/4$. The normalised production matrix defines the density matrix of the $t\bar t$ system,
\begin{equation}
\rho^I=\frac{R^I}{\mathrm{Tr}[R^I]},
\label{eq: normalised matrices}
\end{equation}
which has the form of Eq.~\eqref{eq:rho2_qbit}, with coefficients
\begin{equation}
B_i^I=\frac{\tilde{B}_i^I}{\tilde{A}^I},
\quad\quad
\bar{B}_j^{I}=\frac{\tilde{\bar{B}}_j^{I}}{\tilde{A}^I},
\quad\quad
C_{ij}^I=\frac{\tilde{C}_{ij}^I}{\tilde{A}^I}.
\label{eq:normalised_fano}
\end{equation}

The matrix elements of $R^I$ are computed using the Bouchiat-Michel (BM)
formulae, whose main ingredients are summarised in App.~\ref{app: BM formulae}. This procedure
naturally yields the production density matrix in the helicity basis, where the
spin of each particle is quantised along its own direction of motion.

\begin{figure}[t]
    \centering
    \includegraphics[width=0.3\linewidth]{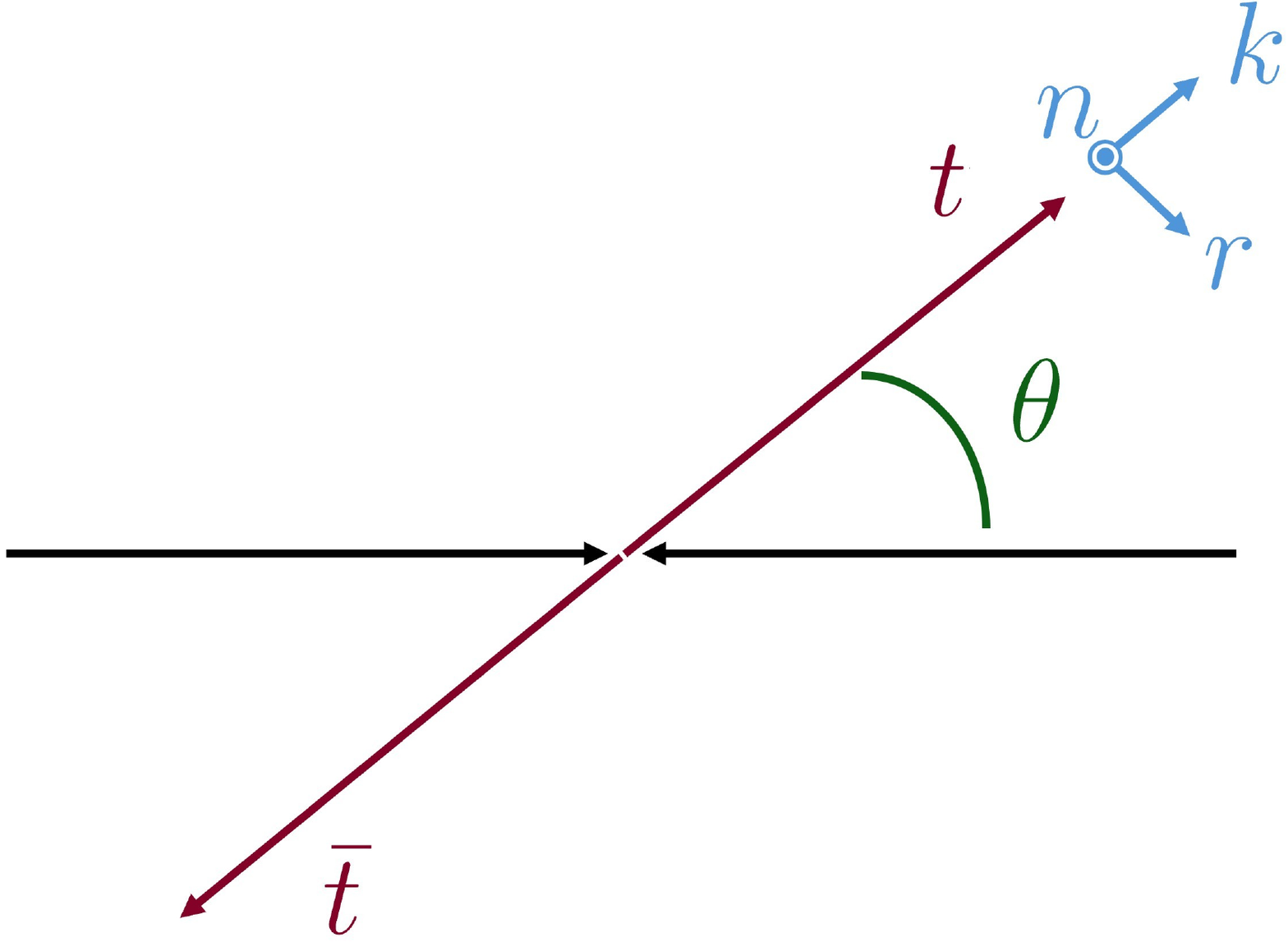}
    \caption{Helicity basis $\{\hat{n}, \hat{r}, \hat{k}\}$ in  $t\bar t$ ZMF.}
    \label{fig: hrf}
\end{figure}

In the zero-momentum frame (ZMF) of the $t\bar t$ pair, the top quark and
antitop are produced back-to-back. It is then useful to introduce a common spin
basis for the two particles. We choose the common axes to coincide with the
top-quark helicity triad $\{\hat n,\hat r,\hat k\}$, with $\hat k$ along the
top momentum. For a $2\rightarrow2$ scattering process, these three directions can be defined in terms of the kinematics of the hard process as
\begin{equation}
    \hat{k}=\hat{t}, \qquad
\hat{r}= \frac{\hat{p}-\cos\theta\hat{k}}{\sin\theta}, \qquad
\hat{n}=\hat{r}\wedge\hat{k}, 
    \label{eq: helicity basis}
\end{equation}
\begin{equation}
\cos\theta=\hat{p}\cdot\hat{k} \,,
  \label{eq:costheta}
\end{equation}
where $\hat{p}$ denotes the direction of one of the incoming particles participating in the hard scattering, while $\hat{k}$ is the direction of motion of the top quark (Fig.~\ref{fig: hrf}). By choosing common axes, the antitop spin is therefore quantised with respect to the same
axes, rather than with respect to its own helicity triad
$\{\hat n,-\hat r,-\hat k\}$. Equivalently, starting from the helicity basis,
this convention is obtained by reversing the antitop helicity assignment. We
refer to this convention as the spin basis. Unless otherwise stated, all
Fano--Bloch coefficients below are given in this spin basis.
The transformation of the density matrix, together with the corresponding $B_i$, $\bar{B}_j$ and $C_{ij}$ coefficients, from the helicity basis to the spin basis is given in App.~\ref{app: production density matrix BM}.

A basis ordered in the same way will also be used for the $1\rightarrow2$ decay process. In that case, the only distinguished direction in the $t\bar t$ ZMF is $\hat{k}$, and there remains a freedom in the definition of the directions $\hat{n}$ and $\hat{r}$ within the plane orthogonal to $\hat{k}$.


\section{Observables}
\label{sec:observables}

In this section we introduce the set of observables used to diagnose CP violation in the production of a $t\bar t$ pair. Our strategy is to exploit as much as possible the information contained in the spin density matrix, rather than selecting only a small number of spin-correlation coefficients. Since the density matrix is fully parametrised by the Fano--Bloch coefficients $\mathbf{B}$, $\mathbf{\bar{B}}$ and $\mathbf C$, any CP-violating effect in production must appear as a violation of the CP symmetry relations obeyed by these coefficients. We therefore consider two complementary classes of observables. The first class consists of direct CP markers constructed from the Fano--Bloch coefficients themselves, such as polarisation asymmetries and antisymmetric spin-correlation observables. The second class consists of quantum information observables, such as discord, concurrence, magic and trace-distance measures, which probe different global properties of the same density matrix. In this way, we can compare conventional CP-sensitive quantities with more global quantum observables and assess whether the latter provide competitive sensitivity to CP-violating interactions.
\subsection{CP transformation properties of the Fano--Bloch coefficients}\label{sec:CP-transformation}

Working in the spin basis introduced in Sec.~\ref{sec:general_decomposition}, the density matrix of the $t\bar t$ pair is written as in Eq.~\eqref{eq:rho2_qbit}. In this basis, CP invariance imposes definite symmetry conditions on the Fano--Bloch coefficients. As shown in App.~\ref{app: Density matrix under CP transformations }, these conditions are\footnote{The conditions in Eq.~\eqref{eq: CP invariance on rho} are specific to a density matrix expressed in the spin basis. The corresponding CP-invariance conditions for the helicity basis read $\mathbf{\bar{B}}=R\,\mathbf B$ and $\mathbf C=R\,\mathbf C^{T}R$, with $R=\mathrm{diag}(1,-1,-1)$.}
\begin{equation}
    \text{CP invariance}
    \quad \Longleftrightarrow \quad
   \mathbf{ B}=\mathbf{\bar{B}} ,
    \qquad
  \mathbf  C= \mathbf C^{T} ,
    \label{eq: CP invariance on rho}
\end{equation}
where the superscript $T$ denotes matrix transposition. Therefore, CP violation in production can be identified with two independent structures: a mismatch between the top and antitop polarisation vectors, and an antisymmetric component of the spin-correlation matrix.
These two relations can be combined into a single symmetry condition by defining $\sigma^0=\mathbb{I}$ and rewriting the density matrix as
\begin{equation}
\rho= \frac{1}{4}\sum_{a, b=0}^3F_{ab}(\sigma^a\otimes\sigma^b)\,.
\end{equation}
In this notation, CP conservation is equivalent to
\begin{equation}
    \text{CP invariance } \Longleftrightarrow F_{ab}=F_{ba}.
    \label{eq: CP on F}
\end{equation}
Thus, CP-violating effects are encoded in the antisymmetric part of $F_{ab}$. We stress once more that all the relations discussed in this subsection are expressed in the spin basis. The explicit relations among the Fano--Bloch coefficients are therefore basis dependent, as they rely on the chosen quantisation axes for the two spins. However, the underlying CP-invariance condition itself is basis independent, and any physical statement about CP violation does not depend on this choice.

It is then useful to decompose the correlation matrix into its symmetric and antisymmetric parts,
\begin{equation}
  \mathbf  C= \mathbf C^S+ \mathbf C^A, \qquad
    \begin{cases}
      \mathbf  C^S=  1/2(\mathbf C+ \mathbf C^T) \\
      \mathbf  C^A=  1/2(\mathbf C- \mathbf C^T) 
    \end{cases} .
    \label{eq: decomposition C in S,A}
\end{equation}
In the CP-conserving limit, the antisymmetric component vanishes,
\(\mathbf C^A=0\). Conversely, \(\mathbf C^A\neq0\) provides a direct
CP-violation marker in the spin-correlation sector, subject to the spin-basis
and CP-related-kinematics qualifications stated above. In the following sections, we will present the full analytical expressions of the Fano--Bloch coefficients for each production mechanism, organizing the results in terms of the symmetric and antisymmetric parts of the correlation matrix. This decomposition makes the separation between CP-even and CP-odd contributions manifest and allows for a transparent identification of the sources of CP violation in each process.
\subsubsection{Polarisation and correlation asymmetries}
\label{sec:asymmtries}

The first class of CP-sensitive observables is obtained directly from the violation of Eq.~\eqref{eq: CP invariance on rho}. In the polarisation sector, this naturally leads to the definition of the CP-odd vector
\begin{equation}
    \Delta \mathbf{B}
    =
    \frac{1}{2}\left(\mathbf B-\mathbf{\bar{B}}\right) \, ,
    \label{eq: deltaB}
\end{equation}
whose components provide CP-sensitive observables in the chosen spin basis. They encode both the magnitude and the directional structure of CP violation in the polarisation sector.

It is also useful to construct a scalar quantity that captures the overall size of this effect, namely the Euclidean distance
\begin{equation}
    \lVert\Delta \mathbf{B}\rVert
    =
    \frac{1}{2}\sqrt{
    \sum_i
    \left(B_i-\bar{B}_i\right)^2
    } \, .
    \label{eq: norm delta B}
\end{equation}
It is easy to see that $\lVert\Delta \mathbf{B}\rVert\leq 1$.
This quantity provides a single measure of the total amount of CP violation present in the polarisation sector. While the individual components of $\Delta \mathbf{B}$ identify the spin directions in which the effect appears, the norm aggregates this information into a basis-independent magnitude under rotations acting identically on the two spins.

We stress, however, that the quantities $\Delta \mathbf{B}$ and $\lVert\Delta \mathbf{B}\rVert$ are not invariant under more general changes of basis that act differently on the two subsystems, such as the transformation between the spin and helicity bases. In particular, while common rotations of the spin quantisation axes preserve the Euclidean norm, the transformation relating the spin and helicity bases involves a non-trivial redefinition of the antitop spin only, and therefore does not leave $\lVert\Delta \mathbf{B}\rVert$ invariant. For this reason, all results presented in this work are to be understood within the spin-basis convention defined in Sec.~\ref{sec:general_decomposition}.

An analogous construction can be applied to the antisymmetric part of the correlation matrix. Since any real antisymmetric $3\times3$ matrix can be mapped onto a three-vector, it is convenient to introduce
\begin{equation}
    \mathrm{a}_i = \frac{1}{2}\,\epsilon_{ijk}\,C_{jk} \, ,
\end{equation}
which isolates the antisymmetric component of the correlation matrix, $C^A_{ij} = \epsilon_{ijk} \mathrm{a}_k \,
$. In the basis $\{\hat n,\hat r,\hat k\}$, the components of $\mathbf{a}$ are given by
\begin{equation}
    \mathbf{a}
    =
    \left(
    \frac{C_{rk}-C_{kr}}{2},
    \frac{C_{kn}-C_{nk}}{2},
    \frac{C_{nr}-C_{rn}}{2}
    \right) ,
    \label{eq:a vector}
\end{equation}
which explicitly display the antisymmetric combinations of the correlation coefficients.

As in the polarisation case, it is useful to define a scalar measure of this CP-odd contribution,
\begin{equation}
    \lVert\mathbf{a}\rVert
    =
    \sqrt{\sum_i (\mathrm{a}_i)^{2}} \, ,
    \label{eq:norm a}
\end{equation}
which quantifies the overall magnitude of CP violation in the correlation sector. It is easy to see that $\lVert\mathbf{a}\rVert\leq 1$.

While the components of $\mathbf{a}$ retain information on the orientation of the CP-violating effects, the norm provides a compact measure of their total size. The two quantities $\lVert\Delta \mathbf{B}\rVert$ and $\lVert\mathbf{a}\rVert$ therefore constitute complementary probes of CP violation in the full spin density matrix.

Besides these scalar quantities, the individual components of $\Delta \mathbf{B}$ and $\mathbf{a}$ can themselves be regarded as CP-sensitive observables, as they encode the directional structure of the underlying effects. In a given spin basis, these coefficients can in principle be accessed through a tomographic reconstruction of the density matrix. Such a reconstruction allows one to establish a direct correspondence between the Fano--Bloch coefficients, the CP-odd angular structures and triple-product correlations commonly employed in phenomenological analyses of $t\bar t$ spin correlations~\cite{Bernreuther:2013aga, Bernreuther:2015yna, CMS:2019nrx, Bernreuther:2024ltu}. 

In the present work, however, we do not make this correspondence explicit, as our focus is on the characterisation of CP violation at the level of the production density matrix. A detailed discussion of the tomographic reconstruction will be presented in a companion work~\cite{Lamba:2026sni}.\\
\subsubsection{Trace distance}
\label{sec:tracedistance}

The trace distance provides a measure of the distance between two quantum states and quantifies their distinguishability under quantum measurements~\cite{Helstrom:1976,Nielsen_Chuang_2010}. For two density matrices $\rho_1$ and $\rho_2$, it is defined as
\begin{equation}
D_T(\rho_1,\rho_2)
=
\frac12
\|\rho_1-\rho_2\|,
\end{equation}
where $\|A\|=\mathrm{Tr}\sqrt{A^\dagger A}$. The trace distance satisfies
\begin{equation}
0\leq D_T(\rho_1,\rho_2)\leq 1,
\end{equation}
with $D_T(\rho_1,\rho_2)=0$ if and only if the two density matrices are identical. Operationally, the trace distance quantifies the maximal distinguishability between two quantum states under arbitrary measurements.

In the present context, the trace distance can be used to quantify CP violation by comparing the density matrix $\rho$ with its CP-transformed counterpart $\rho^{\rm CP}$~\cite{Fabbrichesi:2025igr}. We therefore define the CP-sensitive trace-distance observable as follows
\begin{equation}
D_T^{\rm CP}(\rho)
=
\frac12
\|
\rho-\rho^{\rm CP}
\|.
\label{eq:traceCP}
\end{equation}
This quantity vanishes for CP-conserving states and becomes non-zero whenever the spin density matrix violates the CP symmetry relations discussed in Sec.~\ref{sec:CP-transformation}. Unlike observables constructed from individual Fano--Bloch coefficients, the trace distance provides a global measure of CP violation sensitive to the full structure of the density matrix.
\subsection{Quantum information observables}\label{sec: quantum observables}

Beyond the direct CP-sensitive observables introduced in the previous subsection, it is useful to characterise the produced $t\bar t$ state through quantum information measures constructed from the full spin density matrix. These observables probe complementary properties of the bipartite quantum state and therefore provide additional information on the structure of CP-violating interactions.

In this work we consider three classes of quantum observables that have also recently been introduced in collider studies:

\begin{itemize}
    \item \textbf{Discord}~\cite{Ollivier:2001fdq}, which quantifies general quantum correlations present in the bipartite state, including correlations that are not captured by entanglement measures;

    \item \textbf{Concurrence}~\cite{Hill:1997pfa} and related separability measures, which characterise the degree of entanglement between the top- and antitop-spin subsystems and determine whether the state is separable or genuinely entangled;

    \item \textbf{Magic}~\cite{Bravyi:2004isx}, which quantifies the non-stabilizer nature of the quantum state, namely its departure from the stabilizer-state polytope, and therefore probes genuinely non-classical quantum resources beyond standard entanglement.

\end{itemize}
All these quantities are computed from the same production density matrix, yet they can probe different aspects of the spin correlations induced by CP-even and CP-odd interactions. Comparing their behaviour therefore allows us to identify which genuinely quantum features are most sensitive to CP violation in $t\bar t$ production.
\subsubsection{Quantum Discord}\label{sec:discord}

Quantum discord was introduced by Ollivier and Zurek~\cite{Ollivier:2001fdq},
and closely related measures of classical and quantum correlations were developed
independently by Henderson and Vedral~\cite{Henderson:2001wrr}.
It is a measure of non-classical correlations in a bipartite quantum system and extends the notion of quantum correlations beyond entanglement. Unlike entanglement measures, discord can remain non-vanishing even for separable mixed states, thereby probing a broader class of genuinely quantum correlations encoded in the density matrix. In the context of $t\bar t$ production, this observable is particularly interesting because CP-violating interactions can modify the structure of spin correlations without necessarily inducing large entanglement effects. The origin of quantum discord can be understood from the notion of mutual information. In classical information theory~\cite{it-book}, the mutual information between two subsystems $A$ and $B$ admits two equivalent definitions,
\begin{equation}
\mathcal I(A:B)=H(A)+H(B)-H(A,B),
\end{equation}
and
\begin{equation}
\mathcal J(A:B)=H(A)-H(A|B),
\end{equation}
where $H(X)$ denotes the Shannon entropy and $H(A|B)$ is the conditional entropy. The equivalence between these two expressions follows from Bayes' theorem and reflects the fact that classical measurements do not disturb the system.

In quantum mechanics, however, the definition of conditional entropy becomes measurement dependent because the act of measuring one subsystem generically modifies the state of the other. For a bipartite density matrix $\rho$, the quantum mutual information is defined as
\begin{equation}
\mathcal I(A:B)=S(\rho^A)+S(\rho^B)-S(\rho),
\label{eq:QmutualI}
\end{equation}
where $ S(\rho)=-\mathrm{Tr}\left(\rho \log_2 \rho\right)$ is the von Neumann entropy, and $\rho^A=\mathrm{Tr}_B[\rho]$, $\rho^B=\mathrm{Tr}_A[\rho]$ are the reduced density matrices of the two subsystems.

To define the quantum analogue of the conditional entropy, one considers a local projective measurement $\{b_k\}$ acting on subsystem $B$. Following a measurement outcome labelled by $k$, the state of subsystem $A$ becomes 
\begin{equation}
\rho^{A|b_k}
=
\frac{(\mathbb I\otimes b_k)\rho(\mathbb I\otimes b_k)}
{p_k},
\end{equation}
with probability $p_k=
\mathrm{Tr}\left[(\mathbb I\otimes b_k)\rho(\mathbb I\otimes b_k)\right]$. The measurement-dependent conditional entropy is then defined as
\begin{equation}
S(A|\{b_k\})
=
\sum_k p_k\,S(\rho^{A|b_k}),
\end{equation}
which defines the measurement-dependent correlation measure
\begin{equation}
\mathcal J_B(A:B)
=
S(\rho^A)
-
\min_{\{b_k\}}
S(A|\{b_k\}).
\end{equation}

Quantum discord is defined as the mismatch between the two quantum generalisations of mutual information~\cite{Ollivier:2001fdq},
\begin{equation}
\mathcal D_B(\rho)
=
\mathcal I(A:B)-\mathcal J_B(A:B).
\label{eq:discorddef}
\end{equation}
Unlike the classical case, the two quantities are not generally equivalent because local measurements disturb the quantum state. As a consequence~\cite{PhysRevA.81.052318},
\begin{equation}
\mathcal D_B(\rho)\geq 0,
\end{equation}
with equality holding only for states possessing purely classical correlations. Quantum discord therefore quantifies genuinely quantum correlations that are not necessarily associated with entanglement.

An important feature of quantum discord is that it is generally asymmetric under the exchange of the two subsystems,
\begin{equation}
\mathcal D_A(\rho)\neq \mathcal D_B(\rho),
\end{equation}
since the conditional entropy depends on which subsystem is measured. In the context of the $t\bar t$ spin density matrix, this asymmetry is particularly relevant because CP transformations interchange the top- and antitop-spin subsystems. Differences between left and right discord can therefore probe CP-violating structures in the density matrix.

The computation of quantum discord requires an optimisation over all possible local measurements and is therefore analytically challenging for a generic mixed state. However, for bipartite qubit systems with maximally mixed marginals, corresponding to vanishing polarisation vectors,
\begin{equation}
\mathbf B=\mathbf{\bar{B}}=0,
\end{equation}
an analytic expression is available~\cite{PhysRevA.77.042303}. In this case, the discord is $\mathcal D_A(\rho)=\mathcal D_B(\rho)$, given by
\begin{equation}
{\cal D(\rho)}
=
2+\sum_i \lambda_i\log_2\lambda_i
-\left(\frac{1+c}{2}\right)\log_2(1+c)
-\left(\frac{1-c}{2}\right)\log_2(1-c),
\label{eq:discord}
\end{equation}
where $\lambda_i$ are the eigenvalues of the density matrix and
\begin{equation}
c=\max\{|c_1|,|c_2|,|c_3|\}
\end{equation}
is the largest singular value of the correlation matrix $ \mathbf C$.

The eigenvalues can be expressed in terms of the singular values $c_i$ as
\begin{align}
\lambda_1&=\frac14(1-c_1-c_2-c_3), \nonumber\\
\lambda_2&=\frac14(1-c_1+c_2+c_3), \nonumber\\
\lambda_3&=\frac14(1+c_1-c_2+c_3), \nonumber\\
\lambda_4&=\frac14(1+c_1+c_2-c_3).
\end{align}

For states with non-vanishing polarisation vectors, the analytical evaluation of the full quantum discord becomes considerably more involved. In such cases, we will instead employ the geometric discord discussed in the following subsection, which admits a closed analytical form directly in terms of the Fano--Bloch coefficients. 
\subsubsection{Geometric Discord}\label{se:Gdiscord}

In order to study states with non-vanishing polarisation vectors in a fully analytic manner, we employ the geometric discord introduced in Ref.~\cite{Dakic:2010xfz}. Geometric discord provides a geometric measure of non-classical correlations by quantifying the distance between a given quantum state and the set of zero-discord states. For a bipartite density matrix $\rho$, the left geometric discord is defined as
\begin{equation}
{\cal D}_{G}^{A}(\rho)
=
{\cal D}_{G}^{L}(\rho)
=
\min_{\rho_0\in\Omega_0}
\left\|
\rho-\rho_0
\right\|^2,
\end{equation}
where $\Omega_0$ denotes the set of zero-discord states and $
\left\|
\rho-\rho_0
\right\|^2
=
\mathrm{Tr}\left[(\rho-\rho_0)^2\right]$ is the square of the Hilbert-Schmidt norm.

As in the case of ordinary quantum discord, geometric discord depends on which subsystem is measured and is therefore generally asymmetric under the exchange of the two qubits. This leads to the definitions of left and right geometric discord, corresponding respectively to measurements performed on subsystem $A$ or $B$. For a general two-qubit state written in the Fano--Bloch decomposition, the analytical expressions for the left and right geometric discord are~\cite{Dakic:2010xfz}
\begin{align}
{\cal D}^{L}_{G}(\rho)
&=
\frac14
\left[
\|\mathbf{B}\|^2
+
\|\mathbf{C}\|^2
-
k_{\max}
\right],
\\[2mm]
{\cal D}^{R}_{G}(\rho)
&=
\frac14
\left[
\|\mathbf{\bar{B}}\|^2
+
\|\mathbf{C}\|^2
-
\bar{k}_{\max}
\right],
\end{align}
where
\ba
k_{\max}
&=&
\text{max eigenvalue of }
\left(
\mathbf B \mathbf B^{T}
+
\mathbf C \mathbf C^{T}
\right),
\nonumber\\
\bar{k}_{\max}
&=&
\text{largest eigenvalue of }
\left(
\mathbf{\bar{B}}\mathbf{\bar{B}}^{T}
+
\mathbf C^{T} \mathbf C
\right).
\ea
The geometric discord is positive semidefinite, and for two-qubit systems its values lie in the interval
\begin{equation}
0\leq {\cal D}^{L,R}_{G}(\rho)\leq \frac12.
\end{equation} 
An important feature of geometric discord in the present context is its sensitivity to the asymmetry between the two subsystems. Since CP transformations interchange the top and antitop spins in the spin basis, the difference between left and right geometric discord provides a natural CP-sensitive observable,
\begin{equation}
\Delta {\cal D}_G(\rho)
=
{\cal D}^{L}_{G}(\rho)-{\cal D}^{R}_{G}(\rho).
\end{equation}
Using the expressions above, one obtains
\begin{equation}
\Delta {\cal D}_G(\rho)
=
\frac14
\left[
\|\mathbf{B}\|^2
-
\|\mathbf{\bar{B}}\|^2
-
k_{\max}
+
\bar{k}_{\max}
\right].
\end{equation}
For states with vanishing polarisation vectors, $\mathbf B=\mathbf{\bar{B}}=0$, one has $k_{\max}=\bar{k}_{\max}$,
and therefore $\Delta {\cal D}_G(\rho)=0$. Consequently, a non-vanishing asymmetry between left and right geometric discord can arise only in the presence of non-trivial polarisation structures.

Like ordinary quantum discord, geometric discord probes non-classical correlations beyond entanglement and can remain non-zero even for separable mixed states. It therefore provides a complementary characterisation of the quantum structure of the $t\bar t$ spin density matrix and its modification by CP-violating interactions.
\subsubsection{Concurrence}\label{sec:concurrence}

To quantify the degree of entanglement between the top and antitop  spin subsystems, we employ the concurrence introduced in Ref.~\cite{Wootters:1997id}. Concurrence is a standard entanglement measure for bipartite two-qubit systems and provides a quantitative criterion for distinguishing separable and entangled mixed states. For a given two-qubit density matrix $\rho$, the concurrence is defined as
\begin{equation}
\mathcal{C}(\rho)
=
\max
\left(
0,\,
\lambda_1-\lambda_2-\lambda_3-\lambda_4
\right),
\end{equation}
where $\lambda_i$ $(\lambda_i\geq\lambda_j$ for $i<j)$ denote the eigenvalues of the matrix
\begin{equation}
\mathcal{R}
=
\sqrt{
\sqrt{\rho}\,
\tilde{\rho}\,
\sqrt{\rho}
},
\end{equation}
ordered in decreasing magnitude. Here
\begin{equation}
\tilde{\rho}
=
(\sigma_y\otimes\sigma_y)\,
\rho^{*}\,
(\sigma_y\otimes\sigma_y)
\end{equation}
is the spin-flipped density matrix, with $\rho^{*}$ denoting complex conjugation in the computational basis and $\sigma_y$ the Pauli-$y$ matrix. The eigenvalues of $\mathcal{R}$ are real and non-negative by construction. The concurrence satisfies
\begin{equation}
0\leq \mathcal{C}(\rho)\leq 1.
\end{equation}
A vanishing concurrence corresponds to a separable (non-entangled) state, while $\mathcal{C}(\rho)=1$ characterises a maximally entangled two-qubit state.

Unlike quantum discord, concurrence captures only entanglement correlations and therefore vanishes for separable mixed states even when non-classical correlations remain present. Comparing concurrence with discord therefore allows one to distinguish genuinely entangled CP-induced spin correlations from more general quantum correlations encoded in the $t\bar t$ density matrix.
\subsubsection{Magic}\label{sec:magic}

Besides entanglement and discord, it is also useful to characterise the non-stabilizer structure of the $t\bar t$ spin state. In quantum information theory, the corresponding resource is known as magic or non-stabilizerness and quantifies the genuinely non-classical features required for universal quantum computation~\cite{Leone:2021rzd,Turkeshi:2023lqu}.

The importance of this notion follows from the Gottesman-Knill theorem~\cite{Gottesman:1997qd,Gottesman:1997zz}, which states that quantum computations restricted to stabilizer states and Clifford operations can be efficiently simulated on a classical computer. Since stabilizer states may already possess a large amount of entanglement, entanglement alone is not sufficient to guarantee a genuine quantum computational advantage. The additional resource enabling such an advantage is the non-stabilizer content of the quantum state, commonly referred to as magic. 

For an $n$-qubit system, the Pauli strings are defined as tensor products of Pauli operators,
\begin{equation}
{\cal P}_n
=
A_1\otimes A_2\otimes \cdots \otimes A_n,
\qquad
A_i\in
\left\{
\mathbb I,
\sigma_x,
\sigma_y,
\sigma_z
\right\}.
\end{equation}
The Clifford group is then defined as the set of unitary transformations preserving the Pauli group under conjugation. Pure stabilizer states are obtained by acting with Clifford operations on computational-basis states, while mixed stabilizer states correspond to convex combinations of pure stabilizer states~\cite{Gottesman:1998hu,Chitambar:2018rnj }.

To quantify the amount of magic in the produced density matrix, we employ the stabilizer R\'enyi entropy of order two (SRE$_2$) introduced in Ref.~\cite{Leone:2021rzd}. For mixed states, the quantity used below corresponds to the mixed-state extension of the stabilizer 2-R\'enyi entropy and should therefore be regarded as a diagnostic of non-stabilizerness rather than a fully faithful magic monotone. For a two-qubit density matrix $\rho$, the mixed-state stabilizer 2-R\'enyi entropy is defined as
\begin{equation}
M_2(\rho)
=
-\log_2
\left[
\frac{
\sum_{P\in {\cal P}_2}
\left(
\mathrm{Tr}[\rho P]
\right)^4
}{
\sum_{P\in {\cal P}_2}
\left(
\mathrm{Tr}[\rho P]
\right)^2
}
\right].
\label{eq:M2mixed}
\end{equation}
Using the Fano--Bloch decomposition introduced in Eq.~\eqref{eq:rho2_qbit}, the stabilizer 2-R\'enyi entropy can be written directly in terms of the Fano--Bloch coefficients as
\begin{equation}
M_2(\rho)
=
-\log_2
\left[
\frac{
1+\sum_i B_i^4+\sum_j {\bar B}_j^4+\sum_{ij}C_{ij}^4
}{
1+\sum_i B_i^2+\sum_j {\bar B}_j^2+\sum_{ij}C_{ij}^2
}
\right].
\label{eq:magicnumber}
\end{equation}
The stabilizer 2-R\'enyi entropy satisfies $M_2(\rho)\geq0$, with equality holding for stabilizer states. For two-qubit systems, the maximal value is given by~\cite{Liu:2025frx}
\begin{equation}
M_2^{\mathrm{max}}
=
-\log_2\frac{7}{16}
\simeq
1.1926.
\end{equation}

Unlike concurrence and quantum discord, the stabilizer 2-R\'enyi entropy is not invariant under generic local unitary transformations. Consequently, its numerical value depends on the choice of spin quantisation axes used to define the density matrix. In the present work, all magic observables are evaluated in the spin basis introduced in Sec.~\ref{sec:general_decomposition}.

The quantity $M_2(\rho)$ provides information complementary to both concurrence and discord. While concurrence probes entanglement and discord measures more general quantum correlations, the stabilizer 2-R\'enyi entropy characterises the non-stabilizer structure of the density matrix and therefore probes genuinely non-classical resources beyond standard correlation measures.


\section{$t \bar t $ production channels}
\label{sec:production_channels}

In this section we derive the production density matrices for the \(t\bar t\) system in the benchmark channels relevant to present and future colliders. Building on the general decomposition introduced previously, our goal is to make explicit how the dynamics of each production mechanism is reflected in the Bloch vectors and spin-correlation matrix, and to identify the entries that can carry CP-odd information. We start by considering the decay of a heavy scalar, as it is the simplest example and already provides several useful insights. In the SM, the Higgs boson cannot decay into an on-shell $t\bt$ pair. However, $h\to \tau^+\tau^-$ is an important decay channel and it can be analysed in exactly the same way, see e.g.,~\cite{Altakach:2022ywa,Fabbrichesi:2022ovb}. We then consider  \(e^+e^-\to t\bar t\), \(\gamma\gamma\to t\bar t\),  \(q\bar q\to t\bar t\), \(gg\to t\bar t\),  presenting the results in a unified framework that will be used later to assess the sensitivity of quantum observables to CP violation.
\subsection{$S \rightarrow t\bar{t}$} \label{sect: Stt production}

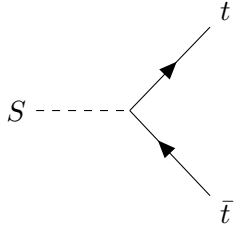
\begin{figure}[t]
    \centering
    \scalebox{1}{
    \begin{tikzpicture}
\begin{feynman}[medium]
\vertex (a) {$S$};
\vertex [right=1.5 cm of a] (b);
\vertex [above right= of b] (f1) {$t$};
\vertex [below right=of b] (c) {$\bar{t}$};
\diagram* {
(a) -- [scalar] (b) -- [fermion] (f1),
(b) -- [anti fermion] (c),
};
\end{feynman}
\end{tikzpicture}}
\caption{Feynman diagram describing the decay of a scalar particle to a $t\bar t$ pair.}
    \label{fig: stt}
\end{figure}

We consider the production of a $t\bt$ pair from the decay of a heavy spin-zero particle $S$ of mass $M_S$ as shown in Fig.~\ref{fig: stt}. The interaction is parametrised as in Ref.~\cite{Altakach:2022ywa}, allowing for a generic admixture of scalar and pseudoscalar couplings, weighted by trigonometric functions of a mixing angle $\delta\in[-\pi/2,\pi/2]$:

\be
    \mathcal{L}_{St\bt}= -k\,S\,\bar{t}\,(\cos\delta+ i \gamma^5\sin\delta)\,t.
    \label{eq: Lagrangian Stt}
\ee
We assume that both the top and antitop quarks are produced on shell, imposing $p_t^2= \bp_t^2= m_{t}^2$, where $m_t$ denotes the top-quark mass. Consequently, the centre-of-mass energy must satisfy $\sqrt{s}=M_S>2m_{t}$.

Applying the BM formulae of App.~\ref{app: BM formulae}, with spin bases
$\{s^a\}$ and $\{\bar s^b\}$ for the $t$ and $\bt$, respectively, the entries of
the production matrix can be expressed in terms of scalar products and Levi-Civita contractions.\footnote{The
relevant scalar products are $p_t\cdot \bp_t$, $p_t\cdot \bs^b$,
$\bp_t\cdot s^a$, and $s^a\cdot\bs^b$.
The only non-vanishing contractions are
$\epsilon(p_t,\bp_t,s^a,\bs^b)=
\epsilon_{\mu\nu\rho\sigma}p_t^\mu \bp_t^\nu s^{a\rho}\bs^{b\sigma}$.
We use the convention $\epsilon^{0123}=1$, as in {\sc FeynCalc}
\cite{Mertig:1990an}.}

In the conventional Lorentz-invariant description, the CP-violating
spin-correlation terms are proportional to the pseudoscalar contraction
$\epsilon(p_t,\bar p_t,s^a,\bar s^b)$. Their coefficient scales as
$\sin(2\delta)$ and therefore vanishes for
$\delta=0,\pm\pi/2$, corresponding respectively to purely scalar or purely
pseudoscalar interactions. CP violation in production thus requires a genuine
scalar-pseudoscalar admixture. Its magnitude is controlled by the mixing
angle $\delta$ and is maximal for $\delta=\pm\pi/4$, where the scalar and
pseudoscalar couplings have equal absolute weight.

To reconstruct the $t\bt$ spin state, namely to build the production matrix and
extract the spin polarisations and correlations, we choose the rest frame of the
scalar resonance, which coincides with the $t\bt$ zero-momentum frame:
$p_S=(M_S,0,0,0)$. In this frame the top and antitop are back-to-back. Since the
decay of a spin-zero particle defines no preferred spatial direction other than
the top momentum, we align $\vec p_t$ with the $\hat k$ axis, while the two
transverse axes can be chosen arbitrarily.\footnote{An alternative choice is
used in Ref.~\cite{Altakach:2022ywa}, where the direction of motion of $S$
selects a preferred spatial axis.} The momenta and spin four-vectors are then
those of Eq.~\ref{eq:kinematics_eett}, with the replacement
$p_1+p_2\to p_S=(M_S,0,0,0)$.

With these conventions, the spin density matrix $R^S$ is constructed according
to Eq.~\eqref{eq:Rmatrix}. Its trace is
\begin{equation}
\mathrm{Tr}[R^S]
=
4\tilde A^S
=
2 k^2 M_S^2
\left(
\beta^2\cos^2\delta+\sin^2\delta
\right).
\end{equation}
The corresponding normalised production density matrix is 
 \be
\rho^{S}
= \frac{1}{2}\begin{pmatrix} 0 & 0 & 0 & 0 \\
      0 & 1 &e^{-i\xi} & 0\\
      0 & e^{i\xi} & 1 & 0\\
      0& 0 & 0 & 0
    \end{pmatrix}, \qquad \begin{cases}
        \xi= \xi{(\delta, \beta)}= 2\tan^{-1}\left[\frac{\tan\delta}{\beta}\right]\\\\
        e^{i\xi(\delta, \beta)}= \frac{\beta \cos\delta+i \sin\delta}{\beta \cos\delta-i  \sin\delta} 
    \end{cases}\,.
    \label{eq: production density matrix stt}
\ee
The structure of $\rho^S$ describes the following configurations for the two particles: 
\be 
\mathbf B= \mathbf{\bar{B}}=0
\ee
\be
\mathbf{C}=\begin{pmatrix}
        &\cos\xi & \purple{-\sin\xi}& 0\\
        & \purple{\sin\xi} & \cos\xi&0 \\
        & 0&0 &-1 \\
    \end{pmatrix}=\begin{pmatrix}
        &\frac{\beta^2 \cos^2\delta - \sin^2\delta}{\beta^2 \cos^2\delta + \sin^2\delta} & \frac{ -\beta \sin2\delta}{\beta^2 \cos^2\delta + \sin^2\delta}& 0\\
        & \frac{ \beta \sin2\delta}{\beta^2 \cos^2\delta + \sin^2\delta} & \frac{\beta^2 \cos^2\delta - \sin^2\delta}{\beta^2 \cos^2\delta + \sin^2\delta}&0 \\
        & 0&0 &-1 \\
    \end{pmatrix}, 
\label{eqs: B, b' and C dor S-tt}
\ee
where the antisymmetric entries are highlighted in purple.
The normalised density matrix in Eq.~\eqref{eq: production density matrix stt}
describes the pure state
\begin{equation}
\ket{\psi}_S
=
\frac{
\ket{+-}
+
e^{i\xi(\delta,\beta)}
\ket{-+}
}{
\sqrt2
}.
\label{eq:scalarstate}
\end{equation}
The relative phase $\xi(\delta,\beta)$ depends on both the CP-mixing angle
$\delta$ and the kinematic parameter $\beta$. However, it can be removed by a
local unitary transformation acting separately on the top and antitop spin
subspaces. Hence, $\ket{\psi}_S$ is locally equivalent to one of the Bell states
\begin{equation}
\ket{\Psi^\pm}
=
\frac{
\ket{+-}\pm\ket{-+}
}{
\sqrt2
}.
\end{equation}
It follows that a $t\bar t$ pair produced in the decay of a heavy spin-zero
resonance is always in a pure maximally entangled state, independently of
$\delta$ and $\beta$. Consequently, all quantum observables invariant under
local unitary transformations, such as concurrence and discord, are insensitive
to the CP-mixing angle and attain their maximal values for any choice of
$\delta$ and $\beta$. 

In this scenario the single-spin polarisation vectors vanish identically,
$\mathbf B=\mathbf{\bar{B}}=0$, as expected for the isotropic decay of a spin-zero
resonance. Consequently, the observable built from the polarisation asymmetry,
\begin{equation}
\Delta\mathbf{B}=0,
\end{equation}
also vanishes identically. The CP structure of the scalar coupling can therefore
be probed only through observables depending on the spin-correlation matrix:
the norm of its antisymmetric component, the CP-sensitive trace distance, and magic.

\begin{figure}[t]
\centering
\includegraphics[width=0.49\textwidth]{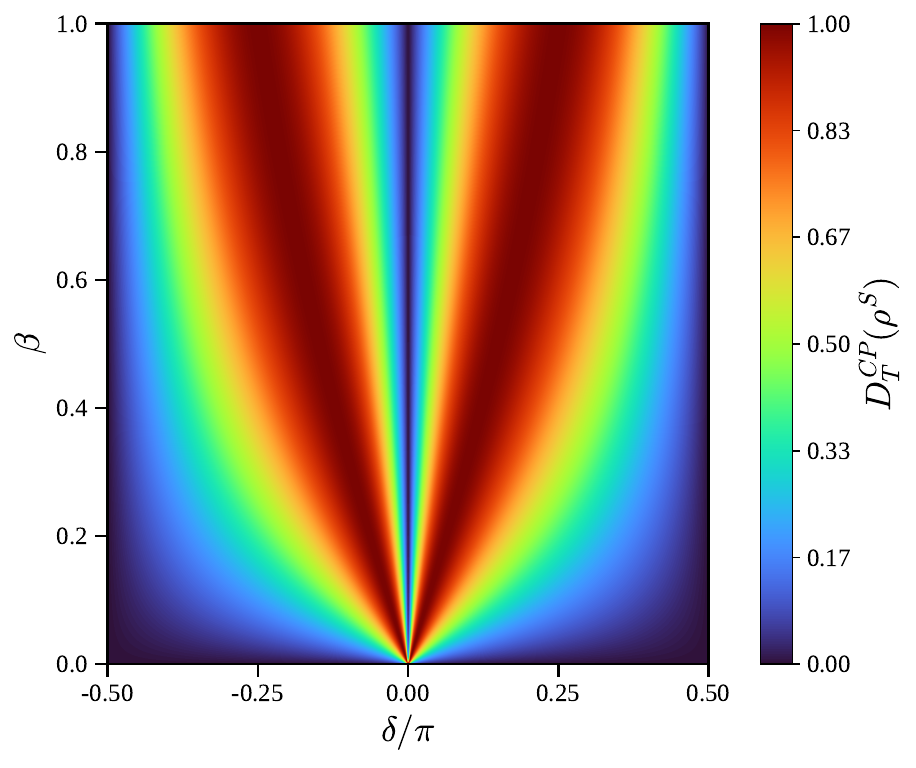}
\includegraphics[width=0.49\textwidth]{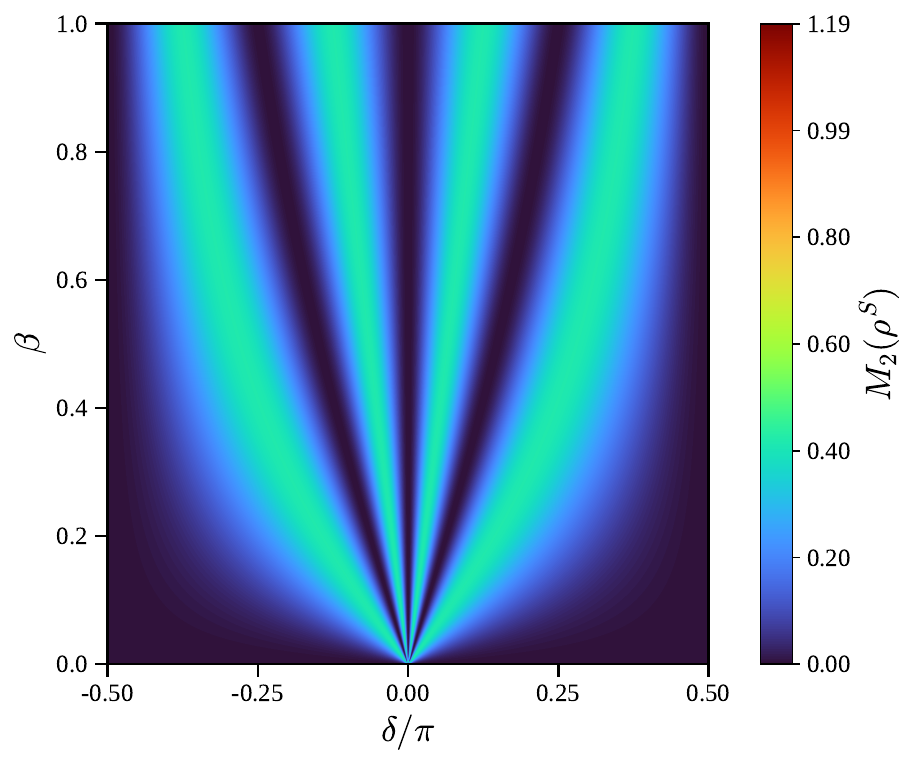}
\caption{
CP-sensitive observables as functions of the CP-mixing angle $\delta$ and the
top-quark velocity $\beta$. Left: CP-sensitive trace distance, equivalently the
norm of the antisymmetric component of the spin-correlation matrix.
Right: Magic.
}
\label{fig: obs stt senistive to delta}
\end{figure}

The first such observable is the norm of the antisymmetric part of the
spin-correlation matrix. In the present case, it coincides exactly with the
CP-sensitive trace distance defined in Eq.~\eqref{eq:traceCP},
\begin{equation}
\|\mathbf{a}^S\|
=
D_T^{CP}(\rho^S)
=
|\sin\xi|\, .
\label{eq:scalarCPobs}
\end{equation}
Its dependence on the CP-mixing angle $\delta$ and on the top-quark velocity
$\beta$ is shown in the left panel of
Fig.~\ref{fig: obs stt senistive to delta}. As expected, the
CP-sensitive trace distance vanishes in the CP-conserving limits
$\delta=0,\pm\pi/2$, corresponding to purely scalar or purely pseudoscalar
interactions. Its size is controlled by the scalar-pseudoscalar interference
and reaches its maximum when $|\sin\xi|=1$, namely for $\xi=\pm\pi/2$. Since
$\xi=\xi(\delta,\beta)$, the corresponding values of $\delta$ depend on the
top-quark velocity. In the relativistic limit, $\beta=1$, the maximum occurs at
$\delta=\pm\pi/4$, where the scalar and pseudoscalar couplings have equal
absolute weight.

The second observable with a non-trivial dependence on the CP phase is magic,
in agreement with Ref.~\cite{Altakach:2026fpl}. In this case, the sensitivity
to $\delta$ originates from the phase structure of the spin-correlation matrix
and survives despite the state being locally equivalent to a Bell state:
\begin{equation}
   M_2(\rho^S)=
   -\log_2
   \left(1-
   \frac{\sin^2 2 \xi }{4}
   \right).
   \label{eq:scalarmagic}
\end{equation}

The right panel of
Fig.~\ref{fig: obs stt senistive to delta} displays magic $M_2(\rho^S)$. Unlike the trace distance,
magic probes the stabilizer structure of the quantum state and therefore
shows a qualitatively different behaviour across the parameter space. In the
CP-conserving regions the density matrix corresponds to a stabilizer state, and
magic vanishes. As the scalar-pseudoscalar admixture is turned on, the
state generally develops non-stabilizer features, leading to a non-zero value of
$M_2(\rho^S)$. However, for specific combinations of $\delta$ and $\beta$, the
magic returns to zero even when the CP-sensitive trace distance is maximal. In
particular, for $\beta=1$, the maximally CP-violating configurations
$\delta=\pm\pi/4$ again correspond to stabilizer states. Equivalently, the
magic vanishes whenever the relative phase takes one of the stabilizer values
$\xi=0,\pm\pi/2,\pm\pi$, which gives rise to the five zero-magic regions
visible in the right panel of Fig.~\ref{fig: obs stt senistive to delta}.

This behaviour shows that CP violation and non-stabilizerness probe distinct
features of the $t\bar t$ quantum state. The antisymmetric correlation norm, or
equivalently the CP-sensitive trace distance, tracks the size of the CP-violating
component of the density matrix. By contrast, magic is sensitive to whether
the state belongs to, or departs from, the stabilizer set. Hence, maximal
CP-odd/even interference does not necessarily imply maximal non-stabilizerness,
even for a pure maximally entangled state. The two observables therefore provide
complementary information: the trace distance identifies the regions where the
CP-violating interference is largest, while magic maps the transition between
stabilizer and non-stabilizer regimes. Taken together, they yield a more refined
characterisation of the CP structure of the produced $t\bar t$ system.

Before concluding this section, let us comment on the physical interpretation of
the interaction in Eq.~\eqref{eq: Lagrangian Stt}. An on-shell SM Higgs boson
cannot decay into a pair of on-shell top quarks because $m_h<2m_t$. Therefore,
Eq.~\eqref{eq: Lagrangian Stt} should be understood as a simplified theoretical
framework, rather than as a process realised within the SM Higgs sector.

The same spin structure can arise whenever a fermion-antifermion pair is produced through the $s$-channel exchange of an intermediate scalar boson, such as the Higgs, irrespective of whether the initial state consists of fermions or vector bosons. In the broken
electroweak phase, the effective Higgs-top interaction can be written as
\begin{equation}
\mathcal{L}_{ht\bar t}
=
h\,\bar t
\left(
a_h+i\gamma^5 b_h
\right)t,
\label{eq:htt}
\end{equation}
where the real coefficients $a_h$ and $b_h$ receive contributions from the SM
Yukawa interaction and from the dimension-six operator $O_{t\phi}$:
\begin{equation}
a_h
=
-\frac{y_t}{\sqrt2}
+
\frac{3v^2}{2\sqrt2\Lambda^2}
\,\operatorname{Re}(C_{t\phi}),
\qquad
b_h
=
\frac{3v^2}{2\sqrt2\Lambda^2}
\,\operatorname{Im}(C_{t\phi}) .
\end{equation}
At tree level, no additional Lorentz structures appear in the Higgs-fermion
interaction. Hence, the results obtained for the simplified coupling in
Eq.~\eqref{eq: Lagrangian Stt} can be directly extended to any scalar
interaction of the form given in Eq.~\eqref{eq:htt}.

In the present work, however, we do not include diagrams with an $s$-channel off-shell Higgs
exchange in $e^+e^-$ or light-quark scattering. In the limit
$m_e=m_q=0$, the Higgs coupling to the initial-state fermions vanishes, making
such contributions negligible. 

Finally, we stress again that although the decay of an on-shell SM Higgs boson into a $t\bar t$ pair
is kinematically forbidden, the framework discussed in this section applies
directly to any kinematically allowed scalar resonance decay into fermion pairs, such as
$h\to\tau^+\tau^-$.

\subsection{$e^+e^- \rightarrow t\bar{t}$}\label{sec:epluseminus}

\begin{figure}[t]
    \centering
    \scalebox{1}{
    \begin{tikzpicture}
    \begin{feynman}[small]
    \vertex (e1) at (0,1) {$e^-$};
    \vertex (e2) at (0,-1) {$e^+$};
    \vertex (v1) at (1,0);
    \vertex (v2) at (2,0);
    \vertex (t1) at (3,1) {$t$};
    \vertex (t2) at (3,-1) {$\bar t$};

    \diagram* {
        (e1) -- [fermion] (v1),
        (e2) -- [anti fermion] (v1),
        (v1) -- [boson] (v2),
        (v2) -- [fermion] (t1),
        (v2) -- [anti fermion] (t2),
    };
    \node[fill=black, draw=black, circle, minimum size=7pt, inner sep=0pt] at (v2) {};
    \end{feynman}
    \end{tikzpicture}}
\caption{Dominant $s$-channel tree-level diagram for $e^+e^- \to t\bar t$. The intermediate neutral vector boson represents either $\gamma$ or $Z$ exchange. The black dot denotes the $t\bar t\gamma/t\bar tZ$ vertices, where anomalous dipole interactions or other new physics contributions may enter.}
    \label{fig: ee-tt scattering}
\end{figure}
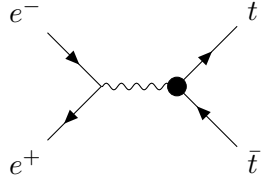

We now turn to the process $e^+e^- \to t\bar t$, which provides the simplest
$2\to2$ scattering benchmark for top-pair production. At tree level in the SM,
the dominant contribution proceeds through $s$-channel $\gamma$ and
$Z$ exchange as shown in Fig.~\ref{fig: ee-tt scattering}. Compared with the scalar decay $S\to t\bar t$, the structure
of the spin density matrix is richer: the produced state depends non-trivially
on the scattering angle, on the top-quark velocity, and on the interference
between the photon- and $Z$-mediated amplitudes. This channel is therefore particularly useful for illustrating how
CP-violating dipole interactions can generate both a difference between the top
and antitop polarisation vectors, $\Delta\mathbf B$, and an antisymmetric
component of the spin-correlation matrix, $\mathbf C^A$, once the kinematic
conventions and spin basis of Sec.~\ref{sec:general_decomposition} and
App.~\ref{app:analyticalexpressions} are adopted.

Quantum observables in $e^+e^-\to t\bar t$ have previously been studied with
polarised lepton beams using a general Lorentz-structure parametrisation of
four-fermion contact interactions~\cite{Altakach:2026fpl}. That analysis does
not include the momentum-dependent electroweak dipole modifications of the
$t\bar t\gamma$ and $t\bar t Z$ vertices considered here. SMEFT effects in
$e^+e^-\to t\bar t$ were also investigated in
Ref.~\cite{Maltoni:2024csn}, including electroweak dipole operators. However,
the dipole Wilson coefficients were restricted to be real, so that only the
CP-conserving dipole contributions were considered. The present analysis
extends these studies by allowing complex dipole coefficients and by examining
the resulting CP-violating components of the spin density matrix.

In the present work we adopt the
SMEFT description of the $t\bar t\gamma$ and $t\bar t Z$ vertices, including
the explicit momentum-dependent electroweak dipole structures proportional to
$\sigma^{\mu\nu}q_\nu$. Defining $q^\nu$ as the momentum carried by the intermediate neutral vector boson exchanged in the $s$ channel, we parametrise the effective $t\bar t V$ vertices, with $V=\gamma,Z$, as
\begin{align}
    Zt\bar{t}&:\quad ie\gamma^\mu \left(\bar{g}_{Vt}-\bar{g}_{At}\gamma^5\right)
    -\frac{v}{\Lambda^2}\sigma^{\mu\nu}q_{\nu}\left(d_ZP_R+ d_Z^*P_L\right)\,, \label{eq: ew dipoles1}
    \\
    \gamma t\bar{t}&:\quad ieQ_t\gamma^\mu
    -\frac{v}{\Lambda^2}\sigma^{\mu\nu}q_{\nu}\left(d_{\gamma}P_R+ d_{\gamma}^*P_L\right)\,.
    \label{eq: ew dipoles2}
\end{align}
We allow for SMEFT corrections only in the final-state $t\bar t V$ vertices, while the initial-state $e^+e^-V$ couplings are kept SM-like:
\begin{align}
Ze^+e^- &:\quad
ie\gamma^\mu \left(\bar{g}_{Ve}-\bar{g}_{Ae}\gamma^5\right),
\qquad
\gamma e^+e^- :\quad ieQ_e\gamma^\mu .
\label{eq:vertices-eeV}
\end{align}
The couplings entering Eqs.~\eqref{eq: ew dipoles1}-\eqref{eq: ew dipoles2} can be expressed in terms of the SMEFT Wilson coefficients of the operators in Tables~\ref{table: CP-odd operators} and \ref{table: CP-even operators}. The mapping is shown in Table~\ref{tab:eett_couplings}. The CP-even operators induce shifts in the SM vector and axial-vector $Zt\bar t$ couplings, while the electroweak dipole operators generate the momentum-dependent structures proportional to $\sigma^{\mu\nu}q_\nu$. The imaginary parts of $d_\gamma$ and $d_Z$ correspond to CP-odd dipole interactions.
\begin{table}[t]
    \centering
    \renewcommand{\arraystretch}{1.7}
    \begin{tabular}{|c|c|}
    \hline
    Effective coefficient & SM limit \\
    \hline
       $\bar{g}_{Vf}=\frac{1}{s_W c_W}\left[g^{\rm SM}_{Vf}+\delta_{ft}\frac{v^2}{4\Lambda^2}\left(C^{3}_{\phi Q}-C^{1}_{\phi Q}-C^{}_{\phi t}\right)\right] $  & $g_{Vf}^{\rm SM}=\frac{T_3^f}{2}-Q_fs^2_W$ \\
       \hline
       $\bar{g}_{Af}=\frac{1}{s_W c_W}\left[g^{\rm SM}_{Af}+\delta_{ft}\frac{v^2}{4\Lambda^2}\left(C^{3}_{\phi Q}- C^{1}_{\phi Q}+C^{}_{\phi t
       }\right)\right] $  & $g_{Af}^{\rm SM}=\frac{T_3^f}{2}$\\
       \hline
       $d_{\gamma}= \sqrt{2}(c_W C_{tB}+ s_W C_{tW})$ & 0\\
       \hline
      $ d_Z= \sqrt{2}(-s_W C_{tB}+ c_W C_{tW})$ &0 \\
       \hline
    \end{tabular}
  \caption{
Effective couplings entering the $t\bar t\gamma$ and $t\bar t Z$ vertices.
We use the shorthand $s_W\equiv\sin\theta_W$ and
$c_W\equiv\cos\theta_W$, and define
$C_{tZ}\equiv -s_W C_{tB}+c_W C_{tW}$.
The factor $\delta_{ft}$ indicates that the SMEFT shifts in
$\bar g_{Vf}$ and $\bar g_{Af}$ are included only for $f=t$; hence the
electron couplings are kept at their SM values.}
\label{tab:eett_couplings}
\end{table}
The four-fermion operators $\mathcal O_{\ell equ}^{(1)}$ and
$\mathcal O_{\ell equ}^{(3)}$ can also contribute to $e^+e^-\to t\bar t$.
However, their interference with the tree-level SM amplitude vanishes at linear
order~\cite{Altakach:2026fpl}. A non-zero CP-violating contribution can arise
only from the interference between amplitudes induced by these two operators,
which enters at quadratic order in the EFT expansion. Since these effects first
appear at order $\Lambda^{-4}$, we do not include them in the following
analysis. To simplify the analytical expressions collected in App.~\ref{app: spin density matrix eett}, we use the decomposition\footnote{Using
\begin{equation}
    d_i P_R+d_i^\ast P_L
    =
    \mathrm{Re}(d_i)+i\gamma^5\mathrm{Im}(d_i),
    \qquad i=\gamma,Z,
\end{equation}
the real parts of $d_i$ correspond to CP-even magnetic-dipole-type
interactions, while the imaginary parts correspond to CP-odd electric-dipole-type
interactions.}
\begin{equation}
    d_\gamma = r_\gamma e^{i\eta_\gamma},
    \qquad
    d_Z = r_Z e^{i\eta_Z}.
    \label{eq:dipole_phase_param}
\end{equation}
Using the vertices in Eqs.~\eqref{eq: ew dipoles1}-\eqref{eq:vertices-eeV},
we construct the unnormalised production matrix $R^{ee}$ and define the
normalised spin density matrix as in Eq.~\eqref{eq: normalised matrices}.
The explicit entries of the Fano--Bloch coefficients of $R^{ee}$ are collected in
App.~\ref{app: spin density matrix eett}. In evaluating $R^{ee}$, we keep both the linear interference terms and the
quadratic contributions from the dimension-six operators,
\begin{equation}
    R^{ee}
    =
    R^{ee}_{\rm SM}
    +
    R^{ee}_{\rm int}
    +
    R^{ee}_{\rm quad},
    \qquad
    R^{ee}_{\rm int}=\mathcal O(\Lambda^{-2}),
    \qquad
    R^{ee}_{\rm quad}=\mathcal O(\Lambda^{-4}) .
    \label{eq:R_ee_expansion}
\end{equation}
The quadratic terms are formally of the same order as possible interference
terms involving dimension-eight operators and therefore should not be interpreted
as a complete EFT prediction at order $\Lambda^{-4}$. In the present analysis
they are retained in order to construct a positive semi-definite density matrix
for finite benchmark values of the Wilson coefficients, which is required for a
consistent definition of quantum observables such as concurrence, discord,
magic, and trace-distance measures.

Using the corresponding Fano--Bloch coefficients, the polarisation vectors
and spin-correlation matrix in the spin basis exhibit the following CP
structure:
\begin{align}
   \mathbf B+\mathbf{\bar{B}}
    &=
    \begin{pmatrix}
           \bullet\\
           \bullet\\
           \bullet
    \end{pmatrix},
    \qquad
    \Delta \mathbf B
    =
    \begin{pmatrix}
          \purple{\bullet} \\
          \purple{\bullet} \\
          \purple{\bullet}
    \end{pmatrix},
 \qquad
  \mathbf  C
    =
    \begin{pmatrix}
        \bullet & \bullet\,\purple{\bullet} & \bullet\,\purple{\bullet}\\
        \bullet\,\purple{\bullet} & \bullet & \bullet\,\purple{\bullet}\\
        \bullet\,\purple{\bullet} & \bullet\,\purple{\bullet} & \bullet
    \end{pmatrix}.
\end{align}
Here, black bullets denote components that may be non-zero in the CP-conserving limit, whereas purple bullets identify components generated by CP-violating interactions. In the spin-correlation matrix, the purple entries belong to its antisymmetric part, $\mathbf C^A$, while in the single-spin sector they contribute to the polarisation difference $\Delta\mathbf B$. These CP-violating contributions vanish for
$\operatorname{Im}(d_\gamma)=\operatorname{Im}(d_Z)=0$. It is worth noting that, unlike all the other production mechanisms considered in this work, $e^+e^-\to t\bar t$ can produce top and antitop quarks that are individually polarised, in addition to exhibiting non-trivial spin correlations. This makes it possible to investigate how CP-violating interactions modify the individual polarisation vectors and, consequently, the observables introduced in Sec.~\ref{sec:observables}. Non-vanishing single-spin polarisations may also provide additional information in a tomographic reconstruction of the $t\bar t$ spin state and enhance sensitivity to CP-violating effects in the subsequent top-quark decay vertices.

\begin{figure}[t]
\centering
\includegraphics[scale=0.335]{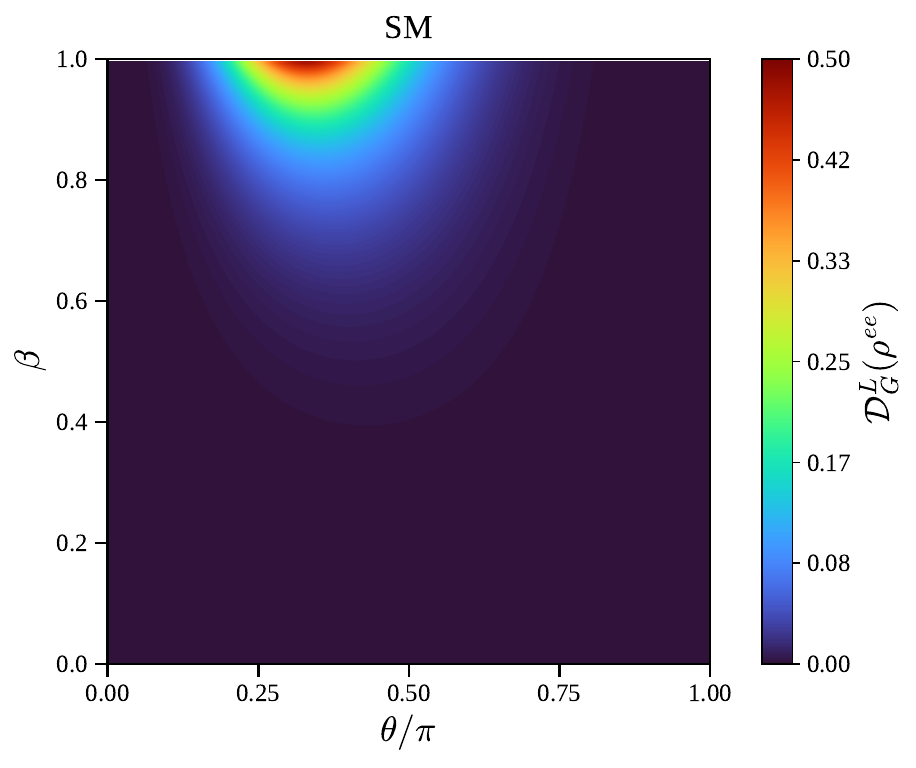}
\includegraphics[scale=0.335]{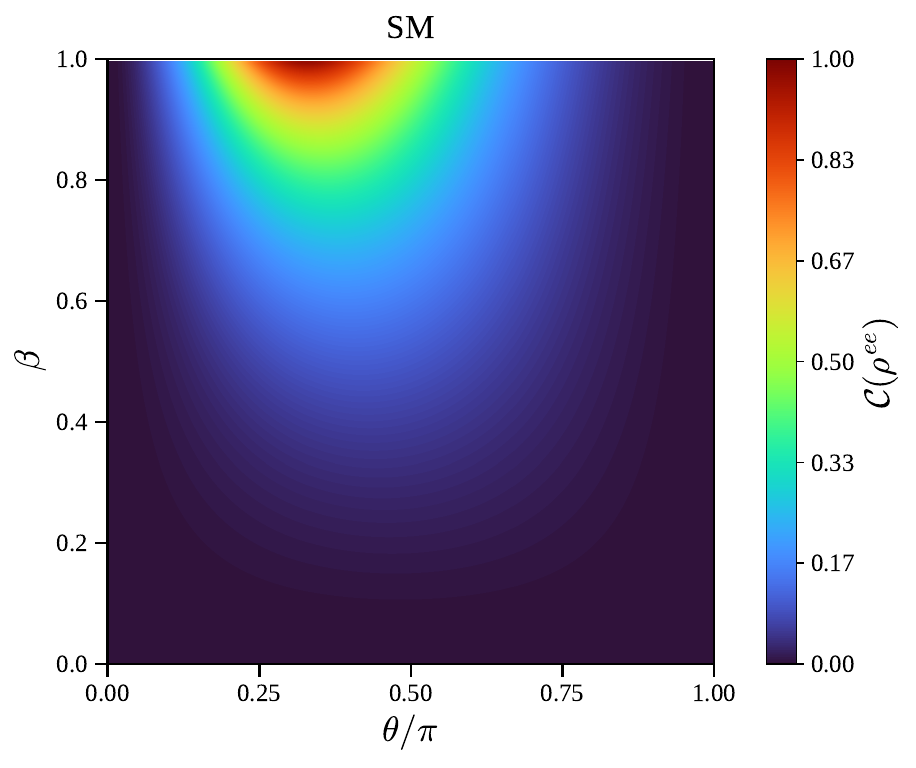}
\includegraphics[scale=0.335]{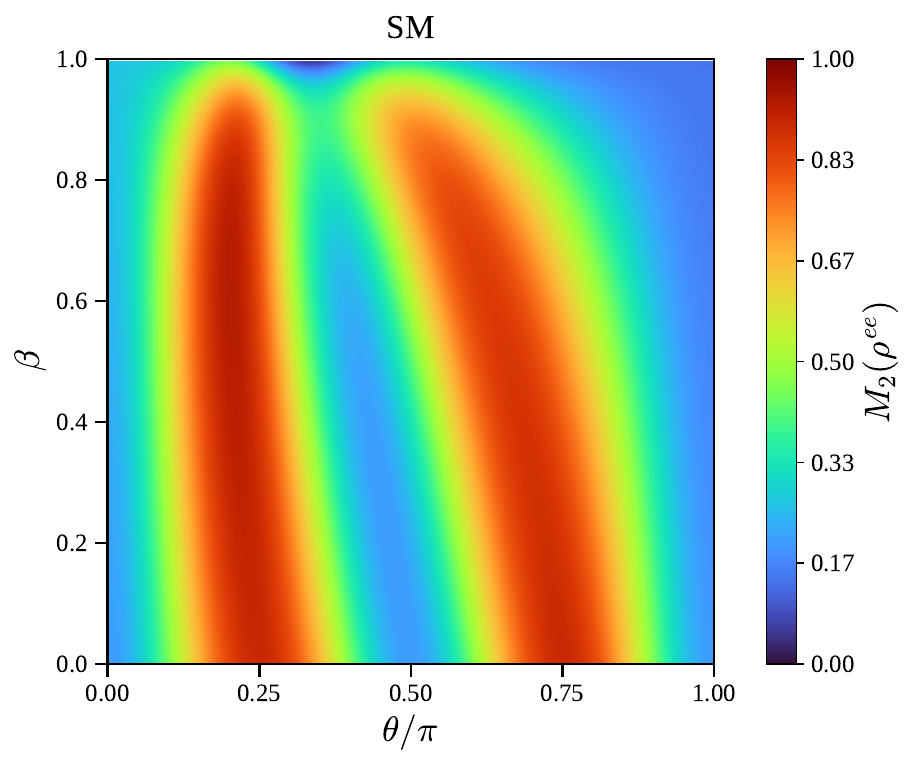}
\caption{
SM predictions for the quantum information observables as functions of the scattering angle $\theta/\pi$ and the top-quark velocity $\beta$. From left to right, we show geometric discord, concurrence, and magic.\label{fig:ee_SM_plots}
}
\end{figure}

\begin{figure}[t]
\centering
\includegraphics[scale=0.335]{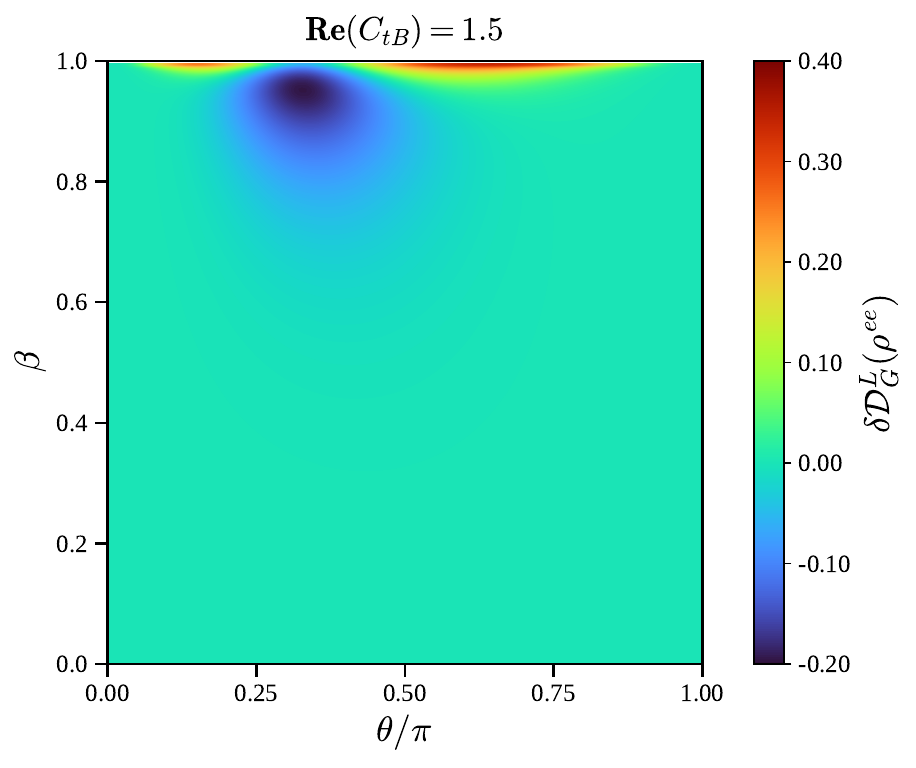}
\includegraphics[scale=0.335]{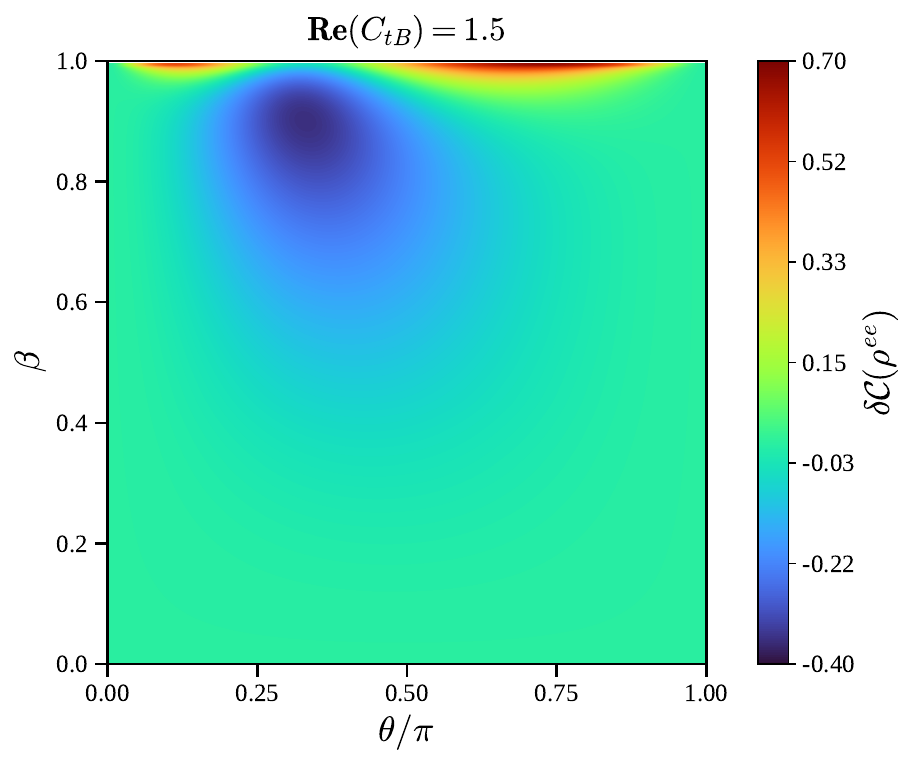}
\includegraphics[scale=0.335]{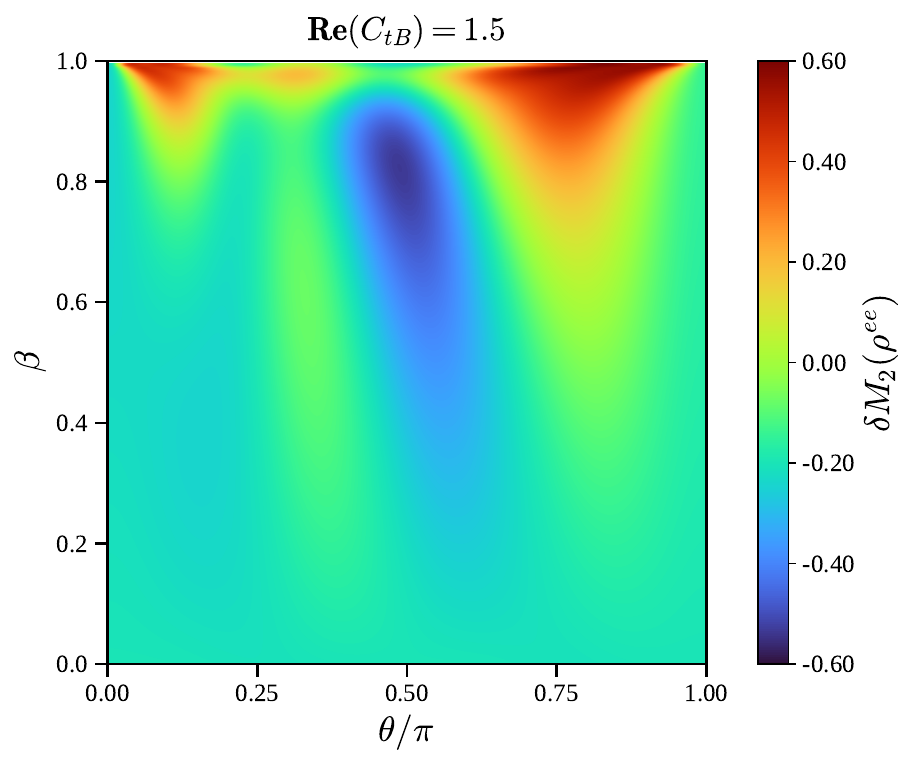}
\includegraphics[scale=0.335]{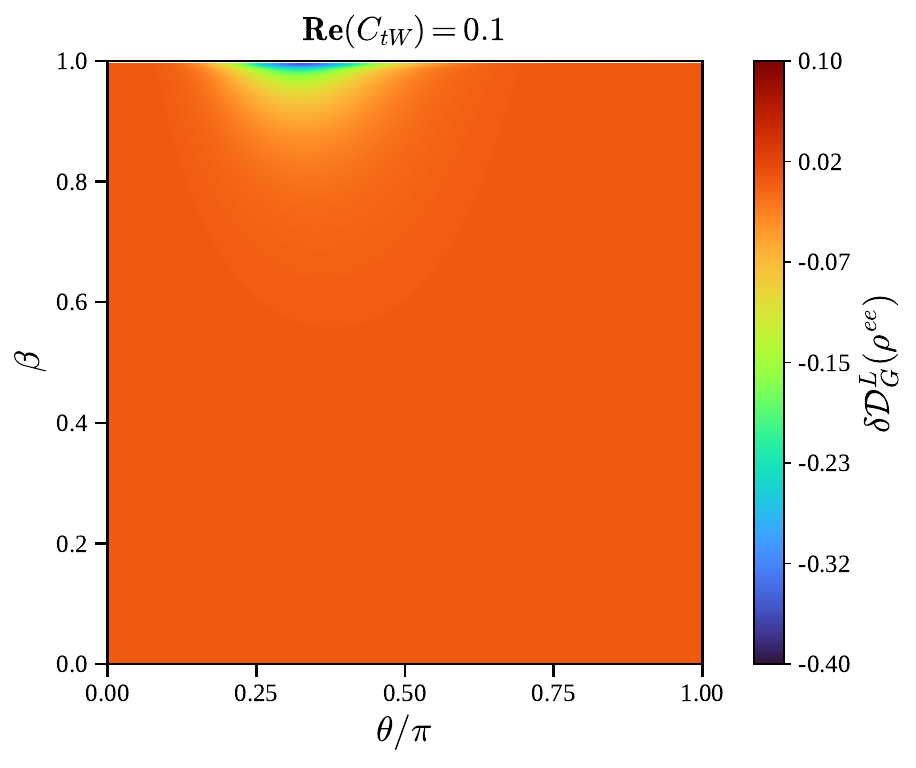}
\includegraphics[scale=0.335]{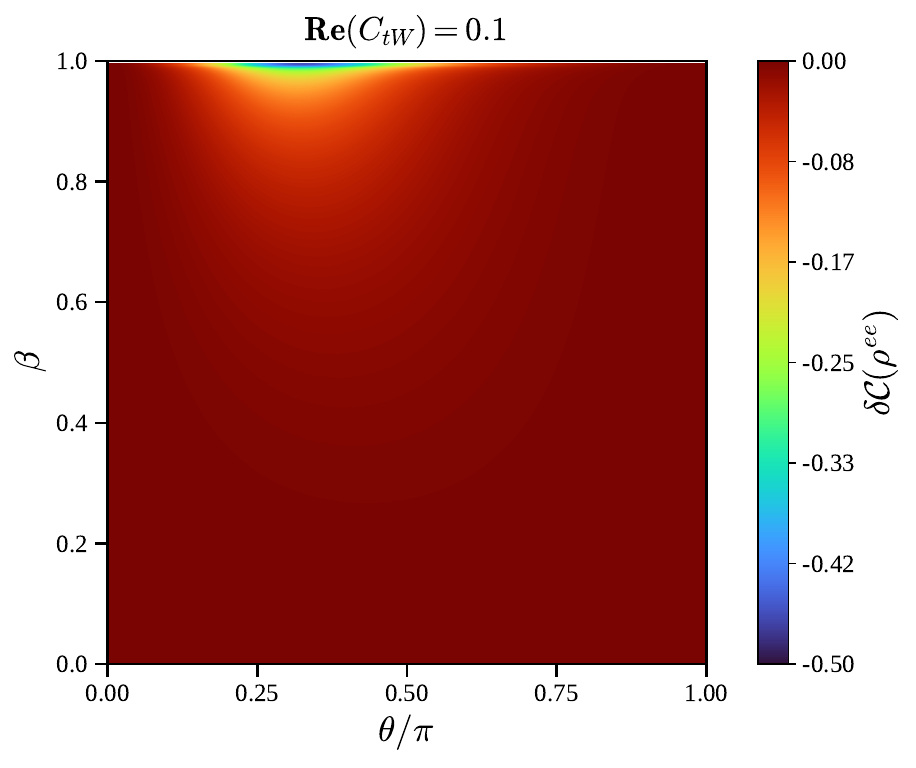}
\includegraphics[scale=0.335]{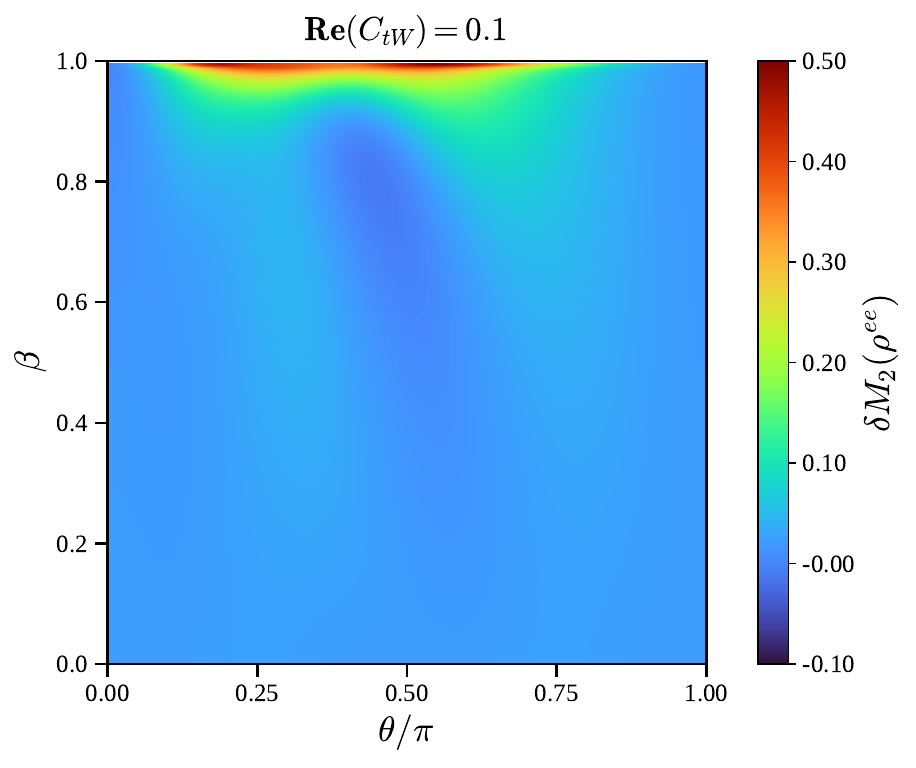}
\caption{
Changes in the quantum information observables induced by the CP-even dipole
coefficients $\operatorname{Re}(C_{tB})$ (upper row) and
$\operatorname{Re}(C_{tW})$ (lower row), with all other SMEFT coefficients
set to zero. From left to right, we show $\delta \mathcal{D}^L_G$, $\delta\mathcal C$, and
$\delta M_2$, where
$\delta\mathcal O\equiv
\mathcal O_{\mathrm{SMEFT}}-\mathcal O_{\mathrm{SM}}$.\label{fig:ee_ReCtB_ReCtW} }
\end{figure}

\begin{figure}[t]
\centering
\includegraphics[scale=0.335]{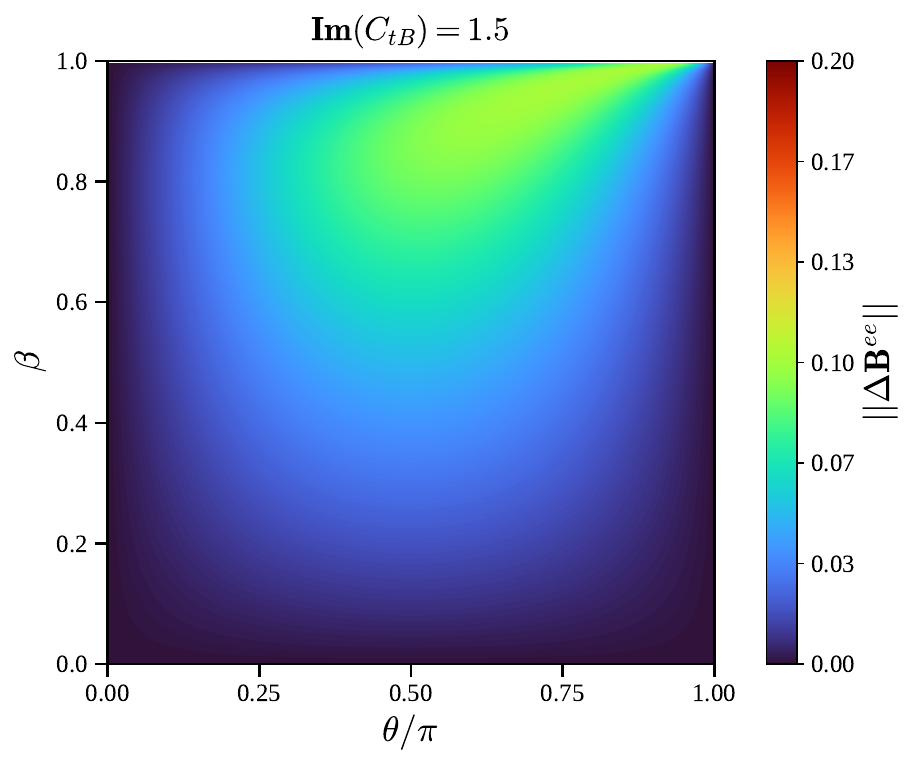}
\includegraphics[scale=0.335]{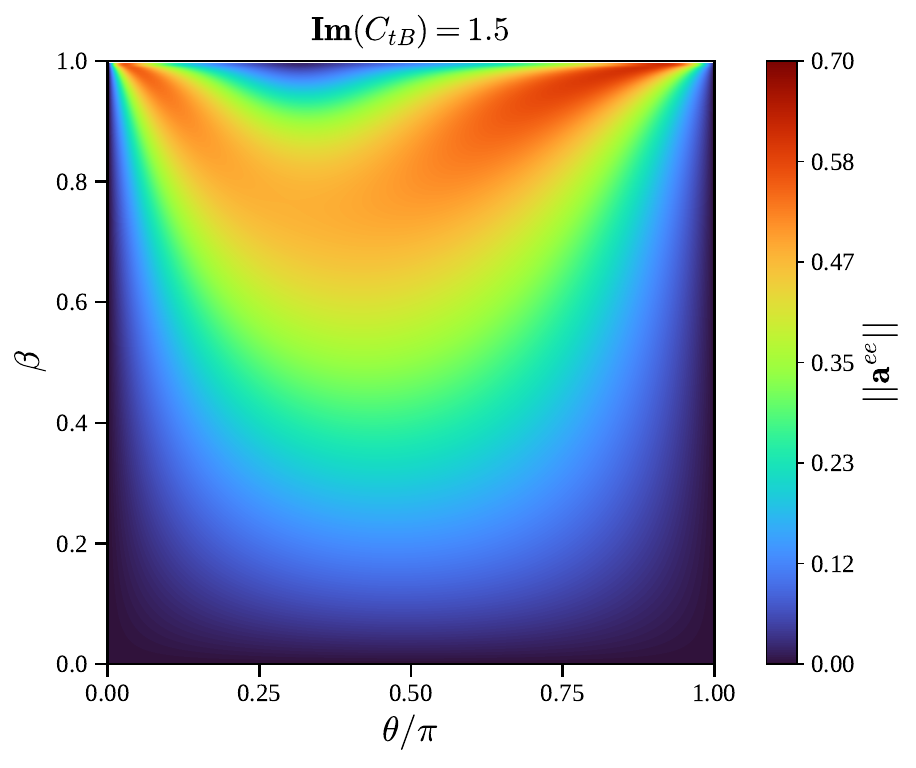}
\includegraphics[scale=0.335]{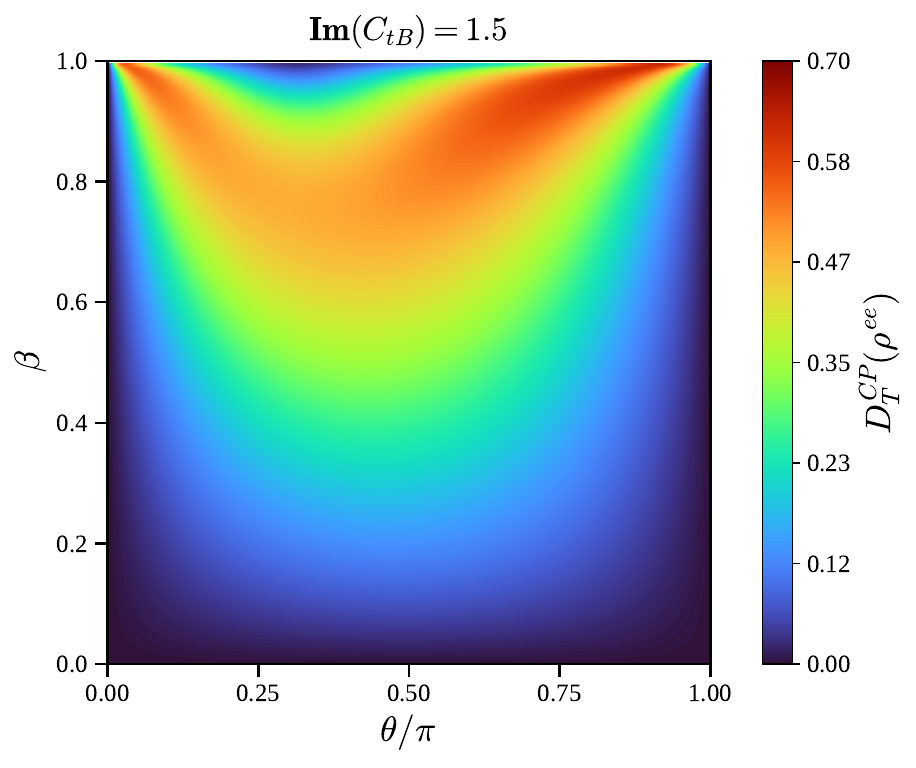}
\includegraphics[scale=0.335]{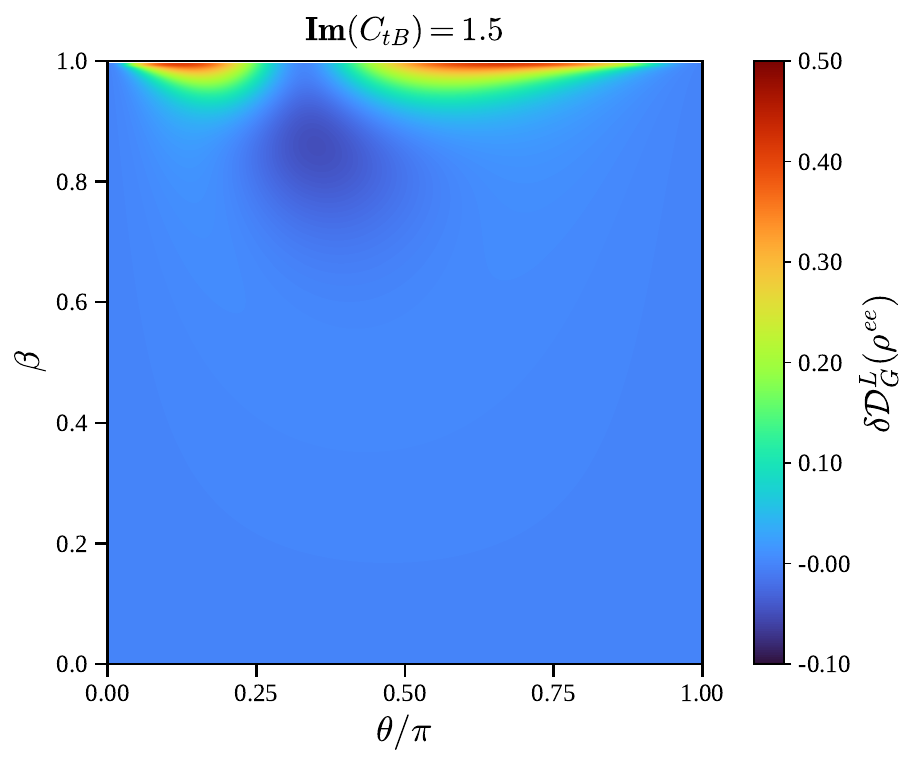}
\includegraphics[scale=0.335]{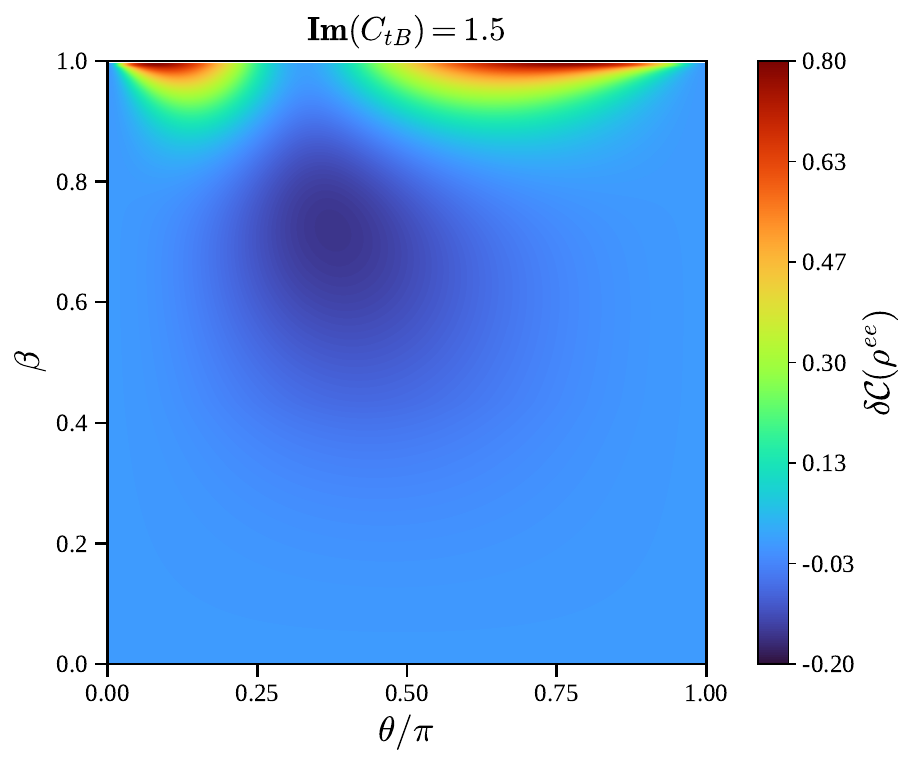}
\includegraphics[scale=0.335]{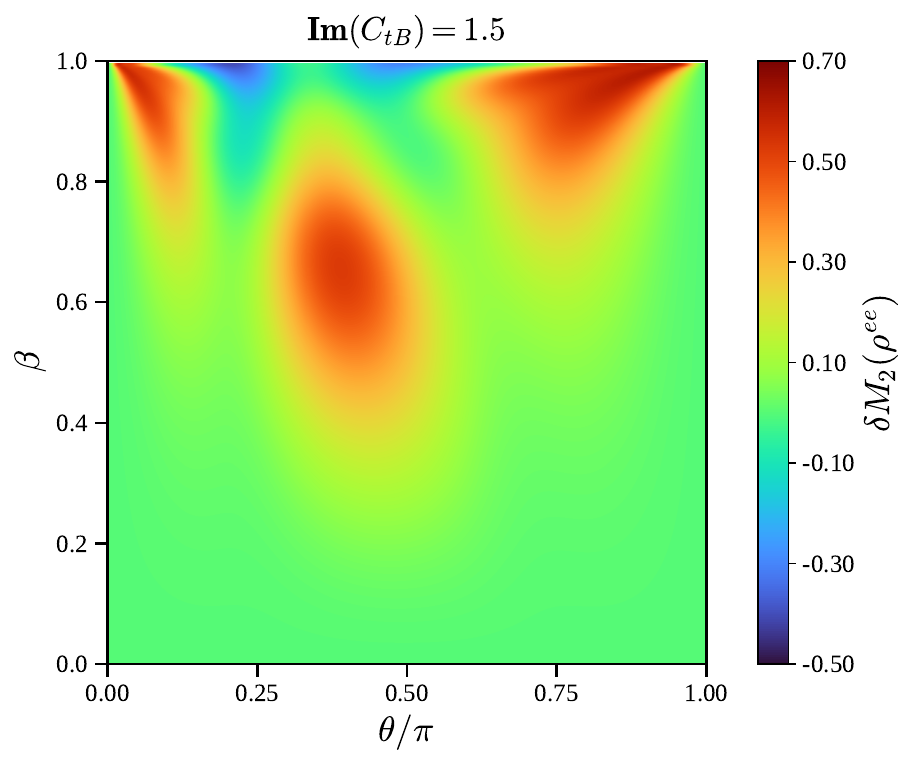}
\caption{The observables for the benchmark
$\operatorname{Im}(C_{tB})=1.5$, with all other SMEFT coefficients set to
zero. The upper panels show, from left to right, the norm of the top-antitop
polarisation difference, the norm of the
antisymmetric spin-correlation matrix, and the CP
trace distance. The lower panels show the
corresponding changes $\delta \mathcal D^L_G$, $\delta\mathcal C$, and
$\delta M_2$ relative to the SM predictions.\label{fig:ee_ImCtB}
}
\end{figure}

\begin{figure}[t!]
\centering
\includegraphics[scale=0.335]{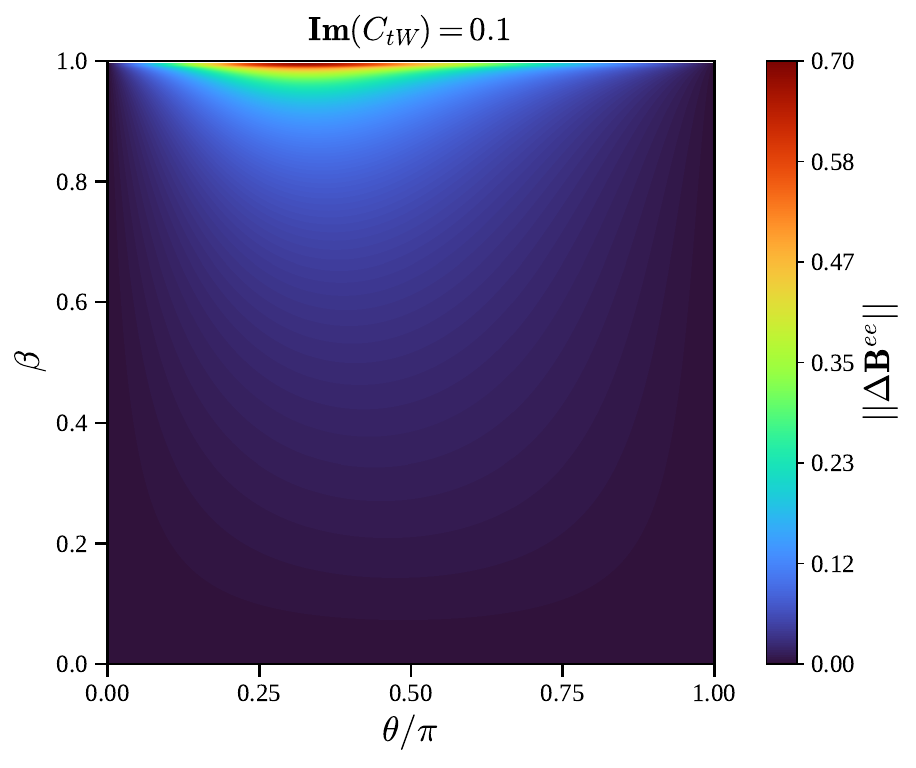}
\includegraphics[scale=0.335]{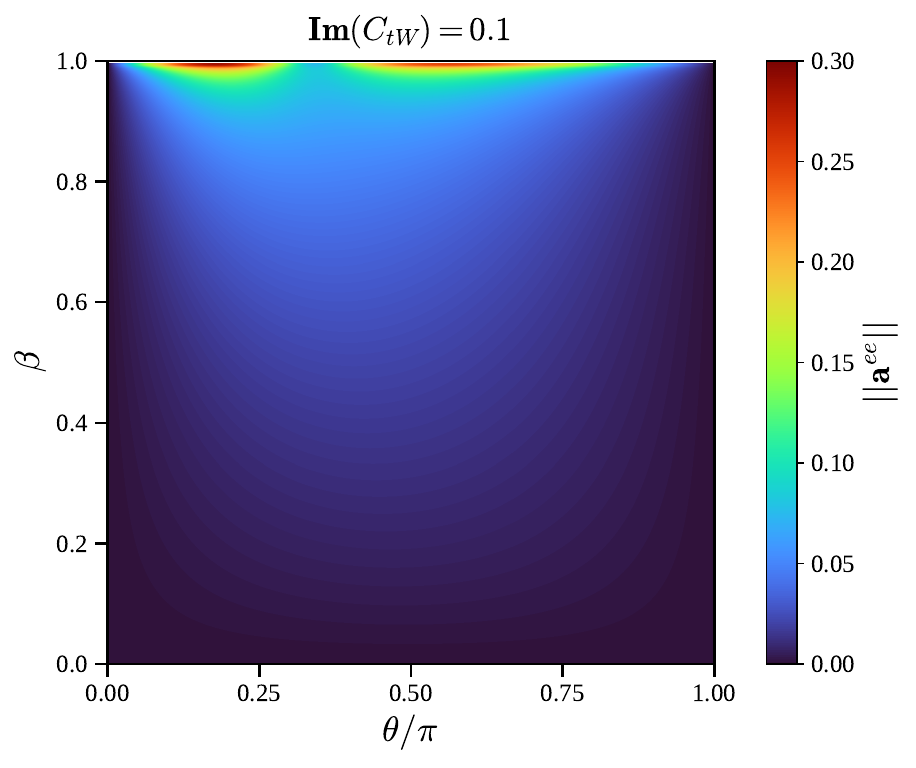}
\includegraphics[scale=0.335]{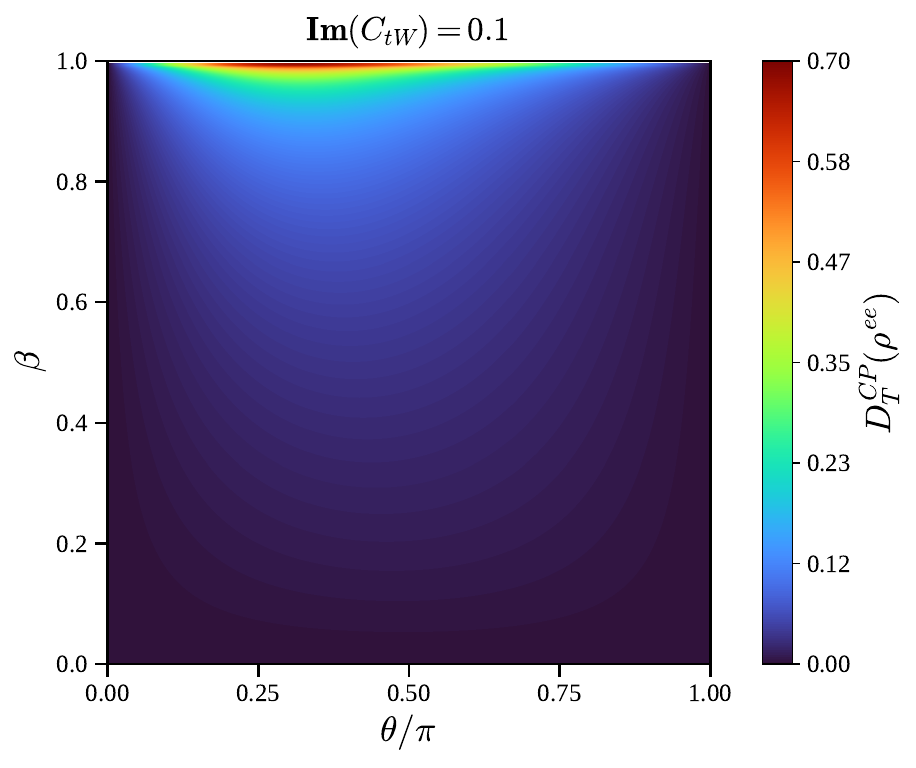}
\includegraphics[scale=0.335]{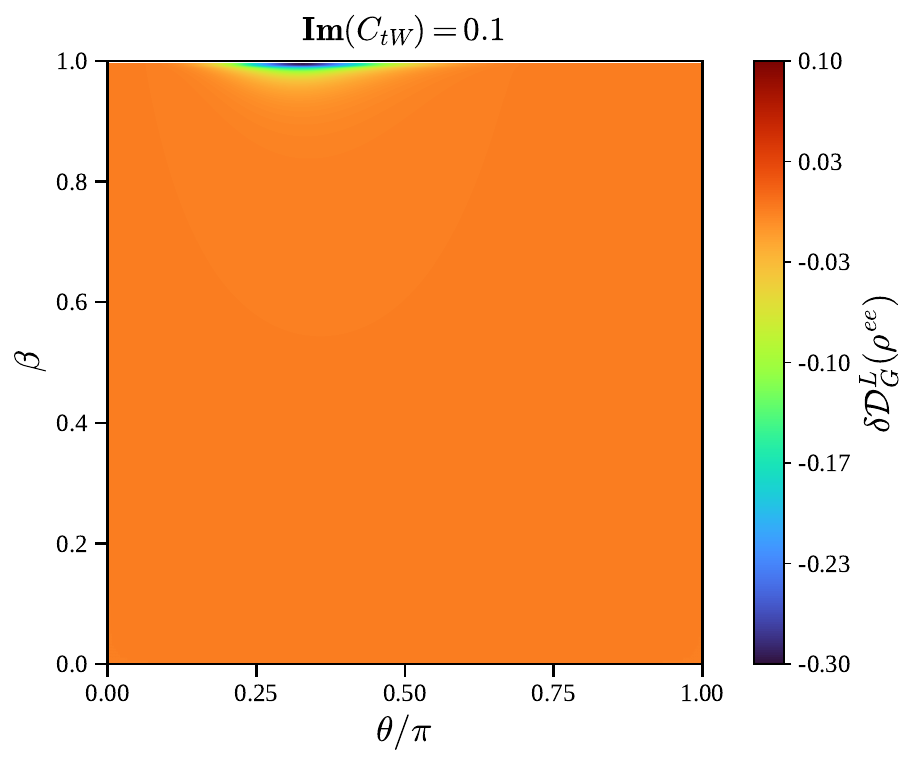}
\includegraphics[scale=0.335]{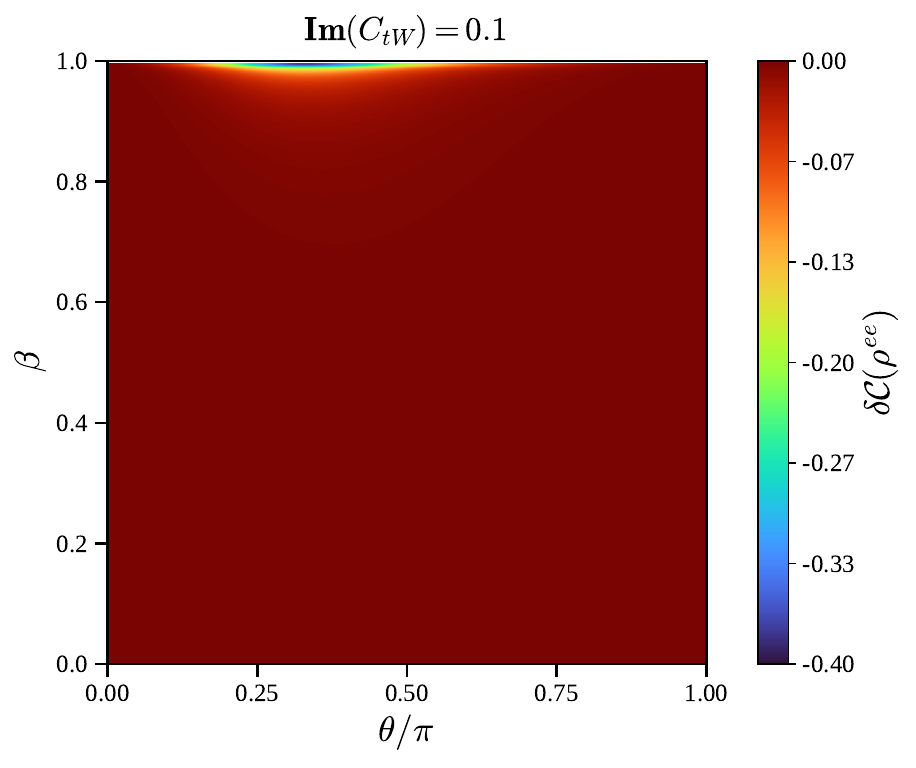}
\includegraphics[scale=0.335]{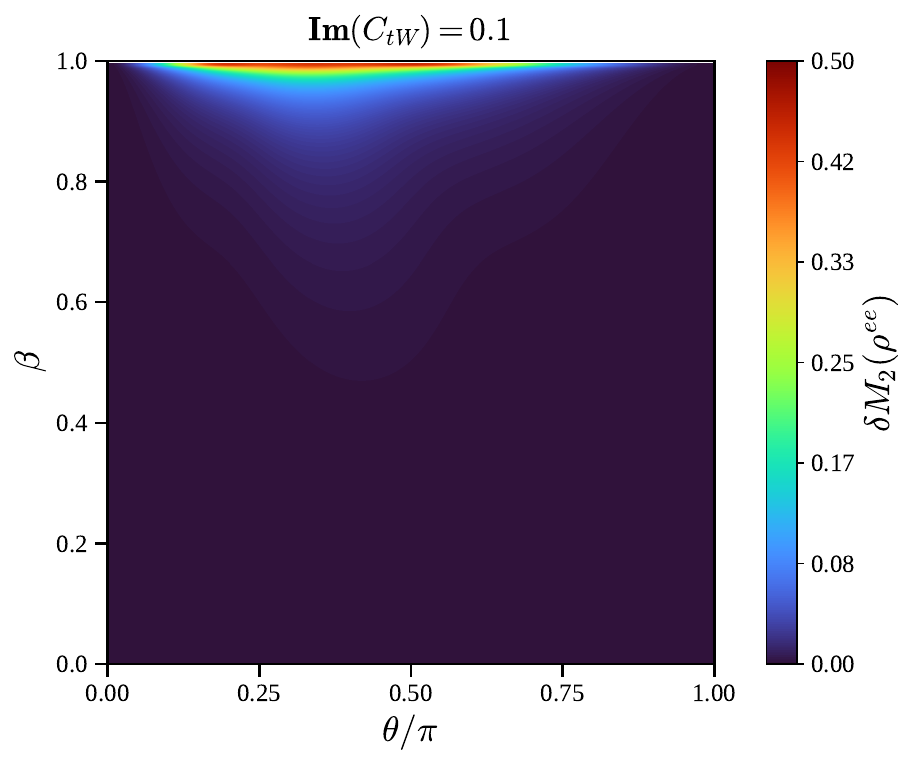}
\caption{The observables for the benchmark
$\operatorname{Im}(C_{tW})=0.1$, with all other SMEFT coefficients set to
zero. The upper panels show, from left to right, the norm of the top-antitop
polarisation difference, the norm of the
antisymmetric spin-correlation matrix, and the CP
trace distance. The lower panels show the
corresponding changes $\delta \mathcal D^L_G$, $\delta\mathcal C$, and
$\delta M_2$ relative to the SM predictions.\label{fig:ee_ImCtW}}
\end{figure}

Figures~\ref{fig:ee_SM_plots}-\ref{fig:ee_ImCtW} show the dependence of the observables defined in Sec.~\ref{sec:observables} on the scattering angle $\theta$ and the top-quark velocity $\beta$. The new physics scale is fixed to $\Lambda=1~\mathrm{TeV}$, and all SM input parameters are set to their reference values. The benchmark Wilson coefficients are chosen within the allowed ranges reported in Tab.~\ref{table:SMEFT limits}. In each benchmark scenario, a single Wilson coefficient is varied while all other SMEFT coefficients are set to zero. We have also evaluated the difference between the left and right geometric
discords for all benchmark scenarios considered below. Its absolute value
remains below $10^{-4}$ throughout the scanned phase space. We therefore do
not distinguish between the two quantities in the following and display only
the geometric discord measure.

We first consider the SM prediction, shown in
Fig.~\ref{fig:ee_SM_plots}. In this case, the top and antitop polarisation vectors are
non-zero, and we therefore use the geometric discord to quantify the
non-classical correlations of the full spin state. Since the SM contribution
is CP conserving, the polarisation and spin-correlation asymmetries,
$\Delta\mathbf B^{ee}$ and $( \mathbf C^A)^{ee}$, vanish, and consequently the CP trace distance is also zero.

Both the geometric discord ${\cal D}^L_G(\rho^{ee})$ and the concurrence
$\mathcal C(\rho^{ee})$ are suppressed near threshold and increase towards the
relativistic regime, reaching their largest values in a restricted angular
region. Their distributions are asymmetric under
$\theta\to\pi-\theta$ because of the parity-violating $Z$-boson contribution
and its interference with the photon-mediated amplitude. The magic
$ M_2(\rho^{ee})$ exhibits a more structured dependence on
$(\theta,\beta)$ and, unlike the concurrence, remains non-zero in the
threshold region. Its maxima do not coincide with those of the concurrence. This confirms that geometric discord, concurrence, and magic probe
complementary properties of the $t\bar t$ spin state.

The effects of the CP-even dipole interactions are displayed in
Fig.~\ref{fig:ee_ReCtB_ReCtW}, where we plot the differences
\begin{equation}
\delta\mathcal O
=
\mathcal O_{\mathrm{SMEFT}}-\mathcal O_{\mathrm{SM}},
\qquad
\mathcal O\in
\left\{
\mathcal D^L_G,\mathcal C, M_2
\right\}.
\label{eq:relativedef}
\end{equation}

For the benchmark $\operatorname{Re}(C_{tB})=1.5$, the largest modifications
of the geometric discord and concurrence occur in the high-$\beta$ region. Both observables exhibit regions of enhancement and suppression:
they are reduced where the corresponding SM values are largest, while positive
shifts appear in other parts of the relativistic region,
$\beta\to1$, except near the
forward and backward limits.

Over the remaining phase space, the deviations are considerably smaller.
The modification of  $\delta M_2$ exhibits a more
pronounced angular dependence, with alternating regions of enhancement and
suppression.
In the relativistic limit, magic is enhanced over most of the angular
range, except near the forward and backward directions, while a suppression
appears in the central angular region at intermediate values of $\beta$.
Thus, a CP-conserving dipole interaction can substantially modify the
quantum information content of the final state.

For the benchmark $\operatorname{Re}(C_{tW})=0.1$, the relative modifications
of the geometric discord and concurrence remain below approximately $\sim 0.1$
over most of the phase space. Larger deviations occur close to $\beta=1$,
where the shifts are predominantly negative and can reach approximately
$\sim-0.4$ and $\sim-0.5$, respectively. By contrast, the modification of magic is small over most of the
phase space and becomes positive in the relativistic region.

We next consider the imaginary parts of the dipole coefficients, shown in
Figs.~\ref{fig:ee_ImCtB} and~\ref{fig:ee_ImCtW}. In addition to the shifts
in the quantum information observables, we display the quantities constructed
directly from the CP-violating Fano--Bloch coefficients,
$\lVert\Delta\mathbf B\rVert$ and $\lVert\mathbf a\rVert$, together with the
CP trace distance $D_T^{\rm CP}(\rho^{ee})$. All three observables vanish
identically in the CP-conserving limit and therefore provide null tests of CP
violation. Once an imaginary dipole coefficient is switched on, these
quantities become non-zero, although their magnitudes depend strongly on the
kinematic region. Consequently, within the assumptions of the present
framework, a non-zero measurement of any of these observables would provide
direct evidence for CP-violating contributions to the production density
matrix.

For the benchmark $\operatorname{Im}(C_{tB})=1.5$, shown in
Fig.~\ref{fig:ee_ImCtB}, the corresponding dipole phases are
$\eta_\gamma=\pi/2$ and $\eta_Z=-\pi/2$. The CP-violating observables are
strongly suppressed near threshold and in the extreme forward and backward
limits, but grow rapidly with increasing $\beta$. The polarisation difference
$\lVert\Delta\mathbf B\rVert$ remains below approximately $\sim0.1$ over the
displayed phase space and exhibits a pronounced angular dependence. Its
distribution is not symmetric under $\theta\to\pi-\theta$. The norm
$\lVert\mathbf a\rVert$, associated with the antisymmetric part of the
spin-correlation matrix, is sizeable over a broader angular range and reaches
its largest values $\sim0.7$ in the relativistic regime, close to the forward and
backward regions. The CP trace distance displays a similar overall dependence,
showing that the two states related by CP become increasingly distinguishable
at high energies.

The lower panels of Fig.~\ref{fig:ee_ImCtB} show that the imaginary dipole
coefficient also modifies the quantum information observables. The geometric
discord and concurrence exhibit a pattern broadly similar to that obtained for
$\operatorname{Re}(C_{tB})=1.5$: both are reduced in the phase-space region
where their SM values are largest and enhanced over other parts of the
relativistic region. However, the locations and magnitudes of these
modifications are shifted in both the $\theta$ and $\beta$ directions. In addition, around
$\beta\simeq0.9$, the effect of the real coefficient is negligible, whereas
the imaginary coefficient produces a shift of approximately $\delta\mathcal O\simeq 0.2$.
The high-$\beta$ region therefore provides kinematic configurations in which
the effects of the real and imaginary dipole contributions can be
distinguished.

The behaviour of magic measure differs more substantially from that
induced by the real coefficient. For
$\operatorname{Im}(C_{tB})=1.5$, magic is suppressed only in restricted
regions of the relativistic regime, approximately around
$\theta\simeq\pi/4$ and $\theta\simeq\pi/2$, while positive shifts occur over
most of the remaining phase space. At intermediate values of $\beta$, the
magic and concurrence can respond in opposite directions: the imaginary
dipole contribution enhances the magic content of the state while reducing
its entanglement. Moreover, in regions where
$\operatorname{Re}(C_{tB})$ suppresses magic, the imaginary coefficient
can instead produce an enhancement. This complementary behaviour indicates
that concurrence and magic are sensitive to different structural changes in
the $t\bar t$ spin density matrix and can therefore help discriminate between
CP-even and CP-violating dipole contributions.

For the benchmark $\operatorname{Im}(C_{tW})=0.1$, the CP-violating
observables are strongly suppressed near threshold and become appreciable
mainly in the relativistic region. The polarisation difference
$\lVert\Delta\mathbf B\rVert$ reaches comparatively large values despite the
small benchmark coefficient, while $\lVert\mathbf a\rVert$ and
$D_T^{\rm CP}(\rho^{ee})$ are enhanced in more restricted regions of the
$(\theta,\beta)$ plane.
The lower panels of Fig.~\ref{fig:ee_ImCtW} show that the geometric discord,
concurrence, and magic are also modified by the imaginary weak-dipole
coefficient. The deviations remain small over most of the phase space and
become more pronounced near $\beta\to1$. In this region, the geometric discord
and concurrence are predominantly reduced, whereas magic receives a
positive correction.

The comparison with the
$\operatorname{Im}(C_{tB})=1.5$ benchmark shows that the sensitivity of each
observable depends on both the Wilson coefficient and the kinematic region.
The distinct magnitude and phase-space dependence of
$\lVert\Delta\mathbf B\rVert$ indicate that the polarisation difference can
also help distinguish whether the CP-violating contribution originates from
the hypercharge or weak-isospin dipole operator.

The same analytical framework can also be applied to
$e^+e^-\to\tau^+\tau^-$ in the presence of electroweak tau dipole operators,
after replacing the top-quark mass, electroweak couplings, and dipole
coefficients by their tau counterparts. The resulting expressions therefore
provide a direct starting point for studying CP-violating tau dipole
interactions.

In summary, $e^+e^-\to t\bar t$ provides simultaneous access to single-spin
polarisations, spin correlations, and quantum information observables. The
real parts of the dipole coefficients modify the discord, concurrence, and
magic while preserving CP, whereas their imaginary parts generate non-zero
$\lVert\Delta\mathbf B\rVert$, $\lVert\mathbf a\rVert$, and $D_T^{\rm CP}$. In particular, the
different responses of the single-spin
sectors provide complementary sensitivity to CP-violating electroweak dipole
interactions.
\subsection{$\gamma\gamma\rightarrow t \bar t $}
\label{sec:gammagamatottbar}

We next consider the process $\gamma\gamma\to t\bar t$, which provides a
particularly clean benchmark for isolating the impact of the electromagnetic
top-quark dipole interaction on the quantum state of the produced pair. In
contrast to the $e^+e^-$ channel, where the interference between photon- and
$Z$-mediated amplitudes generates a richer combination of vector and axial
structures, the $\gamma\gamma$ process depends only on the $t\bar t\gamma$
coupling. Possible CP-violating effects are therefore directly controlled by
the complex coefficient $d_\gamma$ introduced in
Eq.~\eqref{eq:dipole_phase_param}, making this channel particularly
transparent for identifying how the electromagnetic dipole interaction
populates the CP-violating components of the production density matrix.

Quantum entanglement and Bell nonlocality in
$\gamma\gamma\to t\bar t$ have recently been investigated at photon
colliders, including the effects of initial-state photon polarisation
and QCD corrections~\cite{Choi:2026omc,Jiang:2026sfw}. The present analysis
instead focuses on the modifications induced by a complex electroweak
top-quark dipole coefficient and on the resulting discord, concurrence,
magic, and CP-violating spin observables.
\begin{figure}[t]
    \centering
    \scalebox{1}{
        \begin{tikzpicture}
        \begin{feynman}
            \vertex (g1) at (-0.5, 1) {$\gamma$};
            \vertex (g2) at (-0.5,-1) {$\gamma$};
            \vertex (v1) at (1, 1);
            \vertex (v2) at (1,-1);
            \vertex (t1) at (2.5, 1) {$t$};
            \vertex (t2) at (2.5,-1) {$\bar t$};

            \diagram*{
                (g1) -- [boson] (v1),
                (g2) -- [boson] (v2),
                (v1) -- [fermion] (t1),
                (v2) -- [anti fermion] (t2),
                (v2) -- [fermion] (v1),
            };

            \node[
                fill=black,
                draw=black,
                circle,
                minimum size=7pt,
                inner sep=0pt
            ] at (v1) {};
        \end{feynman}
        \end{tikzpicture}

        \hspace{0.5cm}
        \begin{tikzpicture}
        \begin{feynman}
            \vertex (g1) at (-0.5, 1) {$\gamma$};
            \vertex (g2) at (-0.5,-1) {$\gamma$};
            \vertex (v1) at (1, 1);
            \vertex (v2) at (1,-1);
            \vertex (t1) at (2.5, 1) {$t$};
            \vertex (t2) at (2.5,-1) {$\bar t$};

            \diagram*{
                (g1) -- [boson] (v1),
                (g2) -- [boson] (v2),
                (v1) -- [fermion] (t1),
                (v2) -- [anti fermion] (t2),
                (v2) -- [fermion] (v1),
            };

            \node[
                fill=black,
                draw=black,
                circle,
                minimum size=7pt,
                inner sep=0pt
            ] at (v2) {};
        \end{feynman}
        \end{tikzpicture}

   }
    \caption{Representative Feynman diagrams for
    $\gamma\gamma\rightarrow t\bar t$. Diagrams with the photons exchanged are not shown. Double insertions are not included. }
      \label{fig: gamma gamma scattering}
\end{figure}
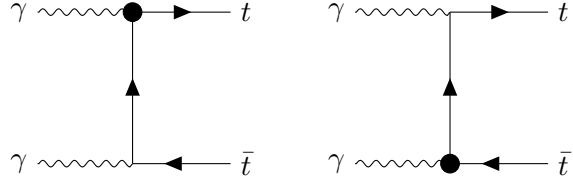
At tree level, the SM amplitude receives contributions from the $t$- and
$u$-channel exchange of a top quark.  The electromagnetic dipole operator modifies the
$t\bar t\gamma$ vertex as given in Eq.~\eqref{eq: ew dipoles2}, and dipole
insertions can occur at either photon-top vertex in both the $t$- and
$u$-channel amplitudes.

In the following, we work in the $t\bar t$ ZMF and adopt the
same spin basis as in the previous subsection, allowing a direct comparison
with the other production channels. The analytical expressions for the
corresponding Fano--Bloch coefficients, including the SM-dipole interference
terms and contributions quadratic in the dimension-six dipole coefficients,
are collected in App.~\ref{app: spin density matrix gagatt}.

In the spin basis, the polarisation vectors and spin-correlation matrix exhibit
the following CP structure:
\begin{align}
   \mathbf B=\mathbf{\bar{B}}&=0,
   &\mathbf C=\left( 
   \begin{array}{ccc}
        \bullet& \textcolor{purple}{\bullet}&\textcolor{purple}{\bullet}\\
         \textcolor{purple}{\bullet}& \bullet&\bullet\\
         \textcolor{purple}{\bullet}&\bullet&\bullet
    \end{array}\right)
    \label{Eq:Cp-sturcture_gaga}
\end{align}
As in the previous subsection, black bullets denote entries that can be
non-zero in the CP-conserving limit, whereas purple bullets identify
contributions generated by CP-violating interactions. For unpolarised initial
photons, the top and antitop polarisation vectors vanish in the present setup,
even after including the electromagnetic dipole interaction. The information
on CP violation is therefore encoded entirely in the spin-correlation matrix.

An important difference with respect to $e^+e^-\to t\bar t$ is that, in
$\gamma\gamma\to t\bar t$, the CP-conserving and CP-violating contributions
populate distinct entries of $\mathbf C$. In particular, the diagonal entries
and the $(2,3)$ and $(3,2)$ components may be non-zero in the CP-conserving
limit, whereas the entries connecting the first spin direction to the second
and third directions are generated only by the CP-violating part of the dipole
interaction. By contrast, in the $e^+e^-$ channel, CP-conserving and
CP-violating contributions can enter the same off-diagonal components of the
spin-correlation matrix. This separation makes the $\gamma\gamma$ channel
especially transparent for identifying CP-violating modifications of the
$t\bar t$ spin state. 

Both the hypercharge and weak dipole operators contribute to the electromagnetic
top-quark dipole coupling. As shown in Tab.~\ref{tab:eett_couplings}, the
coefficient entering the $t\bar t\gamma$ vertex is
\begin{equation}
d_\gamma
=
\sqrt{2}\left(c_W C_{tB}+s_W C_{tW}\right).
\label{eq:EW_dipole}
\end{equation}
Consequently, the process $\gamma\gamma\to t\bar t$ is sensitive to both
$C_{tB}$ and $C_{tW}$, despite involving only the electromagnetic
$t\bar t\gamma$ vertex. In the following, we consider separately the real and
imaginary parts of these two Wilson coefficients. In each benchmark scenario,
only one Wilson coefficient is taken to be non-zero, while all remaining
SMEFT coefficients are set to zero. The benchmark values are chosen within the
allowed ranges reported in Tab.~\ref{table:SMEFT limits}, and the new physics
scale is fixed to $\Lambda=1~\mathrm{TeV}$.
\begin{figure}[t]
\centering
\includegraphics[scale=0.335]{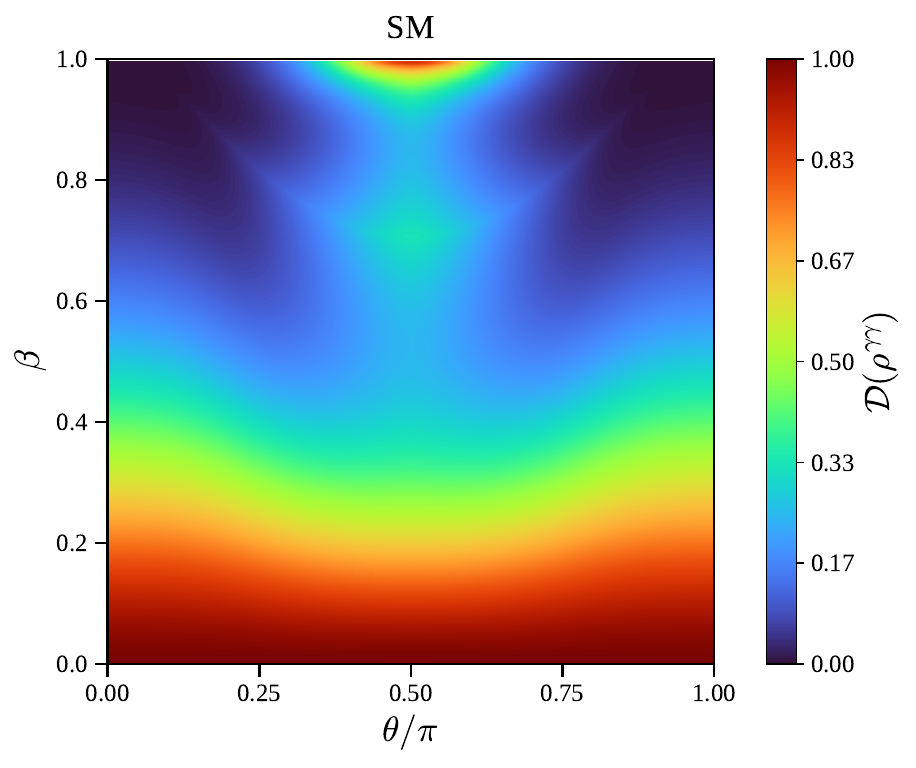}
\includegraphics[scale=0.335]{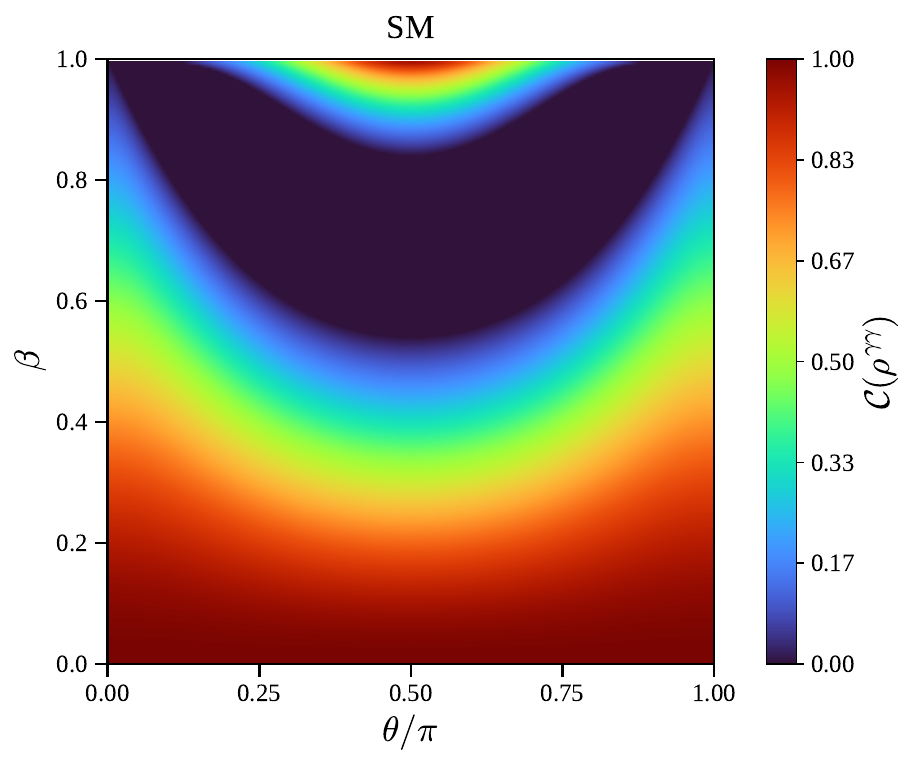}
\includegraphics[scale=0.335]{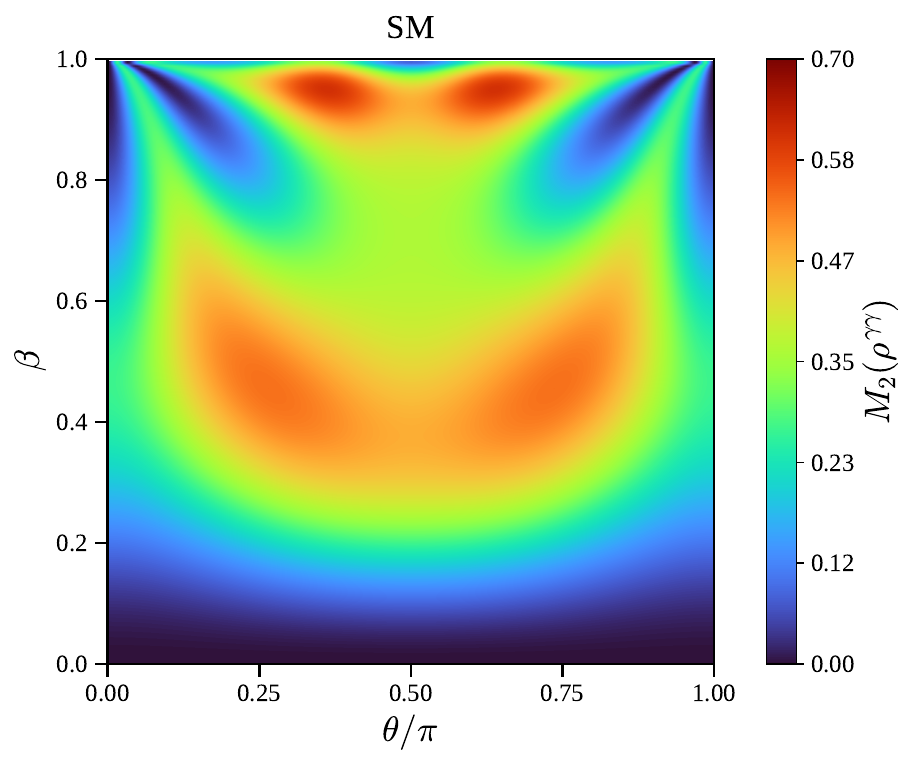}
\caption{SM predictions for the quantum information observables as functions of the scattering angle $\theta/\pi$ and the top-quark velocity $\beta$. From left to right, we show  discord, concurrence, and magic.\label{fig:gammagamma_SM}
}
\end{figure}

We first consider the SM predictions, shown in
Fig.~\ref{fig:gammagamma_SM}. In this channel, the top and antitop
polarisation vectors vanish.
We therefore use the discord $\mathcal D(\rho^{\gamma\gamma})$, defined in
Eq.~\ref{eq:discord}, to quantify the non-classical correlations of the
$t\bar t$ spin state, rather than the geometric discord employed in the
$e^+e^-$ analysis. Since the SM contribution is CP conserving, the
polarisation and spin-correlation asymmetries,
$\Delta\mathbf B^{\gamma\gamma}$ and
$(\mathbf C^A)^{\gamma\gamma}$, vanish, and consequently the CP trace
distance is also zero.

Near threshold, both the discord $\mathcal D(\rho^{\gamma\gamma})$ and the
concurrence $\mathcal C(\rho^{\gamma\gamma})$ attain values close to their
maxima. Their behaviour becomes markedly different as $\beta$ increases. The
concurrence develops a broad band in the $(\theta,\beta)$ plane in which the
$t\bar t$ state is separable. At larger
values of $\beta$, it becomes non-zero again around the central scattering
region and approaches its maximal value near $\theta=\pi/2$ in the
relativistic limit. By contrast, the discord remains finite over much of the intermediate
phase-space region in which the concurrence vanishes, demonstrating that the
$t\bar t$ state can retain non-classical correlations even when it is
separable. Near the relativistic forward and backward limits, however, the
discord is strongly suppressed. In the central angular region at large
$\beta$, it increases again, following the same qualitative trend as the
concurrence.
The magic measure $M_2(\rho^{\gamma\gamma})$ exhibits a qualitatively distinct dependence on the kinematics. It vanishes at threshold, becomes non-zero at intermediate values of $\beta$, and develops a non-trivial angular structure as the relativistic regime is approached. In particular, the phase-space regions in which magic is enhanced do not coincide with those where the concurrence or discord attain their largest values, underscoring the fact that these observables probe complementary features of the $t\bar t$ spin state.

\begin{figure}[t]
\centering
\includegraphics[scale=0.335]{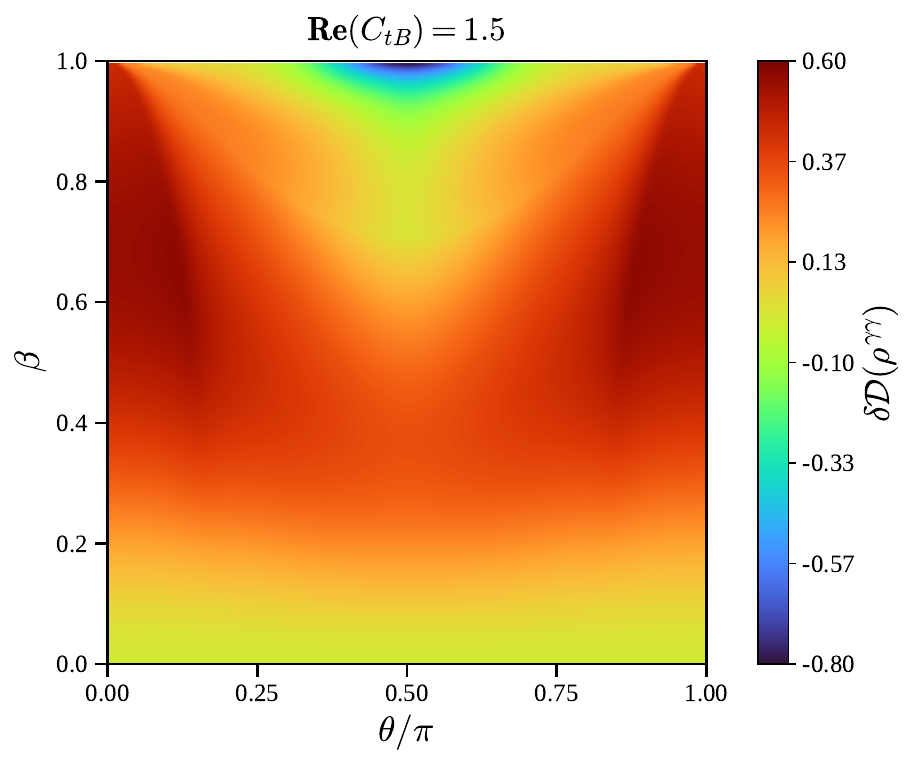}
\includegraphics[scale=0.335]{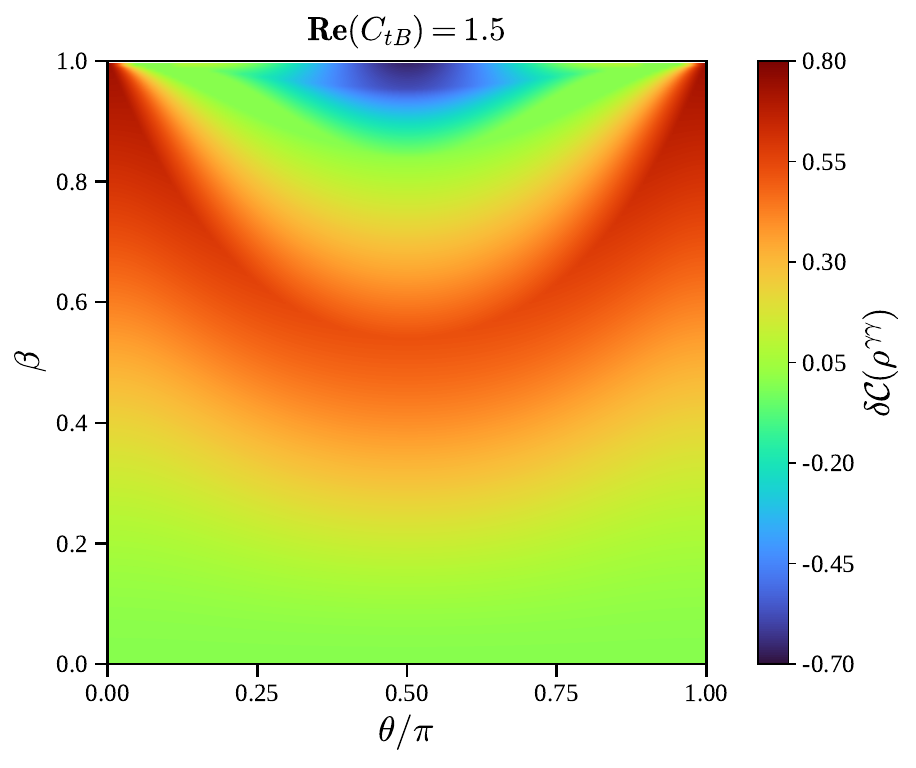}
\includegraphics[scale=0.335]{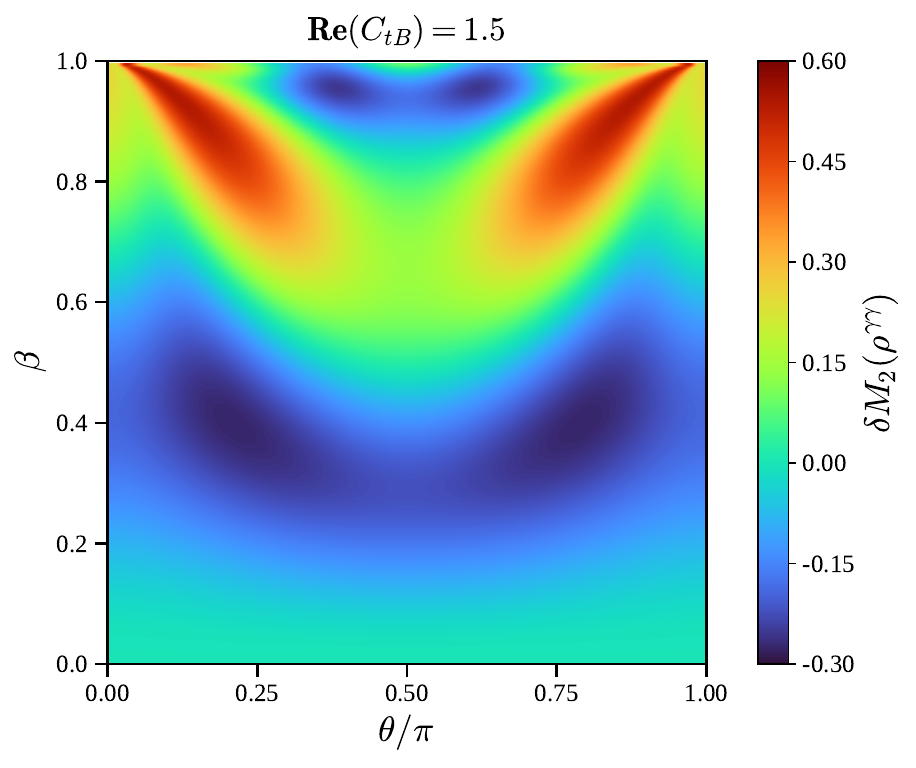}
\includegraphics[scale=0.335]{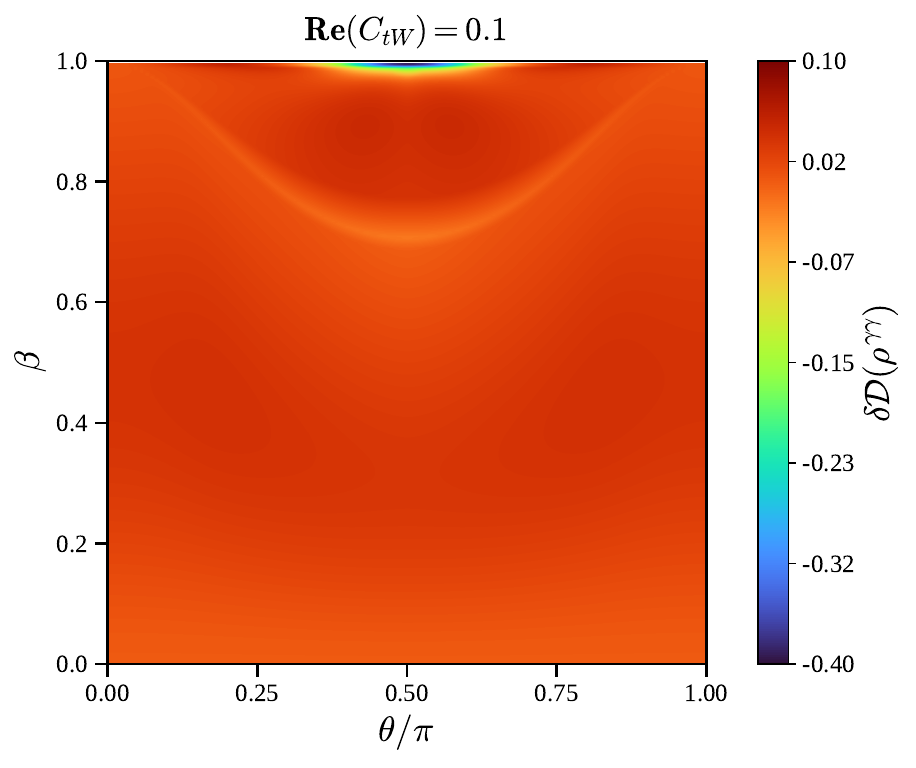}
\includegraphics[scale=0.335]{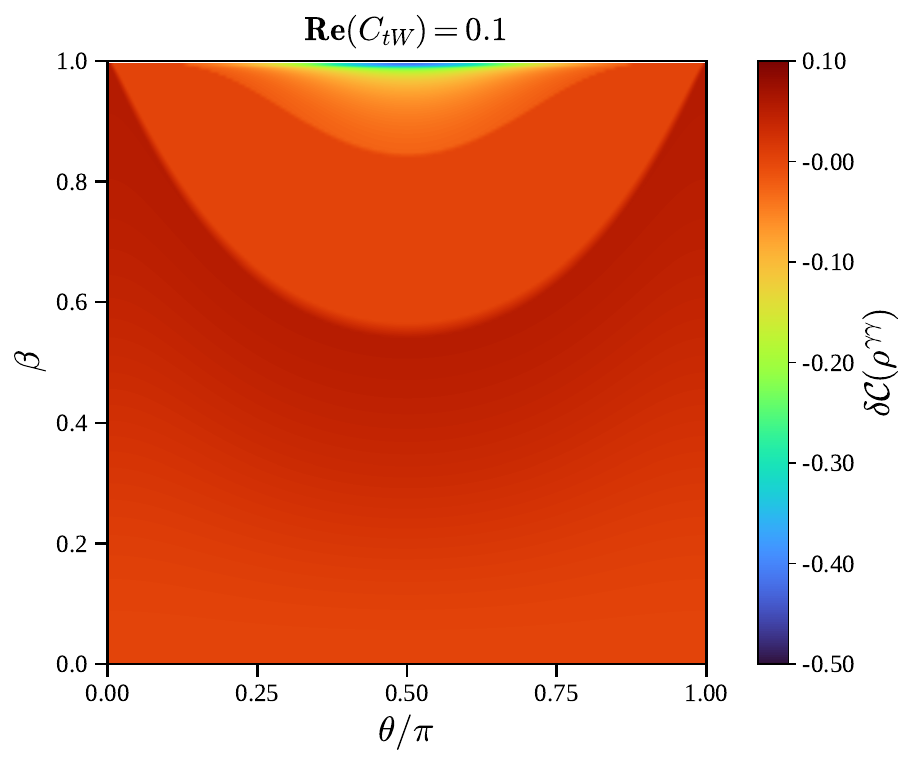}
\includegraphics[scale=0.335]{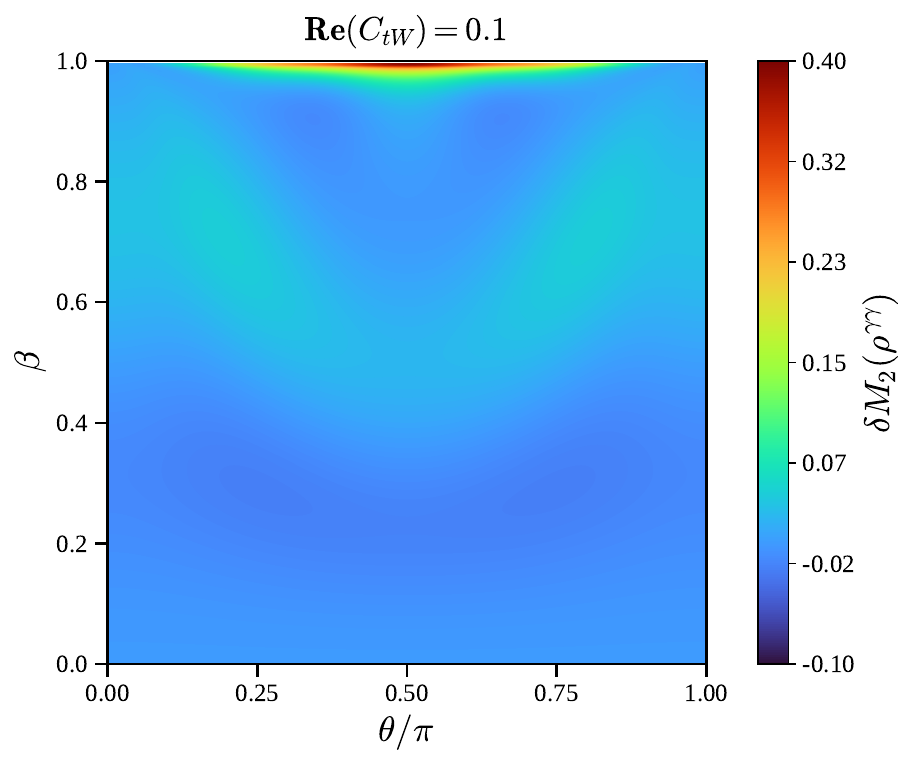}
%
\caption{Changes in the quantum information observables induced by the CP-even dipole
coefficients $\operatorname{Re}(C_{tB})$ (upper row) and
$\operatorname{Re}(C_{tW})$ (lower row), with all other SMEFT coefficients
set to zero. From left to right, we show $\delta \mathcal{D}$, $\delta\mathcal C$, and
$\delta M_2$, where
$\delta\mathcal O\equiv
\mathcal O_{\mathrm{SMEFT}}-\mathcal O_{\mathrm{SM}}$.\label{fig:gammagamma_ReCtB_ReCtW}
}
\end{figure}

The effects of the real parts of the dipole coefficients are shown in
Fig.~\ref{fig:gammagamma_ReCtB_ReCtW}. Since these contributions are CP
conserving, they do not generate the polarisation or spin-correlation
asymmetries, and the CP trace distance remains
zero.

For the benchmark $\operatorname{Re}(C_{tB})=1.5$, the discord and concurrence
receive only small positive corrections near threshold. Their modifications
become substantially larger at intermediate and high values of $\beta$,
particularly in regions where the corresponding SM observables are small or
vanish. Over a broad part of the phase space, both observables are therefore
enhanced. In the relativistic limit, however, they are suppressed around the
central scattering region, $\theta\simeq\pi/2$.

The modification of magic measure exhibits a more complex kinematic
dependence. In general, the dipole contribution tends to suppress magic in
regions where the SM value is already large and to enhance it where the SM
magic is small. An exception occurs for approximately
$0.6\lesssim\beta\lesssim0.9$ around $\theta\simeq\pi/2$, where the SM magic
is already non-negligible and is further enhanced by the dipole interaction.
Near the relativistic boundary, positive shifts develop in angular regions
away from the centre, whereas a broader suppression appears at intermediate
values of $\beta$ around central scattering angles.

For the benchmark $\operatorname{Re}(C_{tW})=0.1$, the discord and concurrence
display a qualitatively similar phase-space dependence, but the deviations
from the SM are considerably smaller except in relativistic limit. This suppression follows from both the
smaller benchmark value and the weaker contribution of $C_{tW}$ to the
electromagnetic dipole coefficient, Eq.~\ref{eq:EW_dipole}. The observables consequently remain
close to their SM values over most of the phase space, with the largest
modifications concentrated near the relativistic boundary. In this region, a suppression appears around $\theta\simeq\pi/2$, while the corrections
elsewhere remain small. In the relativistic region, the magic measure exhibits a qualitatively different pattern. Its modification is predominantly positive, including in regions where the $\operatorname{Re}(C_{tB})$ benchmark produces a suppression.

\begin{figure}[t]
\centering
\includegraphics[scale=0.445]{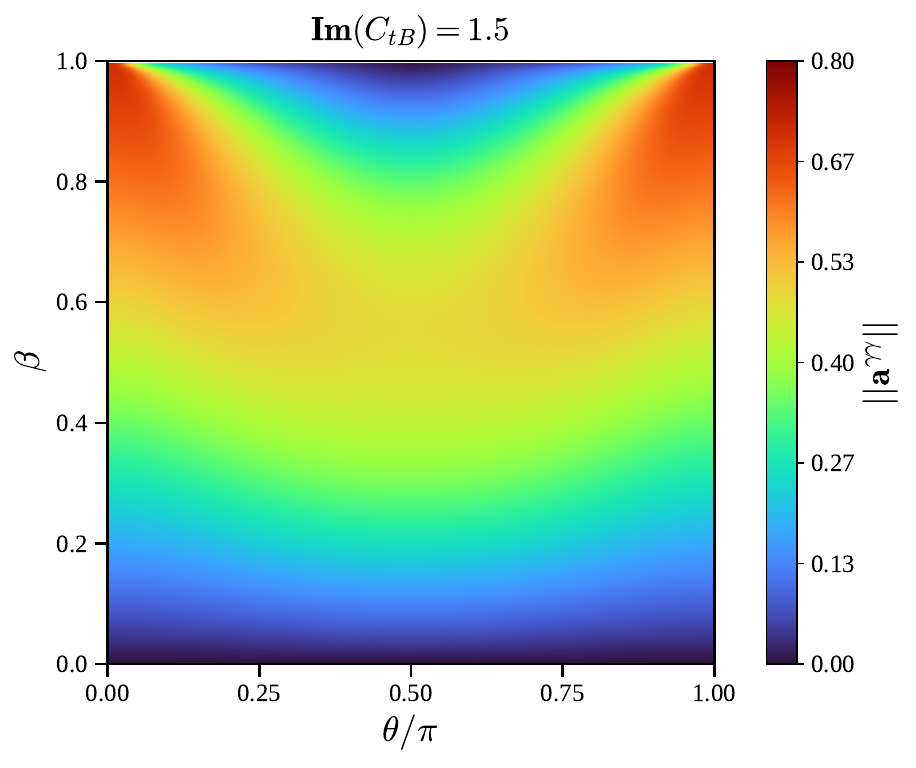}
\includegraphics[scale=0.445]{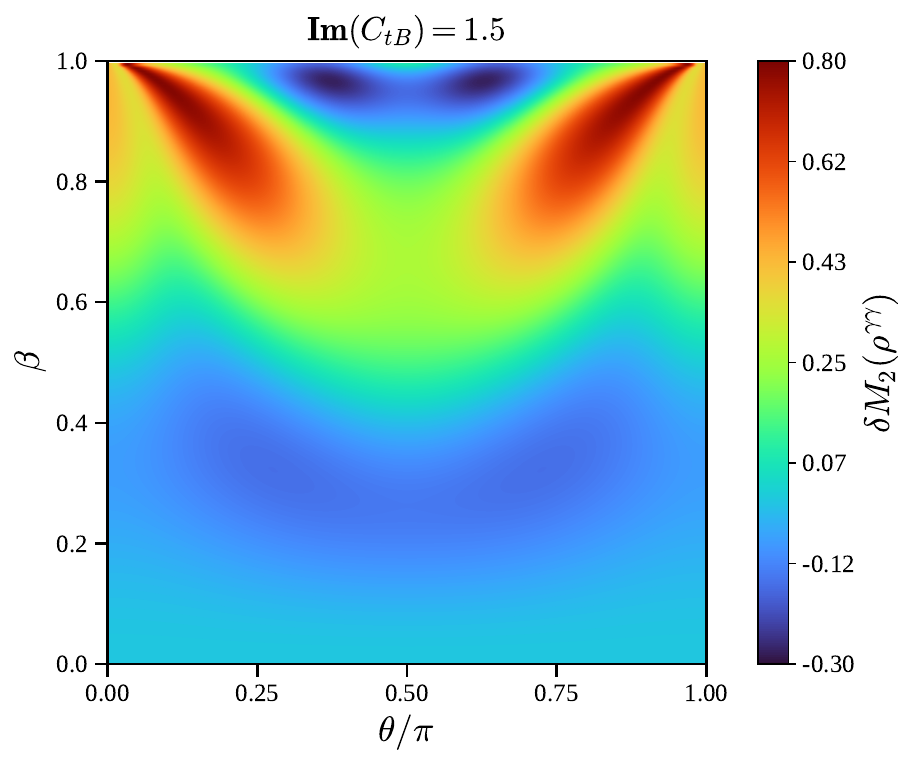}
\includegraphics[scale=0.445]{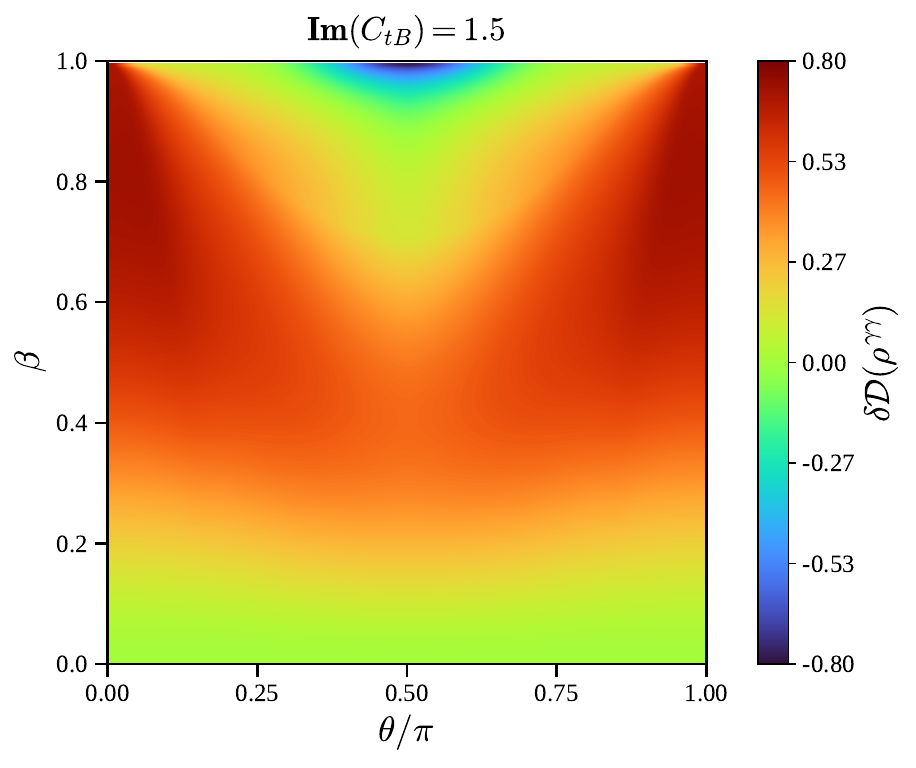}
\includegraphics[scale=0.445]{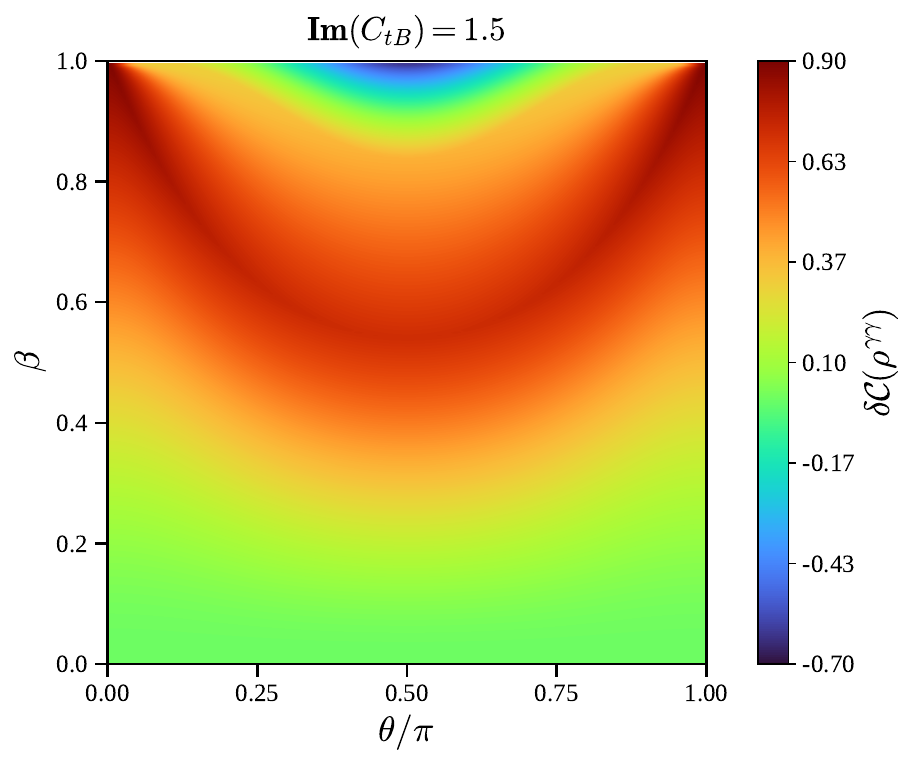}
\caption{\label{fig:gammagamma_ImCtB}The observables for the benchmark
$\operatorname{Im}(C_{tB})=1.5$, with all other SMEFT coefficients set to
zero. The upper panels show, from left to right, the norm of the
antisymmetric spin-correlation matrix, and the
corresponding changes $\delta M_2$ relative to the SM predictions. The lower panels show the
corresponding changes $\delta \mathcal D$, and  $\delta\mathcal C$ relative to the SM predictions.
}
\end{figure}
\begin{figure}[t]
\centering
\includegraphics[scale=0.445]{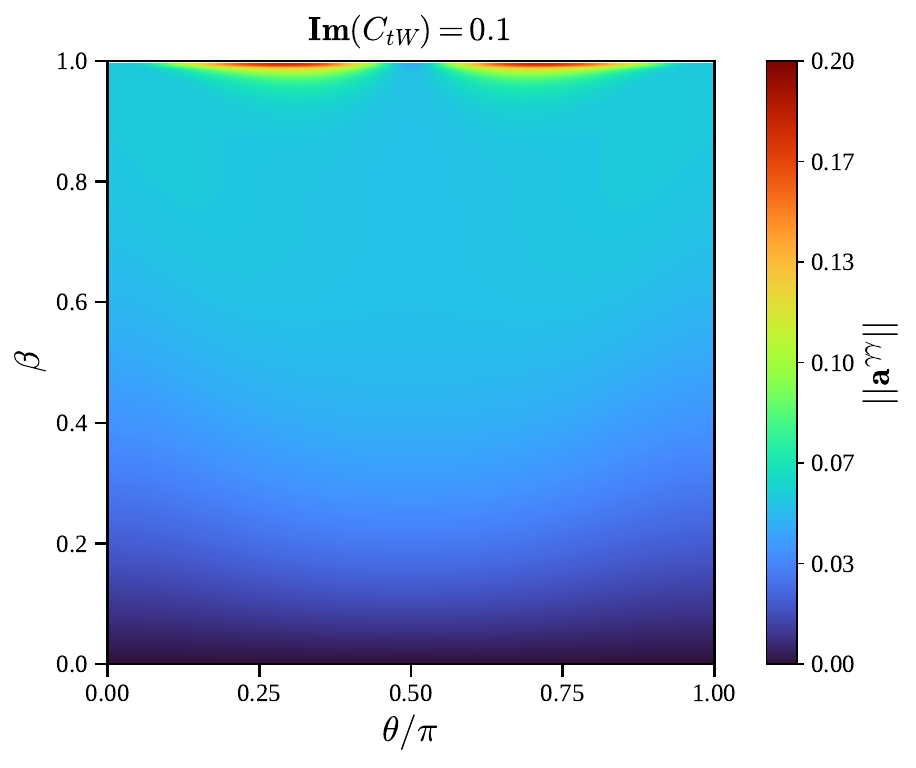}
\includegraphics[scale=0.445]{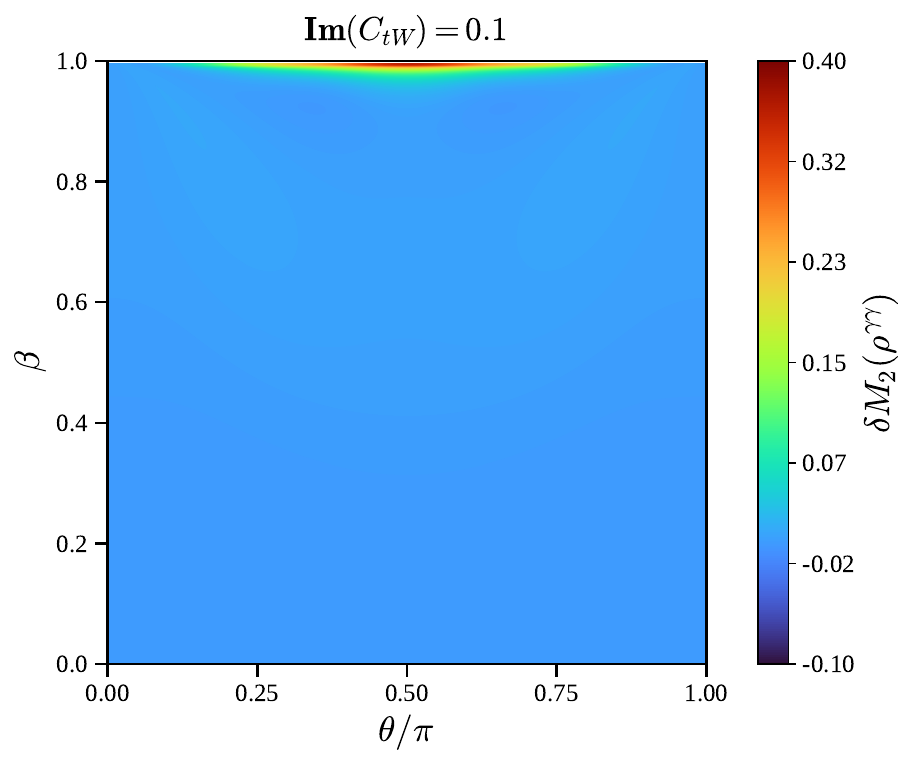}
\includegraphics[scale=0.445]{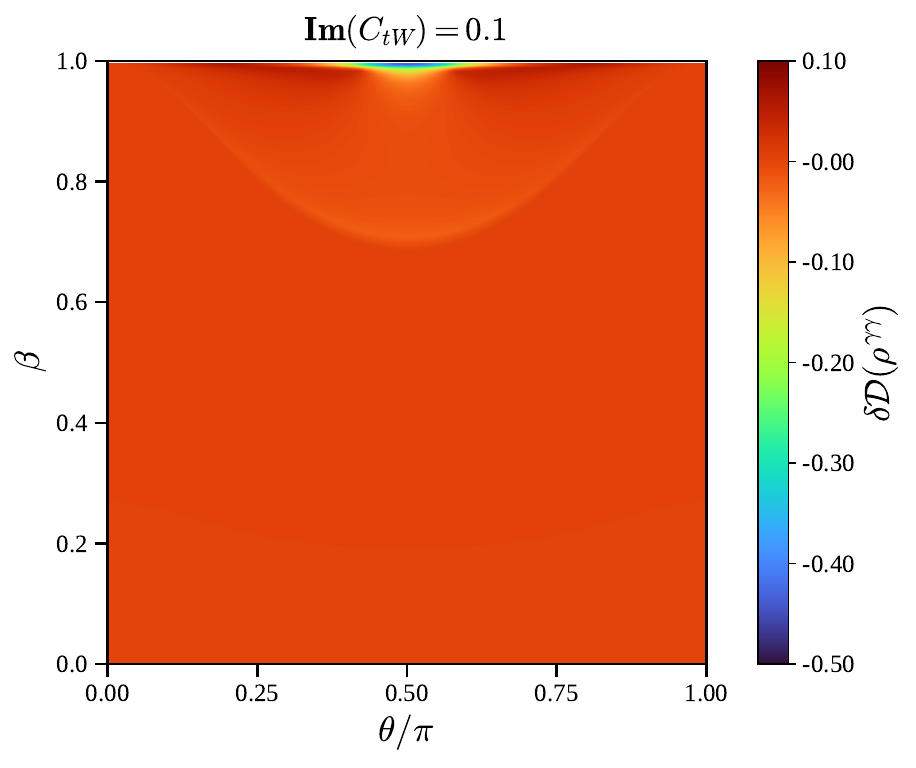}
\includegraphics[scale=0.445]{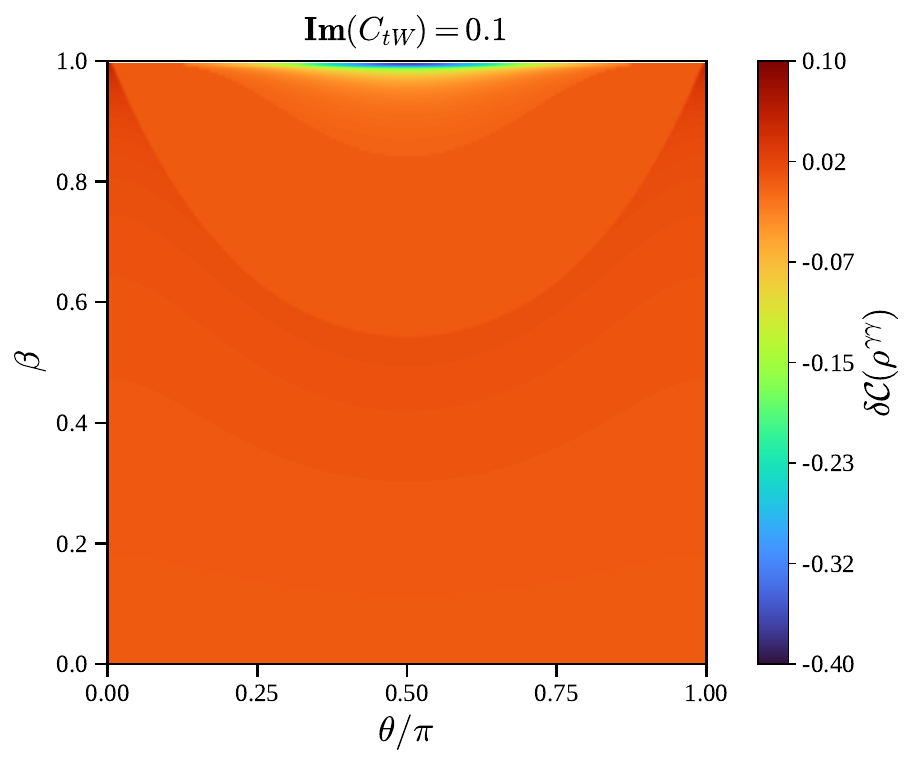}
\caption{\label{fig:gammagamma_ImCtW}The observables for the benchmark
$\operatorname{Im}(C_{tW})=0.1$, with all other SMEFT coefficients set to
zero. The upper panels show, from left to right, the norm of the
antisymmetric spin-correlation matrix, and the
corresponding changes $\delta M_2$ relative to the SM predictions. The lower panels show the
corresponding changes $\delta \mathcal D$, and  $\delta\mathcal C$ relative to the SM predictions.
}
\end{figure}

We next turn to the imaginary parts of the Wilson coefficients, shown in
Figs.~\ref{fig:gammagamma_ImCtB} and
\ref{fig:gammagamma_ImCtW}. Since the polarisation vectors vanish, the
CP-violating information is carried by the entries of the antisymmetric
spin-correlation sector, which we characterise through
$\lVert\mathbf a\rVert$. The trace distance has exactly the same behaviour as the $\lVert\mathbf a\rVert$.

For $\operatorname{Im}(C_{tB})=1.5$,
$\lVert\mathbf a\rVert$ is strongly suppressed near threshold and increases
rapidly with $\beta$. Its largest values occur in two broad angular regions
towards the forward and backward directions, while it remains suppressed near
$\theta\simeq\pi/2$ in the relativistic limit.

The modifications of the discord, concurrence, and magic exhibit a
qualitatively similar dependence on $(\theta,\beta)$ to that obtained for
$\operatorname{Re}(C_{tB})=1.5$. The main difference is quantitative: for the
imaginary coefficient, the positive deviations are enhanced by approximately
$0.2$ in absolute magnitude. In particular, the discord and concurrence
receive positive corrections over broad regions at intermediate and large
values of $\beta$, while both are suppressed near the relativistic limit
around the central scattering angle.

The magic measure follows the same broad pattern as in the real-coefficient
benchmark, with enhancements away from the central angular region and a
suppression near $\theta\simeq\pi/2$ as $\beta\to1$. Its positive shift is,
however, systematically larger for the imaginary coefficient. Thus, the real
and imaginary hypercharge-dipole benchmarks produce similar phase-space
structures in the quantum information observables, but with different overall
magnitudes. In contrast to the quantum information observables, which exhibit qualitatively similar phase-space patterns for the real and imaginary parts of $C_{tB}$, the quantity $\lVert\mathbf a\rVert$ provides a direct discriminator between the two contributions.

For $\operatorname{Im}(C_{tW})=0.1$, the CP-violating correlation measure
$\lVert\mathbf a\rVert$ remains small over most of the phase space and becomes
appreciable only near the relativistic boundary. The modifications of the
discord, concurrence, and magic are quantitatively similar to those obtained
for the benchmark $\operatorname{Re}(C_{tW})=0.1$, both in their overall
magnitudes and in their dependence on $\theta$ and $\beta$. In particular, the
discord and concurrence receive predominantly negative corrections as
$\beta\to1$, especially around the central scattering region, while the change
in the magic measure remains small over most of the phase space and becomes
more pronounced only in the relativistic regime.

In summary, the discord, concurrence, and magic exhibit distinct phase-space
patterns in the SM and receive characteristic modifications from the dipole
operators. These quantum information observables are therefore sensitive to
changes in the structure of the $t\bar t$ spin state and provide complementary
probes of the SMEFT contributions. However, their responses to the real and
imaginary parts of a given Wilson coefficient can be qualitatively, and in
some cases quantitatively, similar. They do not therefore provide, by
themselves, an unambiguous separation between CP-conserving and CP-violating
dipole interactions. By contrast, the correlation asymmetry
$\lVert\mathbf a\rVert$ vanishes for the CP-conserving real coefficients and
becomes non-zero in the presence of an imaginary dipole coefficient.
Consequently, $\lVert\mathbf a\rVert$ provides a null test of CP violation and
a direct discriminator between the CP-conserving and CP-violating
contributions, while the quantum information observables characterise the
resulting changes in the spin-state correlations.

The same analytical framework can be applied to
$\gamma\gamma\to\tau^+\tau^-$ in the presence of electromagnetic tau dipole
interactions. The corresponding production density matrix follows from the
expressions derived here by replacing the top-quark mass, electric charge, and
dipole coefficient by their tau counterparts and by setting the colour factor
to unity. The resulting formulae therefore provide a direct basis for studying
the quantum information and CP-violating spin observables associated with $\tau$
dipole interactions at photon colliders.

\subsection{$pp \rightarrow t\bar{t}$}\label{sec:pptottabr}

We now turn to hadronic top-quark pair production, $pp\to t\bar t$, which is the channel most directly relevant for present collider
phenomenology and, at the same time, the most intricate from the perspective
of the spin density matrix structure. Unlike the leptonic and photonic cases discussed above, the hadronic initial state receives   contributions from two dominant partonic subprocesses, $q\bar q\to t\bar t$ and $gg\to t\bar t$, whose relative importance depends on the collider energy and on the partonic kinematics. This makes the analysis richer, since CP-violating effects can be encoded differently in the two channels and are further intertwined with the non-trivial Lorentz and colour structure of QCD production. 
In the following, we therefore analyse the quark-annihilation and
gluon-fusion subprocesses separately. We work in the $t\bar t$
ZMF and adopt the spin basis introduced in
Section~\ref{sec:general_decomposition}, allowing the CP-violating components of
the polarisation vectors and spin-correlation matrix to be identified and
compared in a common framework. The convolution with the proton-parton
distribution functions, and hence the relative phenomenological weight of
the two subprocesses, will be included later in the hadron-level analysis in Section~\ref{LHCpheno}.

The QCD dipole interactions of the top quark are induced by the
dimension-six operator $O_{tG}$, defined in
Table~\ref{table: CP-odd operators}, with the generally complex Wilson
coefficient $C_{tG}$. Owing to the structure of the non-Abelian gluon
field-strength tensor, this operator modifies the $gt\bar t$ vertex and
simultaneously generates a $ggt\bar t$ contact interaction. In the
conventions adopted here, the corresponding vertices are
\begin{align}
gt\bar t:\qquad
&ig_s\gamma^\mu T^A
-\frac{\sqrt{2}\, v}{\Lambda^2}
\sigma^{\mu\nu}q_\nu
\left(C_{tG} P_R+C_{tG}^{*}P_L\right)T^A ,
\label{eq:ggtvertix}
\\
ggt\bar t:\qquad
&\frac{i\sqrt{2}\,g_s v}{\Lambda^2}
\sigma^{\mu\nu}
\left(C_{tG} P_R+C_{tG}^{*}P_L\right)
f^{ABC}T^A .
\label{eq:ggttvertex}
\end{align}
Here, $q_\nu$ denotes the gluon momentum entering the three-point vertex.
The derivative part of the gluon field-strength tensor
gives rise to the modified three-point interaction, whereas its non-Abelian
part generates the four-point contact interaction. The first term in
Eq.~\eqref{eq:ggtvertix} is the Standard Model QCD contribution, while the
second is induced by $O_{tG}$. The contact interaction in
Eq.~\eqref{eq:ggttvertex} follows from the non-Abelian gauge structure of
QCD and contributes only to the gluon-fusion subprocess.

The effects of dimension-six interactions on the spin density matrix for
hadronic $t\bar t$ production have been investigated in several
complementary settings. The interference contributions linear in the
dimension-six Wilson coefficients to the $gg\to t\bar t$ and
$q\bar q\to t\bar t$ subprocesses, including their effects on top-quark
polarisation and spin correlations, were studied in
Ref.~\cite{Bernreuther:2015yna}. These effects were subsequently revisited
in Ref.~\cite{Bernreuther:2024ltu} through a binned analysis tailored to LHC
phenomenology. In particular, the relevant observables were evaluated in
two-dimensional bins of the $t\bar t$ invariant mass and the top-quark
scattering angle in the $t\bar t$ ZMF. Quadratic dimension-six contributions for CP-even SMEFT
deformations, together with their implications for quantum observables such
as entanglement, were analysed in Ref.~\cite{Aoude:2022imd}. These studies
provide natural points of comparison for our analysis, in which we retain both linear and quadratic terms in the real and imaginary parts of the QCD dipole coefficient. We first consider the quark-annihilation subprocess,
$q\bar q\to t\bar t$, and subsequently turn to the gluon-fusion channel.

\subsubsection{$q\bar q\to t\bar t$}\label{qqtottbar}

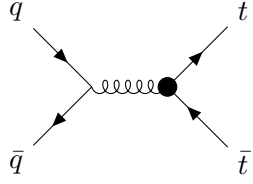
\begin{figure}[t]
    \centering
    \scalebox{1}{
    \begin{tikzpicture}
    \begin{feynman}[small]
    \vertex (e1) at (0,1) {$q$};
    \vertex (e2) at (0,-1) {$\bq$};
    \vertex (v1) at (1,0);
    \vertex (v2) at (2,0);
    \vertex (t1) at (3,1) {$t$};
    \vertex (t2) at (3,-1) {$\bar t$};

    \diagram* {
        (e1) -- [fermion] (v1),
        (e2) -- [anti fermion] (v1),
        (v1) -- [gluon] (v2),
        (v2) -- [fermion] (t1),
        (v2) -- [anti fermion] (t2),
    };
    \node[fill=black, draw=black, circle, minimum size=7pt, inner sep=0pt] at (v2) {};
    \end{feynman}
    \end{tikzpicture}
}
    \caption{Representative tree-level Feynman diagram for the quark-annihilation subprocess
$q\bar q\to t\bar t$ in QCD. The black dot denotes the $gt\bar t$ vertex,
including the contribution induced by the dipole operator $O_{tG}$.  }
    \label{fig:qq_scattering}
\end{figure}

At tree level in QCD, the quark-annihilation subprocess
$q\bar q\to t\bar t$ proceeds through the s-channel exchange of a
gluon, as shown in Fig.~\ref{fig:qq_scattering}. We retain only the dominant
QCD contribution and neglect the subleading electroweak contributions.

The Lorentz structure of the process is analogous to the photon-mediated
contribution to $e^+e^-\to t\bar t$, with the photon replaced by a gluon
and the electromagnetic dipole interaction replaced by the dipole
interaction induced by $O_{tG}$. The additional feature of the present
channel is the colour structure of the QCD amplitude.

We adopt the same spin basis
as in the preceding subsections, allowing a direct comparison with the other
production channels, and we use the kinematic conventions introduced in
Eq.~\eqref{eq:kinematics_eett}. The calculation is performed in
the $t\bar t$ ZMF, which coincides with the partonic
centre-of-mass frame but not, in general, with the proton-proton
centre-of-mass frame. The incoming-quark direction is chosen as the
reference axis for defining the helicity basis and scattering angle. Further details of the frame and kinematic conventions, together with the
analytical expressions for the corresponding Fano--Bloch coefficients, are
provided in App.~\ref{app: spin density matrix qqtt}. The expressions
include both the SM-dipole interference terms and the
contributions quadratic in the dimension-six coefficient.

In this spin basis, the polarisation vectors and spin-correlation matrix
exhibit the following CP structure:
\begin{align}
  \mathbf  B=\mathbf{\bar{B}}&=0,
   & \mathbf C=\left( 
   \begin{array}{ccc}
        \bullet& \textcolor{purple}{\bullet}&\textcolor{purple}{\bullet}\\
         \textcolor{purple}{\bullet}& \bullet&\bullet\\
         \textcolor{purple}{\bullet}&\bullet&\bullet
    \end{array}\right).
\label{eq:symmetries_FANO_qq}
\end{align}
As in the preceding subsections, black bullets denote entries that may be
non-zero in the CP-conserving limit, whereas purple bullets identify
contributions generated by CP-violating interactions. For an unpolarised
$q\bar q$ initial state, the top- and antitop-polarisation vectors vanish
in the present setup, even after including the QCD dipole interaction.
The CP-violating information is therefore encoded entirely in the
spin-correlation matrix. Its CP structure is identical to that obtained for
$\gamma\gamma\to t\bar t$: the CP-conserving and CP-violating
contributions populate the same matrix entries in the two channels.
\begin{figure}[t]
\centering
\includegraphics[scale=0.335]{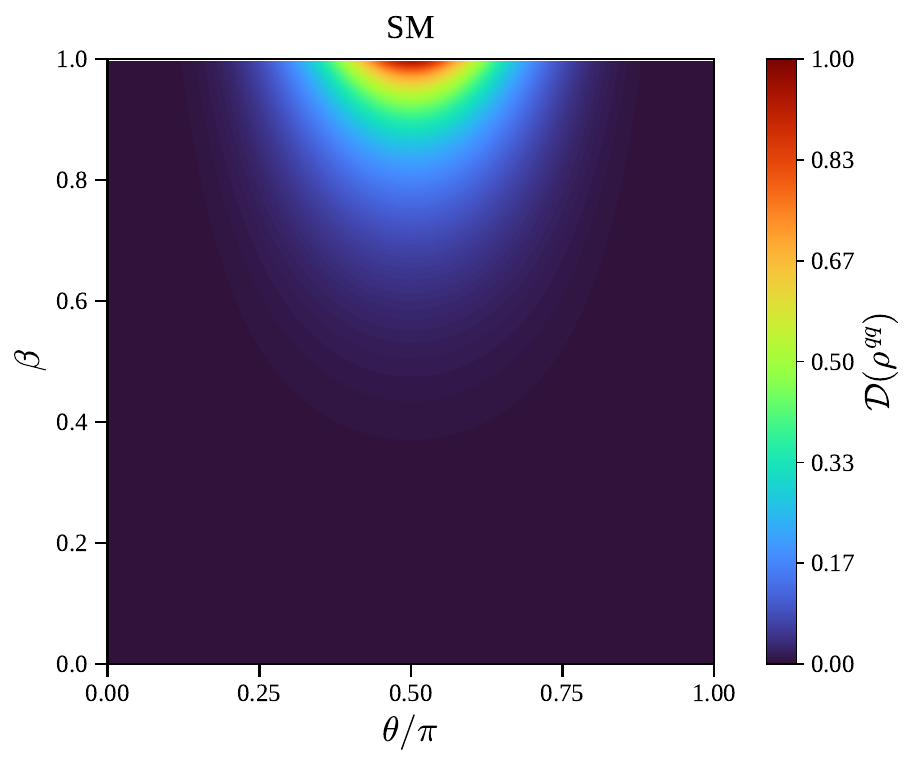}
\includegraphics[scale=0.335]{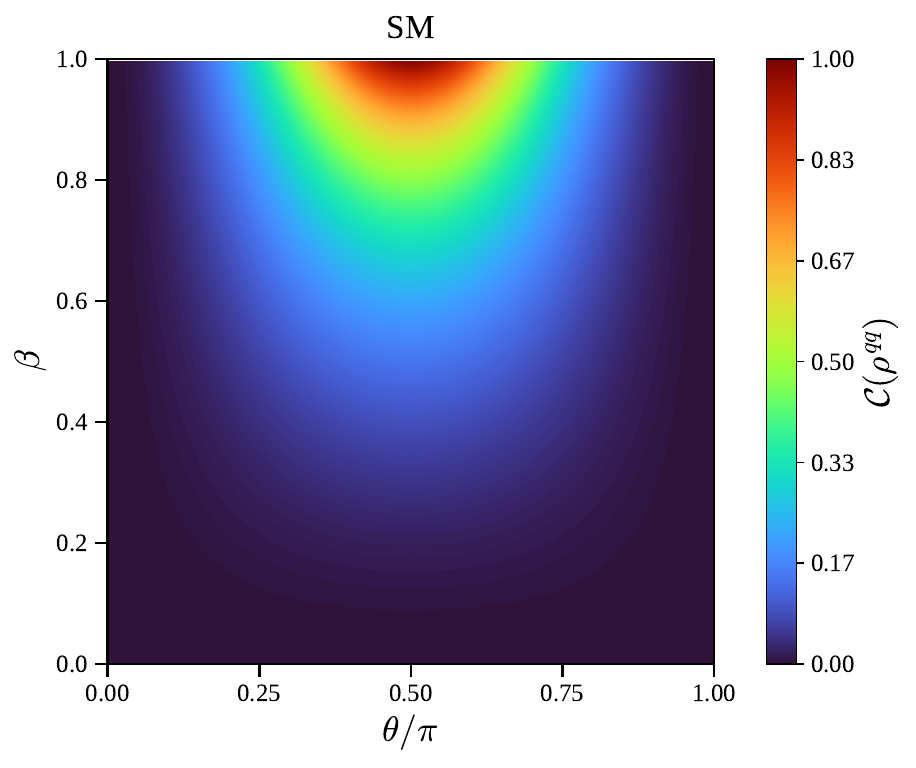}
\includegraphics[scale=0.335]{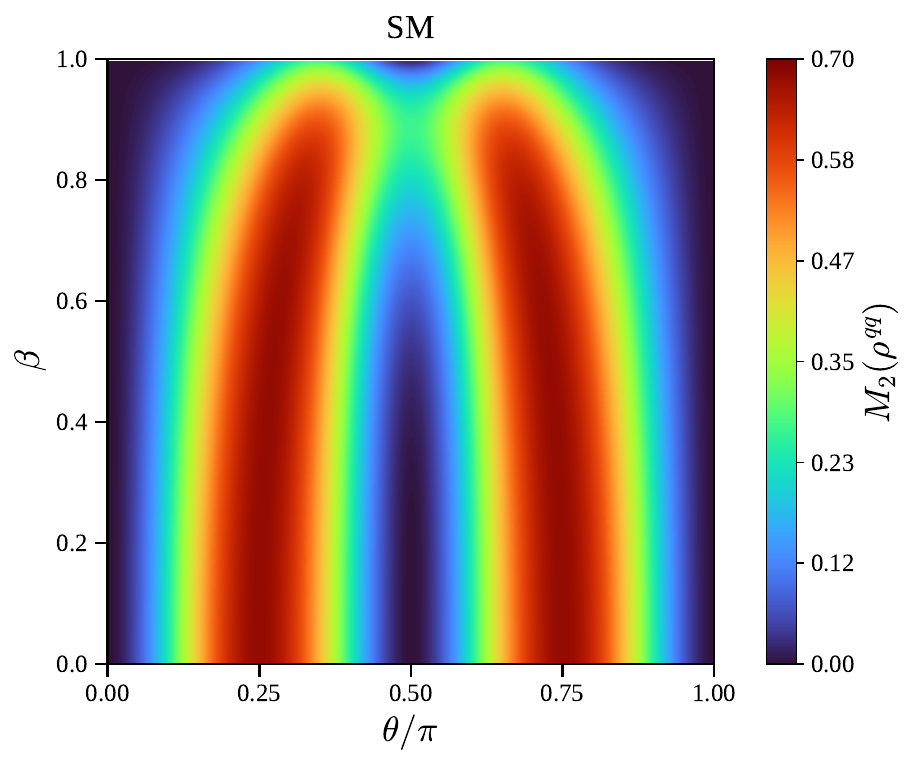}
\caption{SM predictions for the quantum information observables as functions of the scattering angle $\theta/\pi$ and the top-quark velocity $\beta$. From left to right, we show  discord, concurrence, and magic.\label{fig:qq_SM_plots}
}
\end{figure}
We first consider the SM predictions, shown in
Fig.~\ref{fig:qq_SM_plots}. Since the SM contribution is CP conserving, we only display the quantum discord,  concurrence and magic as functions of the top-quark velocity
$\beta$ and the scattering angle $\theta/\pi$.

Both discord and concurrence are largest in the relativistic regime,
$\beta\to1$, and approach their maxima near central production,
$\theta=\pi/2$. Their values are almost negligible near threshold and in the 
forward and backward regions. The observable $M_2$ exhibits a
different angular dependence: it vanishes near central production and is
enhanced in two symmetric angular regions on either side of
$\theta=\pi/2$, with its largest values attained at intermediate
scattering angles.
All three observables are symmetric under
\begin{equation}
\theta\longrightarrow\pi-\theta\,. 
\end{equation}
This reflects the purely vectorial structure of the tree-level QCD
interaction in the unpolarised $q\bar q$ channel. The overall behaviour is
similar to the photon-mediated contribution to
$e^+e^-\to t\bar t$, although the full $e^+e^-$ result does not possess
this symmetry because of the additional axial-vector contribution from
$Z$-boson exchange.

\begin{figure}[t]
\centering
\includegraphics[scale=0.335]{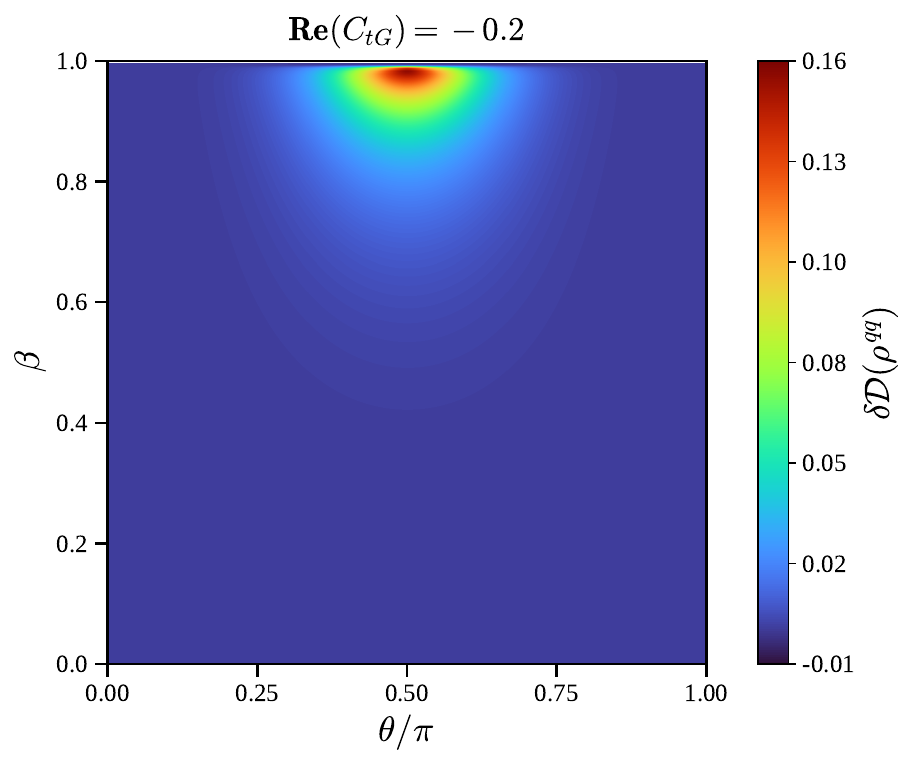}
\includegraphics[scale=0.335]{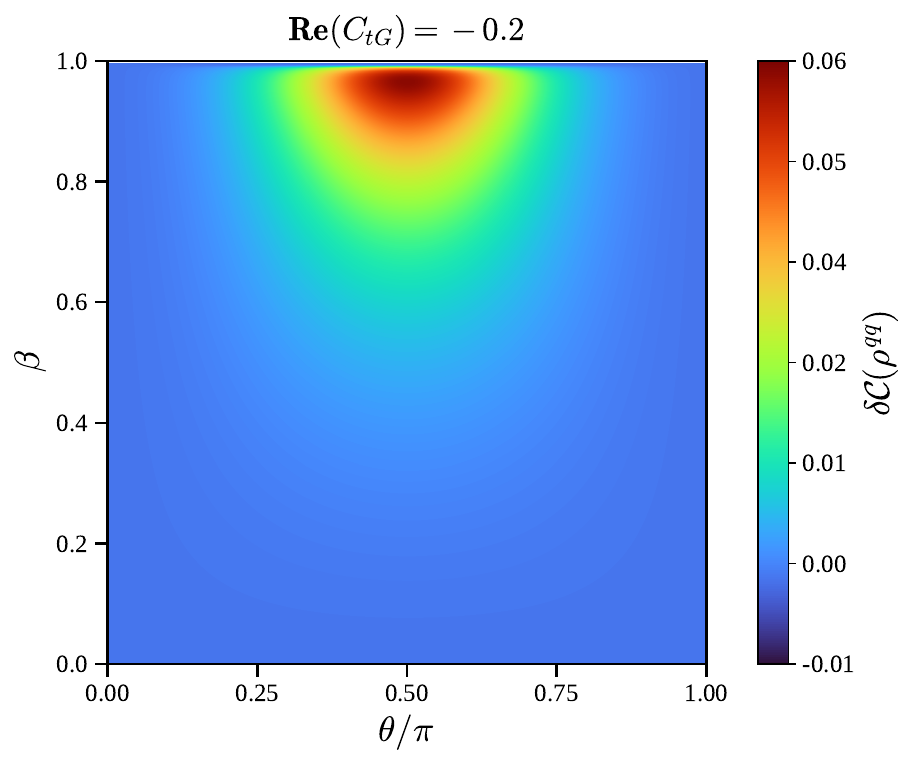}
\includegraphics[scale=0.335]{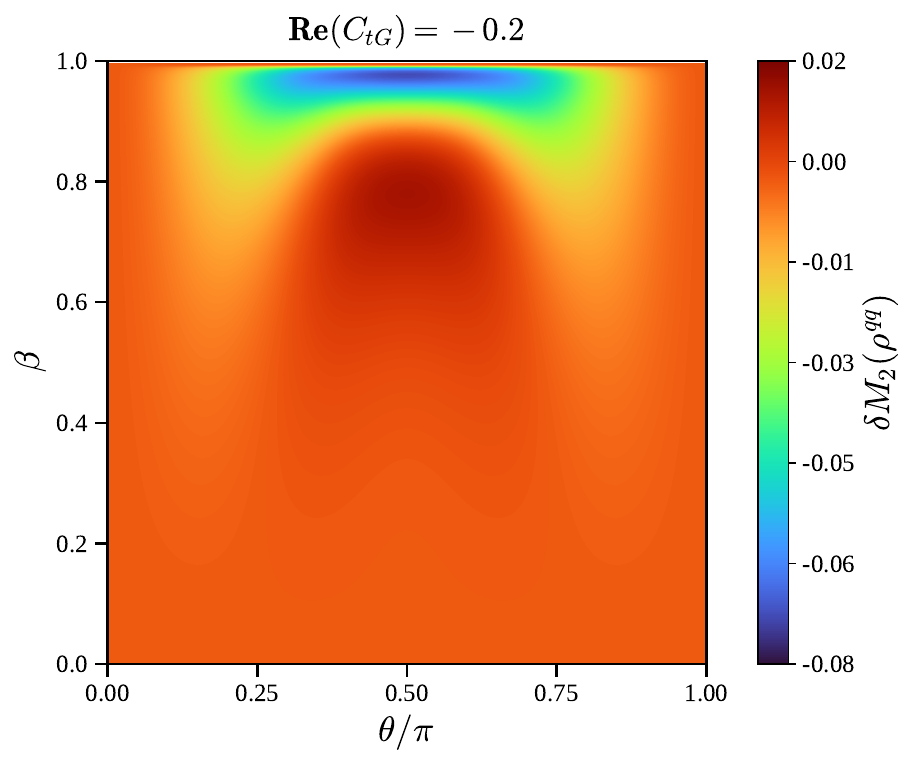}
\caption{Changes in the quantum information observables induced by the CP-even dipole
coefficient $\operatorname{Re}(C_{tG})=-0.2$ , with all other SMEFT coefficients
set to zero. From left to right, we show the
corresponding changes $\delta \mathcal D$, $\delta\mathcal C$, and $\delta M_2$ relative to the SM predictions.\label{fig:qq_ReCtG}
}
\end{figure}
\begin{figure}[t]
\centering
\includegraphics[scale=0.445]{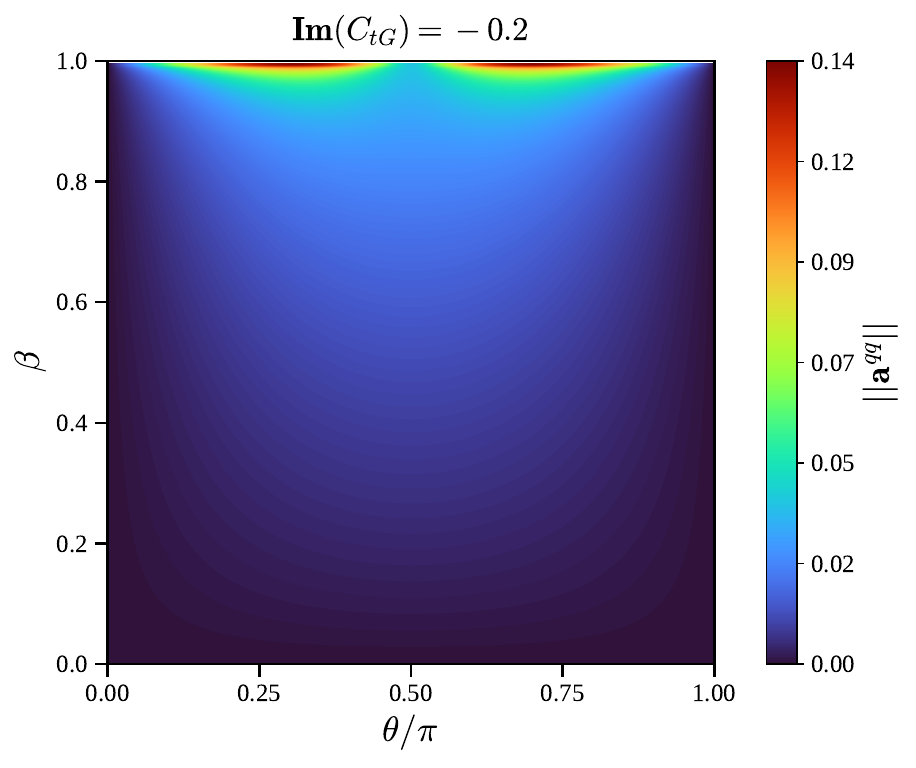}
\includegraphics[scale=0.445]{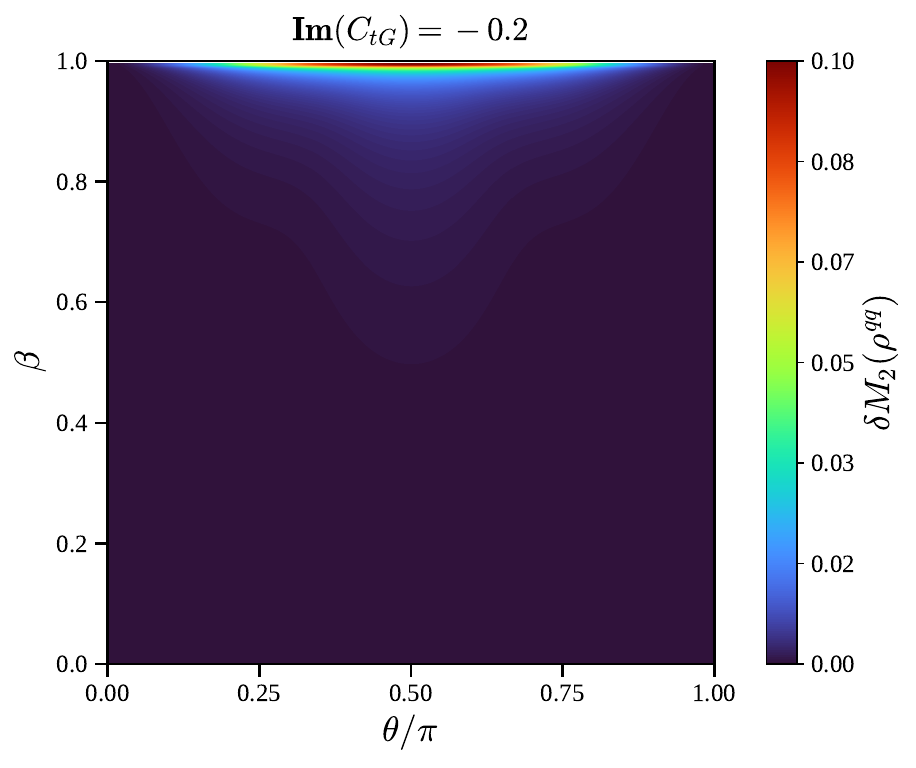}
\includegraphics[scale=0.445]{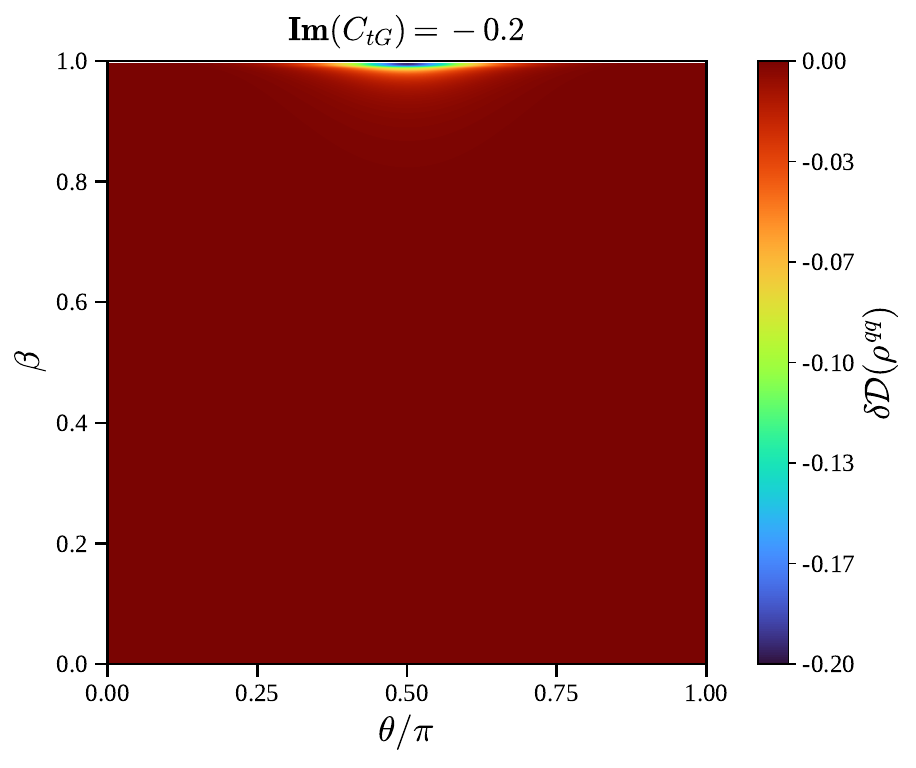}
\includegraphics[scale=0.445]{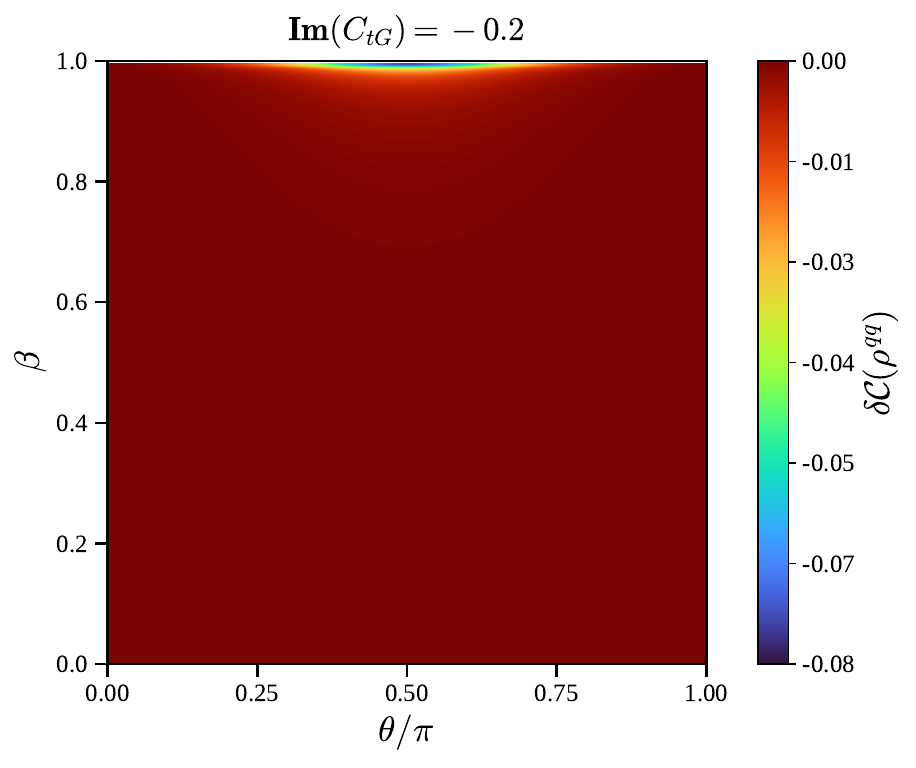}
\caption{\label{fig:qq_ImCtG}
 The observables for the benchmark
$\operatorname{Im}(C_{tG})=-0.2$, with all other SMEFT coefficients set to
zero. The upper panels show, from left to right,
the norm of the antisymmetric spin-correlation matrix and the change
$\delta M_2$ relative to the SM prediction. The lower
panels show the corresponding changes in the discord,
$\delta\mathcal D$, and concurrence, $\delta\mathcal C$.
}
\end{figure}

We next examine the effects of the dipole operator $O_{tG}$, shown in
Figs.~\ref{fig:qq_ReCtG} and \ref{fig:qq_ImCtG}. In each benchmark scenario,
only $C_{tG}$ is taken to be non-zero, while all other SMEFT coefficients
are set to zero. Again, the benchmark values are chosen within the allowed ranges
reported in Table~\ref{table:SMEFT limits}, and the new physics scale is
fixed to $\Lambda=1~\mathrm{TeV}$.

Figure~\ref{fig:qq_ReCtG} shows the changes in the quantum information
observables induced by the CP-even benchmark
$\operatorname{Re}(C_{tG})=-0.2$. The distributions remain symmetric under
$\theta\to\pi-\theta$, as expected for a CP-conserving deformation of the
purely vectorial QCD interaction. For all three observables, the corrections
are negligible over most of the phase space and become sizeable only at
large $\beta$ in the central-production region, before vanishing again at
the kinematic endpoint $\beta=1$.
The discord, $\mathcal D(\rho^{q q})$, and concurrence,
$\mathcal C(\rho^{q q})$, receive positive corrections in this region,
with the effect being more pronounced for the discord. By contrast,
the magic observable, $ M_2(\rho^{q q})$, exhibits a
non-monotonic behaviour: it first receives a small positive correction, then
develops a comparatively larger negative contribution closer to the
ultrarelativistic regime, and finally returns to zero at $\beta=1$. This
behaviour is qualitatively similar to that found in the
$e^+e^-\to t\bar t$ channel for the CP-even dipole benchmark shown in
Fig.~\ref{fig:ee_ReCtB_ReCtW}, although the effects are considerably smaller
in the present case and remain symmetric under
$\theta\to\pi-\theta$.

Figure~\ref{fig:qq_ImCtG} shows the effects of the CP-odd benchmark
$\operatorname{Im}(C_{tG})=-0.2$. In contrast to the CP-even case, the
chromoelectric interaction generates a non-vanishing antisymmetric component
of the spin-correlation matrix. Its norm is negligible over most of the phase
space and becomes sizeable only at very large $\beta$, reaching its largest
values close to the kinematic endpoint $\beta=1$. The distribution remains
symmetric under $\theta\to\pi-\theta$.

The corresponding changes in the quantum information observables are also
strongly localised in the large-$\beta$ region. The correction
$\delta M_2$ is positive and becomes appreciable only close to
$\beta=1$, while remaining small throughout most of the phase space. By
contrast, the discord and concurrence receive negative corrections. These
effects are concentrated around central production, $\theta\simeq\pi/2$,
and close to the ultrarelativistic endpoint, with the reduction being more
pronounced for the discord than for the concurrence.

Thus, although the CP-odd QCD dipole interaction generates a clear
CP-violating contribution to the antisymmetric spin correlations, its effects
on the discord, concurrence, and $M_2$ remain negligible over most
of the kinematic region and become relevant only for highly relativistic
top-quark pairs. It is also noteworthy that the CP-even and CP-odd
QCD dipole contributions affect these observables in opposite ways: the
CP-even benchmark produces positive corrections to the discord and
concurrence and a predominantly negative correction to $ M_2$,
whereas the CP-odd benchmark yields negative corrections to the discord and
concurrence and a positive correction to $ M_2$.
\subsubsection{$gg\rightarrow t\bar{t}$}
\label{sec:ggtottbar}

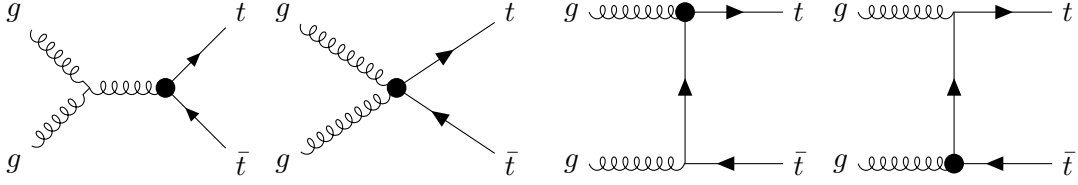
\begin{figure}[t]
    \centering
    \scalebox{1}{
      \begin{tikzpicture}
    \begin{feynman}[small]
    \vertex (g1) at (0,1) {$g$};
    \vertex (g2) at (0,-1) {$g$};
    \vertex (v1) at (1,0);
    \vertex (v2) at (2,0);
    \vertex (t1) at (3,1) {$t$};
    \vertex (t2) at (3,-1) {$\bar t$};
    
    \diagram*{
        (g1) -- [gluon] (v1),
        (g2) -- [gluon] (v1),
        (v1) -- [gluon] (v2),
        (v2) -- [fermion] (t1),
        (v2) -- [anti fermion] (t2),
    };
    \node[fill=black, draw=black, circle, minimum size=7pt, inner sep=0pt] at (v2) {};
    \end{feynman}
    \end{tikzpicture}
    \begin{tikzpicture}
    \begin{feynman}
    \vertex (g1) at (0,1) {$g$};
    \vertex (g2) at (0,-1) {$g$};
    \vertex (v1) at (1.5,0);
    \vertex (t1) at (3,1) {$t$};
    \vertex (t2) at (3,-1) {$\bar t$};
    
    \diagram*{
        (g1) -- [gluon] (v1),
        (g2) -- [gluon] (v1),
        (v1) -- [fermion] (t1),
        (v1) -- [anti fermion] (t2),
    };
    \node[fill=black, draw=black, circle, minimum size=7pt, inner sep=0pt] at (v1) {};
    \end{feynman}
    \end{tikzpicture}
    }
    \scalebox{1}{
    \begin{tikzpicture}
    \begin{feynman}
    \vertex (g1) at (-0.5,1) {$g$};
    \vertex (g2) at (-0.5,-1) {$g$};
    \vertex (v1) at (1,1);
    \vertex (v2) at (1,-1);
    \vertex (t1) at (2.5,1) {$t$};
    \vertex (t2) at (2.5,-1) {$\bar t$};
    
    \diagram*{
        (g1) -- [gluon] (v1) -- [fermion] (t1),
        (g2) -- [gluon] (v2) -- [anti fermion] (t2),
        (v2) -- [fermion] (v1),
    };
    \node[fill=black, draw=black, circle, minimum size=7pt, inner sep=0pt] at (v1) {};
    \end{feynman}
    \end{tikzpicture}
    \begin{tikzpicture}
    \begin{feynman}
    \vertex (g1) at (-0.5,1) {$g$};
    \vertex (g2) at (-0.5,-1) {$g$};
    \vertex (v1) at (1,1);
    \vertex (v2) at (1,-1);
    \vertex (t1) at (2.5,1) {$t$};
    \vertex (t2) at (2.5,-1) {$\bar t$};
    
    \diagram*{
        (g1) -- [gluon] (v1) -- [fermion] (t1),
        (g2) -- [gluon] (v2) -- [anti fermion] (t2),
        (v2) -- [fermion] (v1),
    };
    \node[fill=black, draw=black, circle, minimum size=7pt, inner sep=0pt] at (v2) {};
    \end{feynman}
    \end{tikzpicture}
}
    \caption{Representative tree-level Feynman diagrams for the gluon-fusion
subprocess $gg\to t\bar t$ in QCD. The diagrams include the $s$-channel
gluon-exchange contribution, the $ggt\bar t$ contact interaction induced by
$O_{tG}$, and the $t$-channel  top-quark-exchange contributions ($u$-channel ones are not shown).
For the $t$-channel  diagrams, the black dot denotes a single
dipole insertion at either of the two $gt\bar t$ vertices; diagrams with simultaneous
insertions of new interactions at both vertices are not considered.}
    \label{fig:gg_scattering}
\end{figure}

At leading order in the SM, this process receives contributions from top-quark
exchange in the $t$- and $u$-channels and from an $s$-channel
gluon-exchange diagram. In the presence of the dipole operator
$O_{tG}$, the $gt\bar t$ vertex is modified and an additional four-point
$ggt\bar t$ contact interaction is generated. In the $t$- and $u$-channel
amplitudes, the QCD dipole interaction can be inserted at either of the two
$gt\bar t$ vertices. We include both possible single-insertion
contributions, but do not include amplitudes containing QCD dipole
insertions at both vertices simultaneously. The corresponding representative
diagrams are shown in Fig.~\ref{fig:gg_scattering}.

Although the same QCD dipole operator $O_{tG}$ contributes to both
subprocesses, the gluon-fusion channel has a richer diagrammatic structure.
Besides modifying the $gt\bar t$ vertices in the $s$-, $t$-, and
$u$-channel amplitudes, $O_{tG}$ also generates the $ggt\bar t$ contact
interaction through the non-Abelian part of the gluon field-strength tensor.
We adopt the same frame, spin basis, and kinematic conventions as in the
preceding $q\bar q\to t\bar t$ subsection. The analytical expressions for
the corresponding Fano--Bloch coefficients, including the
SM-dipole interference terms and the contributions
quadratic in the dimension-six coefficient, are collected in
App.~\ref{app: spin density matrix ggtt}.

In this spin basis, the polarisation vectors and spin-correlation matrix
exhibit the following CP structure:
\begin{align}
  \mathbf  B=\mathbf{\bar{B}}&= 0,
   & \mathbf C=\left( 
   \begin{array}{ccc}
        \bullet& \textcolor{purple}{\bullet}&\textcolor{purple}{\bullet}\\
         \textcolor{purple}{\bullet}& \bullet&\bullet\\
         \textcolor{purple}{\bullet}&\bullet&\bullet
    \end{array}\right)
\label{eq:symmetries_FANO_gg}
\end{align}
Black and purple bullets denote CP-conserving and CP-violating contributions,
respectively. For an unpolarised $gg$ initial state, the polarisation vectors
vanish, and the CP-violating information is encoded entirely in the
spin-correlation matrix. Its CP structure is identical to that of the
$q\bar q\to t\bar t$ and $\gamma\gamma\to t\bar t$ channels, although the
coefficients differ in their analytical and kinematic dependence.
\begin{figure}[t]
\centering
\includegraphics[scale=0.335]{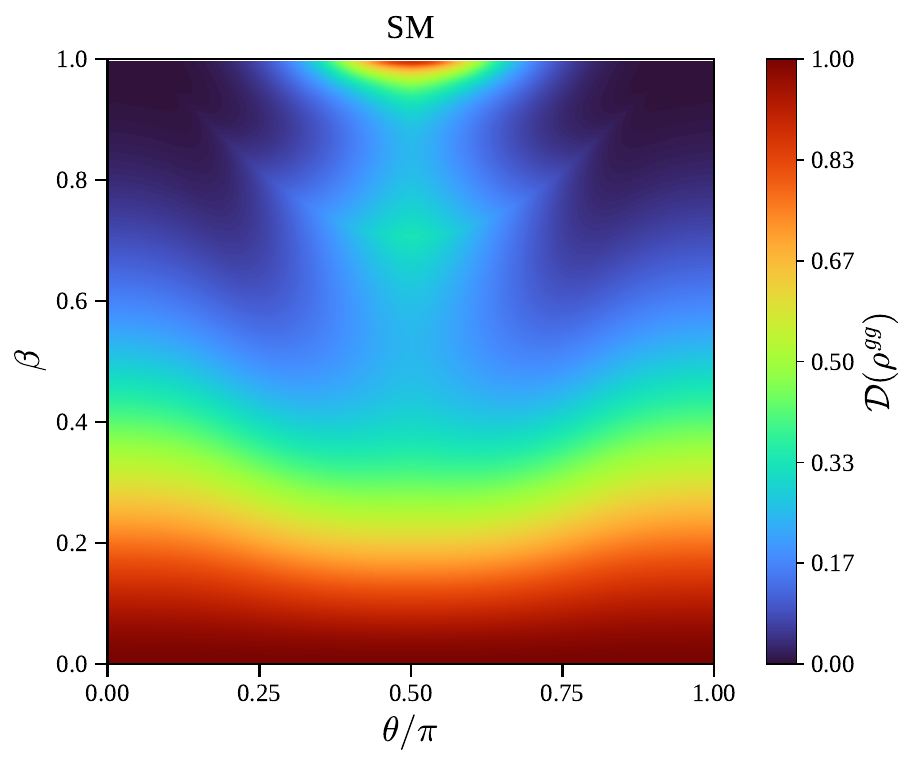}
\includegraphics[scale=0.335]{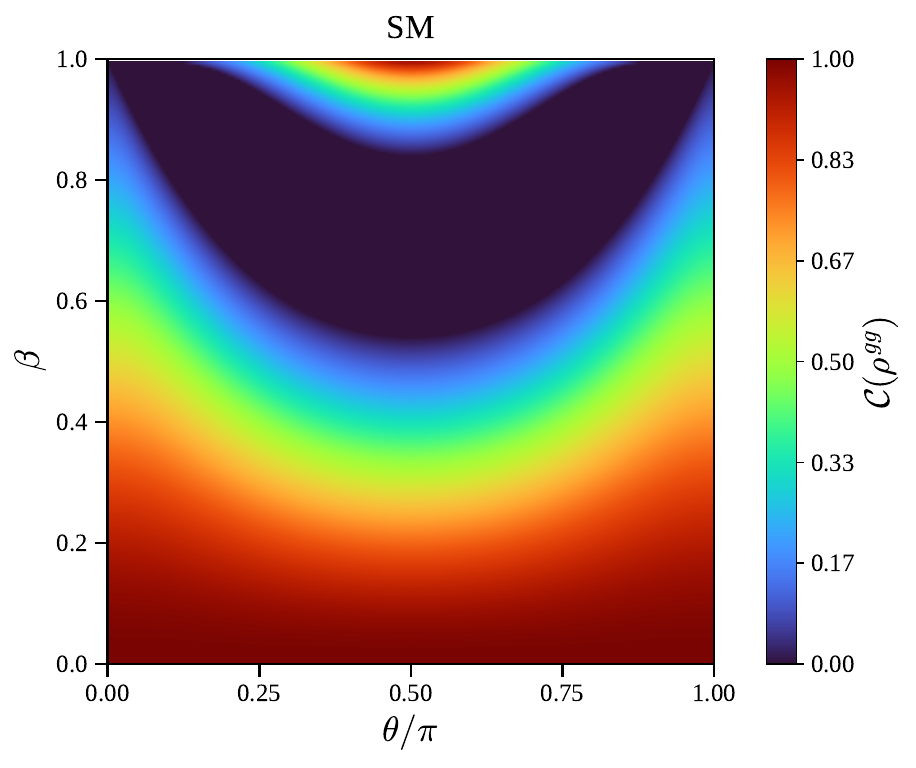}
\includegraphics[scale=0.335]{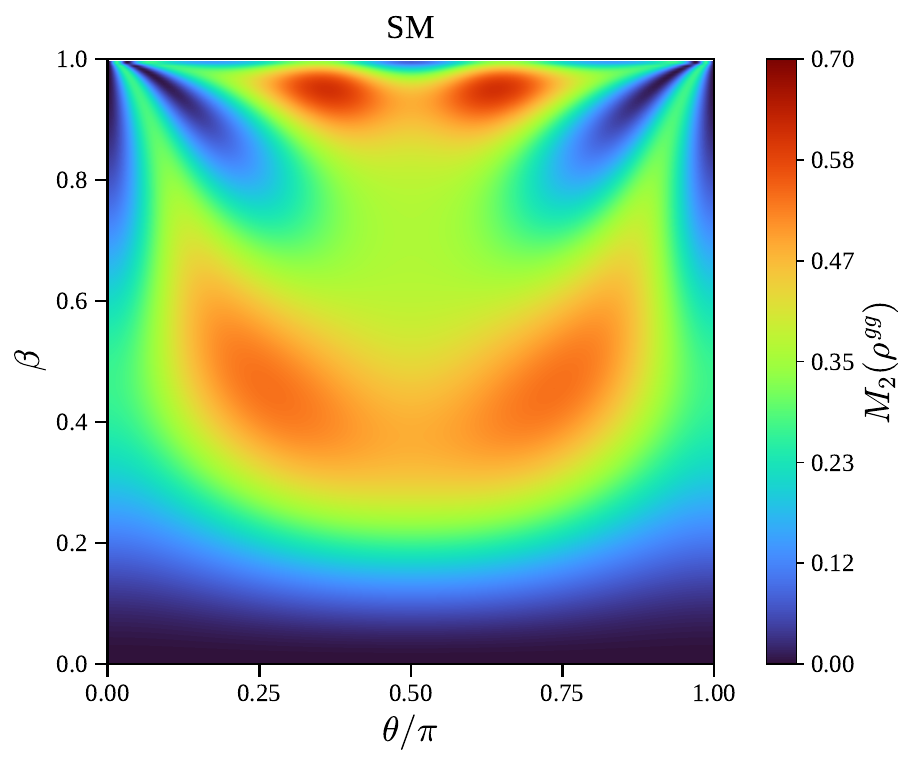}\\
\includegraphics[scale=0.335]{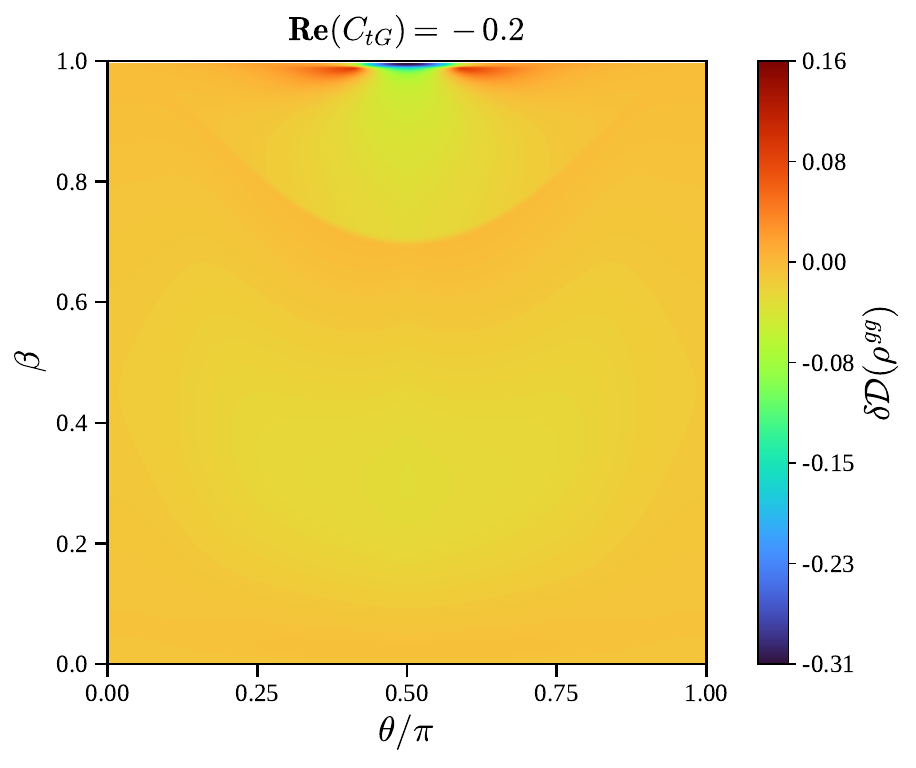}
\includegraphics[scale=0.335]{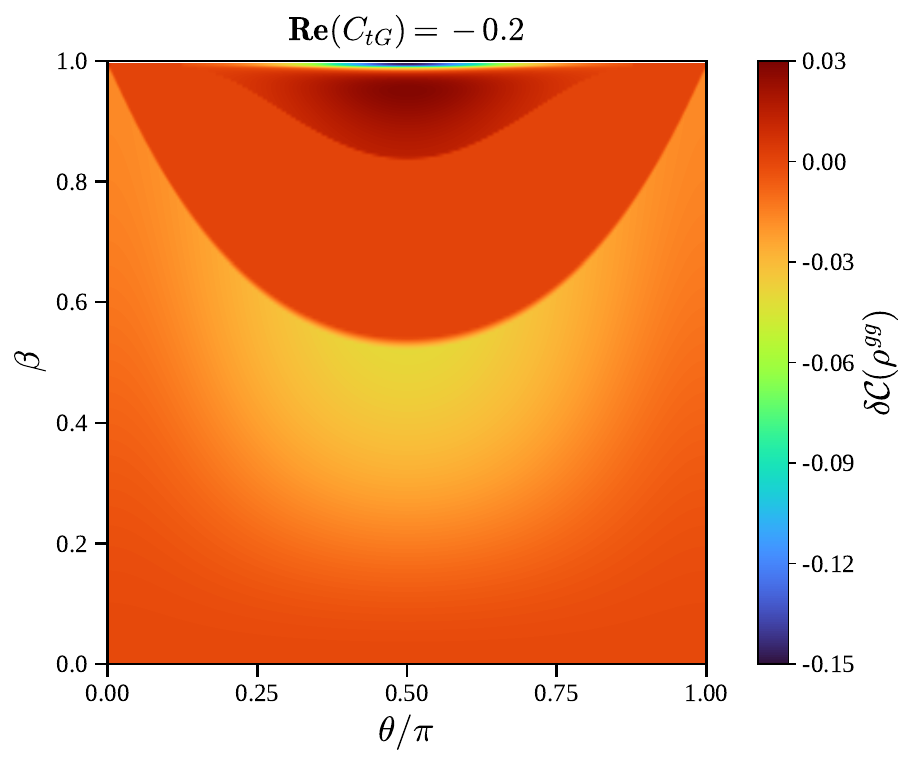}
\includegraphics[scale=0.335]{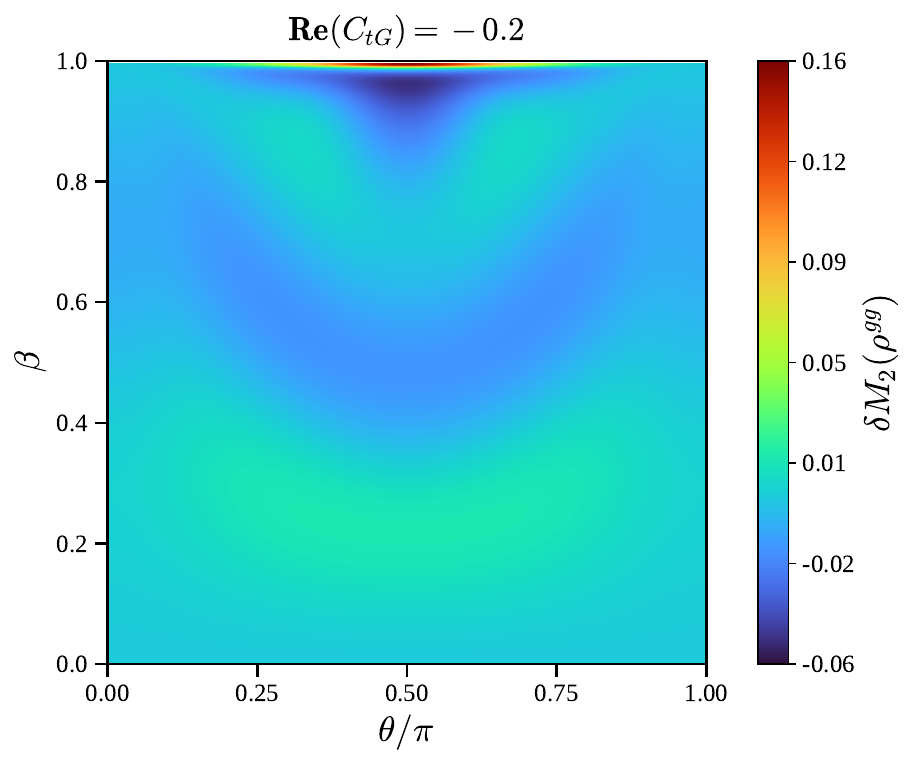}
\caption{SM predictions and SMEFT-induced changes in the quantum-information observables as functions of the scattering angle $\theta/\pi$ and the top-quark velocity $\beta$. The top row shows the SM predictions, while the bottom row shows the corresponding changes induced by the CP-even chromomagnetic dipole coefficient $\operatorname{Re}(C_{tG})=-0.2$, with all other SMEFT coefficients set to zero. From left to right, the panels display the discord, concurrence, and magic in the top row, and their deviations from the SM predictions, $\delta\mathcal D$, $\delta\mathcal C$, and $\delta M_2$, in the bottom row.
}
\label{fig:gg_SM_plots}
\end{figure}
\begin{figure}[t]
\centering
\includegraphics[scale=0.445]{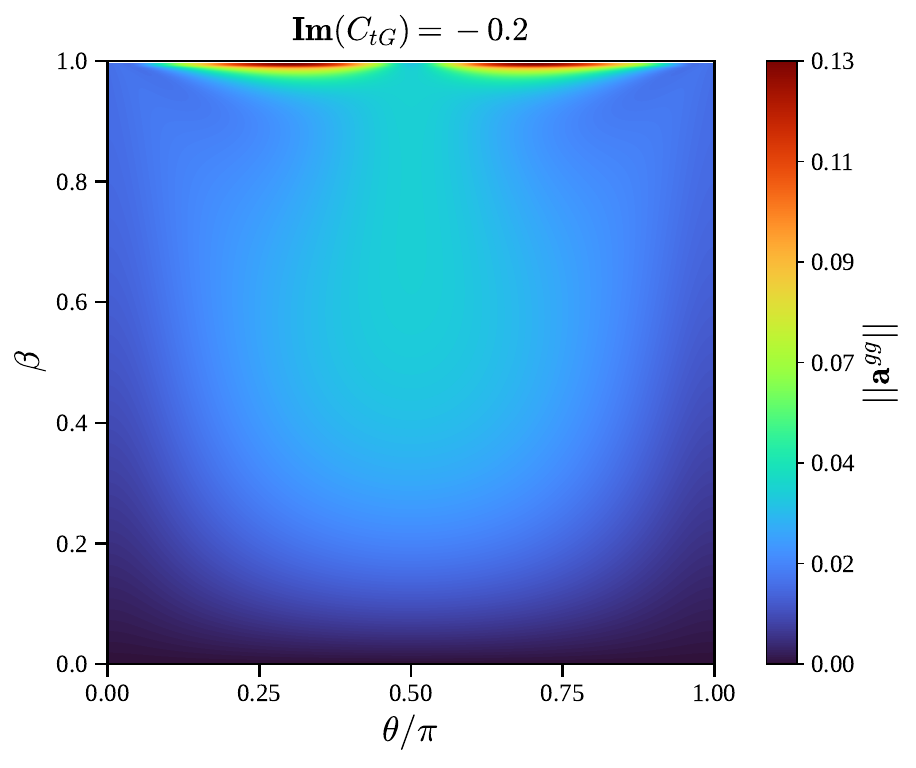}      
\includegraphics[scale=0.445]{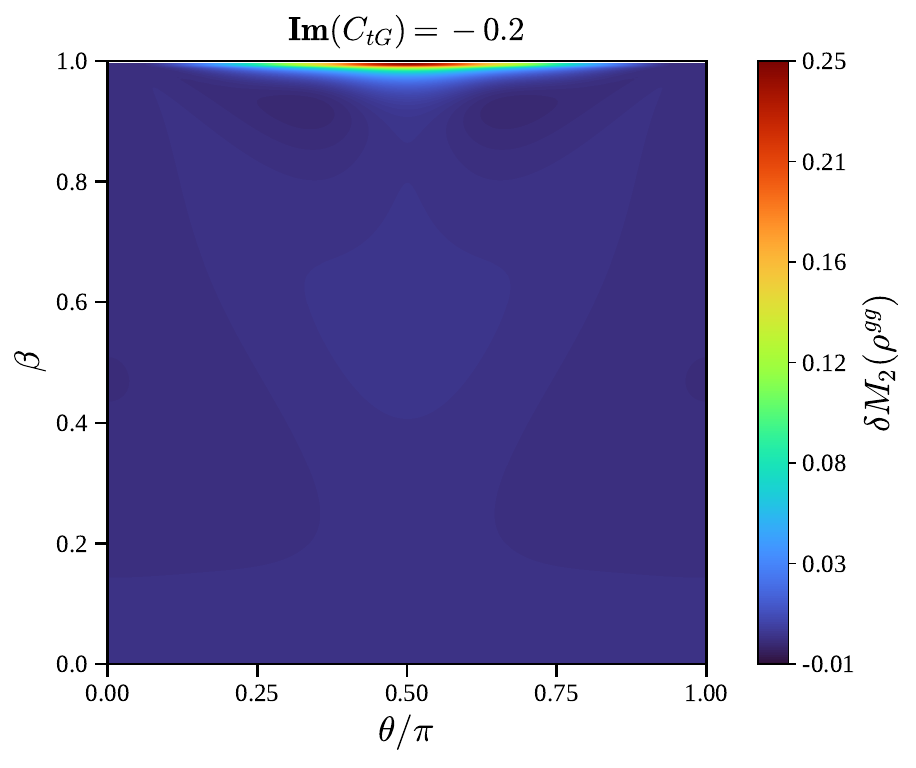}
\includegraphics[scale=0.445]{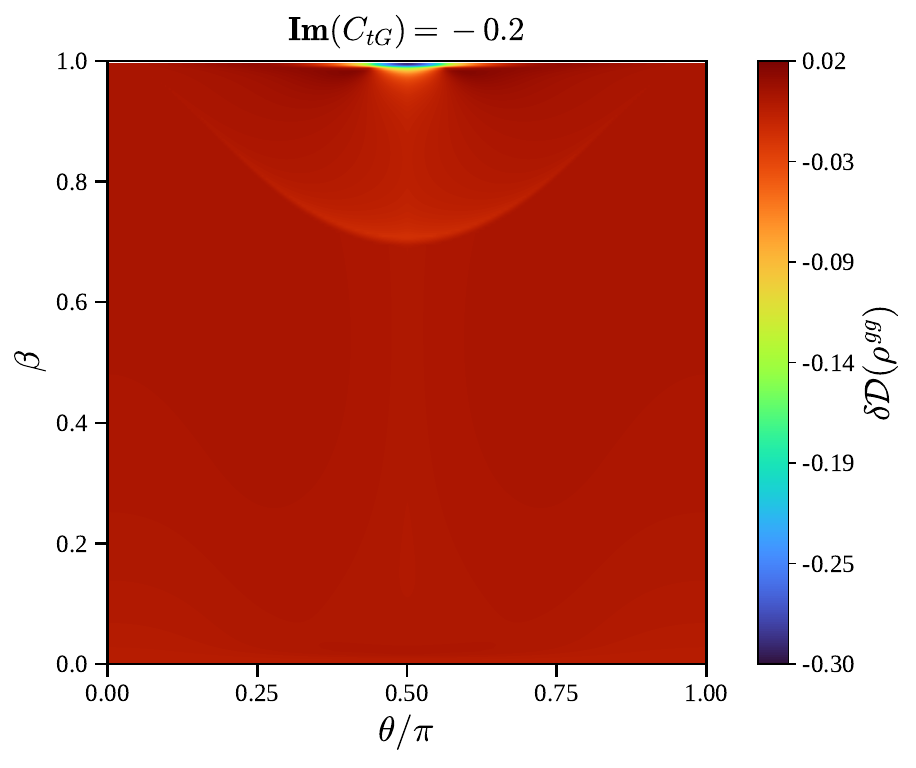}
\includegraphics[scale=0.445]{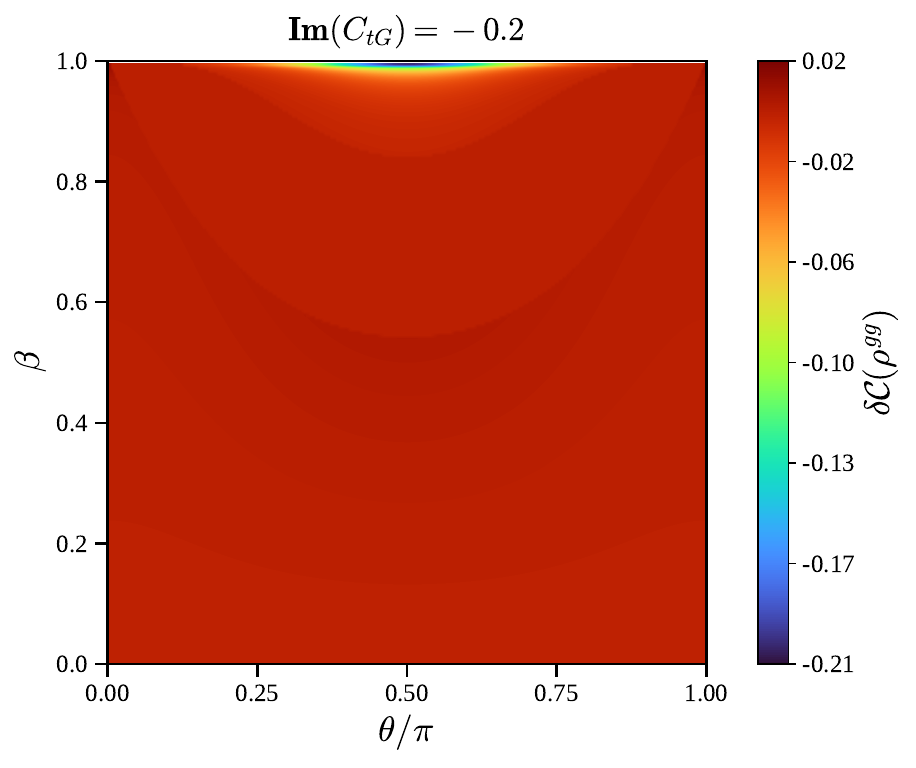}
\caption{\label{fig:gg_ImCtG}
The observables for the benchmark
$\operatorname{Im}(C_{tG})=-0.2$, with all other SMEFT coefficients set to
zero. The upper panels show, from left to right,
the norm of the antisymmetric spin-correlation matrix, $\lVert \mathbf{a}\rVert $, and the change
$\delta M_2$ relative to the SM prediction. The lower
panels show the corresponding changes in the discord,
$\delta\mathcal D$, and concurrence, $\delta\mathcal C$.
}
\end{figure}

The SM predictions are shown in Fig.~\ref{fig:gg_SM_plots}. The
distributions are identical to those obtained for
$\gamma\gamma\to t\bar t$ and discussed in Sec.~\ref{sec:gammagamatottbar}. As
shown in App.~E of Ref.~\cite{Cheng:2024btk}, the colour-symmetric and
colour-antisymmetric parts of the Standard Model $gg\to t\bar t$ amplitude
are proportional to the photon-fusion amplitude at fixed
$(\beta,\theta)$. After summing over the gluon colours, the resulting common
factor which depends on the couplings, colour factors and kinematics cancels upon normalisation, yielding
\begin{equation}
\rho_{gg}^{\rm SM}(\beta,\theta)
=
\rho_{\gamma\gamma}^{\rm SM}(\beta,\theta).
\end{equation}
Consequently, the discord, concurrence, and $M_2$ are identical
functions of $\beta$ and $\theta$ in the two Standard Model channels.

Once the QCD dipole interaction is included, the colour-symmetric part,
which is determined by the symmetric combination of the $t$- and
$u$-channel amplitudes, remains proportional to the corresponding
$\gamma\gamma\to t\bar t$ amplitude. However, this relation no longer holds
for the colour-antisymmetric part, which also receives contributions from
the non-Abelian $s$-channel and contact-interaction diagrams. The full
gluon-fusion production matrix is therefore no longer proportional to the
photon-fusion one. At linear order in the dipole coefficient their traces
remain proportional, although the full matrices, particularly their
off-diagonal entries, differ. Once the quadratic dipole contributions are
included, the proportionality is also lost at the level of the trace. This can be seen in the Fano--Bloch coefficients in App.~\ref{app: spin density matrix ggtt}.

Figure~\ref{fig:gg_SM_plots} shows the changes in the quantum information
observables induced by the CP-even benchmark
$\operatorname{Re}(C_{tG})=-0.2$. Although the corrections remain small over
most of the $(\beta,\theta)$ phase space, each observable exhibits a
characteristic structure. For the discord, $\delta\mathcal D(\rho^{gg})$, the correction is close to
zero at small and intermediate $\beta$. As $\beta$ increases, a broad
negative region develops around central production,
$\theta\simeq\pi/2$, and becomes strongly enhanced close to the endpoint $\beta=1$. This central suppression is accompanied by two narrow positive
regions located symmetrically on either side of $\theta=\pi/2$ near
$\beta=1$. The concurrence, $\delta\mathcal C(\rho^{gg})$, displays a non-monotonic
dependence on $\beta$. A small negative correction first develops around
central production at intermediate $\beta$, followed by a positive
enhancement at larger $\beta$. Very close to $\beta=1$, however, a narrow
negative region reappears around $\theta=\pi/2$. The correction remains
small in the forward, backward and threshold regions. 

The magic observable, $\delta M_2(\rho^{gg})$, receives a small
positive correction at low to intermediate $\beta$, concentrated broadly
around the central angular region. With increasing $\beta$, this contribution
turns negative near $\theta=\pi/2$ and becomes largest close to the
ultrarelativistic endpoint. At large $\beta$, two comparatively small
positive regions also emerge away from central production, symmetrically
placed on either side of $\theta=\pi/2$.

Thus, while the CP-even QCD dipole contribution is modest over most of the
phase space, due to the small benchmark value, the discord, concurrence, and $ M_2$ respond differently
in the highly relativistic region, exhibiting distinct transitions between
positive and negative corrections.

Figure~\ref{fig:gg_ImCtG} shows the effects of the CP-odd benchmark
$\operatorname{Im}(C_{tG})=-0.2$. The norm of the antisymmetric part of the
spin-correlation matrix is non-zero over a broad region of phase space. It
increases with $\beta$ and is generally enhanced around central production,
while the largest values occur close to the ultrarelativistic boundary in
two regions located symmetrically on either side of
$\theta=\pi/2$. Compared
with the $q\bar q\to t\bar t$ channel, the gluon-fusion contribution exhibits
a distinct angular pattern, particularly in the forward and backward
regions and same for $\gamma \gamma \rightarrow t \bar t$. The correction to the magic observable,
$\delta M_2(\rho^{gg})$, exhibits a qualitatively different
kinematic pattern from that induced by the CP-even coefficient. In the
CP-odd case, the effect is small over most of the phase space and is
concentrated in a narrow region close to $\beta=1$, with an enhancement
around central production. By contrast, the CP-even contribution displays
a broader sign-changing band structure extending over a larger range of
$\beta$ and $\theta$.

The corrections to the discord and concurrence are close to zero over most of
the phase space and become negative mainly at large $\beta$. For the discord,
$\delta\mathcal D(\rho^{gg})$, the dominant negative contribution is sharply
localised near $\beta=1$ around central production,
$\theta\simeq\pi/2$. Weaker negative structures extend to lower values of  $\beta$ in
two angular regions located symmetrically on either side of the central
direction. The correction to the concurrence,
$\delta\mathcal C(\rho^{gg})$, exhibits a broader negative region centred at
$\theta\simeq\pi/2$. This region develops already at intermediate $\beta$
and widens with increasing $\beta$, forming a curve. The strongest suppression again occurs close to the
ultra relativistic boundary around central production.

Overall, the CP-odd QCD dipole contribution has a limited impact on the
quantum information observables over most of the phase space, partly because
of the relatively small benchmark value considered here, and becomes
relevant mainly for highly relativistic top-quark pairs. The correction to
$M_2$ exhibits a qualitatively different kinematic pattern from the
CP-even case. Together with the non-vanishing antisymmetric
spin correlations, this provides complementary sensitivity to the CP-odd
QCD dipole interaction in the $(\beta,\theta)$ plane.


\section{Phenomenological analysis}\label{sec:pheno}

In order to assess the sensitivity of the observables discussed in the previous section to new interactions, input from experimental measurements is needed. In this section we will therefore explore the constraints that can be set on CP-even and CP-odd top interactions by considering both current and future measurements at the LHC  and a future lepton collider running above the $t\bar{t}$ threshold. Before presenting our results, we discuss in detail how the theoretical predictions, experimental measurements and projections were extracted, along with their associated uncertainties. An assessment of  different observables and a comparison of the extracted constraints  to the existing ones will also be presented. 
\subsection{LHC sensitivity}\label{LHCpheno}

We will first consider the sensitivity of proton-proton collisions at the LHC at 13 TeV to the QCD dipole interactions. The analytical structure of the spin density matrix for the leading-order partonic scattering processes ($gg$ and $q\bar{q}$) has been computed as an  expansion in the EFT parameter $\Lambda$ (see Apps.~\ref{app: spin density matrix qqtt} and ~\ref{app: spin density matrix ggtt}).  These  ingredients need to be combined with the information of the proton PDFs to extract the spin density matrix for $t\bar{t}$ pairs originating from proton-proton collisions.
Therefore, to perform our sensitivity projections and studies, we used Monte Carlo simulations performed with {\sc MadGraph5$\_$aMC@NLO} \cite{Alwall:2014hca}, recently extended with a framework dedicated to the computation of spin density matrices \cite{Durupt:2025wuk}. 

The primary goal of the simulations is to reconstruct each Fano--Bloch coefficient~\eqref{eq:normalised_fano} with its EFT expansion, defining therefore the analogues of the coefficients in App.~\ref{app:analyticalexpressions} (and their ratio) for $pp$ collisions at the LHC at $\sqrt{s}= 13$ TeV. 
Our Monte Carlo simulations have been performed at leading-order in QCD and our predictions for each Fano--Bloch coefficient have been compared with the experimental measurements presented by the CMS collaboration \cite{CMS:2024zkc}. A $\chi^2$ test is performed to investigate their sensitivity to $O_{tG}$ by deriving the bounds set by each measured observable on the real and imaginary part of its Wilson coefficient $C_{tG}$. The results obtained with this method are collected under the definition ``\textit{current measurements}".

We also extend our study with the same philosophy to what we define the ``\textit{combined measurements}'', where the observables have been obtained by combining the results presented in Ref. \cite{CMS:2024zkc}. In particular, in this class we include the length of the vector identifying the antisymmetric part of the spin-correlation matrix  $\lVert\mathbf{a}\rVert$ and the measure of magic presented in Ref. \cite{Afik:2026pxv}. 

Finally, we include one last class of observables in our sensitivity study, that groups all the quantum observables for which LHC measurements have not yet been presented  (``\textit{projections}''). One representative element of this class is the concurrence, for which the analytical structure is not known in the case where CP violation is present in $t\bt$ production. For this class we can only perform  a projection, assuming that a measurement could be performed in the near future with a $5\%$ relative error. For completeness, we also include a projected result for magic, to compare it with concurrence. 

The procedure for the construction of the $\chi^2$ test  differs slightly for the combined measurements and the projections, and it will be discussed in the relevant section. In building the $\chi^2$ functions throughout this chapter, we made the fundamental simplifying assumption that the uncertainties of the measured observables are uncorrelated.
\subsubsection{Reconstruction of the Fano--Bloch coefficients via Monte Carlo simulations}\label{sec:phenofano}

Before presenting the results of our sensitivity study, we briefly discuss how the predictions for the various observables are extracted through the Monte Carlo simulations. We start by noting that if we allow for the presence of at most one insertion of the $O_{tG}$ operator in the partonic scattering processes occurring in $pp\rightarrow t\bt$ collisions, defining for simplicity $c_{r(i)}$ the real (imaginary) part of the Wilson coefficient $C_{tG}$, we can decompose the production matrix in a given basis (that for us will be the spin basis) as: 
\begin{equation}
    R^{pp}= R^{SM}+ \sum_{m=r, i}c_m R^{m} + \sum_{m, n= i, r}c_mc_nR^{mn}.
\end{equation}
Unlike in the previous sections, where the upper index of $R$ defined the initial state producing the $t\bt$ pair, now $m, n$ mark the different EFT contributions to $R^{pp}$, with $R^m\sim1/\Lambda^2$ and $R^{mn}\sim1/\Lambda^4$. Upon normalising the production matrix, we get the structure of each Fano--Bloch coefficient in the same decomposition. The elements of the spin-correlation matrix, for example, can be decomposed as in Ref. \cite{Maltoni:2024tul, Severi:2022qjy}: 
\begin{equation}
    C_{ab}(c_r,c_i)=  \frac{\displaystyle \sigma_{C_{ab}}^{SM}+ \sum_{m=r,i}c_m\sigma^{m}_{C_{ab}}+\sum_{m,n=r,i}c_mc_n\sigma^{mn}_{C_{ab}} }{\sigma^{SM}+c_r\sigma^{r}+c^2_r\sigma^{rr}+c^2_i\sigma^{ii}}.
    \label{eq:chisq_pptt}
\end{equation}
Each term of the numerator and denominator of Eq.~\eqref{eq:chisq_pptt} can be individually extracted from the simulations, either directly or, for the terms linear in $c_i$, by performing appropriate subtractions between different runs.
The individual polarisations $B_a(c_r, c_i), \bar{B}_b(c_r, c_i)$ can be extracted in a similar manner.

Events were simulated using the model {\sc dim6top$\_$LO$\_$UFO$\_$each$\_$coupling$\_$order} \cite{Aguilar-Saavedra:2018ksv}, setting the value of $\Lambda=1$ TeV and the centre-of-mass energy to $\sqrt{s}=13$ TeV. The PDF selected from the {\sc LHAPDF}  \cite{Buckley:2014ana} was {\sc NNPDF31$\_$nlo$\_$as$\_$0118}, with the renormalisation and factorisation scales set dynamically for each event to the sum of the transverse mass divided by 4.

Following Ref.~\cite{CMS:2024zkc}, we defined four regions in terms of the invariant mass of the $t\bt$ pair ($m_{t\bt}$), each one additionally divided into three subregions classified according to the cosine of the partonic scattering angle $\theta$ ($|\cos\theta|$): 
\begin{equation}
    m_{t\bt}(\text{GeV})\in \{\, [300, 400],\,  [400, 600],\,  [600, 800],\,  > 800\, \}
\end{equation}
 \begin{equation}
    |\cos\theta|\in\{[0, 0.4], [0.4, 0.7], [0.7, 1]\}.
\end{equation}
Those quantities defined our binning (``$m_{t\bt} \rm{\  vs\  } |\cos\theta| $"), where each observable was determined.

The density matrix was extracted for each event in the $t\bt$ ZMF expressed in the top-quark helicity basis $\{\hat{n}, \hat{r}, \hat{k}\}$, using  the code developed in Ref. \cite{Durupt:2025wuk}. We then applied the necessary modifications described in the appendices of \cite{Durupt:2025wuk} as well as those needed to adapt the result to the redefinition of the helicity basis exploited by the CMS collaboration \cite{CMS:2024zkc}: 
 \begin{equation}
     \{\hat{n}, \hat{r}, \hat{k}\}\rightarrow\{\text{sgn}(\cos\theta)\hat{n}, \, \text{sgn}(\cos\theta)\hat{r},\,  \hat{k}\}.
 \end{equation}

We note again that all our Monte Carlo simulations were performed at LO in QCD, providing therefore a physical description of the $t\bt$ system where the two particles are not individually polarised and where the symmetric pattern of the spin-correlation matrix is aligned with Eqs.~\eqref{eq:symmetries_FANO_qq},~\eqref{eq:symmetries_FANO_gg}.\footnote{As a consequence, any observable defined by the individual or combined polarisation vectors did not provide any information in our LHC analysis.} Where possible, when comparing theory predictions and experimental measurements, we employ the best available prediction for the SM contributions, as described in the following section. 
\subsubsection{Current Measurements}\label{sec:currentmeasurement}

\begin{figure}[h]
\centering
\includegraphics[scale=0.55]{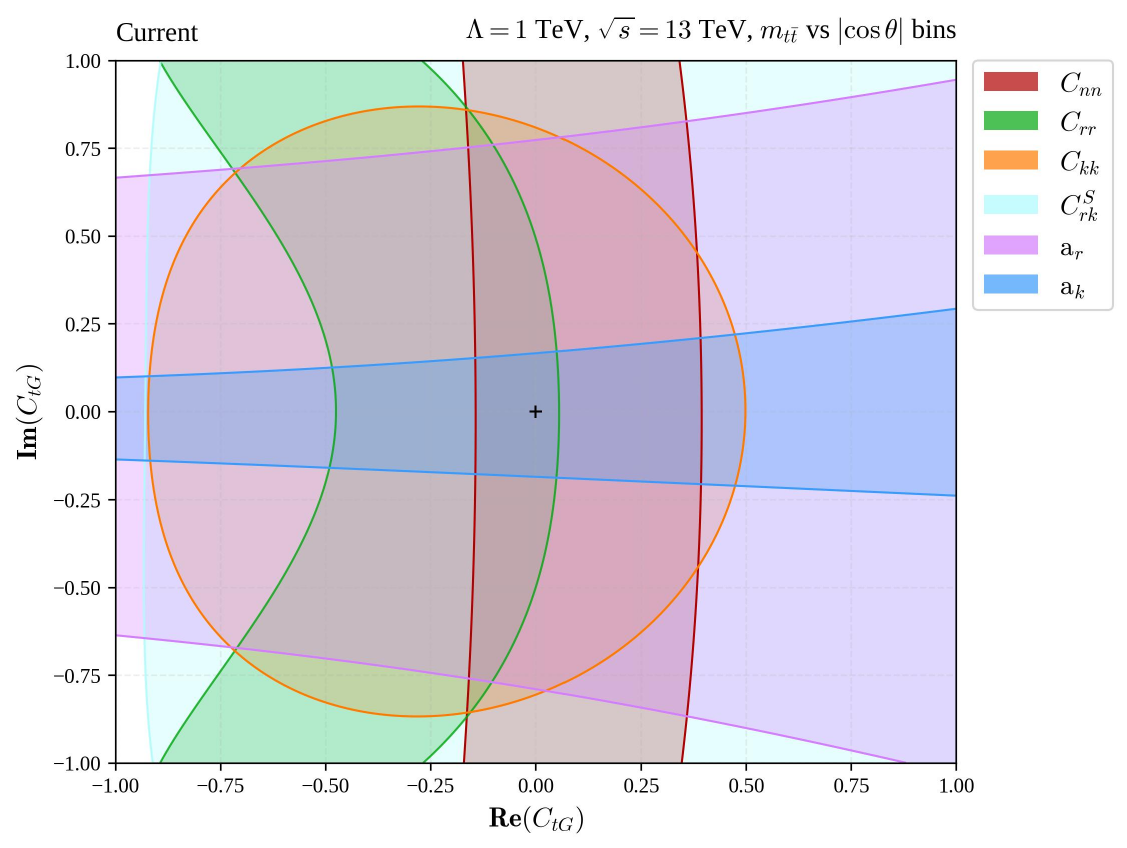}
\caption{2$\sigma$ bounds in the plane of the real and imaginary components of $C_{tG}$, derived by the observables defined as ``\textit{current measurements}", namely the symmetry-defined components of the spin-correlation matrix. The coloured domains represent the regions of the parameter space allowed by the $\chi^2$ fit performed for the corresponding observables. Each fit was realised comparing our reconstructed prediction with the measurement performed by the CMS collaboration \cite{CMS:2024zkc}.}
\label{fig:LHC-current}
\end{figure}

We start by analysing the bounds that can be provided by the Fano--Bloch coefficients, which form the building blocks of the quantum state and allow access to the quantum observables. As we previously mentioned, the defined symmetry structure of the spin-correlation matrix arising from the LO $pp$ scattering suggests the redefinition of the spin-correlation matrix in terms of its symmetric and antisymmetric components, that can be directly compared with the measurements performed by the CMS collaboration \cite{CMS:2024zkc}. This procedure also allows us to define as observables the individual elements of the vector $\mathbf{a}$~\eqref{eq:a vector}, which are direct probes of CP violation.

In Fig.~\ref{fig:LHC-current} we show $2\sigma$ bounds derived from the $\chi^2$ test performed on the set of relevant observables that enter the class of ``\textit{current measurements}". 
To reduce the possibility of misinterpreting  the effect of missing higher-order SM corrections as new physics, in our predictions for each $C_{ab}(c_r, c_i)$ we replaced our $C^{SM}_{ab}$ by the corresponding NNLO QCD prediction obtained within the {\sc MiNNLO+P8} framework as reported by the CMS collaboration  \cite{hepdata.153301}. The sign of each element of the spin-correlation matrix was adequately converted to switch from CMS convention to ours.

As theoretically expected, the globally tightest bound on $\operatorname{Im}(C_{tG})$ is offered by the antisymmetric part of the spin-correlation matrix, which, on the contrary, does not show any sensitivity to $\operatorname{Re}(C_{tG})$. In particular, the most promising observable that allows to bound the effects of CP violation introduced by the imaginary part of the Wilson coefficients is $\mathrm{a}_k$, that is able to define a confidence region that is smaller compared to the experimental bound reported in Tab.~\ref{table:SMEFT limits}. Considering the other Fano--Bloch coefficients, we notice that the most constraining for $\operatorname{Im}(C_{tG})$ is $C_{kk}$, whilst $C_{nn}$ and $C_{rr}$ show better sensitivity to $\operatorname{Re}(C_{tG})$. A proper combination of all Fano--Bloch coefficients, also taking into account the correlations between the experimental uncertainties, is crucial to further reduce the allowed parameter space.  
\subsubsection{Combined measurements and projections}\label{sec:combinedmeasure}

\begin{figure}[h]
\centering
\includegraphics[scale=0.55]{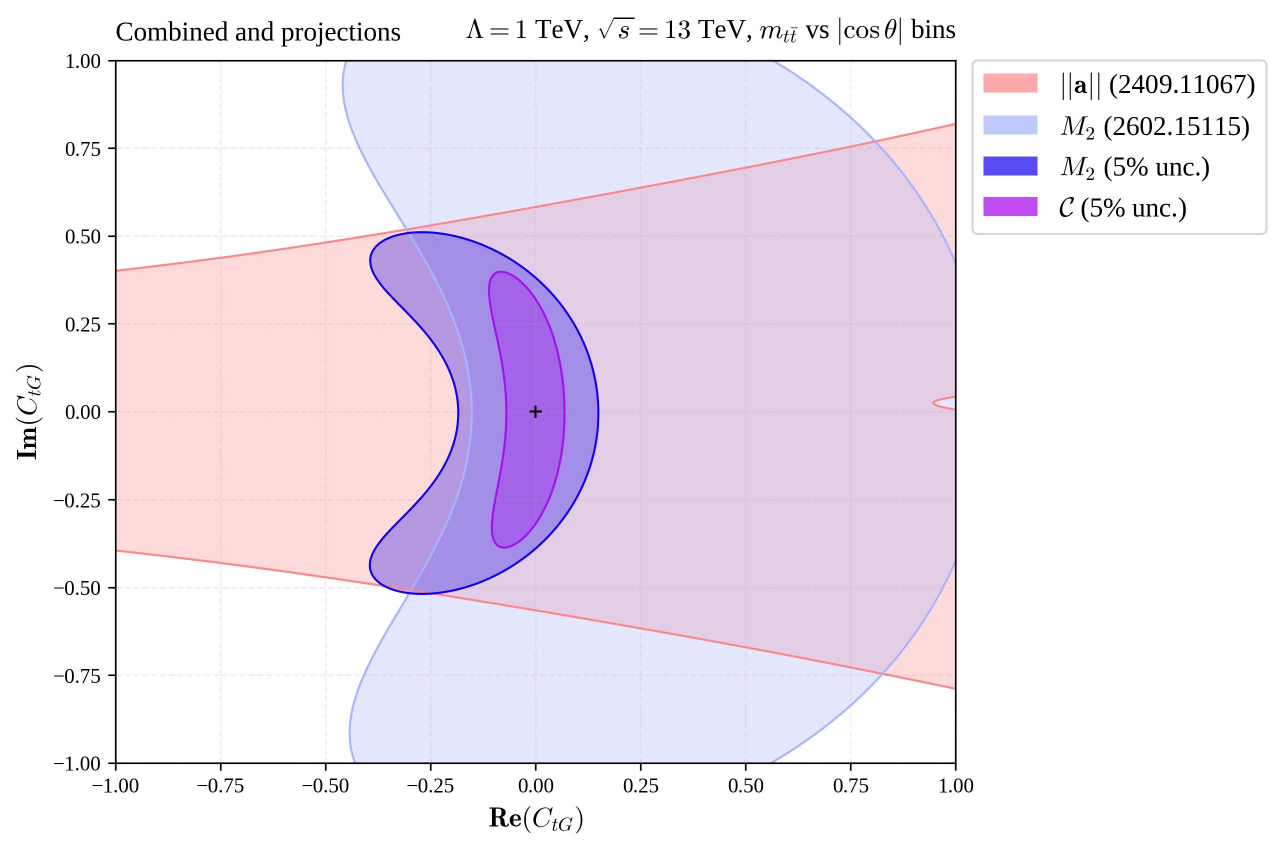}
\caption{2$\sigma$ bounds in the plane of the real and imaginary components of $C_{tG}$, derived by the observables defined as ``\textit{combined measurements}" and ``\textit{projections}". The coloured domains represent the regions of the parameter space allowed by the $\chi^2$ fit performed for the corresponding observables. For $\lVert\mathbf{a}\rVert$, our predictions have been compared with the experimental values that can be derived from Ref.~\cite{CMS:2024zkc}. For $M_2$, we show the results obtained by comparing the predictions both with the experimental data from Ref.~\cite{Afik:2026pxv} and with the SM predicted values, for which a $5\%$ total uncertainty is assumed. For the concurrence, only the latter method was used, and we included in the fit only the contribution from those bins where $\mathcal{C}_{SM}\neq0$.}
\label{fig:LHC-projections}
\end{figure}
A second set of constraints can be derived by combining the information carried by the Fano--Bloch coefficients to build the physical density matrix of $t\bt$ pairs, from which the quantum observables can be evaluated.

Before providing illustrative examples of how these observables could yield bounds on the space of the components of $C_{tG}$, it is natural to focus on the CP-sensitive norm $\lVert\mathbf{a}\rVert$. The definition of this quantity and its associated fit do not require, in principle, the construction of the four-dimensional density matrix, constrained by its physicality requirements. The norm of the vector can be evaluated combining the single components $\mathrm{a}_i$, individually defined following the procedure highlighted in the previous section, with the SM prediction for $C_{ij}$ set by the {\sc MiNNLO+P8} results \cite{hepdata.153301} and the experimental measurements as  recorded in Ref.~\cite{CMS:2024zkc}. The uncertainty was derived from the experimentally reported uncertainties by applying the rule of error propagation, naively neglecting any correlation among them.
The $2\sigma$ bound defined by $\lVert\mathbf{a}\rVert$ is shown in Fig.~\ref{fig:LHC-projections}: comparing it with the results shown in Fig.~\ref{fig:LHC-current}, we see that the bound set by the quantity $\mathrm{a}_k$ is stronger compared to the one provided by $\lVert\mathbf{a}\rVert$. The latter, indeed, combines the constraining information of $\mathrm{a}_k$ with the other two less sensitive components, e.g. with $\mathrm{a}_n$, which carries no sensitivity, effectively diluting the constraining power. 

What is finally the sensitivity offered by the quantum observables? Addressing this question is a hard task, since  most of them still lack a direct experimental measurement. Our goal is to illustrate the projected bounds set by two representative observables, assuming that a possible measurement would align with the SM value and would be defined with a fixed relative uncertainty of $5\%$. The first observable that we selected is magic,  for which, indeed, a differential measurement at the LHC was extracted in Ref.~\cite{Afik:2026pxv}, and that therefore offers the possibility to compare our projection method with the canonical one, involving real data. 

The second observable is concurrence, as a representative entanglement marker. Although several measurements of the entanglement markers D have been performed \cite{ATLAS:2023fsd}, a comprehensive measurement of concurrence is still lacking. This is crucial, especially in a scenario where CP-violating interactions can yield a spin-correlation matrix with modified symmetry properties. Therefore, it is rather informative to consider concurrence as an observable and establish its sensitivity to the CP-odd interactions. 

The extraction of the quantum observables involves the reconstruction of the density matrix in each bin, obtained according to definition~\eqref{eq:rho2_qbit}: 
\begin{equation}
\rho(c_r, c_i) = \frac{1}{4}\left[
\mathbb{I}\otimes\mathbb{I}
+ B_i(c_r, c_i)\,\sigma^i\otimes\mathbb{I}
+ \bar{B}_j(c_r, c_i)\,\mathbb{I}\otimes\sigma^j
+ C_{ij}(c_r, c_i)\,\sigma^i\otimes\sigma^j
\right]. 
\label{eq:rho2_qbit-RECO}
\end{equation}
This step, that could in principle be avoided for magic, is necessary for the concurrence, for which an analytic formula is not known for  our physical scenario where CP can be violated. The same complexity  applies also in assigning an uncertainty for the quantum observables. Whilst uncertainties can be propagated for magic, for which the structure is known (as done in Ref.~\cite{Afik:2026pxv}), for comparison purposes we also show results assigning a fixed relative uncertainty to both $M_2$ and $\mathcal{C}$. 

To guarantee the physicality of the reconstructed density matrix in Eq.~\eqref{eq:rho2_qbit-RECO}, we derived the structure of each coefficient at LO in QCD according to Eq.~\eqref{eq:chisq_pptt}, maintaining our LO predictions also for the SM terms. For this scenario, while the unit value of $\Tr[\rho(c_r, c_i)]$ was guaranteed by definition, we confirmed the positivity of its eigenvalues. 

The $2\sigma$ confidence regions defined by $M_2$ using
experimental data and those derived using projections for $M_2$ and $\mathcal{C}$ are depicted in Fig.~\ref{fig:LHC-projections}. We first note that our SM LO prediction for magic agrees within the $2\sigma$ level with the experimentally extracted value. For $M_2$, we find the projection to give a significantly stronger bound, as the 5\% relative uncertainty applied to each bin is significantly lower than some of the experimental ones in Ref.~\cite{Afik:2026pxv}.  

The tightest bound from the observables in Fig.~\ref{fig:LHC-projections} is provided by the concurrence, that also overcomes the limits set by the spin-correlation matrix for $\operatorname{Re}(C_{tG})$ in Fig.~\ref{fig:LHC-current}. If we only focus on the experimentally  measured quantities, the overall best bound on $\operatorname{Im}(C_{tG})$ remain the one set by $\mathrm{a}_k$. 

We conclude this section  by stressing again that a consistent combination of all measured observables is expected to provide the best constraints, but such a combination is beyond the scope of this work. 
\subsection{Lepton collisions at $\sqrt{s}=365$ GeV}\label{sec:phecnoee}

In addition to the LHC measurements it is particularly interesting to consider future lepton colliders. As we have discussed in the previous sections, the fundamentally different production mode in lepton collisions provides new possibilities in exploring top interactions. We will focus on a lepton collider running at 365 GeV, as envisioned for the FCC-ee \cite{FCC:2025lpp, FCC:2025uan}.

We restrict the phenomenological analysis to a scenario in which
$O_{tB}$ is the only retained operator, which modifies $t\bar t$ production
without affecting the $Wtb$ decay vertex. By contrast, $O_{tW}$ affects
both production and decay and modifies the spin-analysing powers of the
final-state particles~\cite{Aguilar-Saavedra:2010ljg,Severi:2022qjy,
Lamba:2026sni}. The disentanglement of new physics in production and decay,
including its CP character, is addressed in the companion
paper~\cite{Lamba:2026sni}.

Allowing the investigation of the electroweak dipole, an electron--positron collider is complementary to the LHC. In addition, the parity-violating electroweak interaction produces non-trivially polarised top quarks already in the SM, and CP violation effects will induce the additional asymmetry vector $\Delta \mathbf{B}$, that is the counterpart of $\mathbf{a}$.

 The framework we will consider in this section is the FCC-ee proposal  running scenario, with a pair of unpolarised electron-positron beams with an energy of $\sqrt{s}=365$ GeV and an expected integrated luminosity of $\mathcal{L}=3.12$ ab$^{-1}$ \cite{Selvaggi:2025kmd, Armadillo:2026mvp}. As no experimental  measurement is available,  all predictions have been based exclusively on theoretical computations, both for the production and the decay of the $t\bt$ system. In contrast with proton-proton collisions, for a lepton collider and  the tree-level scattering process $e^+e^-\rightarrow t \bt$, the observables can be directly derived from the structure of $\rho^{ee}=R^{ee}/\Tr[R^{ee}]$ in App.~\ref{app: spin density matrix eett}.

  Similarly to the LHC analysis of the previous section, we defined two classes of observables: the first  consists of the individual Fano--Bloch coefficients and their algebraic combinations, which are the quantities most directly reconstructed experimentally. The second class comprises the CP-sensitive combinations, $\lVert\Delta \mathbf{B}\rVert$ and $ \lVert\mathbf{a}\rVert$, and two representative quantum observables, $M_2$ and $\mathcal{C}.$
  We assumed that possible measurements on any of those observables would align with the SM prediction, with an associated uncertainty that varies depending on the type of the observable. 
  
  For $M_2$ and $\mathcal{C}$, we assumed a fixed relative uncertainty of $5\%$, following the ideas presented for the LHC analysis. For the remaining observables, we employed the experimental scenario where spin polarisations and correlations for $t\bt$ pairs could be measured in two final-state channels (F=$f_1f_2$), the dilepton channel ($\ell\ell$) and the lepton+jet channel ($\ell j$), with $\ell=e, \mu$. We assumed that the measurement was performed using  a tomography procedure, based on the angular distribution of the final state fermions $f_1, f_2$ \cite{Baumgart:2012ay, Tweedie:2014yda, Bernreuther:2015yna}, a procedure that  actively depends on the spin-analysing power of those particles. 
  
  Since we are not introducing any modification to the decay vertices of the top quarks, the spin-analysing power of the lepton produced by the top quark decay has been set to its maximum value $\alpha_{\ell}=1$. For the hadronically decaying top quarks, we considered the weighted sum of the two light-quark directions that define the optimal hadronic polarimeter \cite{Tweedie:2014yda}, yielding a spin-analysing power of $\alpha_{j}=0.64$. For each event, we assumed a global reconstruction efficiency of $\epsilon_r=40\%$, as adopted in previous literature \cite{Maltoni:2024csn}. 
  
  Finally, for comparison purposes, we included among the observables the expected number of reconstructed events:
  \begin{equation}
        N^{\text{F}}= \sigma\cdot\mathcal{L}\cdot\epsilon_{r}\cdot Br(t\bt\rightarrow F).
        \label{eq:PHENO_nevens}
  \end{equation} 

  Within this framework, we assigned a statistical uncertainty to the individual Fano--Bloch coefficients following Refs.~\cite{Tweedie:2014yda, Maltoni:2024csn, Cao:2025xnp}:
 \begin{align}
      &\delta B_i^{f_1,\text{F}}\sim \delta \bar{B}_j^{f_1,\text{F}}\sim\frac{1}{|\alpha_{f_1}|}\sqrt{\frac{3}{N_{\text{NLO}}^\text{F}}} &\delta C^{\text{F}}_{ij}\sim\frac{3}{|\alpha_{f_1}\alpha_{f_2}|}\sqrt{\frac{1}{N_{\text{NLO}}^\text{F}}},
      \label{eq:stst_unc_pheno_eett}
 \end{align}
where $f_i$ indicates the particle that is selected as analyser for the decay of the (anti)top quark, and $N_{\text{NLO}}^\text{F}$ is the expected number of events in each channel by approximating the cross section to its SM NLO QCD value $\sigma^{\text{SM}}_{\text{NLO}}\sim 750$ fb \cite{Maltoni:2024csn}. For the number of events, we assumed both a statistical and a relative $1\%$ systematic uncertainty, with:
\begin{equation}
    \delta N^{\text{F}}_{\text{stat}}\sim \sqrt{N^{\text{F}}_{\text{NLO}}}.
\end{equation} For the Fano--Bloch coefficients, no  systematic uncertainty was included in our computation: for the observables of interest, the error that we would get by assuming a $1\%$ systematic uncertainty on the individual coefficients was numerically negligible compared to the statistical one. 
 For the algebraic combinations of Fano coefficients, we adopted the rule of error propagation to define the statistical uncertainty and, as we did for the LHC analysis, we naively assumed that no correlations exist between any of our observables.

\begin{figure}[]
\centering
\includegraphics[scale=0.55]{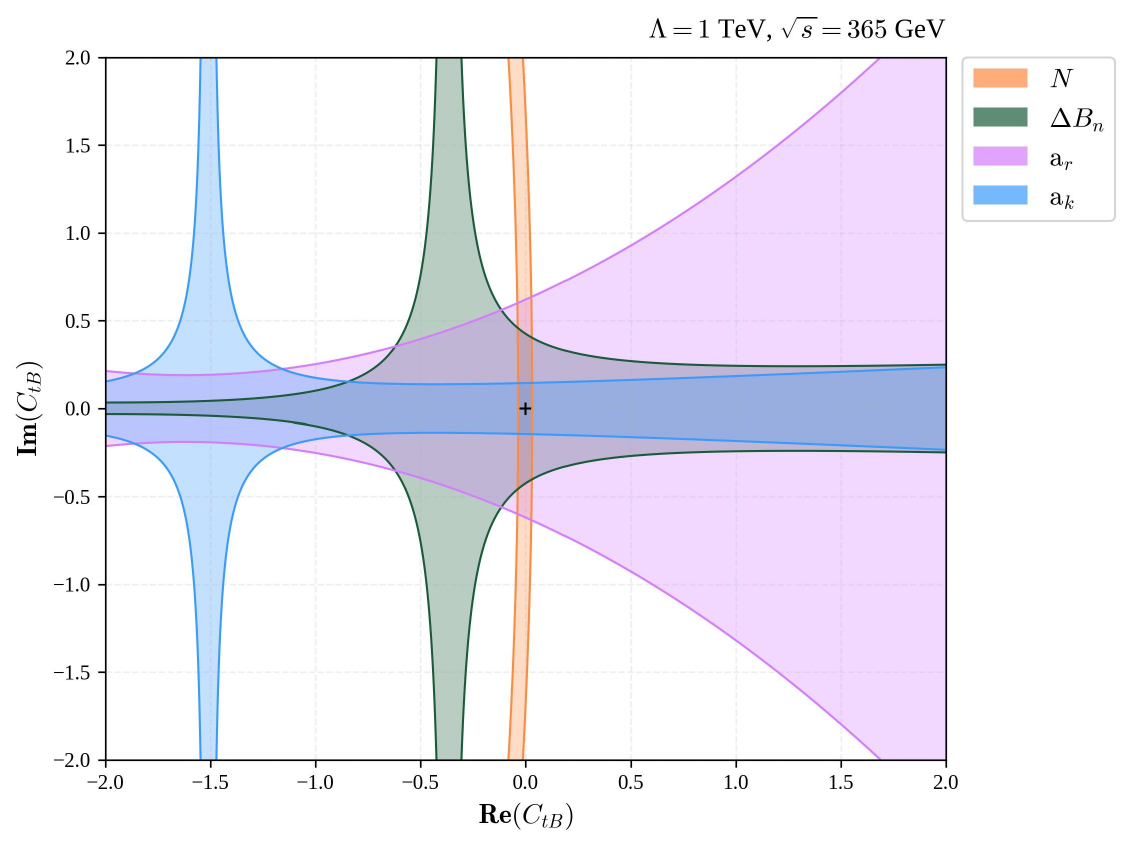}
\caption{2$\sigma$ bounds in the plane of the real and imaginary components of $C_{tB}$, derived by the CP-sensitive combinations of Fano--Bloch coefficients and by the number of events $N$. The coloured domains represent the regions of the parameter space allowed by the $\chi^2$ fit for the corresponding observables. Each fit was realised using our theoretical prediction for each observable, assuming that the measured value would align with the SM prediction.}
\label{fig:ee-asymmetries}
\end{figure}

\begin{figure}[]
\centering
\includegraphics[scale=0.55]{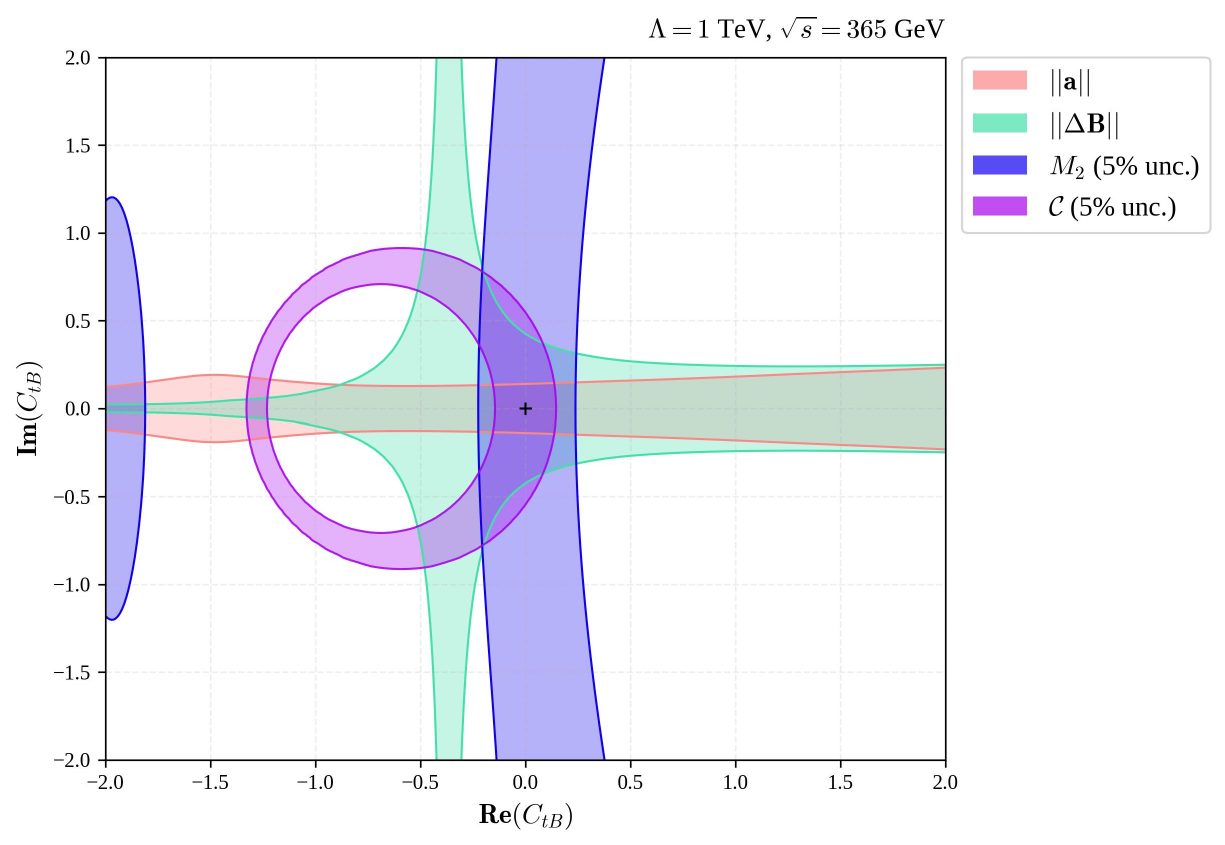}
\caption{2$\sigma$ bounds in the plane of the real and imaginary components of $C_{tB}$, derived by the CP-sensitive lengths $\lVert\Delta\mathbf{B}\rVert, \lVert\mathbf{a}\rVert$ and the quantum observables $M_2, \mathcal{C}$. The coloured domains represent the regions of the parameter space allowed by the $\chi^2$ fit for the corresponding observables. Each fit was realised using our theoretical prediction for each observable, assuming that the measured value would align with the SM prediction.}
\label{fig:ee-combproj}
\end{figure}

In Figs.~\ref{fig:ee-asymmetries} and~\ref{fig:ee-combproj} we show the results for the $\chi^2$ fit for each observable. For those that require experimental reconstruction, we combined the contributions from the two final-state channels F, assuming the absence of correlations between any performed measurements. 
In Fig.~\ref{fig:ee-asymmetries}  we show the $2\sigma$ confidence regions allowed by each CP-sensitive combination of Fano coefficients: we decided to restrict the analysis only to those combinations that would show a better sensitivity to $\operatorname{Im}(C_{tB})$. Each Fano--Bloch coefficient, here integrated over the production angle, has been derived from the unnormalised version presented in App.~\ref{app: spin density matrix eett} following the usual definition in Eq.~\eqref{eq:normalised_fano}: 
\begin{equation}
    B^{}_i= \frac{\int d\cos\theta\  \widetilde{B}^{}_i}{\int d\cos\theta \ \widetilde{A}}, \qquad \bar B^{}_j= \frac{\int d\cos\theta\  \widetilde{\bar B}^{}_j}{\int d\cos\theta \ \widetilde{A}}, \qquad C_{ij}= \frac{\int d\cos\theta \ \widetilde{C}_{ij}}{\int d\cos\theta \  \widetilde{A}}.
\end{equation}

The $2\sigma$ confidence region defined by the number of events $N$ is also presented in Fig.~\ref{fig:ee-asymmetries}, where the cross section appearing in Eq.~\eqref{eq:PHENO_nevens} was computed from $R^{ee}$, replacing our SM prediction with the higher-order value $\sigma^{\text{SM}}_{\text{NLO}}$: 
\begin{equation}
    \sigma= \sigma^{\text{SM}}_{\text{NLO}}+\sigma^{\text{EFT}}_{\text{LO}}.
\end{equation}
In the range $\operatorname{Re}(C_{tB})\in [-1, 1]$, we find that the globally best constraint on the imaginary component of the Wilson coefficient is provided by $\mathrm{a}_k$, following the same pattern observed for the LHC scenario. In this region the bounds are even tighter than the experimentally allowed values. As expected, none of the selected combinations $\mathrm{a}_k,\ \mathrm{a}_r $ and $ \Delta B_n$ show sensitivity to $\operatorname{Re}(C_{tB})$.

We observe that $\Delta B_n$ and $\mathrm{a}_k$ exhibit a peculiar behaviour around
$\operatorname{Re}(C_{tB})\simeq -0.35$ and
$\operatorname{Re}(C_{tB})\simeq -1.5$, respectively, where their sensitivity to
$\operatorname{Im}(C_{tB})$ is lost. This follows from the analytic structure of
the observables: near these values, the contributions to the numerator at
$\mathcal{O}(\Lambda^{-2})$ and $\mathcal{O}(\Lambda^{-4})$ become comparable in
magnitude and opposite in sign, leading to an approximate cancellation.\footnote{The dependence of $\Delta B_n$ and $\mathrm{a}_k$ on the real and
imaginary components of the Wilson coefficient $C_{tB}$ is given by: \[
\Delta B_n= \frac{\ \operatorname{Im}(C_{tB}) \cdot ( \ 1.36523 + 3.72481 \ \operatorname{Re}(C_{tB})\ )}{43.6976 + 37.4784 \ \operatorname{Re}(C_{tB}) + 11.5811 \ \operatorname{Re}(C_{tB})^2+  0.41982 \ \operatorname{Im}(C_{tB})^2 },
\]
\[ \mathrm{a}_k= -\frac{\ \operatorname{Im}(C_{tB})\cdot (\ 3.97813 +2.63678 \ \operatorname{Re}(C_{tB})\ )}{43.6976 + 37.4784 \ \operatorname{Re}(C_{tB}) + 11.5811 \ \operatorname{Re}(C_{tB})^2+  0.41982 \ \operatorname{Im}(C_{tB})^2 } .
\]}

Conversely, the number of events highly constrains $\operatorname{Re}(C_{tB})$, overcoming the current experimental limits, without significantly constraining the CP violation parameter. This is expected from the structure of the EFT contributions to the cross section, where $\operatorname{Im}(C_{tB})$ could enter only via the modulus $|C_{tB}|^2$, inducing modifications starting at the $\Lambda^{-4}$ order. 

The structure of $R^{ee}$ suggests also why the terms $\Delta B_r,\  \Delta B_k$ and $ \mathrm{a}_n$ are not able to strongly constrain the $C_{tB}$ components: these terms, that arise from the $\gamma-Z$ interference, are proportional to the imaginary part of the $Z$-boson propagator, which is expected to be suppressed for our energy scenario. The same behaviour is exhibited by the SM contribution to the normal polarisation $B_n$ \cite{Arens:1992wh}  and the spin correlation coefficients $ C_{nk}, C_{nr}$, that only arise in the presence of absorptive contributions to the scattering amplitudes  \cite{Bernreuther:2015yna}. Due to their suppressed size, the systematic uncertainty associated with these observables is negligible compared to the statistical one.

Finally, the plot in Fig.~\ref{fig:ee-combproj} shows the $2\sigma$ bounds obtained by the CP-sensitive lengths, $\lVert\Delta\mathbf{B}\rVert$ and $\lVert\mathbf{a}\rVert$, and the two quantum observables $M_2$ and $ \mathcal{C}$. As we have found for the LHC,  for these projections the bounds extracted from the concurrence on $\operatorname{Im}(C_{tB})$ are stronger than those offered by magic, that for $|\operatorname{Re}(C_{tB})| < 0.25$ is completely  insensitive to the magnitude of the imaginary part. The CP-violating  $\operatorname{Im}(C_{tB})$ is still mostly constrained by the antisymmetric part of the spin-correlation matrix, whose information is complementary to $M_2$, that conversely tightly bounds the real component of the Wilson coefficient. We note, however, the significant difference in the contour shapes observed for $M_2$ and  $\mathcal{C}$ in the LHC and lepton collider cases, which is related to the difference in the production mechanisms.  


\section{Conclusions}
\label{sec:conclusions}

In this work, we have investigated how CP violation in top-quark interactions
is encoded in the quantum state of a produced \(t\bar t\) pair. The short
lifetime of the top quark allows its spin information to be retained by its
decay products, making the \(t\bar t\) system a distinctive bipartite
spin-\(1/2\) system at colliders. Its spin density matrix, therefore, provides a
natural framework in which to connect CP-violating interactions to definite
structures in the quantum state.

For the unpolarised production processes and kinematic conventions considered
here, after expressing the top and antitop spins in a common basis and
comparing CP-related configurations, CP invariance imposes simple symmetry
relations on the Fano--Bloch coefficients. In particular, CP violation in
production is encoded in two independent structures,
\begin{equation}
    \Delta\mathbf{B}
    =
    \frac{1}{2}\left(\mathbf B-\mathbf{\bar{B}}\right),
    \qquad
   \mathbf C^{A}
    =
    \frac{1}{2}\left(\mathbf C-\mathbf C^{T}\right),
\end{equation}
corresponding respectively to a difference between the top and antitop
polarisation vectors and an antisymmetric component of the spin-correlation
matrix. These quantities identify where CP-odd information is stored in the
density matrix and provide direct production-level markers of CP violation.
The trace distance between the state and its CP transform gives a complementary
global measure of the same symmetry breaking.

We have employed the SMEFT framework to parametrise possible new sources of CP
violation in top-quark interactions. The relevant dimension-six operators have
been identified and their contributions mapped onto the anomalous couplings
entering the effective \(t\bar t\gamma\), \(t\bar t Z\), and \(t\bar t g\)
vertices, as well as onto the scalar--pseudoscalar structure of the \(t\bar t\)
Yukawa interaction.
We have then derived analytic expressions for the
production density matrix in a set of benchmark channels relevant to present
and future colliders: the decay of a spin-zero state, \(e^+e^-\to t\bar t\),
\(\gamma\gamma\to t\bar t\), and the partonic processes
\(q\bar q\to t\bar t\) and \(gg\to t\bar t\). These results provide a unified
description of the CP-even and CP-odd spin structures generated by scalar,
electroweak, photonic, and QCD interactions.

The different channels illustrate that the location of the CP-odd information
in the density matrix is strongly process dependent. In the spin-zero decay,
the produced state is pure and maximally entangled for any value of the
scalar--pseudoscalar admixture. Entanglement is therefore completely
insensitive to the CP-mixing angle in this example, whereas the antisymmetric
part of the correlation matrix directly tracks the scalar--pseudoscalar
interference. In the unpolarised photonic and QCD processes, the CP-odd
information is also predominantly carried by antisymmetric spin correlations.
By contrast, the parity-violating electroweak interaction in
\(e^+e^-\to t\bar t\) permits CP violation to appear both in the polarisation
difference \(\Delta\mathbf{B}\) and in the antisymmetric correlation matrix.

We have studied direct CP-sensitive quantities together with a broad class of
quantum information observables, including quantum discord, geometric discord,
concurrence, and magic.
Their response to CP-violating interactions is
highly observable- and channel-dependent. Quantities constructed directly from
\(\Delta\mathbf{B}\) and \(\mathbf C^{A}\) provide the most transparent and
unambiguous probes of CP violation. Global quantum observables can nevertheless
carry complementary information about the deformation of the state. In
particular, magic diagnoses changes in its non-stabilizer structure, while
concurrence and discord probe different forms of quantum correlation. These
quantities should not, however, be interpreted as direct measures of CP
violation: a CP-violating interaction need not produce a large change in them,
and a sizeable modification can also arise from CP-conserving new physics.

Our phenomenological study demonstrates how these observations can be
exploited in collider measurements. At the LHC, we reconstructed the EFT
dependence of the Fano--Bloch coefficients in \(pp\to t\bar t\) and compared
them with the differential spin-correlation measurements performed by
CMS~\cite{CMS:2024zkc}. Among the currently measured quantities, the component
\(\mathrm{a}_k\) of the antisymmetric correlation vector provides the strongest
individual constraint on \(\operatorname{Im}(C_{tG})\), improving on the
representative existing interval quoted in
Table~\ref{table:SMEFT limits}
. The symmetric coefficients
\(C_{nn}\) and \(C_{rr}\) are instead mainly sensitive to
\(\operatorname{Re}(C_{tG})\), while \(C_{kk}\) retains some sensitivity to its
imaginary part. Combining the three antisymmetric components into the norm
\(\lVert\mathbf{a}\rVert\) does not improve the constraint, because the less
sensitive components dilute the information carried by \(\mathrm{a}_k\).

Among the global quantum observables, a projected measurement of the
concurrence with a \(5\%\) relative uncertainty gives the strongest prospective
constraints and can improve the sensitivity to the CP-even component relative
to the individual spin-correlation coefficients. For the CP-odd component,
however, the best constraint based on presently available measurements remains
that obtained directly from \(\mathrm{a}_k\). Our results for magic were also
compared with its recent extraction from LHC data~\cite{Afik:2026pxv}.

At a future \(e^+e^-\) collider operating at
\(\sqrt{s}=365~\mathrm{GeV}\), as envisaged for the FCC-ee
programme,
we have studied the
electroweak dipole operator \(O_{tB}\), which modifies production without
altering the \(Wtb\) decay vertex. In the region
\(\operatorname{Re}(C_{tB})\in[-1,1]\), the antisymmetric correlation
\(\mathrm{a}_k\) again provides the strongest projected sensitivity to
\(\operatorname{Im}(C_{tB})\), reaching beyond the representative present
bounds. The total event rate is instead primarily
sensitive to \(\operatorname{Re}(C_{tB})\), since the imaginary part contributes
to the inclusive rate only quadratically in the Wilson coefficient. The
polarisation difference supplies an additional CP-sensitive structure that is
absent in the leading-order LHC analysis.

As at the LHC, a projected concurrence measurement is more sensitive to the
CP-odd direction than magic, whereas magic and the event rate provide
complementary sensitivity to the CP-even component. The different shapes of
the allowed regions at hadron and lepton colliders reflect the different
production mechanisms and demonstrate the complementarity of the two
environments.

The numerical projections presented here should be interpreted as an
illustration of the information contained in the different observables rather
than as a complete global analysis. They are based on leading-order production
predictions, one-operator scenarios, a fixed \(5\%\) relative uncertainty for
the projected quantum observables, and the simplifying assumption that the
experimental uncertainties are uncorrelated. We also retain quadratic
dimension-six contributions in order to construct positive-semidefinite
density matrices for finite values of the Wilson coefficients. These terms
should not be interpreted as a complete EFT prediction at
\(\mathcal{O}(\Lambda^{-4})\), since contributions from dimension-eight
operators at the same formal order have not been included. A statistically
consistent combination of the full set of Fano--Bloch coefficients, quantum
observables, rates, and their correlations is expected to provide the strongest
constraints.

Several extensions of this work are natural. The first is the subject of a
companion paper~\cite{Lamba:2026sni}, in which we address the tomographic
reconstruction of the \(t\bar t\) density matrix and make explicit the
connection between the production-level CP markers introduced here and
experimentally accessible angular observables in the decay products. This will
also allow production and decay contributions to CP violation to be
disentangled. A second important direction is the inclusion of QCD radiative corrections and
absorptive contributions, both to test the stability of the leading-order
density-matrix structures and to quantify their impact on collider
sensitivities~\cite{Severi:2022qjy}. Finally, it would be interesting to extend
this framework to electric and magnetic dipole interactions in other fermionic
systems, in particular \(\tau^+\tau^-\) production. Existing studies of
collider constraints on tau dipoles~\cite{Buttazzo:2026amk}, together with
tau-spin and quantum-information analyses
\cite{Rouge:2005iy,Fabbrichesi:2022ovb,Altakach:2022ywa,Ehataht:2023zzt,
Fabbrichesi:2024xtq,Han:2025ewp,Zhang:2025mmm,Altakach:2026fpl}, provide a
natural point of contact.


\section*{Acknowledgements}

We are grateful to Pasquale Calabrese for insightful conversations on quantum observables. F.M. and O.M. thank the CERN Theory Department for hospitality when the final part of this work was performed. O.M. has received partial support from the European Union’s Horizon 2020 research and innovation programme under the Marie Skłodowska-Curie Staff Exchange grant agreement No 101086085-ASYMMETRY. P.L. and F.M. are partially supported by the F.R.S.–FNRS (Belgian Fund for Scientific Research) through the IISN convention 4.4517.08, ``Theory of fundamental interactions''. P.L. also acknowledges the support by the Fonds Spécial de Recherche at UCLouvain.  E.V. is supported by the European Research Council (ERC) under the European Union’s Horizon 2020 research and innovation programme (Grant agreement No. 949451).


\appendix

\section{Details of the spin-density matrix calculation}\label{sec:appA}
\subsection{Density matrix calculation with the Bouchiat-Michel formulae} \label{app: BM formulae}

To construct the production or decay density matrix of a massive spin-$\tfrac{1}{2}$ particle, it is convenient to use spin projection operators rather than evaluate each matrix element separately. For a particle with four-momentum $p^\mu=(E,\vec p)$ and definite helicity $\lambda=\pm1$, the corresponding spin four-vector is
\begin{equation}
s^\mu=\lambda\left(\frac{|\vec p|}{m},\frac{E}{m}\hat p\right),
\qquad
\hat p=\frac{\vec p}{|\vec p|},
\label{eq: spin vector}
\end{equation}
which satisfies $s\cdot p=0$ and $s^2=-1$. The usual helicity projector then reads
\begin{equation}
u(p,\lambda)\bar u(p,\lambda)
=
\frac{1}{2}(\slashed p+m)\left(1+\gamma^5\slashed s\right),
\end{equation}
and similarly for antiparticles. This expression gives the diagonal entries of the density matrix in the helicity basis.

To retain also the off-diagonal helicity components, one uses the Bouchiat-Michel (BM) formulae from Refs.~\cite{Bouchiat:1958yui,Haber:1994pe}. They are based on three orthonormal spin four-vectors $s^{a\mu}$ ($a=1,2,3$), satisfying
\begin{equation}
p\cdot s^a=0,
\qquad
s^a\cdot s^b=-\delta^{ab},
\qquad
\sum_{a=1}^{3}s^a_\mu s^a_\nu
=
-g_{\mu\nu}+\frac{p_\mu p_\nu}{m^2}.
\end{equation}
$\{s^1, s^2, s^3, p/m\}$ are therefore an orthonormal set of four-vectors. If we want the helicity amplitudes, we can define $s^{3\mu}$ to be the spin vector \eqref{eq: spin vector} with positive helicity, while the other two can be conveniently built by means of two spatial unit vectors $\hat{\imath}, \hat{\jmath}$ so that $\{\hat{\imath}, \hat{\jmath}, \hat{p}\}$ are an orthonormal basis in $\mathbb{R}^3$: 
\be
\begin{cases}
    s^{1\mu}= (0, \hat{\imath})  \\
    s^{2\mu}= (0, \hat{\jmath}) \\
    s^{3\mu}= \left(\frac{|\Vec{p}|}{m}, \frac{E}{m} \hat{p}\right).
\end{cases}
\ee
Once the four-vectors $s^{a\mu}$ have been defined,the Bouchiat-Michel formulae can be written:
\begin{align}
u(p,\lambda)\bar u(p,\lambda')
&=
\frac{1}{2}
\left(
\delta_{\lambda'\lambda}
+\gamma^5\slashed{s}^{\,a}\sigma^a_{\lambda'\lambda}
\right)(\slashed p+m),\\
v(p,\lambda')\bar v(p,\lambda)
&=
\frac{1}{2}
\left(
\delta_{\lambda'\lambda}
+\gamma^5\slashed{s}^{\,a}\sigma^a_{\lambda'\lambda}
\right)(\slashed p-m),
\end{align}
where $\sigma^a$ are the Pauli matrices in helicity space. These relations provide a compact way to compute the full density matrix.

\subsection{Relation between helicity basis and spin basis coefficients}
\label{app: production density matrix BM}

Applying the Bouchiat-Michel formulae to the production of a $t(\lambda)\bar t(\mu)$ pair, with helicities $\lambda,\mu=\pm$, the production matrix takes the form
\begin{equation}
R^{(\mathrm{BM})}_{\lambda\lambda'\mu\mu'}
=
\tilde A_{\mathrm{BM}}\,\delta_{\lambda\lambda'}\delta_{\mu\mu'}
+
\delta_{\mu\mu'}\,\sigma^a_{\lambda\lambda'}\,s^a_\alpha\,\tilde B^\alpha_{\mathrm{BM}}
+
\delta_{\lambda\lambda'}\,\sigma^b_{\mu\mu'}\,\bar s^b_\beta\,\tilde{\bar{B}}_{\mathrm{BM}}^{\beta}
+
\sigma^a_{\lambda\lambda'}\sigma^b_{\mu\mu'}\,s^a_\alpha \bar s^b_\beta\,\tilde C_{\mathrm{BM}}^{\alpha\beta},
\label{eq: production rho from BM formulae}
\end{equation}
where $s^a$ and $\bar s^b$ are the spin four-vectors defining the orthonormal bases of the top and antitop quarks, respectively. The quantities $\tilde A_{\mathrm{BM}}$, $\tilde B^\alpha_{\mathrm{BM}}$, $\tilde{\bar{B}}_{\mathrm{BM}}^{\beta}$ and $\tilde C_{\mathrm{BM}}^{\alpha\beta}$ depend on the process kinematics, with Lorentz indices contracted with the corresponding spin vectors.

After performing the Lorentz contractions and normalizing the production matrix, one obtains the density matrix in the helicity basis,
\begin{equation}
\rho^{(\mathrm{BM})}
=
\frac{1}{4}\left[
\mathbb{I}\otimes\mathbb{I}
+
(B_{\mathrm{BM}})_i\,\sigma^i\otimes\mathbb{I}
+
(\bar{B}_{\mathrm{BM}})_j\,\mathbb{I}\otimes\sigma^j
+
(C_{\mathrm{BM}})_{ij}\,\sigma^i\otimes\sigma^j
\right],
\label{eq: rho 2qubit with BM basis}
\end{equation}
with
\begin{equation}
(B_{\mathrm{BM}})_i=
\frac{s^i_\alpha\,\tilde B^\alpha_{\mathrm{BM}}}{\tilde A_{\mathrm{BM}}},
\qquad
(\bar{B}_{\mathrm{BM}})_j=
\frac{\bar s^j_\beta\,\tilde{\bar{B}}_{\mathrm{BM}}^{\beta}}{\tilde A_{\mathrm{BM}}},
\qquad
(C_{\mathrm{BM}})_{ij}=
\frac{s^i_\alpha \bar s^j_\beta\,\tilde C_{\mathrm{BM}}^{\alpha\beta}}{\tilde A_{\mathrm{BM}}}.
\label{eq: coeff of rho 2qubit with BM basis}
\end{equation}

In this representation, each particle is quantised along its own direction of motion. In the $t\bar t$ ZMF, it is often convenient to adopt instead a common quantisation axis, chosen as the top-quark momentum direction: this defines the spin basis used in this work. Operationally, the change of basis amounts to reversing the helicity assignment of the antitop, which is implemented by the Pauli matrix $\sigma^1$,
\begin{equation}
\sigma^1|\pm\rangle=|\mp\rangle .
\end{equation}
The density matrix in the spin basis is therefore
\begin{equation}
\rho = U\,\rho^{(\mathrm{BM})}\,U^\dagger,
\qquad
U=\mathbb{I}\otimes\sigma^1 .
\label{eq:basischangeunitarity}
\end{equation}

As a consequence, the coefficients entering Eq.~\eqref{eq:rho2_qbit} are related to those of Eq.~\eqref{eq: coeff of rho 2qubit with BM basis} through
\begin{equation}
B_i=h_i(B_{\mathrm{BM}})_i,
\qquad
\bar{B}_j=\bar h_j(\bar{B}_{\mathrm{BM}})_j,
\qquad
C_{ij}=h_i\bar h_j(C_{\mathrm{BM}})_{ij},
\label{eq: conversion from BM to usual rho 2 qubit}
\end{equation}
where
\begin{equation}
h=(1,1,1),
\qquad
\bar{h}=(1,-1,-1).
\label{eq: h vectors}
\end{equation}
The signs in $\bar{h}$ reflect the reversal of the antitop spin-quantisation axis when passing from the helicity basis to the spin basis.
\section{Density matrix under CP transformations}
\label{app: Density matrix under CP transformations }

In this appendix we derive the constraints imposed by CP invariance on the
$t\bar t$ spin density matrix. Throughout this appendix we consider the $t\bar t$ spin/helicity density matrix in the ZMF of the pair and assume unpolarised initial states. In the ZMF, parity reverses the three-momenta of the top and antitop, while charge conjugation exchanges the particle labels. The combined CP transformation, therefore, maps the system back to the same kinematic configuration. The explicit form of the CP transformation depends
on the choice of basis used to define the spin degrees of freedom. A generic two-qubit density matrix written in the ordered basis $(|++\rangle, |+-\rangle, |-+\rangle, |--\rangle)$ is
\begin{equation}
\rho=
\begin{pmatrix}
\rho_{11} & \rho_{12} & \rho_{13} & \rho_{14} \\
\rho_{12}^* & \rho_{22} & \rho_{23} & \rho_{24} \\
\rho_{13}^* & \rho_{23}^* & \rho_{33} & \rho_{34} \\
\rho_{14}^* & \rho_{24}^* & \rho_{34}^* & \rho_{44}
\end{pmatrix}.
\end{equation}
Depending on the chosen basis, the labels $\pm$ denote either spin eigenvalues
($s=\pm $) or helicities ($h=\pm $). Under a CP transformation the density matrix transforms as
\begin{equation}
\rho \xrightarrow{CP}
\rho^{CP}
=
U_{CP}\,\rho\,U_{CP}^{\dagger},
\end{equation}
where $U_{CP}$ denotes the unitary representation of CP in the corresponding
basis.
\subsection{CP transformation in the spin basis}\label{sec:CPtransformationspin}

In the zero-momentum frame of the $t\bar t$ pair, the spin basis is defined by quantizing the spins of both particles along the same set of axes. Under parity, the three-momenta of the top and antitop are reversed, while the spin projections remain unchanged since spin is an axial vector. Charge conjugation then exchanges particle and antiparticle. As a result, the combined CP transformation acts on the spin states by exchanging the two subsystems,
\begin{equation}
CP\,|s_1,s_2\rangle
=
\eta_{CP}\,|s_2,s_1\rangle ,
\label{eq:CPspinstate}
\end{equation}
up to an overall phase $\eta_{CP}$. The corresponding unitary operator is therefore given by the SWAP operator,
\begin{equation}
U_{CP}^{\rm spin}
=
\eta_{CP}\,{\rm SWAP},
\end{equation}
with
\begin{equation}
{\rm SWAP}
=
\begin{pmatrix}
1 & 0 & 0 & 0 \\
0 & 0 & 1 & 0 \\
0 & 1 & 0 & 0 \\
0 & 0 & 0 & 1
\end{pmatrix}.
\end{equation}
CP invariance of the density matrix requires
\begin{equation}
\rho
=
U_{CP}^{\rm spin}\,
\rho\,
\left(U_{CP}^{\rm spin}\right)^\dagger ,
\label{eq:CPconditionSpin}
\end{equation}
where $U_{CP}$ denotes the unitary representation of the CP operator in the chosen basis. In terms of the matrix elements, this condition implies
\begin{equation}
\rho_{12}=\rho_{13},
\qquad
\rho_{22}=\rho_{33},
\qquad
\rho_{24}=\rho_{34},
\qquad
\rho_{23}=\rho_{23}^* .
\label{eq:spinmatrixconstraints}
\end{equation}

Using the Pauli decomposition of Eq.~\eqref{eq:rho2_qbit}, these relations translate into the compact conditions
\begin{equation}
\boxed{
\mathbf B=\mathbf{\bar{B}},
\qquad
\mathbf C=\mathbf C^{T}.
}
\label{eq: condition CP in spin basis}
\end{equation}

Therefore, in the spin basis, CP invariance of the density matrix requires identical polarisation vectors for the top and antitop and a symmetric spin-correlation matrix.

\subsection{CP transformation in the helicity basis}\label{sec:Cptranfhelicity}

In the helicity basis, the spin of each particle is quantised along its own
direction of motion. Under parity, the momentum is reversed while the spin remains unchanged, leading to a reversal of the helicity quantum number,
\begin{equation}
P:\ |h\rangle \rightarrow |-h\rangle .
\end{equation}
Charge conjugation exchanges particle and antiparticle. In the zero-momentum
frame of the $t\bar t$ pair, the combined CP transformation therefore acts on
the helicity states as
\begin{equation}
CP\,|h_1,h_2\rangle
=
\eta_{CP}\,|-h_2,-h_1\rangle ,
\label{eq:CPhelicitystate}
\end{equation}
up to an overall phase $\eta_{CP}$. The corresponding unitary operator can be written as
\begin{equation}
U_{CP}^{\rm helicity}
=
\eta_{CP}\,
(\sigma^1\otimes\sigma^1)\,
{\rm SWAP},
\end{equation}
where the Pauli matrix $\sigma^1$ implements the helicity-sign reversal $|h\rangle \to |-h\rangle$, while the SWAP operator exchanges the two subsystems. Explicitly,
\begin{equation}
U_{CP}^{\rm helicity}
=
\eta_{CP}
\begin{pmatrix}
0 & 0 & 0 & 1 \\
0 & 1 & 0 & 0 \\
0 & 0 & 1 & 0 \\
1 & 0 & 0 & 0
\end{pmatrix}.
\end{equation}
CP invariance of the density matrix requires
\begin{equation}
\rho
=
U_{CP}^{\rm helicity}\,
\rho\,
\left(U_{CP}^{\rm helicity}\right)^\dagger .
\label{eq:CPconditionHelicity}
\end{equation}
In terms of the density-matrix elements, this condition implies
\begin{equation}
\rho_{11}=\rho_{44},
\qquad
\rho_{12}=\rho_{24}^*,
\qquad
\rho_{13}=\rho_{34}^*,
\qquad
\rho_{14}=\rho_{14}^* .
\label{eq:helicitymatrixconstraints}
\end{equation}
Using the decomposition of Eq.~\eqref{eq:rho2_qbit}, these relations
translate into
\begin{equation}
\boxed{
\mathbf{\bar{B}}=R\,\mathbf B,
\qquad
\mathbf C=R\,\mathbf C^{T}R ,
}
\label{eq: condition CP in helicity basis}
\end{equation}
where
\begin{equation}
R=\bar{h}=\mathrm{diag}(1,-1,-1).
\end{equation}

Therefore, in the helicity basis, the CP invariance of the density matrix relates
the top and antitop polarisation vectors through the matrix $R$, while the
correlation matrix satisfies the generalised symmetry condition
$\mathbf C=R\,\mathbf C^{T}R$.

The conditions in Eqs.~\eqref{eq: condition CP in spin basis}
and~\eqref{eq: condition CP in helicity basis} are equivalent and are related
through the transformation between the spin and helicity basis Fano--Bloch
coefficients discussed in App.~\ref{app: production density matrix BM}.
\section{Analytic Expressions for the Fano Coefficients}
\label{app:analyticalexpressions}

In this appendix, we present the analytic expressions for the unnormalised Fano--Bloch coefficients of the spin-production density matrices for $t\bar{t}$ pairs produced from different initial states $I$. The unnormalised coefficients are extracted from the production matrix decomposition introduced in Eq.~\eqref{eq: R matrix decomposition} using the following traces
\begin{equation}
 \tilde{A}^I=\frac{1}{4}\operatorname{Tr}[R^I], \quad
 \tilde{B}^I_i=\frac{1}{4}\operatorname{Tr}\!\left[R^I(\sigma_i\otimes \mathbf{1})\right], \quad
 \tilde{\bar{B}}^{I}_j=\frac{1}{4}\operatorname{Tr}\!\left[R^I(\mathbf{1}\otimes\sigma_j)\right], \quad
 \tilde{C}^I_{ij}=\frac{1}{4}\operatorname{Tr}\!\left[R^I(\sigma_i\otimes\sigma_j)\right].
 \label{eq:unnormfano}
\end{equation}
The spin-density matrix can be computed from the helicity amplitudes using Eq.~\eqref{eq:Rmatrix} together with the BM formalism summarised in App.~\ref{app: BM formulae}. Throughout this appendix, all results are given for unpolarised initial states.

The expressions reported below include the SMEFT contributions from the operators listed in Tables~\ref{table: CP-odd operators}-\ref{table: CP-even operators}. For each production channel, we provide the SM tree-level contribution, the SM-EFT interference terms, and the pure EFT contributions. The results are kept fully general and allow for complex Wilson coefficients.

The computation is performed in the ZMF of the $t\bar t$ system. The relevant four-momenta and spin four-vectors are given by:
\begin{align}
&\begin{cases}
   p_{1}=\dfrac{\sqrt{s}}{2}\left(1,0,-\sin\theta,-\cos\theta\right),\\[4pt]
   p_{t}=\dfrac{\sqrt{s}}{2}\left(1,0,0,\beta\right),\\[4pt]
   s^{1}=(0,1,0,0),\\
   s^{2}=(0,0,1,0),\\
   s^{3}=\dfrac{\sqrt{s}}{2m_t}\left(\beta,0,0,1\right),
\end{cases}
\qquad
\begin{cases}
   p_{2}=\dfrac{\sqrt{s}}{2}\left(1,0,\sin\theta,\cos\theta\right),\\[4pt]
   \bar p_{t}=\dfrac{\sqrt{s}}{2}\left(1,0,0,-\beta\right),\\[4pt]
   \bar s^{1}=(0,1,0,0),\\
   \bar s^{2}=(0,0,-1,0),\\
   \bar s^{3}=\dfrac{\sqrt{s}}{2m_t}\left(\beta,0,0,-1\right),
\end{cases}
\label{eq:kinematics_eett}
\end{align}
where $p_1$ and $p_2$ denote the momenta of the initial-state particles, $s=(p_1+p_2)^2$ is the centre-of-mass energy squared, and
\begin{equation}
\beta=\sqrt{1-\frac{4m_t^2}{s}}
\end{equation}
is the top-quark velocity in the $t\bar t$ ZMF.

The spin vectors $s^i$ and $\bar s^i$ are defined in the helicity frames of the top and antitop, respectively. Hence, the density matrix obtained from the BM construction is first expressed in the helicity basis. The corresponding density matrix in the spin basis is obtained by applying the unitary transformation given in Eq.~\eqref{eq:basischangeunitarity}.

Since our primary interest lies in CP-violating effects, we separately present the symmetric and antisymmetric components of the unnormalised spin-correlation matrix: 
\begin{equation}
   \widetilde{\mathbf C} =\widetilde{\mathbf C}^S+ \widetilde{\mathbf C}^A , \qquad
    \begin{cases}
        \widetilde{\mathbf C}^S=  1/2(\widetilde{\mathbf C}+\widetilde{\mathbf C}^T) \\
       \widetilde{\mathbf C}^A=  1/2(\widetilde{\mathbf C}-  \widetilde{\mathbf C}^T) 
    \end{cases} .
\end{equation}
For the individual polarisations, we report each element of the top quark unnormalised polarisation vector $\widetilde{\mathbf{B}}$ and for the difference vector:
\begin{equation}
\Delta\widetilde{\mathbf{B}}
=\frac12 \left(
\widetilde{\mathbf{B}}
-
\widetilde{\mathbf{\bar{B}}}\right).
\label{eq.Bdiff}
\end{equation}
In the CP-conserving limit, the antisymmetric components of the spin-correlation matrix vanish, and the top and antitop quarks are equally polarised. Therefore, a non-vanishing component either of $\widetilde{\mathbf C}^{A}$ or $\Delta\widetilde{\mathbf{B}}$   constitutes a direct signal of CP violation. Throughout this appendix, terms that provide unambiguous evidence of CP violation are highlighted in purple.

\subsection{$e^+e^-\rightarrow t\bar{t}$} \label{app: spin density matrix eett}

In this appendix we present the structure of the spin production density
matrix for a \(t\bar t\) pair produced in lepton-antilepton annihilation,
focusing for definiteness on the process \(e^+e^- \to t\bar t\). The
extension to a generic \(\ell^+\ell^-\) initial state is straightforward.
At leading order the process receives two \(s\)-channel contributions in the SM,
mediated by a photon and a \(Z\) boson, as shown in
Fig.~\ref{fig: ee-tt scattering}. Possible dipole interactions are included as
modifications of the corresponding \(\gamma t\bar t\) and \(Zt\bar t\)
vertices, while the effects of the remaining operators are encoded as shifts of
the SM vector and axial-vector couplings.

The analytic expressions for the Fano coefficients are lengthy. We therefore
decompose the spin production matrix according to the gauge boson exchanged in
the amplitude and in its complex conjugate:
\begin{equation}
   R^{ee}= {}^{(\gamma)}R^{ee}+{}^{(Z)}R^{ee}+{}^{(\gamma Z)}R^{ee},
\end{equation}
with
\begin{equation}
{}^{(\gamma)}R^{ee}= \mathcal{M}_\gamma\mathcal{M}_\gamma^{\dagger},
\qquad
{}^{(Z)}R^{ee}= \mathcal{M}_Z\mathcal{M}_Z^{\dagger},
\qquad
{}^{(\gamma Z)}R^{ee}
=
\mathcal{M}_Z\mathcal{M}_\gamma^{\dagger}
+
\mathcal{M}_\gamma\mathcal{M}_Z^{\dagger}.
\end{equation}
The corresponding unnormalised Fano coefficients are then obtained from
Eq.~\eqref{eq:unnormfano} separately for each contribution. For notational
simplicity, we suppress the superscript \(ee\) on the Fano coefficients in the
following expressions.

We use the kinematic conventions defined in Eq.~\eqref{eq:kinematics_eett},
with \(p_1=p_{e^-}\) and \(p_2=p_{e^+}\). In order to keep the expressions
compact, we introduce the shorthand notation
\begin{equation}
    s_\alpha \equiv \sin\alpha,
    \qquad
    c_\alpha \equiv \cos\alpha,
\end{equation}
and parameterise the real and imaginary parts of the \(Z\)-boson propagator as
\begin{equation}
    \Delta_Z c_{\delta_Z}
    \equiv
    \operatorname{Re}
    \left[
    \frac{1}{s-m_Z^2+i\Gamma_Z m_Z}
    \right],
    \qquad
    \Delta_Z s_{\delta_Z}
    \equiv
    \operatorname{Im}
    \left[
    \frac{1}{s-m_Z^2+i\Gamma_Z m_Z}
    \right].
\end{equation}
We also define
\begin{equation}
    \tilde{\gamma}\equiv \frac{1}{\sqrt{1-\beta^2}},
    \qquad
    F_{ee}\equiv \frac{N_c}{4},
\end{equation}
where \(N_c\) denotes the number of colours.

In the following, we list only the non-vanishing unnormalised Fano
coefficients. The pure photon-exchange contribution is given by
\begin{align*}
{}^{(\gamma)}\tilde{A}&=F_{ee} \Bigl[ \Bigl(\frac{1}{2} e^4 Q_{e}^2 Q_{t}^2 \left(\beta^2 c_{2\theta}-\beta^2+4\right)\Bigr)+\mathbf{\Lambda^{-2}}\Bigl(8 e^3  m_{t} Q_{e}^2 Q_{t} r_{\gamma} v c_{\eta_{\gamma}}\Bigr)\\
&-\mathbf{\Lambda^{-4}}\Bigl(2e^2 m_{t}^2 Q_{e}^2 r_{\gamma}^2 v^2 \left(\tilde{\gamma}^2 \left(\beta^2 c_{2\theta}+\beta^2-2\right)-2 c_{2 \eta_{\gamma}}\right)  \Bigr) \Bigr] \\
{}^{(\gamma)}\tilde{C}_{nn}&=F_{ee}\Bigr[-\Bigl(\beta^2 e^4 Q_{e}^2 Q_{t}^2 s^2_{\theta}\Bigr)+\mathbf{\Lambda^{-4}}\Bigl(4 \beta^2 \tilde{\gamma}^2 e^2  m_{t}^2 Q_{e}^2 r_{\gamma}^2 v^2 c_{2 \eta_{\gamma}} s^2_{\theta}\Bigr)\Bigr] \\
{}^{(\gamma)}\tilde{C}_{rr}&=F_{ee}\Bigr[ -\Bigl(\left(\beta^2-2\right) e^4 Q_{e}^2 Q_{t}^2 s^2_{\theta}\Bigr)+\mathbf{\Lambda^{-2}}\Bigl(8 e^3  m_{t} Q_{e}^2 Q_{t} r_{\gamma} v c_{\eta_{\gamma}} s^2_{\theta}\Bigr)\\
 &+\mathbf{\Lambda^{-4}} \Bigl(4 e^2  m_{t}^2 Q_{e}^2 r_{\gamma}^2 v^2 s^2_{\theta} \left(\tilde{\gamma}^2 c_{2 \eta_{\gamma}}+1\right)\Bigr)\Bigr]\\
{}^{(\gamma)}\tilde{C}_{kk}&=F_{ee}\Bigr[ \Bigl(\frac{1}{2} e^4 Q_{e}^2 Q_{t}^2 \left(-\left(\beta^2-2\right) c_{2\theta}+\beta^2+2\right)\Bigr)+\mathbf{\Lambda^{-2}}\Bigl(8 e^3 m_{t} Q_{e}^2 Q_{t} r_{\gamma} v c_{\eta_{\gamma}} c^2_{\theta}\Bigr)\\
&+\mathbf{\Lambda^{-4}}\Bigl(e^2  m_{t}^2 Q_{e}^2 r_{\gamma}^2 v^2 \left(2 \tilde{\gamma}^2 \left(-2 \beta^2+c_{2\theta}+1\right)+4 c_{2 \eta_{\gamma}} c^2_{\theta}\right)\Bigr)\Bigr]  \\ 
{}^{(\gamma)}\tilde{C}^{S}_{rk}&=F_{ee}\Bigr[\bigl(e^4 Q_{e}^2 Q_{t}^2 s_{2\theta}\tilde{\gamma}^{-1}\Bigr)-\mathbf{\Lambda^{-2}}\Bigl(2\left(\beta^2-2\right) \tilde{\gamma} e^3 m_{t} Q_{e}^2 Q_{t} r_{\gamma} v c_{\eta_{\gamma}} s_{2\theta}\Bigr)\\
 &+\mathbf{\Lambda^{-4}}\Bigl(8 \tilde{\gamma} e^2  m_{t}^2 Q_{e}^2 r_{\gamma}^2 v^2 c^2_{\eta_{\gamma}} s_{\theta} c_{\theta}\Bigr)\Bigr] \\
 \textcolor{purple}{{}^{(\gamma)}\tilde{C}^{A}_{rn}}&=F_{ee}\Bigr[\mathbf{\Lambda^{-2}}\Bigl(4 \beta e^3  m_{t} Q_{e}^2 Q_{t} r_{\gamma} v s_{\eta_{\gamma}} s^2_{\theta}\Bigr)+\mathbf{\Lambda^{-4}}\Bigl(8 \beta \tilde{\gamma}^2 e^2  m_{t}^2 Q_{e}^2 r_{\gamma}^2 v^2 s_{\eta_{\gamma}} c_{\eta_{\gamma}} s^2_{\theta}\Bigr)\Bigr]\\
\textcolor{purple}{{}^{(\gamma)}\tilde{C}^{A}_{kn}}&=F_{ee}\Bigr[\mathbf{\Lambda^{-2}}\Bigl(4 \beta \tilde{\gamma} e^3  m_{t} Q_{e}^2 Q_{t} r_{\gamma} v s_{\eta_{\gamma}} s_{\theta} c_{\theta}\Bigr)+\mathbf{\Lambda^{-4}}\Bigl(2 \beta \tilde{\gamma} e^2  m_{t}^2 Q_{e}^2 r_{\gamma}^2 v^2 s_{2 \eta_{\gamma}} s_{2\theta}\Bigr)\Bigr] 
\refstepcounter{equation}\tag{\theequation}
\end{align*}

The structure of the photon-exchange contribution to $\tilde A$ is consistent with Ref.~\cite{Haberl:1995ek}.

The pure $Z$-exchange contribution reads
\begin{align*}
 {}^{(Z)}\tilde{A}^{} &=F_{ee}\Bigl[\Big( 8 \tilde{\gamma}^4 \Delta_{Z}^2 e^4 m_{t}^4 \\         &*\Big(\left(\bar{g}_{Ae}^2+\bar{g}_{Ve}^2\right) \left(3 \bar{g}_{At}^2 \beta^2+\beta^2 \left(\bar{g}_{At}^2+\bar{g}_{Vt}^2\right) c_{2\theta}-\left(\left(\beta^2-4\right) \bar{g}_{Vt}^2\right)\right)-16 \bar{g}_{Ae} \bar{g}_{At} \beta \bar{g}_{Ve} \bar{g}_{Vt} c_{\theta}\Big)\Bigr)\\
 &+\mathbf{\Lambda^{-2}}\Bigl(128 \tilde{\gamma}^4 \Delta_{Z}^2 e^3  m_{t}^5 r_{Z} v c_{\eta_{Z}} \left(\bar{g}_{Vt} \left(\bar{g}_{Ae}^2+\bar{g}_{Ve}^2\right)-2 \bar{g}_{Ae} \bar{g}_{At} \beta \bar{g}_{Ve} c_{\theta}\right)\Bigr)\\
 &-\mathbf{\Lambda^{-4}}\Bigl(32  \left(\tilde{\gamma}^4 \Delta_{Z}^2 e^2 m_{t}^6 r_{Z}^2 v^2 \left(\bar{g}_{Ae}^2+\bar{g}_{Ve}^2\right) \left(\tilde{\gamma}^2 \left(\beta^2 c_{2\theta}+\beta^2-2\right)-2 c_{2 \eta_{Z}}\right)\right)\Bigr)   \Bigr] \\ 
 {}^{(Z)}\tilde{B}^{ }_n &=F_{ee}\Bigl[ \mathbf{\Lambda^{-2}}\Bigl(  64 e^3 m_{t}^5 r_{Z} v \beta \tilde{\gamma}^5 \Delta_{Z}^2 \left(\bar{g}_{At} \left(\bar{g}_{Ae}^2+\bar{g}_{Ve}^2\right) \beta c_{\theta}-2 \bar{g}_{Ae} \bar{g}_{Ve} \bar{g}_{Vt}\right) s_{\eta_{Z}} s_{\theta}\Bigr) \\
&-\mathbf{\Lambda^{-4}} \Bigl(128 \left(\bar{g}_{Ae} e^2 m_{t}^6 r_{Z}^2 v^2 \bar{g}_{Ve} \beta \tilde{\gamma}^5 \Delta_{Z}^2 s_{2 \eta_{Z}} s_{\theta}\right) \Bigr) \Bigr]   \\ 
{}^{(Z)}\tilde{B}_r&=F_{ee}\Bigl[-\Bigl(32 e^4 m_{t}^4 \bar{g}_{Vt} \tilde{\gamma}^3 \Delta_{Z}^2 \left(\bar{g}_{At} \left(\bar{g}_{Ae}^2+\bar{g}_{Ve}^2\right) \beta c_{\theta}-2 \bar{g}_{Ae} \bar{g}_{Ve} \bar{g}_{Vt}\right) s_{\theta}\Bigr)\\
&-\mathbf{\Lambda^{-2}}\Bigl(64 e^3 m_{t}^5 r_{Z} v \tilde{\gamma}^3 \Delta_{Z}^2 c_{\eta_{Z}} \left(\bar{g}_{At} \left(\bar{g}_{Ae}^2+\bar{g}_{Ve}^2\right) \beta \tilde{\gamma}^2 c_{\theta}-2 \bar{g}_{Ae} \bar{g}_{Ve} \bar{g}_{Vt} \left(\tilde{\gamma}^2+1\right)\right) s_{\theta}\Bigr) \\
&+\mathbf{\Lambda^{-4}}\Bigl(256 \bar{g}_{Ae} e^2 m_{t}^6 r_{Z}^2 v^2 \bar{g}_{Ve} \tilde{\gamma}^5 \Delta_{Z}^2 c^2_{ \eta_{Z}} s_{\theta} \Bigr)   \Bigr]  \\
{}^{(Z)}\tilde{B}_k&=F_{ee}\Bigl[- \Bigl(16e^4 m_{t}^4 \tilde{\gamma}^4 \Delta_{Z}^2 \left(\bar{g}_{At} \left(\bar{g}_{Ae}^2+\bar{g}_{Ve}^2\right) \bar{g}_{Vt} \beta (c_{2\theta}+3)-4 \bar{g}_{Ae} \bar{g}_{Ve} \left(\bar{g}_{Vt}^2+\bar{g}_{At}^2 \beta^2\right) c_{\theta}\right)\Bigr)\\
&- \mathbf{\Lambda^{-2}}\Bigl(32e^3 m_{t}^5 r_{Z} v \tilde{\gamma}^4 \Delta_{Z}^2 c_{\eta_{Z}} \left(\bar{g}_{At} \left(\bar{g}_{Ae}^2+\bar{g}_{Ve}^2\right) \beta (c_{2\theta}+3)-8 \bar{g}_{Ae} \bar{g}_{Ve} \bar{g}_{Vt} c_{\theta}\right)\Bigr) \\
&+\mathbf{\Lambda^{-4}} \Bigl(256 \bar{g}_{Ae} e^2 m_{t}^6 r_{Z}^2 v^2 \bar{g}_{Ve} \tilde{\gamma}^4 \Delta_{Z}^2 c^2_{ \eta_{Z}} c_{\theta} \Bigr)     \Bigr]\\
\textcolor{purple}{{}^{(Z)}\Delta\widetilde{B}_n}&=F_{ee}\Bigl[  \mathbf{\Lambda^{-2}}\Bigl( 64 e^3 m_{t}^5 r_{Z} v \beta \tilde{\gamma}^5 \Delta_{Z}^2 \left(\bar{g}_{At} \left(\bar{g}_{Ae}^2+\bar{g}_{Ve}^2\right) \beta c_{\theta}-2 \bar{g}_{Ae} \bar{g}_{Ve} \bar{g}_{Vt}\right) s_{\eta_{Z}} s_{\theta} \Bigr)\\
&-\mathbf{\Lambda^{-4}}\Bigl(128 \bar{g}_{Ae} e^2 m_{t}^6 r_{Z}^2 v^2 \bar{g}_{Ve} \beta \tilde{\gamma}^5 \Delta_{Z}^2 s_{2 \eta_{Z}} s_{\theta}\Bigr)     \Bigr] \refstepcounter{equation}\tag{\theequation}
\end{align*}

\begin{align*}
 {}^{(Z)}\tilde{C}_{nn}&=F_{ee}\Bigl[ \Bigl(16 \beta^2 \tilde{\gamma}^4 \Delta_{Z}^2 e^4 m_{t}^4 \left(\bar{g}_{Ae}^2+\bar{g}_{Ve}^2\right) \left(\bar{g}_{At}^2-\bar{g}_{Vt}^2\right) s^2_{\theta}\Bigr)\\
 &+\mathbf{\Lambda^{-4}}\Bigl(64 \beta^2 \tilde{\gamma}^6 \Delta_{Z}^2 e^2  m_{t}^6 r_{Z}^2 v^2 \left(\bar{g}_{Ae}^2+\bar{g}_{Ve}^2\right) c_{2 \eta_{Z}} s^2_{\theta} \Bigr)\Bigr] \\
 {}^{(Z)}\tilde{C}_{rr}&=F_{ee}\Bigl[-\Bigl(16\tilde{\gamma}^4 \Delta_{Z}^2 e^4 m_{t}^4 s^2_{\theta}\left(\bar{g}_{Ae}^2+\bar{g}_{Ve}^2\right)  \left(\bar{g}_{At}^2 \beta^2+\left(\beta^2-2\right) \bar{g}_{Vt}^2\right)\Bigr)\\
 &+\mathbf{\Lambda^{-2}}\Bigl(128 \tilde{\gamma}^4 \Delta_{Z}^2 e^3  m_{t}^5 r_{Z} v \bar{g}_{Vt} \left(\bar{g}_{Ae}^2+\bar{g}_{Ve}^2\right) c_{\eta_{Z}} s^2_{\theta}\Bigr)\\
 &+\mathbf{\Lambda^{-4}}\Bigl(64 \tilde{\gamma}^6 \Delta_{Z}^2 e^2  m_{t}^6 r_{Z}^2 v^2  s^2_{\theta}\left(\bar{g}_{Ae}^2+\bar{g}_{Ve}^2\right) \left(-\beta^2+c_{2 \eta_{Z}}+1\right)\Bigr) \Bigr] \\
 {}^{(Z)}\tilde{C}_{kk}&=F_{ee}\Bigl[- \Bigl(8\tilde{\gamma}^4 \Delta_{Z}^2 e^4 m_{t}^4 \Bigl(\left(\bar{g}_{Ae}^2+\bar{g}_{Ve}^2\right) \Big(-3 \bar{g}_{At}^2 \beta^2+c_{2\theta} \left(\left(\beta^2-2\right) \bar{g}_{Vt}^2-\bar{g}_{At}^2 \beta^2\right)\\
 &-\left(\left(\beta^2+2\right) \bar{g}_{Vt}^2\right)\Big)+16 \bar{g}_{Ae} \bar{g}_{At} \beta \bar{g}_{Ve} \bar{g}_{Vt} c_{\theta}\Bigr)\Bigr)\\
 &+\mathbf{\Lambda^{-2}}\Bigl(128 \tilde{\gamma}^4 \Delta_{Z}^2 e^3  m_{t}^5 r_{Z} v c_{\eta_{Z}} c_{\theta} \left(\bar{g}_{Vt} \left(\bar{g}_{Ae}^2+\bar{g}_{Ve}^2\right) c_{\theta}-2 \bar{g}_{Ae} \bar{g}_{At} \beta \bar{g}_{Ve}\right)\Bigr)\\
 &- \mathbf{\Lambda^{-4}} \Bigl(16\tilde{\gamma}^6 \Delta_{Z}^2 e^2 m_{t}^6 r_{Z}^2 v^2 \left(\bar{g}_{Ae}^2+\bar{g}_{Ve}^2\right)\Big(\beta^2 c_{2 (\eta_{Z}+\theta )}+\left(\beta^2-1\right) c_{2 (\eta_{Z}-\theta )}+2 \left(\beta^2-1\right) c_{2 \eta_{Z}}\\
 &+4 \beta^2-c_{2 (\eta_{Z}+\theta )}-2 c_{2\theta}-2\Big)\Bigr) \Bigr] \\ 
 {}^{(Z)}\tilde{C}^{S}_{rk}&=F_{ee}\Bigl[\Bigl(32 \tilde{\gamma}^3 \Delta_{Z}^2 e^4 m_{t}^4 \bar{g}_{Vt} s_{\theta} \left( \bar{g}_{Vt} \left(\bar{g}_{Ae}^2+\bar{g}_{Ve}^2\right) c_{\theta}-2 \bar{g}_{Ae} \bar{g}_{At} \beta  \bar{g}_{Ve}\right)\Bigr)\\
 &+\mathbf{\Lambda^{-2}}\Bigl(64 \tilde{\gamma}^3 \Delta_{Z}^2 e^3  m_{t}^5 r_{Z} v c_{\eta_{Z}} s_{\theta} \left(\left(\tilde{\gamma}^2+1\right) \bar{g}_{Vt} \left(\bar{g}_{Ae}^2+\bar{g}_{Ve}^2\right) c_{\theta}-2 \bar{g}_{Ae} \bar{g}_{At} \beta \tilde{\gamma}^2 \bar{g}_{Ve}\right)\Bigr)\\
 &+\mathbf{\Lambda^{-4}}\Bigl(64 \tilde{\gamma}^5 \Delta_{Z}^2 e^2  m_{t}^6 r_{Z}^2 v^2 \left(\bar{g}_{Ae}^2+\bar{g}_{Ve}^2\right) c^2_{ \eta_{Z}} s_{2\theta} \Bigr)\Bigr]  \\
\textcolor{purple}{{}^{(Z)} \tilde{C}^{A}_{rn}}&= F_{ee}\Bigl[\mathbf{\Lambda^{-2}}\Bigl(64 \beta \tilde{\gamma}^4 \Delta_{Z}^2 e^3  m_{t}^5 r_{Z} v \bar{g}_{Vt} \left(\bar{g}_{Ae}^2+\bar{g}_{Ve}^2\right) s_{\eta_{Z}} s^2_{\theta}\Bigr)\\
&+\mathbf{\Lambda^{-4}}\Bigl(64 \beta \tilde{\gamma}^6 \Delta_{Z}^2 e^2  m_{t}^6 r_{Z}^2 v^2 \left(\bar{g}_{Ae}^2+\bar{g}_{Ve}^2\right) s_{2 \eta_{Z}} s^2_{\theta} \Bigr)\Bigr] \\
\textcolor{purple}{{}^{(Z)}\tilde{C}^{A}_{kn}}&= F_{ee}\Bigl[ \mathbf{\Lambda^{-2}}\Bigl(64 \beta \tilde{\gamma}^5 \Delta_{Z}^2 e^3  m_{t}^5 r_{Z} v s_{\eta_{Z}} s_{\theta} \left(\bar{g}_{Vt} \left(\bar{g}_{Ae}^2+\bar{g}_{Ve}^2\right) c_{\theta}-2 \bar{g}_{Ae} \bar{g}_{At} \beta \bar{g}_{Ve}\right)\Bigr)\\
&+\mathbf{\Lambda^{-4}}\Bigl(32 \beta \tilde{\gamma}^5 \Delta_{Z}^2 e^2  m_{t}^6 r_{Z}^2 v^2 \left(\bar{g}_{Ae}^2+\bar{g}_{Ve}^2\right) s_{2 \eta_{Z}} s_{2\theta} \Bigr)\Bigr] \refstepcounter{equation}\tag{\theequation}
\end{align*}
Finally, the \(\gamma-Z\) interference contribution is
\begin{align*}
{}^{(\gamma Z)}\tilde{A}^{ } &=F_{ee}\Bigl[\Bigl( 4 \tilde{\gamma}^2 \Delta_{Z} e^4 m_{t}^2 Q_{e} Q_{t} c_{\delta_{Z}} \left(\bar{g}_{Ve} \bar{g}_{Vt} \left(\beta^2 c_{2\theta}-\beta^2+4\right)-4 \bar{g}_{Ae} \bar{g}_{At} \beta c_{\theta}\right)\Bigr)\\
&+\mathbf{\Lambda^{-2}} \Bigl(32 \tilde{\gamma}^2 \Delta_{Z} e^3 m_{t}^3 Q_{e} v c_{\delta_{Z}} (r_{\gamma} c_{\eta_{\gamma}} (\bar{g}_{Ve} \bar{g}_{Vt}-\bar{g}_{Ae} \bar{g}_{At} \beta c_{\theta})+Q_{t} r_{Z} \bar{g}_{Ve} c_{\eta_{Z}})\Bigr)\\
&-\mathbf{\Lambda^{-4}}\Bigl(8\tilde{\gamma}^4 \Delta_{Z} e^2 m_{t}^4 Q_{e} r_{\gamma} r_{Z} v^2 \bar{g}_{Ve} c_{\delta_{Z}} \left(2 \left(\beta^2 c_{2\theta}+\beta^2-2\right) c_{(\eta_{\gamma}-\eta_{Z})}+4 \left(\beta^2-1\right) c_{(\eta_{\gamma}+\eta_{Z})}\right)\Big) \Bigr] 
\refstepcounter{equation}\tag{\theequation}
\end{align*}
\begin{align*}
{}^{(\gamma Z)}\tilde{B}^{}_n &=F_{ee}\Bigl[   -8 \Bigl(\bar{g}_{Ae} \bar{g}_{At} e^4 m_{t}^2 Q_{e} Q_{t} \beta \tilde{\gamma} \Delta_{Z} s_{\delta_{Z}} s_{\theta}\Bigr)-\mathbf{\Lambda^{-2}} \bigl(16 e^3 m_{t}^3 Q_{e} v \beta \tilde{\gamma}^3 \Delta_{Z} (Q_{t} r_{Z} \bar{g}_{Ve} \beta c_{\eta_{Z}} c_{\theta} s_{\delta_{Z}}\\
&+r_{\gamma} c_{\eta_{\gamma}} (\bar{g}_{Ae} \bar{g}_{At}-\bar{g}_{Ve} \bar{g}_{Vt} \beta c_{\theta}) s_{\delta_{Z}}+c_{\delta_{Z}} (r_{\gamma} (\bar{g}_{Ae} \bar{g}_{Vt}-\bar{g}_{At} \bar{g}_{Ve} \beta c_{\theta}) s_{\eta_{\gamma}}+\bar{g}_{Ae} Q_{t} r_{Z} s_{\eta_{Z}})) s_{\theta}\bigr) \\
&-\mathbf{\Lambda^{-4}} \Bigl(32\bar{g}_{Ae} e^2 m_{t}^4 Q_{e} r_{\gamma} r_{Z} v^2 \beta \tilde{\gamma}^3 \Delta_{Z} c_{\delta_{Z}} s_{(\eta_{\gamma}+\eta_{Z})} s_{\theta}\Bigr)    \Bigr]  \\ 
{}^{(\gamma Z)}\tilde{B}^{ }_r&=F_{ee}\Bigl[-\Bigl(8e^4 m_{t}^2 Q_{e} Q_{t} \tilde{\gamma} \Delta_{Z} c_{\delta_{Z}} (\bar{g}_{At} \bar{g}_{Ve} \beta c_{\theta}-2 \bar{g}_{Ae} \bar{g}_{Vt}) s_{\theta}\Bigr)\\
&-\mathbf{\Lambda^{-2}}\Bigl(16e^3 m_{t}^3 Q_{e} v \tilde{\gamma}^3 \Delta_{Z} (c_{\delta_{Z}} (\bar{g}_{Ae} \left(\beta^2-2\right) (r_{\gamma} \bar{g}_{Vt} c_{\eta_{\gamma}}+Q_{t} r_{Z} c_{\eta_{Z}})+\bar{g}_{At} r_{\gamma} \bar{g}_{Ve} \beta c_{\eta_{\gamma}} c_{\theta})\\
&+\beta s_{\delta_{Z}} (r_{\gamma} (\bar{g}_{Ae} \bar{g}_{At} \beta-\bar{g}_{Ve} \bar{g}_{Vt} c_{\theta}) s_{\eta_{\gamma}}+Q_{t} r_{Z} \bar{g}_{Ve} c_{\theta} s_{\eta_{Z}})) s_{\theta}\Bigr)\\
&+\mathbf{\Lambda^{-4}}\Bigl(32 e^2 m_{t}^4 Q_{e} r_{\gamma} r_{Z} v^2 \tilde{\gamma}^3 \Delta_{Z} (2 \bar{g}_{Ae} c_{\delta_{Z}} c_{\eta_{\gamma}} c_{\eta_{Z}}+\bar{g}_{Ve} \beta c_{\theta} s_{\delta_{Z}} s_{(\eta_{\gamma}-\eta_{Z})}) s_{\theta}    \Bigr)\Bigr]  \\ 
{}^{(\gamma Z)}\tilde{B}^{ }_k&=F_{ee}\Bigl[- \Bigl(4e^4 m_{t}^2 Q_{e} Q_{t} \tilde{\gamma}^2 \Delta_{Z} c_{\delta_{Z}} (\bar{g}_{At} \bar{g}_{Ve} \beta (c_{2\theta}+3)-4 \bar{g}_{Ae} \bar{g}_{Vt} c_{\theta})\Bigr)\\
&-\mathbf{\Lambda^{-2}}\Bigl(8e^3 m_{t}^3 Q_{e} v \tilde{\gamma}^2 \Delta_{Z} (2 \bar{g}_{Ve} \beta s_{\delta_{Z}} (r_{\gamma} \bar{g}_{Vt} s_{\eta_{\gamma}}-Q_{t} r_{Z} s_{\eta_{Z}}) s^2_{\theta}\\
&+c_{\delta_{Z}} (r_{\gamma} c_{\eta_{\gamma}} (\bar{g}_{At} \bar{g}_{Ve} \beta (c_{2\theta}+3)-4 \bar{g}_{Ae} \bar{g}_{Vt} c_{\theta})-4 \bar{g}_{Ae} Q_{t} r_{Z} c_{\eta_{Z}} c_{\theta}))\Bigr) \\
&- \mathbf{\Lambda^{-4}} \Bigl(32 e^2 m_{t}^4 Q_{e} r_{\gamma} r_{Z} v^2 \tilde{\gamma}^2 \Delta_{Z} \left(\bar{g}_{Ve} \beta \tilde{\gamma}^2 s_{\delta_{Z}} s_{(\eta_{\gamma}-\eta_{Z})} s^2_{\theta}-2 \bar{g}_{Ae} c_{\delta_{Z}} c_{\eta_{\gamma}} c_{\eta_{Z}} c_{\theta}\right)\Bigr)      \Bigr] \\
\textcolor{purple}{{}^{(\gamma Z)}\Delta\widetilde{B}^{}_n} &=F_{ee}\Bigl[   - \mathbf{\Lambda^{-2}}\Bigl(16e^3 m_{t}^3 Q_{e} v \beta \tilde{\gamma}^3 \Delta_{Z} c_{\delta_{Z}} (r_{\gamma} (\bar{g}_{Ae} \bar{g}_{Vt}-\bar{g}_{At} \bar{g}_{Ve} \beta c_{\theta}) s_{\eta_{\gamma}}+\bar{g}_{Ae} Q_{t} r_{Z} s_{\eta_{Z}}) s_{\theta}\Bigr) \\
&-\mathbf{\Lambda^{-4}}\Bigl(32\bar{g}_{Ae} e^2 m_{t}^4 Q_{e} r_{\gamma} r_{Z} v^2 \beta \tilde{\gamma}^3 \Delta_{Z} c_{\delta_{Z}} s_{(\eta_{\gamma}+\eta_{Z})} s_{\theta}\Bigr)   \Bigr]    \\ 
\textcolor{purple}{{}^{(\gamma Z)}\Delta\widetilde{B}^{ }_r}&=F_{ee}\Bigl[ - \mathbf{\Lambda^{-2}}\Bigl(16e^3 m_{t}^3 Q_{e} v \beta \tilde{\gamma}^3 \Delta_{Z} s_{\delta_{Z}} (r_{\gamma} (\bar{g}_{Ae} \bar{g}_{At} \beta-\bar{g}_{Ve} \bar{g}_{Vt} c_{\theta}) s_{\eta_{\gamma}}+Q_{t} r_{Z} \bar{g}_{Ve} c_{\theta} s_{\eta_{Z}}) s_{\theta}\Bigr) \\
&+\mathbf{\Lambda^{-4}} \Bigl(16 e^2 m_{t}^4 Q_{e} r_{\gamma} r_{Z} v^2 \bar{g}_{Ve} \beta \tilde{\gamma}^3 \Delta_{Z} s_{\delta_{Z}} s_{(\eta_{\gamma}-\eta_{Z})} s_{2\theta} \Bigr)\Bigr]     \\ 
\textcolor{purple}{{}^{(\gamma Z)}\Delta\widetilde{B}^{ }_k}&=F_{ee}\Bigl[ \mathbf{\Lambda^{-2}}\Bigl(16 e^3 m_{t}^3 Q_{e} v \bar{g}_{Ve} \beta \tilde{\gamma}^2 \Delta_{Z} s_{\delta_{Z}} (Q_{t} r_{Z} s_{\eta_{Z}}-r_{\gamma} \bar{g}_{Vt} s_{\eta_{\gamma}}) s^2_{\theta} \Bigr)\\
&- \mathbf{\Lambda^{-4}} \Bigl(32e^2 m_{t}^4 Q_{e} r_{\gamma} r_{Z} v^2 \bar{g}_{Ve} \beta \tilde{\gamma}^4 \Delta_{Z} s_{\delta_{Z}} s_{(\eta_{\gamma}-\eta_{Z})} s^2_{\theta}\Bigr) \Bigr]  \refstepcounter{equation}\tag{\theequation} 
\end{align*}
\begin{align*}
 {}^{(\gamma Z)}\tilde{C}_{nn}&= F_{ee}\Bigl[ - \Bigl(8\beta^2 \tilde{\gamma}^2 \Delta_{Z} e^4 m_{t}^2 Q_{e} Q_{t} \bar{g}_{Ve} \bar{g}_{Vt} c_{\delta_{Z}} s^2_{\theta}\Bigr)\\
 &+\mathbf{\Lambda^{-4}}\Bigl(32 \beta^2 \tilde{\gamma}^4 \Delta_{Z} e^2  m_{t}^4 Q_{e} r_{\gamma} r_{Z} v^2 \bar{g}_{Ve} c_{\delta_{Z}} s^2_{\theta} c_{(\eta_{\gamma}+\eta_{Z})}\Bigr)\Bigr] \\
 {}^{(\gamma Z)} \tilde{C}_{rr}&= F_{ee}\Bigl[ -\Bigl(8\left(\beta^2-2\right) \tilde{\gamma}^2 \Delta_{Z} e^4 m_{t}^2 Q_{e} Q_{t} \bar{g}_{Ve} \bar{g}_{Vt} c_{\delta_{Z}} s^2_{\theta}\Bigr)\\
 &+\mathbf{\Lambda^{-2}} \Bigl(32 \tilde{\gamma}^2 \Delta_{Z} e^3 m_{t}^3 Q_{e} v \bar{g}_{Ve} c_{\delta_{Z}} s^2_{\theta} (Q_{t} r_{Z} c_{\eta_{Z}}+r_{\gamma} \bar{g}_{Vt} c_{\eta_{\gamma}})\Bigr)\\
  &+\mathbf{\Lambda^{-4}}\Bigl(32 \tilde{\gamma}^4 \Delta_{Z} e^2  m_{t}^4 Q_{e} r_{\gamma} r_{Z} v^2 \bar{g}_{Ve} c_{\delta_{Z}} s^2_{\theta} \left(c_{(\eta_{\gamma}+\eta_{Z})}-\left(\beta^2-1\right) c_{(\eta_{\gamma}-\eta_{Z})}\right) \Bigr)\Bigr] \\
{}^{(\gamma Z)} \tilde{C}_{kk}&= F_{ee}\Bigl[ -\Bigl(4\tilde{\gamma}^2 \Delta_{Z} e^4 m_{t}^2 Q_{e} Q_{t} c_{\delta_{Z}} \left(4 \bar{g}_{Ae} \bar{g}_{At} \beta c_{\theta}+\bar{g}_{Ve} \bar{g}_{Vt} \left(\left(\beta^2-2\right) c_{2\theta}-\beta^2-2\right)\right)\Bigr)\\
 &+\mathbf{\Lambda^{-2}}\Bigl(32 \tilde{\gamma}^2 \Delta_{Z} e^3  m_{t}^3 Q_{e} v c_{\delta_{Z}} c_{\theta} (r_{\gamma} c_{\eta_{\gamma}} (\bar{g}_{Ve} \bar{g}_{Vt} c_{\theta}-\bar{g}_{Ae} \bar{g}_{At} \beta)+Q_{t} r_{Z} \bar{g}_{Ve} c_{\eta_{Z}} c_{\theta})\Bigr)\\
 &-\mathbf{\Lambda^{-4}} \Bigl(8\tilde{\gamma}^4 \Delta_{Z} e^2 m_{t}^4 Q_{e} r_{\gamma} r_{Z} v^2 \bar{g}_{Ve} c_{\delta_{Z}} \left(4 \beta^2 s_{\eta_{\gamma}} s_{\eta_{Z}} s^2_{\theta}+2 c_{\eta_{\gamma}} c_{\eta_{Z}} \left(\left(\beta^2-2\right) c_{2\theta}+3 \beta^2-2\right)\right)\Bigr)\Bigr] \\
{}^{(\gamma Z)} \tilde{C}^{S }_{rn}&= F_{ee}\Bigl[ - \Bigl(8\bar{g}_{At} \beta \tilde{\gamma}^2 \Delta_{Z} e^4 m_{t}^2 Q_{e} Q_{t} \bar{g}_{Ve} s_{\delta_{Z}} s^2_{\theta}\Bigr)-\mathbf{\Lambda^{-2}} \Bigl(16\bar{g}_{At} \beta \tilde{\gamma}^2 \Delta_{Z} e^3 m_{t}^3 Q_{e} r_{\gamma} v \bar{g}_{Ve} s_{\delta_{Z}} c_{\eta_{\gamma}} s^2_{\theta}\Bigr)\Bigr] \\
 {}^{(\gamma Z)}\tilde{C}^{S }_{rk}&=F_{ee}\Bigl[ \Bigl( 8 \tilde{\gamma} \Delta_{Z} e^4 m_{t}^2 Q_{e} Q_{t} c_{\delta_{Z}} s_{\theta} (2 \bar{g}_{Ve} \bar{g}_{Vt} c_{\theta}-\bar{g}_{Ae} \bar{g}_{At} \beta)\Bigr)\\
 &-\mathbf{\Lambda^{-2}}\Bigl(16 \tilde{\gamma}^3 \Delta_{Z} e^3  m_{t}^3 Q_{e} v c_{\delta_{Z}} s_{\theta}\Bigl(\left(\beta^2-2\right) Q_{t}r_{Z} \bar{g}_{Ve} c_{\eta_{Z}} c_{\theta}+r_{\gamma} c_{\eta_{\gamma}}\Bigl(\bar{g}_{Ae} \bar{g}_{At} \beta+\left(\beta^2-2\right) \\
 &*\bar{g}_{Ve} \bar{g}_{Vt} c_{\theta}\Bigr)\Bigr)\Bigr)+\mathbf{\Lambda^{-4}}\Bigl(32 \tilde{\gamma}^3 \Delta_{Z} e^2  m_{t}^4 Q_{e} r_{\gamma} r_{Z} v^2 \bar{g}_{Ve} c_{\delta_{Z}} c_{\eta_{\gamma}} c_{\eta_{Z}} s_{2\theta}\Bigr)\Bigr] \\
{}^{(\gamma Z)} \tilde{C}^{S }_{kn}&= F_{ee}\Bigl[-\Bigl(4\bar{g}_{At} \beta \tilde{\gamma} \Delta_{Z} e^4 m_{t}^2 Q_{e} Q_{t} \bar{g}_{Ve} s_{\delta_{Z}} s_{2\theta}\Bigr)\\
&- \mathbf{\Lambda^{-2}} \Bigl(16\beta \tilde{\gamma}^3 \Delta_{Z} e^3 m_{t}^3 Q_{e} v s_{\delta_{Z}} s_{\theta} (c_{\eta_{\gamma}} (\bar{g}_{At} r_{\gamma} \bar{g}_{Ve} c_{\theta}-\bar{g}_{Ae} \beta r_{\gamma} \bar{g}_{Vt})+\bar{g}_{Ae} \beta Q_{t} r_{Z} c_{\eta_{Z}})\Bigr) \Bigr] \\
 \textcolor{purple}{{}^{(\gamma Z)}\tilde{C}^{A }_{rn}}&= F_{ee}\Bigl[ \mathbf{\Lambda^{-2}}\Bigl(16 \beta \tilde{\gamma}^2 \Delta_{Z} e^3  m_{t}^3 Q_{e} v \bar{g}_{Ve} c_{\delta_{Z}} s^2_{\theta} (Q_{t} r_{Z} s_{\eta_{Z}}+r_{\gamma} \bar{g}_{Vt} s_{\eta_{\gamma}})\Bigr)\\
 &+\mathbf{\Lambda^{-4}}\Bigl(32 \beta \tilde{\gamma}^4 \Delta_{Z} e^2  m_{t}^4 Q_{e} r_{\gamma} r_{Z} v^2 \bar{g}_{Ve} c_{\delta_{Z}} s^2_{\theta} s_{(\eta_{\gamma}+\eta_{Z})}\Bigr)\Bigr] \\
\textcolor{purple}{{}^{(\gamma Z)}\tilde{C}^{A }_{rk}}&= F_{ee}\Bigl[- \mathbf{\Lambda^{-2}} \Bigl(16\beta \tilde{\gamma}^3 \Delta_{Z} e^3 m_{t}^3 Q_{e} v s_{\delta_{Z}} s_{\theta} (s_{\eta_{\gamma}} (\bar{g}_{At} \beta r_{\gamma} \bar{g}_{Ve} c_{\theta}-\bar{g}_{Ae} r_{\gamma} \bar{g}_{Vt})+\bar{g}_{Ae} Q_{t} r_{Z} s_{\eta_{Z}})\Bigr)\\
&+\mathbf{\Lambda^{-4}}\Bigl(32 \bar{g}_{Ae} \beta \tilde{\gamma}^3 \Delta_{Z} e^2  m_{t}^4 Q_{e} r_{\gamma} r_{Z} v^2 s_{\delta_{Z}} s_{\theta} s_{(\eta_{\gamma}-\eta_{Z})} \Bigr)\Bigr] \\
 \textcolor{purple}{{}^{(\gamma Z)}\tilde{C}^{A }_{kn}}&= F_{ee}\Bigl[- \mathbf{\Lambda^{-2}} \Bigl(16\beta \tilde{\gamma}^3 \Delta_{Z} e^3 m_{t}^3 Q_{e} v c_{\delta_{Z}} s_{\theta} (r_{\gamma} s_{\eta_{\gamma}} (\bar{g}_{Ae} \bar{g}_{At} \beta-\bar{g}_{Ve} \bar{g}_{Vt} c_{\theta})-Q_{t} r_{Z} \bar{g}_{Ve} s_{\eta_{Z}} c_{\theta})\Bigr)\\
 &+\mathbf{\Lambda^{-4}}\Bigl(16 \beta \tilde{\gamma}^3 \Delta_{Z} e^2  m_{t}^4 Q_{e} r_{\gamma} r_{Z} v^2 \bar{g}_{Ve} c_{\delta_{Z}} s_{2\theta} s_{(\eta_{\gamma}+\eta_{Z})}\Bigr)  \Bigr] \refstepcounter{equation}\tag{\theequation}
\end{align*}
The SM structure has been checked against
Ref.~\cite{Altakach:2026fpl} and verified numerically using the code of
Ref.~\cite{Durupt:2025wuk}.
Terms
proportional to the imaginary part of the $Z$-boson propagator are usually
neglected in the literature. Such terms generate, for instance, the SM contribution to $\tilde B_n$ \cite{Arens:1992wh} and to the
spin-correlation coefficients $\tilde C^S_{kn}$ and $\tilde C^S_{rn}$,
which are non-zero only in the presence of absorptive parts of the
scattering amplitudes \cite{Bernreuther:2015yna} in SM. Since these effects are
controlled by the absorptive part of the propagator, they are expected to be
suppressed away from the $Z$ pole. We nevertheless retain
these terms for completeness.\\

\subsection{$\gamma \gamma \rightarrow t\bar{t}$  } \label{app: spin density matrix gagatt}

In this appendix we present the structure of the spin production density
matrix for a \(t\bar t\) pair produced in photon-photon scattering
\(\gamma\gamma\to t\bar t\). At leading order, this process receives two
contributions, corresponding to top-quark exchange in the \(t\)- and
\(u\)-channels, as shown in Fig.~\ref{fig: gamma gamma scattering}. Unlike
the lepton-antilepton annihilation case, only the dipole interactions enter
the effective description considered here, through modifications of the
\(\gamma t\bar t\) vertex.
We use the same kinematic conventions as in Eq.~\eqref{eq:kinematics_eett}
and the shorthand notation introduced in App.~\ref{app: spin density matrix eett}. For the
\(\gamma\gamma\to t\bar t\) process, we further define
\begin{equation}
F_{\gamma\gamma}
\equiv
\frac{N_c (eQ_t)^2}{8\left(1-\beta^2 c_\theta^2\right)^2}.
\end{equation}
In the following, we list only the non-vanishing unnormalised Fano
coefficients.
\begin{align*}
\tilde{A} & =F_{\gamma\gamma}\Bigl[\Bigl(e^2 Q_t^2(8+8\beta^2-11\B^4+4\B^2(-2+\B^2)c_{2\theta}-\B^4c_{4\theta})\Bigr) \\
&+\mathbf{\Lambda^{-2}}\Big(-32 e Q_t v r_{\gamma} c_{\e_{\gamma}}m_t\nonumber(-2+\B^2+\B^2 c_{2\theta})\Big)\\
&+\mathbf{\Lambda^{-4}}\Big(4 m_t^2 v^2 r_{\gamma}^2 \tilde{\gamma}^2(32-40 \B^2+17 \B^4+\B^2(12(-2+\B^2)c_{2\theta}+3\B^2 c_{4\theta}\nonumber-8(-1+\B^2)s_{\theta}^2 c_{2\eta_{\gamma}}))\Big)\Bigr]
\\
{\tilde{C}_{nn}}&= F_{\gamma\gamma}\Bigl[ \Big(-e^2Q_t^2(8-16\B^2+11\B^4+\B^4(-4c_{2\theta}+c_{4\theta}))\Big)\\
&+\mathbf{\Lambda^{-2}}\Big(32 e Q_t v r_{\gamma}c_{\eta_{\gamma}}m_t(-2+\B^2+\B^2 c_{2\theta})\Big)\\
&+\mathbf{\Lambda^{-4}}\Big(16m_t^2 v^2r_{\gamma}^2\tilde{\gamma}^2((\tilde{\gamma}^{-2}(-8+5\B^2+3\B^2c_{2\theta}))+2\B^4c_{2\eta_\gamma}s_{\theta}^4)\Big)\Bigr]
\\
{\tilde{C}_{rr}}&= F_{\gamma\gamma}\Bigl[ \Big(-e^2Q_t^2(8+11\B^2(-2+\B^2)+\B^2(-2+\B^2)(-4c_{2\theta}+c_{4\theta}))\Big)\\
&+\mathbf{\Lambda^{-2}}\Big(8 e Q_t v r_{\gamma}c_{\eta_{\gamma}}m_t(-8+7\B^2+\B^2 c_{4\theta})\Big)\\
&+\mathbf{\Lambda^{-4}}\Big(4 m_t^2 v^2 r_{\gamma}^2 \tilde{\gamma}^2((\tilde{\gamma}^{-2}(-32+21\B^2+\B^2(8c_{2\theta}+3c_{4\theta})))+8\B^2c_{2\eta_\gamma}s_{\theta}^4)\Big)\Bigr]
\\
{\tilde{C}_{kk}}&= F_{\gamma\gamma}\Bigl[ \Big(e^2Q_t^2(-8+2\B^2+11\B^4-4\B^4c_{2\theta}+\B^2(-2+\B^2)c_{4\theta})\Big)\\
&+\mathbf{\Lambda^{-2}}\Big(-8 e Q_t v r_{\gamma}c_{\eta_{\gamma}}m_t(8-9\B^2+\B^2 c_{4\theta})\Big)
\\&+\mathbf{\Lambda^{-4}}\Big(4 m_t^2 v^2r_{\gamma}^2\tilde{\gamma}^2(-32+47\B^2-24\B^4+\B^2((20-8\B^2)c_{2\theta}-3c_{4\theta}-2(-1+\B^2)c_{2\eta_\gamma}s_{2\theta}^2))\Big)\Bigr]
\\
{\tilde{C}^{S }_{rk}}&= F_{\gamma\gamma}\Bigl[ \Big(16e^2 Q_t^2\B^2\tilde{\gamma}^{-1}c_{\theta}s_{\theta}^3\Big)+\mathbf{\Lambda^{-2}}\Big(8e Q_t v r_{\gamma}c_{\eta_\gamma}m_t\tilde{\gamma}\B^2(\B^2+(-2+\B^2)c_{2\theta})s_{2\theta}\Big)\\
&+\mathbf{\Lambda^{-4}}\Big(32 m_t^2 v^2 r_{\gamma}^2\B^2\tilde{\gamma}c_{\theta}(3+ c_{2\eta_{\gamma}})s_{\theta}^3\Big)\Bigr]
\\
\textcolor{purple}{\tilde{C}^{A }_{rn}}&= F_{\gamma\gamma}\Bigl[ \mathbf{\Lambda^{-2}}\Big(4eQ_t v r_{\gamma}s_{\e_\gamma}\B m_t(-20+15\B^2+4 c_{2\theta}+\B^2 c_{4\theta})\Big)+\mathbf{\Lambda^{-4}}\Big(32 m_t^2 v^2 \B^3r_{\gamma}^2 \tilde{\gamma}^2 s_{2\e_\gamma}s_{\theta}^4
\Big)\Bigr]
\\
 \textcolor{purple}{\tilde{C}^{A }_{kn}}&=F_{\gamma\gamma}\Bigl[\mathbf{\Lambda^{-2}}\Big(-8e Q_t v r_{\gamma}s_{\e_{\gamma}}\B m_t\tilde{\gamma}(2-3\B^2+\B^2 c_{2\theta})s_{2\theta}\Big)+\mathbf{\Lambda^{-4}}\Big(32 m_t^2 v^2 r_{\gamma}^2s_{2\eta_{\gamma}}\B^3\tilde{\gamma} c_{\theta}s_{\theta}^3\Big)\Bigr]\\
 \refstepcounter{equation}\tag{\theequation}
\end{align*}

\subsection{$q\bar{q}\rightarrow t\bar{t}$  } 
\label{app: spin density matrix qqtt}

In this appendix we present the structure of the spin production density
matrix for a \(t\bar t\) pair produced in quark-antiquark annihilation,
\(q\bar q\to t\bar t\). At leading order, this process can receive
\(s\)-channel contributions from gluon, photon, and \(Z\)-boson exchange. In
this subsection, however, we restrict ourselves to the QCD contribution, for
which the partonic amplitude is mediated by an \(s\)-channel gluon, as shown
in Fig.~\ref{fig:qq_scattering}. Possible EFT effects are included through
modifications of the effective \(gt\bar t\) vertex.

We use the same kinematic conventions as in Eq.~\eqref{eq:kinematics_eett}
and the shorthand notation introduced in
App.~\ref{app: spin density matrix eett}. The calculation is performed in
the \(t\bar t\) zero-momentum frame, which coincides with the zero-momentum
frame of the incoming partons, but not in general with the proton-proton
centre-of-mass frame. To define the top-quark helicity coordinate system in the \(t\bar t\)
zero-momentum frame, we choose the reference direction \(\hat p\) appearing in
Eq.~\eqref{eq:costheta} to coincide with the direction of the incoming
quark momentum, $\hat p \equiv \hat p_q $.
With respect to Eq.~\eqref{eq:kinematics_eett}, the incoming momenta are
identified as
\begin{equation}
    p_1=p_{\bar q},
    \qquad
    p_2=p_q.
\end{equation}
For compactness of the analytical expressions in both \(q\bar q\to t\bar t\) and \(gg\to t\bar t\), we introduce the notation
\begin{equation}
d_G
=
r_G e^{i\eta_G}=\sqrt{2}C_{tG}.
\label{eq:dG_definition}
\end{equation}
Here, \(r_G=|d_G|\) denotes the magnitude of the dipole coefficient,
while \(\eta_G=\arg(d_G)\) is its CP phase, controlling the relative
contributions of the CP-even chromomagnetic and CP-odd chromoelectric
interactions. For the \(q\bar q\to t\bar t\) process, we further define
\begin{equation}
    F_{q\bar q}\equiv \frac{g_s^2(N_c^2-1)}{8N_c^2} .
\end{equation}
 In the following, we list only the non-vanishing unnormalised Fano
coefficients.
\begin{align*}
 \tilde{A} & = F_{q\bar q}\Bigl[\Bigl(\frac{1}{4}g_s^2 \left(\beta^2c_{2 \theta}-\beta^2+4\right)\Bigr)+\mathbf{\Lambda^{-2}}\Bigl(4g_s m_t r_G vc_{\eta_G} \Bigr)\\
 &+ \mathbf{\Lambda^{-4}}\Bigl(m_t^2 r_G^2 v^2 \left(\tilde{\gamma}^2(2-\beta^2c_{2 \theta}-\beta^2)+ 2c_{2 \eta_G}\right) \Bigr)\Bigr]
 \\
 \tilde{C}_{nn} &=  F_{q\bar q}\Bigl[-\Bigl(\frac{1}{2} \beta^2g_s^2s ^2_{\theta}\Bigr)+\mathbf{\Lambda^{-4}}\Bigl(2 \beta^2 \tilde{\gamma}^2 m_t^2 r_G^2 v^2c_{2 \eta_G}s ^2_{\theta}\Bigr)\Bigr]\\
 \tilde{C}_{rr} &= F_{q\bar q}\Bigl[ -\Bigl(\frac{1}{2} \left(\beta^2-2\right)g_s^2s ^2_{\theta}\Bigr)+\mathbf{\Lambda^{-2}}\Bigl(4g_s m_t r_G vc_{\eta_G}s ^2_{\theta}\Bigr)+\mathbf{\Lambda^{-4}}\Bigl(2 
  m_t^2 r_G^2 v^2s ^2_{\theta} \left(\tilde{\gamma}^2c_{2 \eta_G}+1\right)\Bigr)\Bigr] \\
 \tilde{C}_{kk} &=  F_{q\bar q}\Bigl[\Bigl(\frac{1}{4}g_s^2 \left(\left(2-\beta^2\right)c_{2 \theta }+\beta^2+2\right)\Bigr)+\mathbf{\Lambda^{-2}}\Bigl(4g_s m_t r_G vc_{\eta_G}c^2_{\theta}\Bigr)\\
 &-  \mathbf{\Lambda^{-4}}\Bigl(\tilde{\gamma}^2m_t^2 r_G^2 v^2 \left(2 \beta^2s ^2_{\eta_G}s ^2_{\theta }+c^2_{\eta_G}
   \left(\left(\beta^2-2\right)c_{2 \theta }+3 \beta^2-2\right)\right)\Bigr)\Bigr]\\
 \tilde{C}^S_{rk} &=  F_{q\bar q}\Bigl[\Bigl(g_s^2s _{\theta}c_{\theta}\tilde{\gamma}^{-1}\Bigr)-\mathbf{\Lambda^{-2}}\Bigl(\left(\beta^2-2\right) \tilde{\gamma}g_s m_t r_G vc_{\eta_G}s_{2 \theta}\Bigr)+\mathbf{\Lambda^{-4}}\Bigl(4 \tilde{\gamma} m_t^2 r_G^2 v^2c^2_{\eta_G}s _{\theta}c_{\theta}\Bigr)\Bigr]\\
 \textcolor{purple}{ \tilde{C}^A_{rn}} &= F_{q\bar q}\Bigl[\mathbf{\Lambda^{-2}}\Bigl(2 \beta g_s m_t r_G vs_{\eta_G}s ^2_{\theta}\Bigr)+\mathbf{\Lambda^{-4}}\Bigl(4 \beta \tilde{\gamma}^2 m_t^2 r_G^2 v^2s_{\eta_G}c_{\eta_G}s ^2_{\theta }\Bigr)\Bigr]\\
\textcolor{purple}{ \tilde{C}^A_{kn}} &= F_{q\bar q}\Bigl[\mathbf{\Lambda^{-2}}\Bigl(\beta \widetilde{\gamma}g_s m_t r_G vs_{\eta_G}s_{2 \theta }\Bigr)+\mathbf{\Lambda^{-4}}\Bigl(\beta \tilde{\gamma} m_t^2
   r_G^2 v^2s_{2 \eta_G}s_{2 \theta }\Bigr)\Bigr]
 \refstepcounter{equation}\tag{\theequation}
\end{align*}
The structure of $\tilde A$ is consistent with Ref.~\cite{Haberl:1995ek}. We have compared our results with Refs.~\cite{Bernreuther:2015yna,Aoude:2022imd}.
Reference~\cite{Aoude:2022imd} provides the SM contribution and the CP-even
EFT corrections up to quadratic order in the corresponding couplings for $N_c=3$. After
translating to our coupling normalisation and spin-basis conventions, we find
complete agreement with their SM expressions. For the EFT contributions, we
also find a good agreement with their results, except, for example, for the spin-correlation coefficient
\(\tilde C_{rk}\), which differs by an overall sign.

Reference~\cite{Bernreuther:2015yna} includes both CP-even and CP-odd
contributions, but only up to linear order in the EFT couplings. After
accounting for the different normalisation of the effective couplings, we find
agreement with their linear results. The expressions presented here extend
these comparisons by retaining both CP-even and CP-odd contributions up to
quadratic order in the EFT couplings.

\subsection{$g g \rightarrow t\bar{t}$  } 
\label{app: spin density matrix ggtt}

In this appendix we present the structure of the spin production density
matrix for a \(t\bar t\) pair produced in gluon fusion,
\(gg\to t\bar t\). At leading order in QCD, this process receives
contributions from top-quark exchange in the \(t\)- and \(u\)-channels and
from an \(s\)-channel gluon-exchange diagram. In the presence of the top
dipole operators, the \(gt\bar t\) vertex is modified and a four-point
\(ggt\bar t\) contact interaction is generated. The corresponding diagrams
are shown in Fig.~\ref{fig:gg_scattering}.

The calculation is performed in the \(t\bar t\) zero-momentum frame, which
coincides with the zero-momentum frame of the incoming partons, but not in
general with the proton-proton centre-of-mass frame. We use the same
kinematic conventions as in Eq.~\eqref{eq:kinematics_eett} and the shorthand
notation introduced in App.~\ref{app: spin density matrix eett}. For the
\(gg\to t\bar t\) process, we further define
\begin{equation}
F_{gg}
\equiv
\frac{g_s^2}{128\,N_c\left(N_c^2-1\right)
\left(1-\beta^2 c_\theta^2\right)^2}.
\end{equation}
In the following, we list only the non-vanishing unnormalised Fano
coefficients.
\begin{align}
\tilde{A} & =F_{gg}\Bigl[\Bigl(g_s^2 (-4 +N_c^2(2+\B^2)+N_c^2 \B^2 c_{2\theta})(8+8\B^2-11\B^4+4\B^2(-2+\B^2)c_{2\theta}-\B^4 c_{4\theta}) \Bigr)\nonumber\\
&+\mathbf{\Lambda^{-2}}\Big(-32 g_s m_t v r_{G} c_{\eta_G} (-2+\B^2+\B^2 c_{2\theta})(-4+N_c^2(2+\B^2)+N_c^2 \B^2 c_{2\theta})\Big)\nonumber\\
&+\mathbf{\Lambda^{-4}}\Big(2 v^2 m_t^2 r_{G}^2\tilde{\gamma}^2(-8(32-40\B^2+17\B^4)+2N_c^2(80-96\B^2+28\B^4+7\B^6)\nonumber\\
&+\B^2(2\B^2(-12+N_c^2(4+\B^2))c_{4\theta}+N_c^2\B^4c_{6\theta}+16\tilde{\gamma}^{-2}(-4+N_c^2(2+\B^2))c_{2\eta_G}s_{\theta}^2\nonumber\\
&+c_{2\theta}(-96(-2+\B^2)+N_c^2(-96+32\B^2+15\B^4)+16N_c^2\B^2\tilde{\gamma}^{-2}c_{2\eta_G}s_{\theta}^2)))\Big)\Bigr]
\end{align}

\begin{align*}
{\tilde{C}_{nn}}&=F_{gg}\Bigl[\Bigl(-g_s^2 (-4 +N_c^2(2+\B^2)+N_c^2 \B^2 c_{2\theta})(8-16\B^2+11\B^4+\B^4(-4 c_{2\theta}+c_{4\theta}))\Bigr) \\
&+\mathbf{\Lambda^{-2}}\Big(64 g_s m_t v r_{G} c_{\eta_G}(N_c^2-2)(-2+\B^2+\B^2 c_{2\theta})\Big)\\
&+\mathbf{\Lambda^{-4}}\Big(32 m_t^2 v^2 \tilde{\gamma}^2 r_{G}^2(-\tilde{\gamma}^{-2}(2(-8+5\B^2)-N_c^2(-6+3\B^2+\B^4)-\B^2(-6+N_c^2(1+\B^2))c_{2\theta})\\
&+\B^4(-4+N_c^2(2+\B^2)+N_c^2\B^2c_{2\theta})c_{2\eta_G}s_{\theta}^4)\Big)\Bigr]
\\
{\tilde{C}_{rr}}&= F_{gg}\Bigl[\Bigl(-g_s^2 (-4 +N_c^2(2+\B^2)+N_c^2 \B^2 c_{2\theta})(8+11\B^2(\B^2-2)+\B^2(\B^2-2)(-4 c_{2\theta}+c_{4\theta}))\Bigr) \\
&+\mathbf{\Lambda^{-2}}\Big(4g_s m_t v r_{G} c_{\eta_G}(64-56\B^2+2N_c^2(-16+14\B^2+\B^4)-N_c^2\B^4 c_{2\theta}\\
&-2\B^2(4+N_c^2(-2+\B^2))c_{4\theta}+N_c^2\B^4c_{6\theta})\Big)+\mathbf{\Lambda^{-4}}\\
&*\Big(2m_t^2 \tilde{\gamma}^2 v^2 r_{G}^2((\tilde{\gamma}^{-2}(256-168\B^2+2N_c^2(26\B^2+9\B^4-64)+(N_c^2(32+15\B^4)-64\B^2)c_{2\theta}\\
&-2\B^2(12+N_c^2(-6+\B^2))c_{4\theta}+N_c^2\B^4c_{6\theta}))-16\B^2(-4+N_c^2(2+\B^2)-N_c^2\B^2 c_{2\theta})c_{2\eta_G}s_{\theta}^4)\Big)\Bigr]
\\
{\tilde{C}_{kk}}&= F_{gg}\Bigl[\Bigl(g_s^2 (-4 +N_c^2(2+\B^2)+N_c^2 \B^2 c_{2\theta})(-8+2\B^2+11\B^4-4\B^4 c_{2\theta}+\B^2(-2+\B^2)c_{4\theta})\Bigr) \\
&+\mathbf{\Lambda^{-2}}\Big(-8g_s m_t vr_{G} c_{\eta_G}(-4+N_c^2(2+\B^2)+N_c^2\B^2 c_{2\theta})(8-9\B^2+\B^2 c_{4\theta})\Big)\\
&+\mathbf{\Lambda^{-4}}\Big(2 m_t^2 v^2\tilde{\gamma}^2 r_{G}^2(8(32-47\B^2+24\B^4)-2N_c^2(64-94\B^2+39\B^4+10\B^6)\\
&+c_{2\theta}(32\B^2(-5+2\B^2)-N_c^2(32-112\B^2+15\B^4+16\B^6)-4N_c^2\B^4(\B^2-1)c_{2\eta_G}s_{2\theta}^2)+\B^2\\
&*(2(12+N_c^2(-2+\B^2)(3+2\B^2))c_{4\theta}-N_c^2\B^2c_{6\theta}-4(\B^2-1)(-4+N_c^2(2+\B^2))c_{2\eta_G}s_{2\theta}^2))\Big)\Bigr]
\\
{\tilde{C}^{S }_{rk}}&= F_{gg}\Bigl[\Bigl(16 g_s^2 \B^2 \tilde{\gamma}^{-1}c_{\theta}s_{\theta}^3(-4+N_c^2(2+\B^2)+N_c^2\B^2c_{2\theta})\Bigr)+\mathbf{\Lambda^{-2}}\\
&*\Big(4 g_s m_t v \tilde{\gamma} r_{G} c_{\eta_G} s_{2\theta}\B^2(-8\B^2-N_c^2(8-14\B^2+\B^4)+(\B^2-2)(4(N_c^2-2)c_{2\theta}+N_c^2\B^2 c_{4\theta}))\Big)
\\
&+\mathbf{\Lambda^{-4}}\Big(-2 m_t^2 v^2 \tilde{\gamma} r_{G}^2 s_{2\theta}(48\B^2-2N_c^2(-16+28\B^2+\B^4)+2\B^2(-24 c_{2\theta}
+N_c^2(12 c_{2\theta}+\B^2 c_{4\theta})\\
&-4(-4+N_c^2(2+\B^2)+N_c^2\B^2 c_{2\theta})c_{2\eta_G}s_{\theta
}^2))\Big)\Bigr]
\\
\textcolor{purple}{\tilde{C}^{A }_{rn}}&=F_{gg}\Bigl[\mathbf{\Lambda^{-2}}\Big(2 g_s m_t v  r_{G} s_{\eta_G}\B(2(80-60\B^2+N_c^2(-32+14\B^2+9\B^4))+(-32+N_c^2(32-32\B^2\\
&+15\B^4))c_{2\theta}-2\B^2(4+N_c^2(-2+\B^2))c_{4\theta}+N_c^2\B^4 c_{6\theta})\Big)\\
&+\mathbf{\Lambda^{-4}}\Big(32 m_t^2 \tilde{\gamma}^2 v^2 r_{G}^2 s_{2\eta_G}\B^3 s_{\theta}^4(-4+N_c^2(2+\B^2)+N_c^2\B^2c_{2\theta}) \Big)\Bigr]
\\
 \textcolor{purple}{\tilde{C}^{A }_{kn}}&=F_{gg}\Bigl[\mathbf{\Lambda^{-2}}\Big(-4m_t v\tilde{\gamma}r_{G} s_{\eta_G}s_{2\theta}\B(8(-2+3\B^2)-N_c^2(-16+20\B^2+\B^4)+4(-2+N_c^2)\B^2c_{2\theta}\\
 &+N_c^2\B^4 c_{4\theta})\Big)+\mathbf{\Lambda^{-4}}\Big(32 m_t^2\tilde{\gamma} v^2  r_G^2 s_{2\eta_G} \B^3 c_{\theta} s_{\theta}^3(-4+N_c^2(2+\B^2)+N_c^2\B^2c_{2\theta})\Big)\Bigr]
   \refstepcounter{equation}\tag{\theequation}
\end{align*}

We have compared our results with Refs.~\cite{Bernreuther:2015yna,Aoude:2022imd}.
Reference~\cite{Aoude:2022imd} provides the SM contribution and the CP-even
EFT corrections up to quadratic order in the corresponding couplings for $N_c=3$. After
translating to our coupling normalisation and spin-basis conventions, we find
complete agreement with their SM expressions. For the EFT contributions, we
also find agreement, except for the spin-correlation coefficient
\(\tilde C_{rk}\), which differs by an overall sign.

Reference~\cite{Bernreuther:2015yna} includes both CP-even and CP-odd
contributions, but only up to linear order in the EFT couplings. After
accounting for the different normalisation of the effective couplings, we find
agreement with their linear results. The expressions presented here extend
these comparisons by retaining both CP-even and CP-odd contributions up to
quadratic order in the EFT couplings.

\bibliographystyle{JHEP}
\bibliography{main}

\providecommand{\href}[2]{#2}\begingroup\raggedright\begin{thebibliography}{100}

\bibitem{Barger:1988jj}
V.D.~Barger, J.~Ohnemus and R.J.N.~Phillips, \emph{{Spin Correlation Effects in the Hadroproduction and Decay of Very Heavy Top Quark Pairs}}, \href{https://doi.org/10.1142/S0217751X89000297}{\emph{Int. J. Mod. Phys. A} {\bfseries 4} (1989) 617}.

\bibitem{Mahlon:1995zn}
G.~Mahlon and S.J.~Parke, \emph{{Angular correlations in top quark pair production and decay at hadron colliders}}, \href{https://doi.org/10.1103/PhysRevD.53.4886}{\emph{Phys. Rev. D} {\bfseries 53} (1996) 4886} [\href{https://arxiv.org/abs/hep-ph/9512264}{{\ttfamily hep-ph/9512264}}].

\bibitem{Stelzer:1995gc}
T.~Stelzer and S.~Willenbrock, \emph{{Spin correlation in top quark production at hadron colliders}}, \href{https://doi.org/10.1016/0370-2693(96)00178-5}{\emph{Phys. Lett. B} {\bfseries 374} (1996) 169} [\href{https://arxiv.org/abs/hep-ph/9512292}{{\ttfamily hep-ph/9512292}}].

\bibitem{Parke:1996pr}
S.J.~Parke and Y.~Shadmi, \emph{{Spin correlations in top quark pair production at $e^{+} e^{-}$ colliders}}, \href{https://doi.org/10.1016/0370-2693(96)00998-7}{\emph{Phys. Lett. B} {\bfseries 387} (1996) 199} [\href{https://arxiv.org/abs/hep-ph/9606419}{{\ttfamily hep-ph/9606419}}].

\bibitem{Mahlon:1996pn}
G.~Mahlon and S.J.~Parke, \emph{{Improved spin basis for angular correlation studies in single top quark production at the Tevatron}}, \href{https://doi.org/10.1103/PhysRevD.55.7249}{\emph{Phys. Rev. D} {\bfseries 55} (1997) 7249} [\href{https://arxiv.org/abs/hep-ph/9611367}{{\ttfamily hep-ph/9611367}}].

\bibitem{Mahlon:1997uc}
G.~Mahlon and S.J.~Parke, \emph{{Maximizing spin correlations in top quark pair production at the Tevatron}}, \href{https://doi.org/10.1016/S0370-2693(97)00987-8}{\emph{Phys. Lett. B} {\bfseries 411} (1997) 173} [\href{https://arxiv.org/abs/hep-ph/9706304}{{\ttfamily hep-ph/9706304}}].

\bibitem{Brandenburg:2002xr}
A.~Brandenburg, Z.G.~Si and P.~Uwer, \emph{{QCD corrected spin analyzing power of jets in decays of polarized top quarks}}, \href{https://doi.org/10.1016/S0370-2693(02)02098-1}{\emph{Phys. Lett. B} {\bfseries 539} (2002) 235} [\href{https://arxiv.org/abs/hep-ph/0205023}{{\ttfamily hep-ph/0205023}}].

\bibitem{D0:2015kta}
{\scshape D0} collaboration, \emph{{Measurement of Spin Correlation between Top and Antitop Quarks Produced in $p\bar{p}$ Collisions at $\sqrt{s} =$ 1.96 TeV}}, \href{https://doi.org/10.1016/j.physletb.2016.03.053}{\emph{Phys. Lett. B} {\bfseries 757} (2016) 199} [\href{https://arxiv.org/abs/1512.08818}{{\ttfamily 1512.08818}}].

\bibitem{CMS:2013roq}
{\scshape CMS} collaboration, \emph{{Measurements of $t\bar{t}$ Spin Correlations and Top-Quark Polarization Using Dilepton Final States in $pp$ Collisions at $\sqrt{s}$ = 7 TeV}}, \href{https://doi.org/10.1103/PhysRevLett.112.182001}{\emph{Phys. Rev. Lett.} {\bfseries 112} (2014) 182001} [\href{https://arxiv.org/abs/1311.3924}{{\ttfamily 1311.3924}}].

\bibitem{ATLAS:2014aus}
{\scshape ATLAS} collaboration, \emph{{Measurements of spin correlation in top-antitop quark events from proton-proton collisions at $\sqrt{s}=7$ TeV using the ATLAS detector}}, \href{https://doi.org/10.1103/PhysRevD.90.112016}{\emph{Phys. Rev. D} {\bfseries 90} (2014) 112016} [\href{https://arxiv.org/abs/1407.4314}{{\ttfamily 1407.4314}}].

\bibitem{CMS:2016piu}
{\scshape CMS} collaboration, \emph{{Measurements of t t-bar spin correlations and top quark polarization using dilepton final states in pp collisions at sqrt(s) = 8 TeV}}, \href{https://doi.org/10.1103/PhysRevD.93.052007}{\emph{Phys. Rev. D} {\bfseries 93} (2016) 052007} [\href{https://arxiv.org/abs/1601.01107}{{\ttfamily 1601.01107}}].

\bibitem{ATLAS:2016bac}
{\scshape ATLAS} collaboration, \emph{{Measurements of top quark spin observables in $ t\overline{t} $ events using dilepton final states in $ \sqrt{s}=8 $ TeV pp collisions with the ATLAS detector}}, \href{https://doi.org/10.1007/JHEP03(2017)113}{\emph{JHEP} {\bfseries 03} (2017) 113} [\href{https://arxiv.org/abs/1612.07004}{{\ttfamily 1612.07004}}].

\bibitem{CMS:2019nrx}
{\scshape CMS} collaboration, \emph{{Measurement of the top quark polarization and $\mathrm{t\bar{t}}$ spin correlations using dilepton final states in proton-proton collisions at $\sqrt{s} =$ 13 TeV}}, \href{https://doi.org/10.1103/PhysRevD.100.072002}{\emph{Phys. Rev. D} {\bfseries 100} (2019) 072002} [\href{https://arxiv.org/abs/1907.03729}{{\ttfamily 1907.03729}}].

\bibitem{ATLAS:2019zrq}
{\scshape ATLAS} collaboration, \emph{{Measurements of top-quark pair spin correlations in the $e\mu$ channel at $\sqrt{s} = 13$ TeV using $pp$ collisions in the ATLAS detector}}, \href{https://doi.org/10.1140/epjc/s10052-020-8181-6}{\emph{Eur. Phys. J. C} {\bfseries 80} (2020) 754} [\href{https://arxiv.org/abs/1903.07570}{{\ttfamily 1903.07570}}].

\bibitem{ATLAS:2023fsd}
{\scshape ATLAS} collaboration, \emph{{Observation of quantum entanglement with top quarks at the ATLAS detector}}, \href{https://doi.org/10.1038/s41586-024-07824-z}{\emph{Nature} {\bfseries 633} (2024) 542} [\href{https://arxiv.org/abs/2311.07288}{{\ttfamily 2311.07288}}].

\bibitem{CMS:2024zkc}
{\scshape CMS} collaboration, \emph{{Measurements of polarization and spin correlation and observation of entanglement in top quark pairs using lepton+jets events from proton-proton collisions at s=13{\,}{\,}TeV}}, \href{https://doi.org/10.1103/PhysRevD.110.112016}{\emph{Phys. Rev. D} {\bfseries 110} (2024) 112016} [\href{https://arxiv.org/abs/2409.11067}{{\ttfamily 2409.11067}}].

\bibitem{CMS:2024pts}
{\scshape CMS} collaboration, \emph{{Observation of quantum entanglement in top quark pair production in proton{\textendash}proton collisions at $\sqrt{s} = 13$ TeV}}, \href{https://doi.org/10.1088/1361-6633/ad7e4d}{\emph{Rept. Prog. Phys.} {\bfseries 87} (2024) 117801} [\href{https://arxiv.org/abs/2406.03976}{{\ttfamily 2406.03976}}].

\bibitem{Fano:1957zz}
U.~Fano, \emph{{Description of States in Quantum Mechanics by Density Matrix and Operator Techniques}}, \href{https://doi.org/10.1103/RevModPhys.29.74}{\emph{Rev. Mod. Phys.} {\bfseries 29} (1957) 74}.

\bibitem{Fano:1983zz}
U.~Fano, \emph{{Pairs of two-level systems}}, \href{https://doi.org/10.1103/RevModPhys.55.855}{\emph{Rev. Mod. Phys.} {\bfseries 55} (1983) 855}.

\bibitem{2003.02280}
Y.~Afik and J.R.M.n.~de~Nova, \emph{{Entanglement and quantum tomography with top quarks at the LHC}}, \href{https://doi.org/10.1140/epjp/s13360-021-01902-1}{\emph{Eur. Phys. J. Plus} {\bfseries 136} (2021) 907} [\href{https://arxiv.org/abs/2003.02280}{{\ttfamily 2003.02280}}].

\bibitem{Fabbrichesi:2021npl}
M.~Fabbrichesi, R.~Floreanini and G.~Panizzo, \emph{{Testing Bell Inequalities at the LHC with Top-Quark Pairs}}, \href{https://doi.org/10.1103/PhysRevLett.127.161801}{\emph{Phys. Rev. Lett.} {\bfseries 127} (2021) 161801} [\href{https://arxiv.org/abs/2102.11883}{{\ttfamily 2102.11883}}].

\bibitem{Severi:2021cnj}
C.~Severi, C.D.E.~Boschi, F.~Maltoni and M.~Sioli, \emph{{Quantum tops at the LHC: from entanglement to Bell inequalities}}, \href{https://doi.org/10.1140/epjc/s10052-022-10245-9}{\emph{Eur. Phys. J. C} {\bfseries 82} (2022) 285} [\href{https://arxiv.org/abs/2110.10112}{{\ttfamily 2110.10112}}].

\bibitem{Severi:2022qjy}
C.~Severi and E.~Vryonidou, \emph{{Quantum entanglement and top spin correlations in SMEFT at higher orders}}, \href{https://doi.org/10.1007/JHEP01(2023)148}{\emph{JHEP} {\bfseries 01} (2023) 148} [\href{https://arxiv.org/abs/2210.09330}{{\ttfamily 2210.09330}}].

\bibitem{Aoude:2022imd}
R.~Aoude, E.~Madge, F.~Maltoni and L.~Mantani, \emph{{Quantum SMEFT tomography: Top quark pair production at the LHC}}, \href{https://doi.org/10.1103/PhysRevD.106.055007}{\emph{Phys. Rev. D} {\bfseries 106} (2022) 055007} [\href{https://arxiv.org/abs/2203.05619}{{\ttfamily 2203.05619}}].

\bibitem{Afik:2022kwm}
Y.~Afik and J.R.M.n.~de~Nova, \emph{{Quantum information with top quarks in QCD}}, \href{https://doi.org/10.22331/q-2022-09-29-820}{\emph{Quantum} {\bfseries 6} (2022) 820} [\href{https://arxiv.org/abs/2203.05582}{{\ttfamily 2203.05582}}].

\bibitem{Aguilar-Saavedra:2022uye}
J.A.~Aguilar-Saavedra and J.A.~Casas, \emph{{Improved tests of entanglement and Bell inequalities with LHC tops}}, \href{https://doi.org/10.1140/epjc/s10052-022-10630-4}{\emph{Eur. Phys. J. C} {\bfseries 82} (2022) 666} [\href{https://arxiv.org/abs/2205.00542}{{\ttfamily 2205.00542}}].

\bibitem{Afik:2022dgh}
Y.~Afik and J.R.M.~de~Nova, \emph{{Quantum Discord and Steering in Top Quarks at the LHC}}, \href{https://doi.org/10.1103/PhysRevLett.130.221801}{\emph{Phys. Rev. Lett.} {\bfseries 130} (2023) 221801} [\href{https://arxiv.org/abs/2209.03969}{{\ttfamily 2209.03969}}].

\bibitem{Cheng:2023qmz}
K.~Cheng, T.~Han and M.~Low, \emph{{Optimizing fictitious states for Bell inequality violation in bipartite qubit systems with applications to the tt\textasciimacron{} system}}, \href{https://doi.org/10.1103/PhysRevD.109.116005}{\emph{Phys. Rev. D} {\bfseries 109} (2024) 116005} [\href{https://arxiv.org/abs/2311.09166}{{\ttfamily 2311.09166}}].

\bibitem{Han:2023fci}
T.~Han, M.~Low and T.A.~Wu, \emph{{Quantum entanglement and Bell inequality violation in semi-leptonic top decays}}, \href{https://doi.org/10.1007/JHEP07(2024)192}{\emph{JHEP} {\bfseries 07} (2024) 192} [\href{https://arxiv.org/abs/2310.17696}{{\ttfamily 2310.17696}}].

\bibitem{Dong:2023xiw}
Z.~Dong, D.~Gon\c{c}alves, K.~Kong and A.~Navarro, \emph{{Entanglement and Bell inequalities with boosted tt\textasciimacron{}}}, \href{https://doi.org/10.1103/PhysRevD.109.115023}{\emph{Phys. Rev. D} {\bfseries 109} (2024) 115023} [\href{https://arxiv.org/abs/2305.07075}{{\ttfamily 2305.07075}}].

\bibitem{Cheng:2024btk}
K.~Cheng, T.~Han and M.~Low, \emph{{Optimizing Entanglement and Bell Inequality Violation in Top Anti-Top Events}},  \href{https://arxiv.org/abs/2407.01672}{{\ttfamily 2407.01672}}.

\bibitem{Aguilar-Saavedra:2024hwd}
J.A.~Aguilar-Saavedra, \emph{{A closer look at post-decay $t \bar t$ entanglement}}, \href{https://doi.org/10.1103/PhysRevD.109.096027}{\emph{Phys. Rev. D} {\bfseries 109} (2024) 096027} [\href{https://arxiv.org/abs/2401.10988}{{\ttfamily 2401.10988}}].

\bibitem{Aguilar-Saavedra:2023lwb}
J.A.~Aguilar-Saavedra, \emph{{Decay of entangled fermion pairs with post-selection}}, \href{https://doi.org/10.1016/j.physletb.2023.138409}{\emph{Phys. Lett. B} {\bfseries 848} (2024) 138409} [\href{https://arxiv.org/abs/2308.07412}{{\ttfamily 2308.07412}}].

\bibitem{DelGratta:2025qyp}
M.~Del~Gratta, F.~Fabbri, P.~Lamba, F.~Maltoni and D.~Pagani, \emph{{Quantum properties of $H\to VV^*$: precise predictions in the SM and sensitivity to new physics}},  \href{https://arxiv.org/abs/2504.03841}{{\ttfamily 2504.03841}}.

\bibitem{Lamba:2026qnk}
P.~Lamba, M.~Del~Gratta, F.~Fabbri, F.~Maltoni and D.~Pagani, \emph{{Precise Standard Model predictions for quantum properties in $H\rightarrow ZZ^*\rightarrow 4l$}}, \href{https://doi.org/10.1140/epjp/s13360-026-07933-w}{\emph{Eur. Phys. J. Plus} {\bfseries 141} (2026) 710}.

\bibitem{Afik:2025ejh}
Y.~Afik et~al., \emph{{Quantum information meets high-energy physics: input to the update of the European strategy for particle physics}}, \href{https://doi.org/10.1140/epjp/s13360-025-06752-9}{\emph{Eur. Phys. J. Plus} {\bfseries 140} (2025) 855} [\href{https://arxiv.org/abs/2504.00086}{{\ttfamily 2504.00086}}].

\bibitem{A:2026pug}
S.K.~A., A.~Das and S.~Mandal, \emph{{Quantum spin correlations in $Z^\prime$-mediated $t\bar{t}$ production at future lepton colliders}},  \href{https://arxiv.org/abs/2607.15153}{{\ttfamily 2607.15153}}.

\bibitem{LoChiatto:2024dmx}
P.~Lo~Chiatto, \emph{{Interference resurrection of the {\ensuremath{\tau}} dipole through quantum tomography}}, \href{https://doi.org/10.1103/8gtq-twfc}{\emph{Phys. Rev. D} {\bfseries 112} (2025) 015017} [\href{https://arxiv.org/abs/2408.04553}{{\ttfamily 2408.04553}}].

\bibitem{Aguilar-Saavedra:2024fig}
J.A.~Aguilar-Saavedra and J.A.~Casas, \emph{{Entanglement Autodistillation from Particle Decays}}, \href{https://doi.org/10.1103/PhysRevLett.133.111801}{\emph{Phys. Rev. Lett.} {\bfseries 133} (2024) 111801} [\href{https://arxiv.org/abs/2401.06854}{{\ttfamily 2401.06854}}].

\bibitem{Cheng:2024rxi}
K.~Cheng, T.~Han and M.~Low, \emph{{Quantum Tomography at Colliders: With or Without Decays}},  \href{https://arxiv.org/abs/2410.08303}{{\ttfamily 2410.08303}}.

\bibitem{Aguilar-Saavedra:2024vpd}
J.A.~Aguilar-Saavedra, \emph{{Full quantum tomography of top quark decays}}, \href{https://doi.org/10.1016/j.physletb.2024.138849}{\emph{Phys. Lett. B} {\bfseries 855} (2024) 138849} [\href{https://arxiv.org/abs/2402.14725}{{\ttfamily 2402.14725}}].

\bibitem{Han:2024ugl}
T.~Han, M.~Low, N.~McGinnis and S.~Su, \emph{{Measuring quantum discord at the LHC}}, \href{https://doi.org/10.1007/JHEP05(2025)081}{\emph{JHEP} {\bfseries 05} (2025) 081} [\href{https://arxiv.org/abs/2412.21158}{{\ttfamily 2412.21158}}].

\bibitem{Maltoni:2024tul}
F.~Maltoni, C.~Severi, S.~Tentori and E.~Vryonidou, \emph{{Quantum detection of new physics in top-quark pair production at the LHC}}, \href{https://doi.org/10.1007/JHEP03(2024)099}{\emph{JHEP} {\bfseries 03} (2024) 099} [\href{https://arxiv.org/abs/2401.08751}{{\ttfamily 2401.08751}}].

\bibitem{Maltoni:2024csn}
F.~Maltoni, C.~Severi, S.~Tentori and E.~Vryonidou, \emph{{Quantum tops at circular lepton colliders}}, \href{https://doi.org/10.1007/JHEP09(2024)001}{\emph{JHEP} {\bfseries 09} (2024) 001} [\href{https://arxiv.org/abs/2404.08049}{{\ttfamily 2404.08049}}].

\bibitem{Aoude:2025jzc}
R.~Aoude, H.~Banks, C.D.~White and M.J.~White, \emph{{Probing new physics in the top sector using quantum information}},  \href{https://arxiv.org/abs/2505.12522}{{\ttfamily 2505.12522}}.

\bibitem{Fabbrichesi:2025psr}
M.~Fabbrichesi, R.~Floreanini and L.~Marzola, \emph{{Local vs. nonlocal entanglement in top-quark pairs at the LHC}},  \href{https://arxiv.org/abs/2505.02902}{{\ttfamily 2505.02902}}.

\bibitem{Altakach:2026fpl}
M.M.~Altakach, P.~Lamba, F.~Maltoni and K.~Sakurai, \emph{{Quantum properties of heavy-fermion pairs at a lepton collider with polarised beams}},  \href{https://arxiv.org/abs/2601.09558}{{\ttfamily 2601.09558}}.

\bibitem{Choi:2026omc}
S.Y.~Choi, D.W.~Kang, J.S.~Lee and C.B.~Park, \emph{{Quantum entanglement and Bell nonlocality in top-quark pair production at a photon linear collider}},  \href{https://arxiv.org/abs/2603.12830}{{\ttfamily 2603.12830}}.

\bibitem{Arai:2026jtc}
M.~Arai, K.~Mawatari and N.~Okada, \emph{{Disentangling new physics with quantum entanglement in $t\bar{t}$ production at future lepton colliders}},  \href{https://arxiv.org/abs/2604.21332}{{\ttfamily 2604.21332}}.

\bibitem{Fang:2026ddi}
Y.-J.~Fang, A.~Bhoonah, K.~Cheng, T.~Han, Y.~Liu and H.~Zhang, \emph{{Spin Correlation and Quantum Entanglement of Fermion Pairs in Transversely Polarized $e^-e^+$ Collisions}},  \href{https://arxiv.org/abs/2604.11887}{{\ttfamily 2604.11887}}.

\bibitem{Antozzi:2026vdi}
L.~Antozzi, E.~Chalbaud, F.~D{\'e}liot, F.~Fabbri, M.C.N.~Fiolhais, B.~Fuks et~al., \emph{{Extracting a Toponium Signal at the LHC with Spin and Quantum Information Tools}},  \href{https://arxiv.org/abs/2602.23426}{{\ttfamily 2602.23426}}.

\bibitem{Afik:2026pxv}
Y.~Afik, R.~Demina, A.~Herrera, O.~Heinz~Hindrichs, J.R.M.~de~Nova and B.~Ravina, \emph{{Experimental characterization of the hierarchy of quantum correlations in top quark pairs}},  \href{https://arxiv.org/abs/2602.15115}{{\ttfamily 2602.15115}}.

\bibitem{Guo:2026yhz}
Y.-C.~Guo, T.~Han, M.~Low and Y.~Su, \emph{{Quantum Tomography of Fermion Pairs in $e^+e^-$ Collisions: Longitudinal Beam Polarization Effects}},  \href{https://arxiv.org/abs/2602.02719}{{\ttfamily 2602.02719}}.

\bibitem{Aoude:2026eeg}
R.~Aoude, J.M.~Camacho, V.~Durupt, G.~Garc{\'\i}a-Mir, F.~Maltoni, M.~Moreno~Ll{\'a}cer et~al., \emph{{Radiation effects on the entanglement of fermion pairs at colliders}},  \href{https://arxiv.org/abs/2604.16268}{{\ttfamily 2604.16268}}.

\bibitem{Bonnefoy:2021tbt}
Q.~Bonnefoy, E.~Gendy, C.~Grojean and J.T.~Ruderman, \emph{{Beyond Jarlskog: 699 invariants for CP violation in SMEFT}}, \href{https://doi.org/10.1007/JHEP08(2022)032}{\emph{JHEP} {\bfseries 08} (2022) 032} [\href{https://arxiv.org/abs/2112.03889}{{\ttfamily 2112.03889}}].

\bibitem{Kane:1991bg}
G.L.~Kane, G.A.~Ladinsky and C.P.~Yuan, \emph{{Using the Top Quark for Testing Standard Model Polarization and CP Predictions}}, \href{https://doi.org/10.1103/PhysRevD.45.124}{\emph{Phys. Rev. D} {\bfseries 45} (1992) 124}.

\bibitem{Atwood:1991ka}
D.~Atwood and A.~Soni, \emph{{Analysis for magnetic moment and electric dipole moment form-factors of the top quark via e+ e- ---{\ensuremath{>}} t anti-t}}, \href{https://doi.org/10.1103/PhysRevD.45.2405}{\emph{Phys. Rev. D} {\bfseries 45} (1992) 2405}.

\bibitem{Bernreuther:1992be}
W.~Bernreuther, O.~Nachtmann, P.~Overmann and T.~Schr{\"o}der, \emph{{Angular correlations and distributions for searches of CP violation in top quark production and decay}}, \href{https://doi.org/10.1016/0550-3213(92)90545-M}{\emph{Nucl. Phys. B} {\bfseries 388} (1992) 53}.

\bibitem{Atwood:2000tu}
D.~Atwood, S.~Bar-Shalom, G.~Eilam and A.~Soni, \emph{{CP violation in top physics}}, \href{https://doi.org/10.1016/S0370-1573(00)00112-5}{\emph{Phys. Rept.} {\bfseries 347} (2001) 1} [\href{https://arxiv.org/abs/hep-ph/0006032}{{\ttfamily hep-ph/0006032}}].

\bibitem{Zhang:2010dr}
C.~Zhang and S.~Willenbrock, \emph{{Effective-Field-Theory Approach to Top-Quark Production and Decay}}, \href{https://doi.org/10.1103/PhysRevD.83.034006}{\emph{Phys. Rev. D} {\bfseries 83} (2011) 034006} [\href{https://arxiv.org/abs/1008.3869}{{\ttfamily 1008.3869}}].

\bibitem{Aguilar-Saavedra:2008nuh}
J.A.~Aguilar-Saavedra, \emph{{A Minimal set of top anomalous couplings}}, \href{https://doi.org/10.1016/j.nuclphysb.2008.12.012}{\emph{Nucl. Phys. B} {\bfseries 812} (2009) 181} [\href{https://arxiv.org/abs/0811.3842}{{\ttfamily 0811.3842}}].

\bibitem{Gupta:2009wu}
S.K.~Gupta, A.S.~Mete and G.~Valencia, \emph{{CP violating anomalous top-quark couplings at the LHC}}, \href{https://doi.org/10.1103/PhysRevD.80.034013}{\emph{Phys. Rev. D} {\bfseries 80} (2009) 034013} [\href{https://arxiv.org/abs/0905.1074}{{\ttfamily 0905.1074}}].

\bibitem{Bernreuther:2013aga}
W.~Bernreuther and Z.-G.~Si, \emph{{Top quark spin correlations and polarization at the LHC: standard model predictions and effects of anomalous top chromo moments}}, \href{https://doi.org/10.1016/j.physletb.2013.06.051}{\emph{Phys. Lett. B} {\bfseries 725} (2013) 115} [\href{https://arxiv.org/abs/1305.2066}{{\ttfamily 1305.2066}}].

\bibitem{Bernreuther:2015yna}
W.~Bernreuther, D.~Heisler and Z.-G.~Si, \emph{{A set of top quark spin correlation and polarization observables for the LHC: Standard Model predictions and new physics contributions}}, \href{https://doi.org/10.1007/JHEP12(2015)026}{\emph{JHEP} {\bfseries 12} (2015) 026} [\href{https://arxiv.org/abs/1508.05271}{{\ttfamily 1508.05271}}].

\bibitem{Cirigliano:2016nyn}
V.~Cirigliano, W.~Dekens, J.~de~Vries and E.~Mereghetti, \emph{{Constraining the top-Higgs sector of the Standard Model Effective Field Theory}}, \href{https://doi.org/10.1103/PhysRevD.94.034031}{\emph{Phys. Rev. D} {\bfseries 94} (2016) 034031} [\href{https://arxiv.org/abs/1605.04311}{{\ttfamily 1605.04311}}].

\bibitem{Cirigliano:2016njn}
V.~Cirigliano, W.~Dekens, J.~de~Vries and E.~Mereghetti, \emph{{Is there room for CP violation in the top-Higgs sector?}}, \href{https://doi.org/10.1103/PhysRevD.94.016002}{\emph{Phys. Rev. D} {\bfseries 94} (2016) 016002} [\href{https://arxiv.org/abs/1603.03049}{{\ttfamily 1603.03049}}].

\bibitem{Aguilar-Saavedra:2018ksv}
J.A.~Aguilar-Saavedra, C.~Degrande, G.~Durieux, F.~Maltoni, E.~Vryonidou and e.a.~Zhang, C., \emph{{Interpreting top-quark LHC measurements in the standard-model effective field theory}},  \href{https://arxiv.org/abs/1802.07237}{{\ttfamily 1802.07237}}.

\bibitem{deBeurs:2018pvs}
M.~de~Beurs, E.~Laenen, M.~Vreeswijk and E.~Vryonidou, \emph{{Effective operators in $t$-channel single top production and decay}}, \href{https://doi.org/10.1140/epjc/s10052-018-6399-3}{\emph{Eur. Phys. J. C} {\bfseries 78} (2018) 919} [\href{https://arxiv.org/abs/1807.03576}{{\ttfamily 1807.03576}}].

\bibitem{Degrande:2021zpv}
C.~Degrande and J.~Touch{\`e}que, \emph{{A reduced basis for CP violation in SMEFT at colliders and its application to diboson production}}, \href{https://doi.org/10.1007/JHEP04(2022)032}{\emph{JHEP} {\bfseries 04} (2022) 032} [\href{https://arxiv.org/abs/2110.02993}{{\ttfamily 2110.02993}}].

\bibitem{Bernreuther:2024ltu}
W.~Bernreuther, L.~Chen and Z.-G.~Si, \emph{{Binned top quark spin correlation and polarization observables for the LHC at 13.6~TeV}}, \href{https://doi.org/10.1103/PhysRevD.109.116016}{\emph{Phys. Rev. D} {\bfseries 109} (2024) 116016} [\href{https://arxiv.org/abs/2403.04371}{{\ttfamily 2403.04371}}].

\bibitem{deBlas:2025xhe}
J.~de~Blas, A.~Goncalves, V.~Miralles, L.~Reina, L.~Silvestrini and M.~Valli, \emph{{Constraining new physics effective interactions via a global fit of electroweak, Drell-Yan, Higgs, top, and flavour observables}}, \href{https://doi.org/10.1007/JHEP03(2026)013}{\emph{JHEP} {\bfseries 03} (2026) 013} [\href{https://arxiv.org/abs/2507.06191}{{\ttfamily 2507.06191}}].

\bibitem{Armadillo:2026mvp}
T.~Armadillo, E.~Celada, J.~ter Hoeve, F.~Maltoni, L.~Mantani, J.~Rojo et~al., \emph{{New Physics Reach through Precision at Future Colliders: a Multi-Pronged Approach}},  \href{https://arxiv.org/abs/2604.16596}{{\ttfamily 2604.16596}}.

\bibitem{Godbole:2006tq}
R.M.~Godbole, S.D.~Rindani and R.K.~Singh, \emph{{Lepton distribution as a probe of new physics in production and decay of the t quark and its polarization}}, \href{https://doi.org/10.1088/1126-6708/2006/12/021}{\emph{JHEP} {\bfseries 12} (2006) 021} [\href{https://arxiv.org/abs/hep-ph/0605100}{{\ttfamily hep-ph/0605100}}].

\bibitem{Boudjema:2009fz}
F.~Boudjema and R.K.~Singh, \emph{{A Model independent spin analysis of fundamental particles using azimuthal asymmetries}}, \href{https://doi.org/10.1088/1126-6708/2009/07/028}{\emph{JHEP} {\bfseries 07} (2009) 028} [\href{https://arxiv.org/abs/0903.4705}{{\ttfamily 0903.4705}}].

\bibitem{Rahaman:2021fcz}
R.~Rahaman and R.K.~Singh, \emph{{Breaking down the entire spectrum of spin correlations of a pair of particles involving fermions and gauge bosons}}, \href{https://doi.org/10.1016/j.nuclphysb.2022.115984}{\emph{Nucl. Phys. B} {\bfseries 984} (2022) 115984} [\href{https://arxiv.org/abs/2109.09345}{{\ttfamily 2109.09345}}].

\bibitem{Baumgart:2012ay}
M.~Baumgart and B.~Tweedie, \emph{{A New Twist on Top Quark Spin Correlations}}, \href{https://doi.org/10.1007/JHEP03(2013)117}{\emph{JHEP} {\bfseries 03} (2013) 117} [\href{https://arxiv.org/abs/1212.4888}{{\ttfamily 1212.4888}}].

\bibitem{Fischer:2018lme}
M.~Fischer, S.~Groote and J.G.~K{\"o}rner, \emph{{$T$-odd correlations in polarized top quark decays in the sequential decay $t(\uparrow) \to X_b+W^+(\to \ell^+ + \nu_\ell)$ and in the quasi three-body decay $t(\uparrow) \to X_b+ \ell^+ + \nu_\ell$}}, \href{https://doi.org/10.1103/PhysRevD.97.093001}{\emph{Phys. Rev. D} {\bfseries 97} (2018) 093001} [\href{https://arxiv.org/abs/1802.02492}{{\ttfamily 1802.02492}}].

\bibitem{Bernreuther:2017cyi}
W.~Bernreuther, L.~Chen, I.~Garc{\'\i}a, M.~Perell{\'o}, R.~Poeschl, F.~Richard et~al., \emph{{CP-violating top quark couplings at future linear $e^+e^-$ colliders}}, \href{https://doi.org/10.1140/epjc/s10052-018-5625-3}{\emph{Eur. Phys. J. C} {\bfseries 78} (2018) 155} [\href{https://arxiv.org/abs/1710.06737}{{\ttfamily 1710.06737}}].

\bibitem{Lamba:2026sni}
P.~Lamba, F.~Maltoni, O.~Miniati and E.~Vryonidou, \emph{{Quantum detection of CP violation in the $t\bar{t}$ system: tomography}},  \href{https://arxiv.org/abs/2607.25034}{{\ttfamily 2607.25034}}.

\bibitem{Krauss:2016ely}
F.~Krauss, S.~Kuttimalai and T.~Plehn, \emph{{LHC multijet events as a probe for anomalous dimension-six gluon interactions}}, \href{https://doi.org/10.1103/PhysRevD.95.035024}{\emph{Phys. Rev. D} {\bfseries 95} (2017) 035024} [\href{https://arxiv.org/abs/1611.00767}{{\ttfamily 1611.00767}}].

\bibitem{Hirschi:2018etq}
V.~Hirschi, F.~Maltoni, I.~Tsinikos and E.~Vryonidou, \emph{{Constraining anomalous gluon self-interactions at the LHC: a reappraisal}}, \href{https://doi.org/10.1007/JHEP07(2018)093}{\emph{JHEP} {\bfseries 07} (2018) 093} [\href{https://arxiv.org/abs/1806.04696}{{\ttfamily 1806.04696}}].

\bibitem{terHoeve:2025gey}
J.~ter Hoeve, L.~Mantani, J.~Rojo, A.N.~Rossia and E.~Vryonidou, \emph{{Connecting scales: RGE effects in the SMEFT at the LHC and future colliders}}, \href{https://doi.org/10.1007/JHEP06(2025)125}{\emph{JHEP} {\bfseries 06} (2025) 125} [\href{https://arxiv.org/abs/2502.20453}{{\ttfamily 2502.20453}}].

\bibitem{ATLAS:2025adk}
{\scshape ATLAS} collaboration, \emph{{Constraints on effective field theories via quadruple-differential angular decay rates from $t$-channel single-top-quark production at $\sqrt{s}=13$ TeV with the ATLAS detector}},  \href{https://arxiv.org/abs/2510.23372}{{\ttfamily 2510.23372}}.

\bibitem{ATLAS:2023eld}
{\scshape ATLAS} collaboration, \emph{{Inclusive and differential cross-section measurements of $ t\overline{t}Z $ production in pp collisions at $ \sqrt{s} $ = 13 TeV with the ATLAS detector, including EFT and spin-correlation interpretations}}, \href{https://doi.org/10.1007/JHEP07(2024)163}{\emph{JHEP} {\bfseries 07} (2024) 163} [\href{https://arxiv.org/abs/2312.04450}{{\ttfamily 2312.04450}}].

\bibitem{CMS:2023xyc}
{\scshape CMS} collaboration, \emph{{Search for physics beyond the standard model in top quark production with additional leptons in the context of effective field theory}}, \href{https://doi.org/10.1007/JHEP12(2023)068}{\emph{JHEP} {\bfseries 12} (2023) 068} [\href{https://arxiv.org/abs/2307.15761}{{\ttfamily 2307.15761}}].

\bibitem{Helstrom:1976}
C.W.~Helstrom, \emph{Quantum Detection and Estimation Theory}, Academic Press (1976).

\bibitem{Nielsen_Chuang_2010}
M.A.~Nielsen and I.L.~Chuang, \emph{Quantum Computation and Quantum Information: 10th Anniversary Edition}, Cambridge University Press (2010).

\bibitem{Fabbrichesi:2025igr}
M.~Fabbrichesi, R.~Floreanini, E.~Gabrielli and L.~Marzola, \emph{{Measuring CP violation using quantum state tomography}},  \href{https://arxiv.org/abs/2512.07939}{{\ttfamily 2512.07939}}.

\bibitem{Ollivier:2001fdq}
H.~Ollivier and W.H.~Zurek, \emph{{Introducing Quantum Discord}}, \href{https://doi.org/10.1103/PhysRevLett.88.017901}{\emph{Phys. Rev. Lett.} {\bfseries 88} (2001) 017901} [\href{https://arxiv.org/abs/quant-ph/0105072}{{\ttfamily quant-ph/0105072}}].

\bibitem{Hill:1997pfa}
S.~Hill and W.K.~Wootters, \emph{{Entanglement of a pair of quantum bits}}, \href{https://doi.org/10.1103/PhysRevLett.78.5022}{\emph{Phys. Rev. Lett.} {\bfseries 78} (1997) 5022} [\href{https://arxiv.org/abs/quant-ph/9703041}{{\ttfamily quant-ph/9703041}}].

\bibitem{Bravyi:2004isx}
S.~Bravyi and A.~Kitaev, \emph{{Universal quantum computation with ideal Clifford gates and noisy ancillas}}, \href{https://doi.org/10.1103/PhysRevA.71.022316}{\emph{Phys. Rev. A} {\bfseries 71} (2005) 022316} [\href{https://arxiv.org/abs/quant-ph/0403025}{{\ttfamily quant-ph/0403025}}].

\bibitem{Henderson:2001wrr}
L.~Henderson and V.~Vedral, \emph{{Classical, quantum and total correlations}}, \href{https://doi.org/10.1088/0305-4470/34/35/315}{\emph{J. Phys. A} {\bfseries 34} (2001) 6899} [\href{https://arxiv.org/abs/quant-ph/0105028}{{\ttfamily quant-ph/0105028}}].

\bibitem{it-book}
\emph{Entropy, relative entropy, and mutual information},  in \emph{Elements of Information Theory}, pp.~13--55, John Wiley \& Sons, Ltd (2005), \href{https://doi.org/https://doi.org/10.1002/047174882X.ch2}{DOI} [\href{https://arxiv.org/abs/https://onlinelibrary.wiley.com/doi/pdf/10.1002/047174882X.ch2}{{\ttfamily https://onlinelibrary.wiley.com/doi/pdf/10.1002/047174882X.ch2}}].

\bibitem{PhysRevA.81.052318}
A.~Ferraro, L.~Aolita, D.~Cavalcanti, F.M.~Cucchietti and A.~Ac\'{\i}n, \emph{Almost all quantum states have nonclassical correlations}, \href{https://doi.org/10.1103/PhysRevA.81.052318}{\emph{Phys. Rev. A} {\bfseries 81} (2010) 052318}.

\bibitem{PhysRevA.77.042303}
S.~Luo, \emph{Quantum discord for two-qubit systems}, \href{https://doi.org/10.1103/PhysRevA.77.042303}{\emph{Phys. Rev. A} {\bfseries 77} (2008) 042303}.

\bibitem{Dakic:2010xfz}
B.~Daki{\'c}, V.~Vedral and {\v{C}}.~Brukner, \emph{{Necessary and Sufficient Condition for Nonzero Quantum Discord}}, \href{https://doi.org/10.1103/PhysRevLett.105.190502}{\emph{Phys. Rev. Lett.} {\bfseries 105} (2010) 190502} [\href{https://arxiv.org/abs/1004.0190}{{\ttfamily 1004.0190}}].

\bibitem{Wootters:1997id}
W.K.~Wootters, \emph{{Entanglement of formation of an arbitrary state of two qubits}}, \href{https://doi.org/10.1103/PhysRevLett.80.2245}{\emph{Phys. Rev. Lett.} {\bfseries 80} (1998) 2245} [\href{https://arxiv.org/abs/quant-ph/9709029}{{\ttfamily quant-ph/9709029}}].

\bibitem{Leone:2021rzd}
L.~Leone, S.F.E.~Oliviero and A.~Hamma, \emph{{Stabilizer R{\'e}nyi Entropy}}, \href{https://doi.org/10.1103/PhysRevLett.128.050402}{\emph{Phys. Rev. Lett.} {\bfseries 128} (2022) 050402} [\href{https://arxiv.org/abs/2106.12587}{{\ttfamily 2106.12587}}].

\bibitem{Turkeshi:2023lqu}
X.~Turkeshi, A.~Dymarsky and P.~Sierant, \emph{{Pauli spectrum and nonstabilizerness of typical quantum many-body states}}, \href{https://doi.org/10.1103/PhysRevB.111.054301}{\emph{Phys. Rev. B} {\bfseries 111} (2025) 054301} [\href{https://arxiv.org/abs/2312.11631}{{\ttfamily 2312.11631}}].

\bibitem{Gottesman:1997qd}
D.~Gottesman, \emph{{A Theory of fault tolerant quantum computation}}, \href{https://doi.org/10.1103/PhysRevA.57.127}{\emph{Phys. Rev. A} {\bfseries 57} (1998) 127} [\href{https://arxiv.org/abs/quant-ph/9702029}{{\ttfamily quant-ph/9702029}}].

\bibitem{Gottesman:1997zz}
D.~Gottesman, \emph{{Stabilizer codes and quantum error correction}},  \href{https://arxiv.org/abs/quant-ph/9705052}{{\ttfamily quant-ph/9705052}}.

\bibitem{Gottesman:1998hu}
D.~Gottesman, \emph{{The Heisenberg representation of quantum computers}},  in \emph{{22nd International Colloquium on Group Theoretical Methods in Physics}}, pp.~32--43, 7, 1998 [\href{https://arxiv.org/abs/quant-ph/9807006}{{\ttfamily quant-ph/9807006}}].

\bibitem{Chitambar:2018rnj}
E.~Chitambar and G.~Gour, \emph{{Quantum resource theories}}, \href{https://doi.org/10.1103/revmodphys.91.025001}{\emph{Rev. Mod. Phys.} {\bfseries 91} (2019) 025001} [\href{https://arxiv.org/abs/1806.06107}{{\ttfamily 1806.06107}}].

\bibitem{Liu:2025frx}
Q.~Liu, I.~Low and Z.~Yin, \emph{{Maximal Magic for Two-qubit States}},  \href{https://arxiv.org/abs/2502.17550}{{\ttfamily 2502.17550}}.

\bibitem{Altakach:2022ywa}
M.M.~Altakach, P.~Lamba, F.~Maltoni, K.~Mawatari and K.~Sakurai, \emph{{Quantum information and CP measurement in H\textrightarrow{}\ensuremath{\tau}+\ensuremath{\tau}- at future lepton colliders}}, \href{https://doi.org/10.1103/PhysRevD.107.093002}{\emph{Phys. Rev. D} {\bfseries 107} (2023) 093002} [\href{https://arxiv.org/abs/2211.10513}{{\ttfamily 2211.10513}}].

\bibitem{Fabbrichesi:2022ovb}
M.~Fabbrichesi, R.~Floreanini and E.~Gabrielli, \emph{{Constraining new physics in entangled two-qubit systems: top-quark, tau-lepton and photon pairs}}, \href{https://doi.org/10.1140/epjc/s10052-023-11307-2}{\emph{Eur. Phys. J. C} {\bfseries 83} (2023) 162} [\href{https://arxiv.org/abs/2208.11723}{{\ttfamily 2208.11723}}].

\bibitem{Mertig:1990an}
R.~Mertig, M.~Bohm and A.~Denner, \emph{{FEYN CALC: Computer algebraic calculation of Feynman amplitudes}}, \href{https://doi.org/10.1016/0010-4655(91)90130-D}{\emph{Comput. Phys. Commun.} {\bfseries 64} (1991) 345}.

\bibitem{Jiang:2026sfw}
J.~Jiang, Z.-G.~Si, H.~Zhang and X.-Y.~Zhang, \emph{{Spin correlations and quantum entanglement in $\gamma \gamma \to t\bar t$ at polarized photon colliders with NLO QCD corrections}},  \href{https://arxiv.org/abs/2606.21152}{{\ttfamily 2606.21152}}.

\bibitem{Alwall:2014hca}
J.~Alwall, R.~Frederix, S.~Frixione, V.~Hirschi, F.~Maltoni, O.~Mattelaer et~al., \emph{{The automated computation of tree-level and next-to-leading order differential cross sections, and their matching to parton shower simulations}}, \href{https://doi.org/10.1007/JHEP07(2014)079}{\emph{JHEP} {\bfseries 07} (2014) 079} [\href{https://arxiv.org/abs/1405.0301}{{\ttfamily 1405.0301}}].

\bibitem{Durupt:2025wuk}
V.~Durupt, F.~Maltoni and O.~Mattelaer, \emph{{Automated computation of spin-density matrices and quantum observables for collider physics}}, \href{https://doi.org/10.1007/JHEP04(2026)103}{\emph{JHEP} {\bfseries 04} (2026) 103} [\href{https://arxiv.org/abs/2510.17730}{{\ttfamily 2510.17730}}].

\bibitem{Buckley:2014ana}
A.~Buckley, J.~Ferrando, S.~Lloyd, K.~Nordstr{\"o}m, B.~Page, M.~R{\"u}fenacht et~al., \emph{{LHAPDF6: parton density access in the LHC precision era}}, \href{https://doi.org/10.1140/epjc/s10052-015-3318-8}{\emph{Eur. Phys. J. C} {\bfseries 75} (2015) 132} [\href{https://arxiv.org/abs/1412.7420}{{\ttfamily 1412.7420}}].

\bibitem{hepdata.153301}
{CMS Collaboration}, ``{Measurements of polarization and spin correlation and observation of entanglement in top quark pairs using lepton+jets events from proton-proton collisions at $\sqrt{s}$ = 13 TeV}.'' {HEPData (collection)}, 2024.

\bibitem{FCC:2025lpp}
{\scshape FCC} collaboration, \emph{{Future Circular Collider Feasibility Study Report: Volume 1, Physics, Experiments, Detectors}}, \href{https://doi.org/10.1140/epjc/s10052-025-15077-x}{\emph{Eur. Phys. J. C} {\bfseries 85} (2025) 1468} [\href{https://arxiv.org/abs/2505.00272}{{\ttfamily 2505.00272}}].

\bibitem{FCC:2025uan}
{\scshape FCC} collaboration, \emph{{Future Circular Collider Feasibility Study Report: Volume 2, Accelerators, Technical Infrastructure and Safety}}, \href{https://doi.org/10.1140/epjs/s11734-025-01967-4}{\emph{Eur. Phys. J. ST} {\bfseries 234} (2025) 5713} [\href{https://arxiv.org/abs/2505.00274}{{\ttfamily 2505.00274}}].

\bibitem{Aguilar-Saavedra:2010ljg}
J.A.~Aguilar-Saavedra and J.~Bernabeu, \emph{{W polarisation beyond helicity fractions in top quark decays}}, \href{https://doi.org/10.1016/j.nuclphysb.2010.07.012}{\emph{Nucl. Phys. B} {\bfseries 840} (2010) 349} [\href{https://arxiv.org/abs/1005.5382}{{\ttfamily 1005.5382}}].

\bibitem{Selvaggi:2025kmd}
{\scshape FCC} collaboration, M.~Selvaggi, A.~Blondel and J.~Eysermans, eds., \emph{{Prospects in electroweak, Higgs and Top physics at FCC}}, \href{https://doi.org/10.17181/n2emg-43f06}{\emph{{repository.cern/records/n2emg-43f06}} (2025) }.

\bibitem{Tweedie:2014yda}
B.~Tweedie, \emph{{Better Hadronic Top Quark Polarimetry}}, \href{https://doi.org/10.1103/PhysRevD.90.094010}{\emph{Phys. Rev. D} {\bfseries 90} (2014) 094010} [\href{https://arxiv.org/abs/1401.3021}{{\ttfamily 1401.3021}}].

\bibitem{Cao:2025xnp}
H.~Cao and F.~Petriello, \emph{{Single-spin measurements and heavy new physics in the e+e-{\textrightarrow}tt{\textasciimacron} process at an FCC-ee}}, \href{https://doi.org/10.1103/794y-gp3r}{\emph{Phys. Rev. D} {\bfseries 113} (2026) 035033} [\href{https://arxiv.org/abs/2511.01994}{{\ttfamily 2511.01994}}].

\bibitem{Arens:1992wh}
T.~Arens and L.M.~Sehgal, \emph{{Secondary leptons as probes of top quark polarization in e+ e- --{\ensuremath{>}} t anti-t}}, \href{https://doi.org/10.1016/0550-3213(93)90236-I}{\emph{Nucl. Phys. B} {\bfseries 393} (1993) 46}.

\bibitem{Buttazzo:2026amk}
D.~Buttazzo, G.~Levati, Y.~Ma, F.~Maltoni, P.~Paradisi and Z.~Wang, \emph{{Probing $\tau$ lepton dipole moments at future Lepton Colliders}},  \href{https://arxiv.org/abs/2604.14281}{{\ttfamily 2604.14281}}.

\bibitem{Rouge:2005iy}
A.~Rouge, \emph{{CP violation in a light Higgs boson decay from tau-spin correlations at a linear collider}}, \href{https://doi.org/10.1016/j.physletb.2005.05.076}{\emph{Phys. Lett. B} {\bfseries 619} (2005) 43} [\href{https://arxiv.org/abs/hep-ex/0505014}{{\ttfamily hep-ex/0505014}}].

\bibitem{Ehataht:2023zzt}
K.~Ehat\"aht, M.~Fabbrichesi, L.~Marzola and C.~Veelken, \emph{{Probing entanglement and testing Bell inequality violation with e+e-\textrightarrow{}\ensuremath{\tau}+\ensuremath{\tau}- at Belle II}}, \href{https://doi.org/10.1103/PhysRevD.109.032005}{\emph{Phys. Rev. D} {\bfseries 109} (2024) 032005} [\href{https://arxiv.org/abs/2311.17555}{{\ttfamily 2311.17555}}].

\bibitem{Fabbrichesi:2024xtq}
M.~Fabbrichesi and L.~Marzola, \emph{{Dipole momenta and compositeness of the $\tau$ lepton at Belle II}},  \href{https://arxiv.org/abs/2401.04449}{{\ttfamily 2401.04449}}.

\bibitem{Han:2025ewp}
T.~Han, M.~Low and Y.~Su, \emph{{Entanglement and Bell Nonlocality in $\tau^+ \tau^-$ at the BEPC}},  \href{https://arxiv.org/abs/2501.04801}{{\ttfamily 2501.04801}}.

\bibitem{Zhang:2025mmm}
Y.~Zhang, B.-H.~Zhou, Q.-B.~Liu, S.~Li, S.-C.~Hsu, T.~Han et~al., \emph{{Entanglement and Bell Nonlocality in $\tau^+ \tau^-$ at the LHC using Machine Learning for Neutrino Reconstruction}},  \href{https://arxiv.org/abs/2504.01496}{{\ttfamily 2504.01496}}.

\bibitem{Bouchiat:1958yui}
C.~Bouchiat and L.~Michel, \emph{{Mesure de la polarisation des electrons relativistes}}, \href{https://doi.org/10.1016/0029-5582(58)90046-4}{\emph{Nucl. Phys.} {\bfseries 5} (1958) 416}.

\bibitem{Haber:1994pe}
H.E.~Haber, \emph{{Spin formalism and applications to new physics searches}},  in \emph{{21st Annual SLAC Summer Institute on Particle Physics: Spin Structure in High-energy Processes (School: 26 Jul - 3 Aug, Topical Conference: 4-6 Aug) (SSI 93)}}, pp.~231--272, 4, 1994 [\href{https://arxiv.org/abs/hep-ph/9405376}{{\ttfamily hep-ph/9405376}}].

\bibitem{Haberl:1995ek}
P.~Haberl, O.~Nachtmann and A.~Wilch, \emph{{Top production in hadron hadron collisions and anomalous top - gluon couplings}}, \href{https://doi.org/10.1103/PhysRevD.53.4875}{\emph{Phys. Rev. D} {\bfseries 53} (1996) 4875} [\href{https://arxiv.org/abs/hep-ph/9505409}{{\ttfamily hep-ph/9505409}}].

\end{thebibliography}\endgroup

\end{document}